\documentclass[12pt]{article}
\usepackage{mcite}
\usepackage{ifthen}
\usepackage{epsfig,cite}
\usepackage {amsmath}
\usepackage{amssymb}
\usepackage{ulem}
\include{epsf}
\usepackage{color}
\newcount\timecount
\newcount\hours \newcount\minutes  \newcount\temp \newcount\pmhours
\hours = \time \divide\hours by 60 \temp = \hours \multiply\temp by
60 \minutes = \time \advance\minutes by -\temp
\def\hour{\the\hours}
\def\minute{\ifnum\minutes<10 0\the\minutes
            \else\the\minutes\fi}
\def\clock{
\ifnum\hours=0 12:\minute\ AM \else\ifnum\hours<12 \hour:\minute\ AM
      \else\ifnum\hours=12 12:\minute\ PM
            \else\ifnum\hours>12
                 \pmhours=\hours
            \advance\pmhours by -12
                 \the\pmhours:\minute\ PM
                 \fi
            \fi
      \fi
\fi }

\def\monthname{\relax\ifcase\month 0/\or January\or February\or
   March\or April\or May\or June\or July\or August\or September\or
   October\or November\or December\else\number\month/\fi}

\def\bold#1{\setbox0=\hbox{$#1$}%
     \kern-.025em\copy0\kern-\wd0
     \kern.05em\copy0\kern-\wd0
     \kern-.025em\raise.0433em\box0 }

\def\beq{\begin{equation}}
\def\eeq{\end{equation}}

\newlength{\capindent}
\newlength{\capwidth}
\newlength{\figwidth}
\newcommand{\icaption}[2][!*!,!]{\hspace*{\capindent}%
  \begin{minipage}{\capwidth}
    \ifthenelse{\equal{#1}{!*!,!}}%
      {\caption{#2}}%
      {\caption[#1]{#2}}
  \end{minipage}}
\newcommand{\EEGG}{\rm e^+ e^-\rightarrow \gamma\gamma(\gamma)}

\newcommand{\EEEEG}{\rm e^+ e^-\rightarrow e^+e^-(\gamma)}
\newcommand{\LAMP}{ \Lambda_{+}}
\newcommand{\LAMM}{ \Lambda_{-}}

\newcommand{\EEff}{\rm e^+ e^-\rightarrow f^{+}f^{-}}
\newcommand{\sm}{SM}
\newcommand{\invh}{h^{-1}}
\newcommand{\Mp}{{\rm Mpc}}

\begin{document}
\begin{titlepage}
\pagestyle{empty} \baselineskip=21pt
\vskip 0.2in
\begin{center}
{\large{\bf Experimental and Theoretical Evidence for
            Extended Particle Models}}
\end{center}
\begin{center}
\vskip 0.2in {\bf Chih-Hsun Lin}$^1$, 
{\bf J\" urgen~Ulbricht}$^{2}$, {\bf
Jian~Wu}$^{3}$ and  {\bf Jiawei~Zhao}$^{3}$ \vskip 0.1in

{\it
$^1$ Institute of Physics Academia Sinica, Taipei 11, Taiwan     \\
$^2$ Swiss Institute of Technology ETH-Z\"urich, CH-8093 Z\"urich,
Switzerland \\
$^3$ University of Science and Technology of China,
230009 Hefei Anhui, China                                             \\
}

\vskip 0.2in {\bf Abstract}
\end{center}
\baselineskip=18pt \noindent

We review the experimental searches on those interactions where the       
fundamental particles could exhibit a non point-like behavior.    
In particular we focus on the QED reaction    
measuring the differential cross sections for the process $ \EEGG $
at energies from  $\sqrt{s} $=55~GeV to 207 GeV
using the data collected with the VENUS, TOPAZ, ALEPH, DELPHI
L3 and OPAL from 1989 to 2003. The global fit to the data is 
5 standard deviations away from the standard model expectation for the hypothesis
of an excited state of the electron, corresponding to the cut-off scale $\Lambda
=12.5$~TeV. Assuming that this cut-off scale restricts the characteristic size
of QED interaction to $15.7 \times 10^{-18}$~cm, we perform an effort to assign in
a semi-mechanical way all available properties of fundamental particles to a
hypothetical classical object. Such object can be modeled as a classical
gyroscope consisted of a non rotating inner massive kernel surrounded by an
outer rotating massive layer equipped with charged sorted in a way to match the
charge contents for different interactions. The model size
of an electron agrees with $ 1.86 \times 10^{-17} $ cm with the
experiment. The introduction of a particle like structure related
to gravity allows to estimate the inner mass kernel of an electron to
$ 1.7 \times 10^{-19} $ cm and the mass of a scaler to $ 154 $ GeV.
The extension of the model to
electrical charged particle-like structure in nonlinear
electrodynamics coupled to General Relativity confirms the model
in the global geometrical structure of mass and field distribution.

\vspace*{0.2cm}

\end{titlepage}
%
%
\section{ INTRODUCTION }
\label{sec:level1}

The experimental and theoretical success in high energy physics of the
last years allowed to established the standard model ( SM ) a unique
theory to unify our knowledge in particle physics.
The model consists of three families
of point like fundamental particles ( FP ) and three interactions.
The very precise measurement of the total cross section of the $ Z^{o} $
production at the Large Electron Positron Collider ( LEP ) allowed to
determine the number of families of FP to three \cite{Threefamilies}.
The mass of the last missing FP the top quark was predicted by the
very precise experiments from LEP at CERN \cite{massTOPLEP}
and discovered at FERMILAB \cite{massTOPCDF}. The three weak interaction
bosons $ Z^{o} $, $ W^{+} $ and $ W^{-} $ get discovered at the SPS at CERN
\cite{Z-WdiscovereySPS} and their parameters are measured to a high
precision at LEP \cite{parameterZ-WLEP}.

In spite of this success the SM has still drawbacks. The last
cornerstone of the SM the scalar Higgs field is missing. The
Higgs field provides with the approbate coupling the SM fermions
and bosons with mass. Lower experimental limits of
$ m_{Higgs}> 114.4 $ GeV  \cite{HiggsLimitL3} and upper
limits of $ m_{Higgs}< 154 $ GeV
\cite{HiggsLimitIRINA,HiggsLimitLEP} exist.
The model needs at least 19 parameters \cite{SMparameter},
is at energies above a scale of 1 TeV in the radiative corrections
divergent and allows no unification of the three interaction at one energy
\cite{ZicciciPaper}. In particular the weakest interaction the
gravity is missing in the theory.

The above discussed considerations triggered extensive experimental
and theoretical efforts to search for physics beyond the SM. Experiments
at LEP, HERA and CDF searched for a deviation of the point like behavior
of FP \cite{papersNONpointFM}. Such a deviation could be interpreted
as a micro structure or a new composite substructure of FP \cite{Preons}.
Excellent experiments
searching for lepto quarks \cite{leptoquarks}, for an anomalous
magnetic moment of FP \cite{g-2experiment} and an electric dipole moment
of electrons \cite{electricDIPOL}. Very promising to be successfully
in a detection of physics beyond the SM is the search for neutrino
oscillations \cite{NeutrinoOscilations,ExperimentNeutrinoAndre}.

The theoretical efforts is characterized by the
Yang-Mills theory \cite{YangMills1} as gauge theory of quantum 
field theory based on the SU(N) group, the Standard Model and 
String theory.
The string theory combines quantum mechanics and general 
relativity into a quantum theory of gravity \cite{string1}.
It is a candidate for a theory of everything,
a way to describe all the known natural forces
gravitational, electromagnetic, weak and strong together with matter,
quarks and leptons in a mathematically complete system.
The string is in the string theory the fundamental building block
with a size of $10^{-33}$ cm.  Our four dimensional world
can only be described if we add  further six dimension.
Every space-time coordinate of our four dimensional
world is accompanied by six compact not observable
space dimensions. The  string theory  is in conflict
with the from Karl Popper ( 1902-1994 ) required
falsification of a physical theory \cite{string2}.

These theoretical activities introduced
new hypothetical elementary
particles beyond the known fundamental particles.
The Dilaton \cite{dilaton1}
a particle of a scalar field that always comes with gravity.
The sphaleron \cite{Sphaleron1} a static solution to the electroweak
field equations of the Standard Model of particle physics. Geometrically, 
a sphaleron is a saddle point of the electroweak potential energy
much like the saddle point of the surface $ Z = = x^{2} - y^{2} $ in three
dimensional analytic geometry. The Skyrmion \cite{Skyrmions1}
is a mathematical model used to model baryons. Skyrmions
are not individual baryons but coherent states of known
baryons and resonances on a compact manifold associated
with the spin and flavor symmetry group.
The Super Symmetric Models is understood as the low energy
theory of a high energy model. It would be possible to unify all three
interactions at one energy, solve the divergencies of the radiative
corrections and introduce only a soft breaking Higgs mechanism
\cite{SuperSymetries}. A price has to be paid for these advantages,
it is necessary to double the numbers of FP and gauge particles by
introduction of Super Symmetric Partners to the SM fermions and bosons.

In this paper we summerize in a worldwide global fit
all experimental data of direct contact term interaction.
These interaction is sensitive to not point like behavior
of fundamental particle. This is an essential test of the Standard Theory
which describes these particles as points.
We develop an overall scheme to sort all known fundamental
particles in a common scheme. We use for this development
an empirically ansatz guided by the details of the
running coupling constants of the Standard Model,
the very early time development of the of the Big Bang model
and the experimental data of the fundamental particles.
For an electron we use this scheme together  with an
empirically classical ansatz to predict the size of an electron
and compare it with the size of the worldwide global fit
for the electron. Finally we compare the size of an
electron calculated from this empirically classical ansatz
with particle-like structure related to gravity and
electrical charged particle-like structure in Nonlinear
Electrodynamics coupled to General Relativity.

The paper is ordered as follows. We discuss from
Standard Model to contact interactions the experimental status,
the cosmology and the early Universe, Summary about experimental
status of Standard and Big Bang Model, Possible micro
structure of Fundamental Particles,
Particle-like structure related to gravity,
electrical charged particle-like structure in Nonlinear
Electrodynamics coupled to General Relativity and conclude.

\section {From Standard Model to contact interactions: experimental status}

To discuss possible deviations from the SM we first summarize
the main parameters of the model and give an overview of experiments 
which test a possible signatures of non point-like behavior of FP.

\subsection {Standard Model Parameters}
\label{sec:Standard Model Parameters}

The Standard Model of particle physics has been established by a series of 
experiments and theoretical developments over the past century
including:
\begin{itemize}
\item
1897 - The discovery of the electron;
\item
1910 - The discovery of the nucleus;
\item
1930 - The nucleus found to be made of protons and neutrons; neutrino 
postulated;
\item
1936 - The muon discovered;
\item
1947 - Pion and strange particles discovered;
\item
1950's - Many strongly-interacting particles discovered;
\item
1964 - Quarks proposed;
\item
1967 - The Standard Model proposed;
\item
1973 - Neutral weak interactions discovered;
\item
1974 - The charm quark discovered;
\item
1975 - The $\tau$ lepton discovered;
\item
1977 - The bottom quark discovered;
\item
1979 - The gluon discovered;
\item
1983 - The intermediate $W^\pm, Z^0$ bosons discovered;
\item
1989 - Three neutrino species counted;
\item
1994 - The top quark discovered;
\item
1998 - Neutrino oscillations discovered.
\end{itemize}

All the above historical steps, apart from the last (which was made with 
neutrinos from astrophysical sources), fit within the SM, 
and the SM continues to survive all experimental tests at accelerators.

The set of SM spin-1/2 matter (anti-matter)~\footnote{After the 
discovery of the first anti particle the positron
1933 \cite{PositronDis} experiments proved every FP is
accompanied by an anti-partner. Experimentally, an anti-partner looks like the
same as its matter parter apart the sign of its electric
charge~Fig.\ref{fig.1}. However, the nature demonstrates that the matter
anti-matter asymmetry is currently established in the 
Universe~\cite{AntiHeliumAMS01,matter-antimatter} called also baryon asymmetry.
Therefore, it is quite a challenge to produce and keep saved from annihilation
even simple anti-Hydrogen atoms~\cite{AntiHydrogen}. Moreover, since the
conditions required to generate the baryon asymmetry assume a physics beyond
the SM it would be of great interest to discover
anti-nuclei~\cite{AntiHeliumAMS01,AntimatterSakharov} in cosmic rays.} 
particles is shown in Fig.{\ref{fig.1}. We know from experiments at 
CERNs LEP accelerator in 1989 that there can only be three 
neutrinos~\cite{Threefamilies}: $ N_{\nu} = 2.9841 \pm 0.0083 $, which
is a couple of standard deviations below 3, but that cannot be
considered a significant discrepancy. 

\begin{figure}[htbp]
\begin{center}
\rotatebox{90}{
 \includegraphics[width=9.0cm,height=8.0cm]{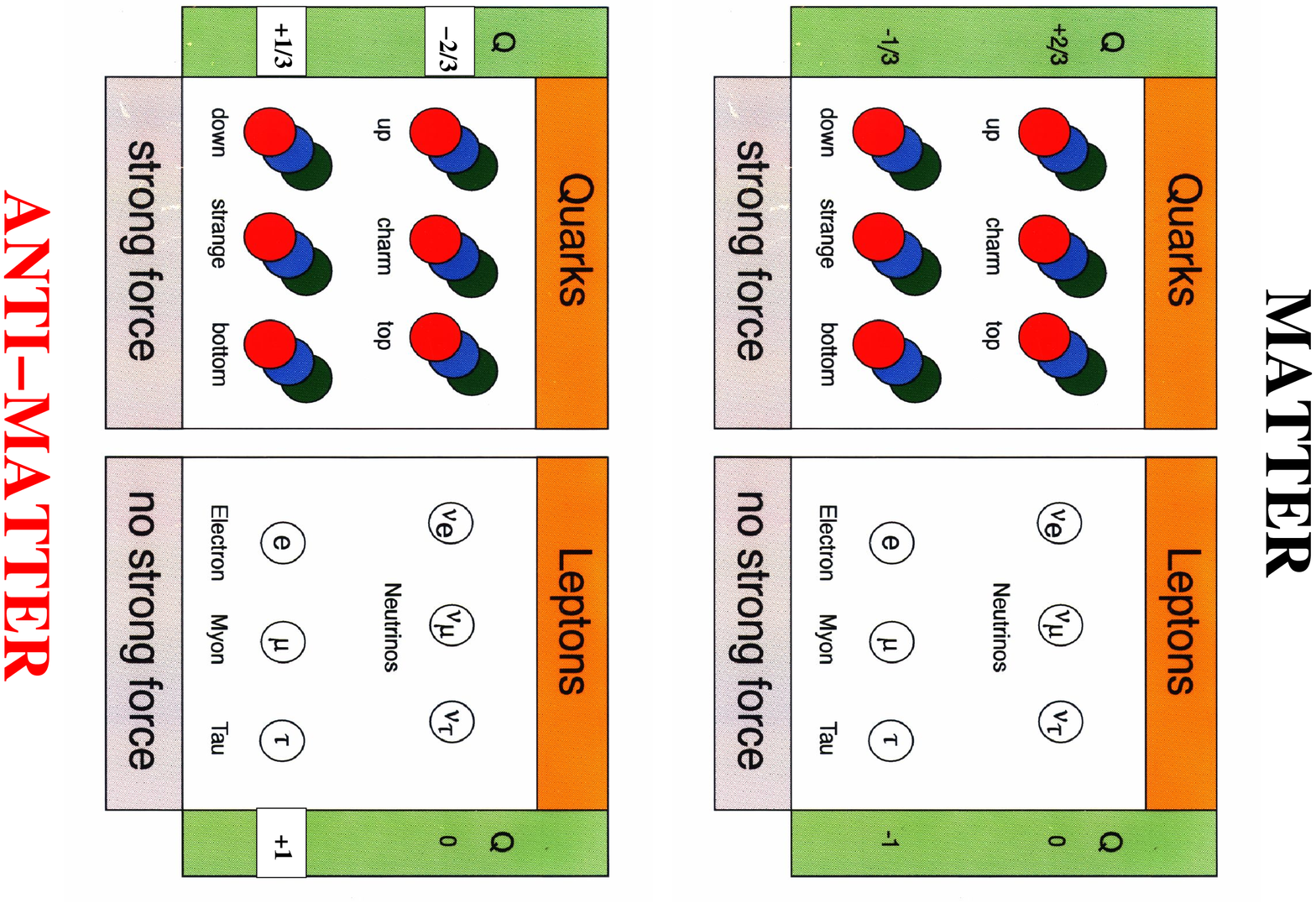}}
\end{center}
\caption{Scheme of fundamental particles and anti particles. }
\label{fig.1}
\end{figure}

The forces between these matter particles are carried by spin-1 bosons:
electromagnetism by the familiar massless photon $\gamma$, the weak
interactions by the massive intermediate $W^\pm$ and $Z^0$ bosons that
weight $\simeq 80, 91$~GeV, respectively, and the strong interactions by
the massless gluon. 

The Standard Model agrees with all confirmed experimental data from
accelerators, but is theoretically very
unsatisfactory. In particular, it does not explain the particle
quantum numbers, such as the electric charge $Q$, weak isospin $I$,
hypercharge $Y$ and colour, and contains at least 19 arbitrary parameters.
These include three independent vector-boson couplings and a possible
CP-violating strong-interaction parameter $ \Theta_{QCD} $, six quark and three
charged-lepton masses, three generalized Cabibbo weak mixing angles and
the CP-violating Kobayashi-Maskawa phase, as well as two independent
masses for weak bosons.

The mass of $ e $ , $\mu $ and
$ \tau $ is measured to high accuracy \cite{Databooklet2005}. The current
experimental limits on the neutrino masses read: $ m_{\nu_{e}} \le 3 $ eV,
$ m_{\nu_{\mu}} \le 0.19 $ MeV and $ m_{\nu_{\tau}} \le 18.2 $ MeV. It is almost
clear, as reviled by atmospheric and 
solar neutrino experiments, that neutrinos posses a non-zero mass. However,
the explanation requires to invoke a beyond standard model physics, which is out
of the scope of the current chapter. Due to the confinement, the bare masses of
up- and down-quarks are measured in ranges given by upper and lower limits
\cite{Databooklet2005}. The mass of the top quark is measured with a high
accuracy,  $m_t{\rm (LEP)}=178.1^{+10.4}_{-8.3} ~{\rm GeV}$
and $m_t{\rm (CDF)}=174.3 \pm 5.1 ~{\rm GeV} $ by LEP
experiments~\cite{massTOPLEP} and CDF collaboration~\cite{massTOPCDF}.
In Tab.~\ref{ParemeterFamily} the three families of FPs defining 
the first group SM parameters are summarized. 
%
\begin{table}
\caption{\it Three families of FPs define the first group SM parameters.}
\label{ParemeterFamily}
\begin{center}
\begin{tabular}{||l|l|l|l||}                               \hline
 name      & family           & family           & family  \\
           & 1                & 2                & 3       \\  \cline{2-4}
           & ~~~ charge ~~mass[MeV] & ~~~ charge ~~mass[MeV] 
           & ~~~ charge ~~mass[MeV]                        \\ \hline
 lepton    & $e$  ~ $-1$ ~~~~~~$0.51099892$  & $\mu$ ~~$-1$ ~~~~~~$105.658369 $
           & $\tau$ ~~$-1$ ~~~~~~$1776.99$                 \\ \hline
 neutrino  & $\nu_{e}$ ~~~0 ~~~~~~0          & $\nu_{\mu}$ ~~~0 ~~~~~~0
           & $\nu_{\tau}$ ~~~0 ~~~~~~0                        \\ \hline
 up-quark  & $u$  ~ $+2/3$     ~~~$1.5-4.0$   & $c$   ~~~$+2/3$ ~~$1150-1350$
           & $t$ ~~ $+2/3$     ~~$174300$                     \\ \hline
 down-quark& $d$ ~~ $-1/3$      ~~$4.0-8.0$   & $s$   ~~~$-1/3$ ~~$80-130$
           & $b$ ~~ $-1/3$     ~~$4100-4400$                  \\ \hline
\end{tabular}
\end{center}
\end{table}
%

%

The masses of the gauge bosons are measured to high accuracy
\cite{Databooklet2005}.
In Fig.{\ref{fig.3}} are displayed three
different type of interactions of FPs which  are governed by
the three
coupling constants, namely  $ \alpha_{s} $ for the strong, $ \alpha $ for the
electro magnetic and $ \alpha_{weak} $ for the weak interaction. The fourth
interaction, the gravitational one being not described by the SM is included for
the completeness.

Among the key objectives of particle physics are attempts to unify these
different interactions (at least 3 of them), and to explain the very different
masses of the various matter particles and spin-1 bosons.
%
\begin{figure}[htbp]
\vspace{5.0mm}
\begin{center}
\rotatebox{00}{
 \includegraphics[width=9.0cm,height=6.0cm]{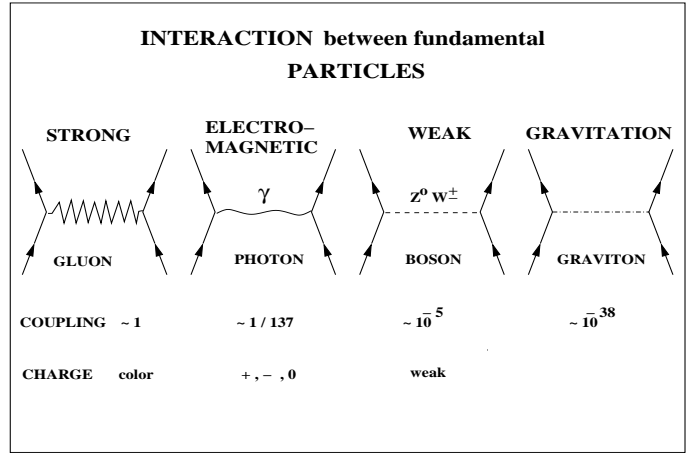}}
\end{center}
\caption{Interactions between fundamental particles.}
\label{fig.3}
\end{figure}

In the last decades, it has been revealed experimentally that coupling (gauge)
constants in the quantum field theories of the SM are not
constant. The couplings, which set the strength for the interactions, change
their value if one probes smaller distances with higher energies. This is due to
contributions of virtual particles that cause a ``running'' of the coupling with
the energy scale. This energy scale is also referred to as the ``sliding
scale''. If one evaluates the necessary Feynman diagrams to compute this effect,
it turns out that the couplings run logarithmically with the sliding scale, and
their slope depends on the particle content. 

This generic property of quantum
field theory has an analogy in classical physics~\cite{UehlingSerber}. Indeed,
consider electric and magnetic phenomena. Let us take some
dielectric medium and put a sample electric charge in it. What
happens is that  the medium is polarized. It contains  electric
dipoles which are arranged in such a way as  to screen the charge. It is a
consequence of the Coulomb law:
attraction of the opposite charges and repulsion of the same ones.
This is the origin of electric screening. The opposite situation occurs  in a
magnetic medium. According to
the Biot-Savart law, electric currents of the same direction are
attracted to each other, while those of the opposite one are
repulsed. This leads to anti screening of
electric currents in a magnetic medium.

In the SM, the role of the medium is played by the vacuum. Vacuum is
polarized due to the presence of virtual pairs of particles in it.
The matter fields and transverse quanta of  vector fields in this
case behave like dipoles in a dielectric medium and cause
screening, while the longitudinal quanta of vector fields behave
like currents and cause anti screening. These two effects compete
each other (see Eq.\ref{regeq1} below).

Thus, the couplings become the functions of a distance or an
energy scale $\mu $, with $ Q $ as test energy.

\begin{equation}
 \alpha_i(\frac{Q^2}{\mu^2}) = \alpha_i(\mbox{distance}),
 \ \ \ \ \ \ \alpha_i \equiv g_i^2/4\pi.
\label{alphadistance}
\end{equation}

This dependence is described by the renormalization group equations and is
confirmed experimentally \cite{ConsitencyGUT}.

In the SM the strong and weak couplings associated with
non-Abelian gauge groups decrease with energy, while the
electromagnetic one associated with the Abelian group on the
contrary increases. Thus, it becomes possible that at some energy
scale they become equal. According to the GUT idea, this equality
is not occasional but is a manifestation of a unique origin of
these three interactions. As a result of spontaneous symmetry
breaking, the unifying group is broken and the unique interaction
is split into three branches which we call strong, weak
 and electromagnetic interactions. This happens at a very
high energy of an order of $10^{15\div 16}$ GeV. Of course, this
energy is out of the range of accelerators; however, some
 crucial predictions follow from the very fact of unification.

After the precise measurement of the $SU(3)\times SU(2) \times
U(1)$ coupling constants, it has become possible to check the
unification numerically.

The three coupling constants to be compared are
 \begin{eqnarray}
\alpha_1&=&(5/3)g^{\prime2}/(4\pi)=5\alpha/(3\cos^2\theta_W),
\nonumber \\ \alpha_2&=& g^2/(4\pi)=\alpha/\sin^2\theta_W, \\
\alpha_3&=& g_s^2/(4\pi) \nonumber
 \end{eqnarray}
where $g',~g$ and $g_s$ are the usual $U(1)$, $SU(2)$ and $SU(3)$
coupling constants and $\alpha$ is the fine structure constant.
The factor of 5/3 in the definition of $\alpha_1$ has been
included for  proper normalization of the generators.

The couplings, when defined as renormalized values including loop
corrections require the specification of a renormalization
prescription for which the modified minimal subtraction
($\overline{MS}$) scheme~\cite{msbar} is used.

In this scheme, the world averaged values of the coup\-lings at
the Z$^0$ energy are obtained from a fit to the LEP and Tevatron
data~\cite{fine},\cite{test-sm},\cite{bethke}:
 \begin{eqnarray}
  \label{worave}
  \alpha^{-1}(M_Z)             & = & 128.978\pm 0.027   \nonumber\\
  \sin^2\theta_{\overline{MS}} & = & 0.23146\pm 0.00017\\
  \alpha_s                     & = & 0.1184\pm 0.0031, \nonumber
 \end{eqnarray}
that gives
 \begin{equation} \alpha_1(M_Z)=0.017 ,\ \ \alpha_2(M_Z)=0.034, \
\ \alpha_3(M_Z)=0.118\pm 0.003.
 \end{equation}
 Assuming that the SM is valid up to the unification
scale, one can then use the known RG equations for the three
couplings. They are the following:
 \begin{equation}
\frac{d\tilde{\alpha}_i}{dt} =  b_i\tilde{\alpha}_i^2, \ \ \ \
\tilde{\alpha}_i=\frac{\alpha_i}{4\pi}, \ \ \ \ \
t=log(\frac{Q^2}{\mu^2}), \label{alpha}
 \end{equation}

where $ \mu $ is the energy scale and for 
the SM the coefficients $b_i$ are

 \begin{equation}
b_i=\left( \begin{array}{r} b_1 \\ b_2 \\b_3 \end{array} \right)
   =
\left( \begin{array}{r}           0    \\
                             - 22 / 3  \\
                                -11    \end{array} \right) +N_{Fam}
\left( \begin{array}{r}         4 / 3  \\
                                4 / 3  \\
                                4 / 3  \end{array} \right) + N_{Higgs}
\left( \begin{array}{r}         1 / 10 \\
                                1 / 6  \\
                                  0    \end{array} \right) .
 \label{regeq1}
 \end{equation}
Here  $N_{Fam}$ is the number of generations of matter multiplets
and $N_{Higgs}$ is the number of Higgs doublets. We use
$N_{Fam}=3$ and $N_{Higgs}=1$ for the minimal SM, which gives
$b_i=(41/10, -19/6, -7)$.

Notice a positive contribution (screening) from the matter
multiplets and negative one (anti screening) from the gauge fields.
For the Abelian group $U(1)$ this contribution is absent due to
the absence of a self-interaction of Abelian gauge fields.

The solution to eq.\ref{alpha} is very simple

 \begin{equation}
\frac{1}{\tilde{\alpha}_i(Q^2)} = \frac{1}{\tilde{\alpha}_i(\mu^2)}-
 b_i log(\frac{Q^2}{\mu^2}). \label{alphasol}
 \end{equation}

The result is demonstrated for example in
Fig.\ref{fig.4} (upper part ) showing the
evolution of the inverse of the couplings as a function of the
logarithm of energy scale $ \mu $ \cite{ConsitencyGUT}.
In this presentation, the evolution becomes a
straight line in first order. The second order corrections are
small and do not cause any visible deviation from a straight line.
Fig.\ref{fig.4} clearly demonstrates that within the SM the
coupling constant unification at a single point is impossible. It
is excluded by more than 8 standard deviations. This result means
that the unification can only be obtained if new physics enters
between the electroweak and the Planck scales!

\begin{figure}
\vspace{-5.0 mm}
\begin{center}
 \epsfig{file=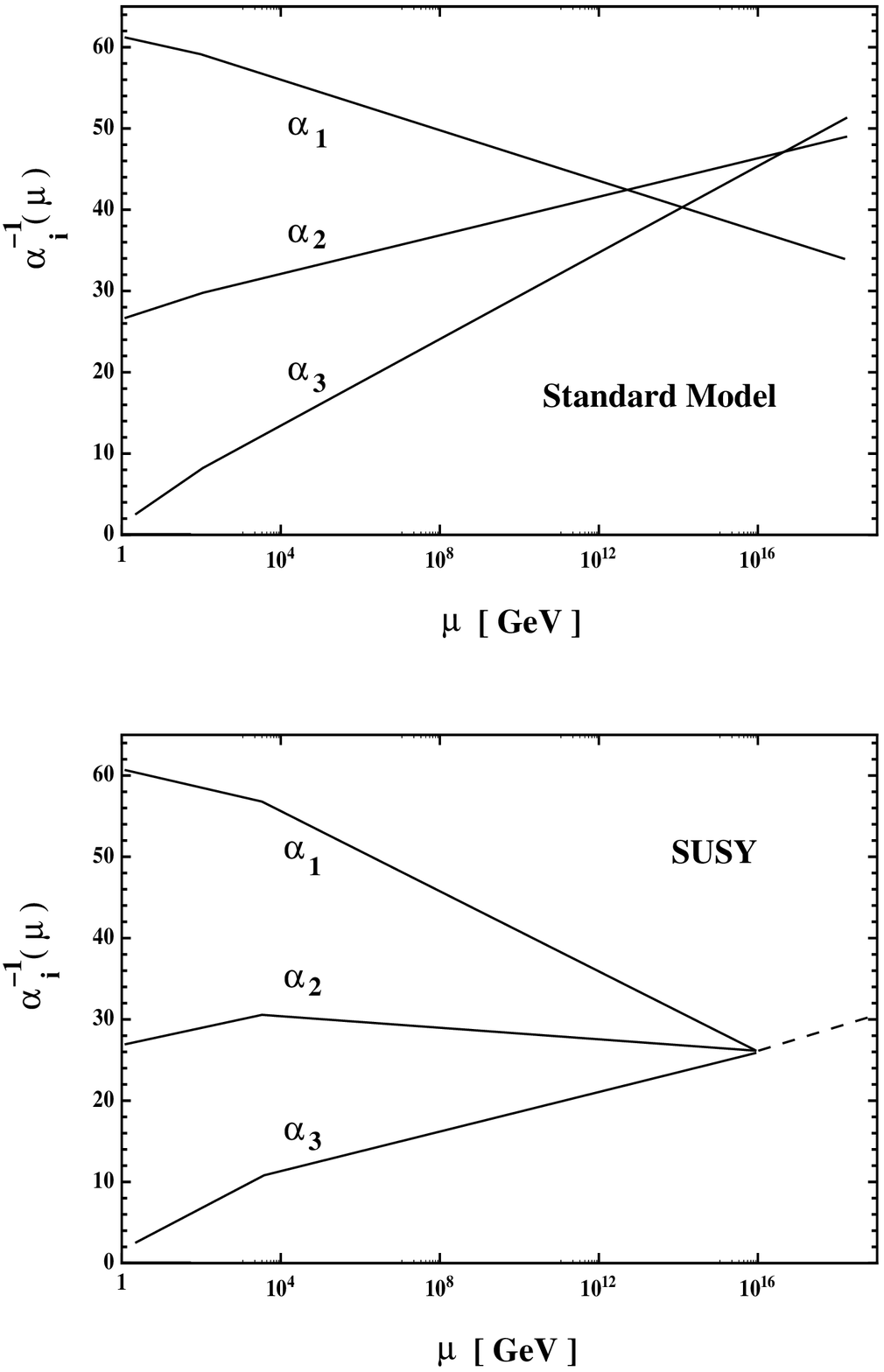,width=10.8cm,height=15.3cm}
\end{center}
\caption{The variation of $ \alpha_{i}$ with
         the scale $ \mu ( GeV )$ for SM and SUSY model.}
\label{fig.4}
\end{figure}

The different
models show a range of unification from
about $ 10^{15}$ GeV $ < M_{X} < 10^{17}$ GeV in the GUT-model
at different scale energies or in the SUSY-model 
Fig.\ref{fig.4} (lower part ) at one scale energy
\cite{UNIFACATIONmodels,PhysikBlaetter1991}. At scales
above $M_{X} \sim 10^{17} $ GeV only on interaction is left over.
This scale is still some magnitudes smaller as the Planck scale of
$ 10^{19} $ GeV where the gravitation has to be taken into account.

Six quarks, three leptons masses and three coupling constants
are the first set of 12 SM parameters. The are summarized
in Tab.~\ref{ParemeterFamily} and the first three lines
of Tab.~\ref{SecondSetParametersSM}.
Further seven parameters are needed to complete the model.

Four are related to the quark decay. Measurements of the quark
decay demonstrate that the quarks $u, c~ and~ t $ with charge
$ 2/3 $ prefer to decay to $ d, s~ and~ b $ with charge $ -1/3 $.
For example in the $ D^{+} $ decay the charm prefers to decays via
$ c \rightarrow s u \bar{d} $ with a spectator $  \bar{d}$.
The quark mass eigenstates $ ( d, s~ and~ b ) $ are not
same as the weak eigenstates $ ( d', s'~ and~ b' ) $.
The matrix relating these bases for the six quarks was
parameterized by Kobayashi and Maskawa \cite{KobayashiMaskawa}.

The mixing is mostly expressed in terms of a $ 3 \times 3 $
unitary matrix $ V $ operating on the $ -e/3 $ quark mass
eigenstates $ ( d, s~ and~ b ) $ Eq.\ref{Mixmatrix}:

\begin{equation}
\begin{pmatrix}
d' \\
s' \\
b'
\end{pmatrix}
=
\begin{pmatrix}
V_{ud} & V_{us} & V_{ub} \\
V_{cd} & V_{cs} & V_{cb} \\
V_{td} & V_{ts} & V_{tb}
\end{pmatrix}
\begin{pmatrix}
d \\
s \\
b
\end{pmatrix}
\label{Mixmatrix}
\end{equation}

This $ 3 \times 3 $ unitary matrix has nine free parameters
but only four of these are physical significant. The
standard parameterization \cite{Databooklet2005} of $ V $ 
utilizes the angles $ \theta_{12} $, $ \theta_{23} $,
$ \theta_{13} $ and the phase $ \delta_{13} $. The matrix
elements of the first row and third column have been directly
measured in decay processes. With $ c_{ij}=\cos \theta_{ij} $,
$ s_{ij}=\sin\theta_{ij} $ and $i,j=1,2,3 $ for the generation label
it is possible to connect the elements of $ V $ direct to the three 
angles under discussion.
It is $ V_{ud}=c_{12} $, $ V_{us}=s_{12} $,
$ V_{ub}=s_{13} e^{-i\delta_{13}} $, $ V_{cb}=s_{23} $ and
$ V_{tb}=c_{23} $ with the angles $ \theta_{12} \sim 13^{o} $,
$ \theta_{23} \sim 2.4^{o} $ and $ \theta_{13} \sim 0.2^{o} $.
Including the loop-level process \cite{Databooklet2005} it is
possible to constrain $ \delta_{13}=60^{o} \pm 14^{o} $.
Tab. \ref{SecondSetParametersSM} summerizes the actual known data.

\begin{table}
\caption{\it The second group of free SM parameters. }
\begin{center}
\begin{tabular}{||l|l||}                                      \hline
 coupling constant $\alpha_{1}(M_{Z})$  & $ 0.017 $
                                                           \\ \hline
 coupling constant $\alpha_{2}(M_{Z})$  & $ 0.034 $
                                                           \\ \hline
 coupling constant $\alpha_{3}(M_{Z})$
                           & $0.118 \pm 0.003 $
                                                            \\ \hline
 CKM - matrix angle $ s_{ij}=\sin \Theta_{ij} $
                           & $ s_{12}=0.2243 \pm 0.0016 $
                                                            \\ \hline
 CKM - matrix angle $ s_{ij}=\sin \Theta_{ij} $
                           & $ s_{23}=0.0413 \pm 0.0015 $
                                                            \\ \hline
 CKM - matrix angle $ s_{ij}=\sin \Theta_{ij} $
                           & $ s_{13}=0.0037 \pm 0.0005 $
                                                            \\ \hline
 CKM - phase  angle        & $ \delta_{13}=1.05 \pm 0.24 $
                                                            \\ \hline
 QCD Vacuum Angle $\Theta_{QCD}$
                           & $ \sim 0 $
                                                            \\ \hline
 Higgs quadratic coupling $\mu_{\star}$
                           & Unknown                        \\ \hline
 Higgs self-coupling coupling strength $\lambda$
                           & Unknown                        \\ \hline
\end{tabular}
\label{SecondSetParametersSM}
\end{center}
\end{table}

The origin of all the masses in the \sm~ is postulated to be a weak 
doublet of scalar Higgs fields, whose kinetic term in the Lagrangian is
\beq
{\cal L}_\phi = -\vert D_\mu \phi\vert^2
\label{onefour}
\eeq
and which has the magic potential:
\beq
{\cal L}_V = -V(\phi ) : V(\phi ) = -\mu^2\phi^{\dagger}\phi +
{\lambda \over 2} (\phi^{\dagger}\phi)^2
\label{onefive}
\eeq
Because of the negative sign for the 
quadratic term in Eq.\ref{onefive}, the
symmetric solution $<0\vert\phi\vert 0> = 0$ is unstable, and if
$\lambda > 0$ the favored solution has a non-zero vacuum expectation
value which we may write in the form:
\beq
<0\vert\phi\vert 0> = <0\vert\phi^{\dagger}\vert 0> = v\begin{pmatrix}
0 \\
\frac{1}{\sqrt{2}}
\end{pmatrix} : v^2 = {\mu^2\over 2\lambda}
\label{onesix}
\eeq
corresponding to spontaneous breakdown of the electroweak symmetry.

Expanding around the vacuum: $\phi = <0\vert\phi\vert 0> + \,\hat\phi$,
the
kinetic term Eq.\ref{onefour} for the Higgs field yields mass terms for
the vector bosons:
\beq
{\cal L}_\phi \ni -{g^2v^2\over 2}~~W^+_\mu ~W^{\mu -} - g^{\prime 2}
~{v^2\over 2}~B_\mu~B^\mu + g~g^\prime  v^2~B_\mu~W^{\mu 3} -
g^2~{v^2 \over 2}~W^3_\mu~W^{\mu 3}
\label{oneseven}
\eeq
corresponding to masses
\beq
m_{W^\pm} = {gv\over 2}
\label{oneeight}
\eeq
for the charged vector bosons.
The neutral vector bosons $(W^3_\mu , B_\mu)$ have a 2$\times$2
mass-squared matrix:
\beq
\begin{pmatrix}
\frac{g^2}{2} & \frac{-gg^\prime}{2}\cr \cr
\frac{-gg^\prime}{2} & \frac{g^{\prime 2}}{2}\\
\end{pmatrix} v^2
\label{onenine}
\eeq
This is easily diagonalizable to yield the mass eigenstates:
\beq
Z_\mu = {gW^3_\mu - g^\prime B_\mu\over \sqrt{g^2+g^{\prime 2}}}~ :~~ m_Z
= {1\over 2} \sqrt{g^2+g^{\prime 2}} v~;~~
A_\mu = {g^\prime W^3_\mu + g B_\mu\over \sqrt{g^2+g^{\prime 2}}} ~:~~
m_A = 0
\label{oneten}
\eeq
that we identify with the massive $Z^0$ and massless $\gamma$, 
respectively. It is useful to
introduce the electroweak mixing angle $\theta_W$ defined by
\beq
\sin\theta_W = {g^\prime\over \sqrt{g^2+g^{\prime 2}}} 
\label{oneeleven}
\eeq
in terms of the  weak SU(2) coupling $g$ and the weak U(1) 
coupling $g^\prime$. Many other
quantities can be expressed in terms of $\sin\theta_W$ \cite{HalsenMartin}:
for example, $m^2_W/ m^2_Z = \cos^2\theta_W$. 

With these boson masses, one indeed obtains charged-current
interactions of the current-current form \cite{HalsenMartin} shown
above, and the neutral currents take the form:
\beq
J^0_\mu \equiv J^3_\mu - \sin^2\theta_W~J^{em}_\mu~,~~G^{NC}_F \equiv
{g^2+g^{\prime 2}\over 8 m^2_Z}
\label{neutralcurrent}
\eeq
The ratio of neutral- and charged-current interaction strengths is often
expressed as
\beq
\rho = {G^{NC}_F\over G_F} = {m^2_W\over m^2_Z \cos^2\theta_W}
\label{oneforteen}
\eeq
which takes the value unity in the \sm, apart from quantum corrections 
(loop effects).

The previous field-theoretical discussion of the Higgs mechanism can be
rephrased in more physical language. It is well known that a massless
vector boson such as the photon $\gamma$ or gluon $g$ has just two
polarization states: $\lambda = \pm 1$. However, a massive vector boson
such as the $\rho$ has three polarization states: $\lambda = 0, \pm 1$.
This third polarization state is provided by a spin-0 field.
In order to make $m_{W^\pm,Z^0} \not= 0$, this should
have non-zero electroweak isospin $I \not= 0$, and the simplest
possibility is a complex iso doublet $(\phi^+,\phi^0)$, as assumed above.
This has four degrees of freedom, three of which are eaten by the $W^\pm$
and $Z^0$ as their third polarization states, leaving us with one physical
Higgs boson $H$. Once the vacuum expectation value $\vert \langle
0\vert\phi\vert 0 \rangle \vert = v/ \sqrt{2}~:~~v = \mu / 
\sqrt{2\lambda}$ is
fixed, the mass of the remaining physical Higgs boson is given by 
\beq
m^2_H = 2\mu_{\star}^2 = 4 \lambda v^2,
\label{onefifteen} 
\eeq 
which is a free parameter in the \sm.

In spite of the big effort of the physics community in the last years
it was not possible to detect a Higgs particle \cite{HiggsLimitLEP}.
So far we know only limits of the Higgs mass. It also means we have
no experimental information about the self coupling $ \lambda $ and the
mass parameter $ \mu_{\star} $. These are the last two free parameters of the
SM in Tab. \ref{SecondSetParametersSM}.

\subsection {Status of experimental limits on the sizes of
             Fundamental Particles}
\label{sec:Status of experimental limits on the sizes of
           Fundamental Particles}

To test the finite size of fundamental particles,
experiments are performed to search for compositeness
or to investigate a non-point-like
behavior in strong, electromagnetic and electroweak interactions.
In the succeeding sections we assume it exist
for each interaction its own characteristic
energy scale $\Lambda$ related to the characteristic size
of interaction region.
\subsubsection {Strong Interaction}
To test  the color charge of the quarks, the entrance
and the exit channels of the reaction
in the scattering experiment should
be dominated by the strong interaction. This condition
is fulfilled by the CDF $ p\bar{p} $
data \cite{CDFDOexitedquark} which exclude excited quarks $ q^{*} $ with
a mass between $200$ and $760$~GeV at 95\%~CL. The
UA2 data \cite{UA2exitedquark} exclude $u^{*}$ and $d^{*}$ quark
masses smaller than $ 288 $~GeV at 90\%~CL.
In this case characteristic energy scale is given by
the mass of the excited quark. Associated characteristic size is
$r_q \sim \hbar/(m_{q}^* c)<3.5\times 10^{-17}$ cm.

\subsubsection { Electromagnetic Interaction }
The purely electromagnetic interaction $ \EEGG $
is ideal to test the QED because
it is not interfered by the $ Z^{o} $ decay. This
reaction proceeds via the exchange of a virtual electron
in the t - and u - channels, while the s - channel is
forbidden due to angular momentum conservation.
Differential cross sections for the process $ \EEGG $,
are measured at energies from  $\sqrt{s} $=55~GeV to 207 GeV
using the data collected with the
VENUS \cite{VENUSdiffCrossSection},
TOPAZ \cite{TOPAZdiffCrossSection},
ALEPH \cite{ALEPHdiffCrossSection},
DELPHI \cite{DELPHIdiffCrossSection},
L3 \cite{L3diffCrossSection} and
OPAL \cite{OPALdiffCrossSection}
detector from 1989 to 2003. The measurements of the
differential cross section from VENUS, TOPAZ, ALEPH, DELPHI
and L3 are summarized in Fig.\ref{VTADL}. The date are marked
in color. For comparison with theory
a QED calculation of the differential cross section
up to radiative effects up to $ O(\alpha^{3}) $ is displayed
in a solid black line. All values are normalized to $ 91.2 $ GeV.
\begin{figure}
\vspace{-5.0 mm}
\begin{center}
 \epsfig{file=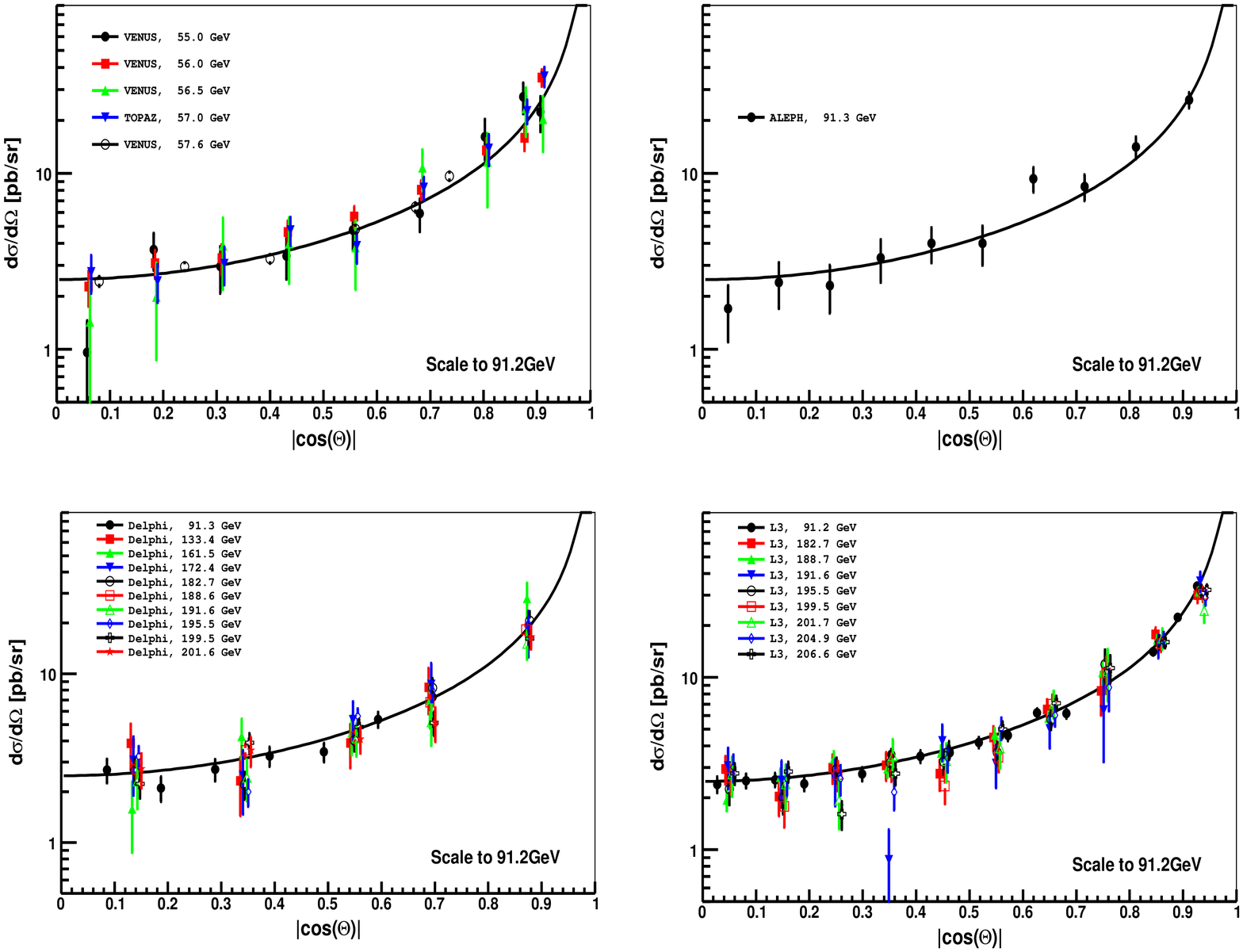,width=17.0cm,height=10.0cm}
\end{center}
\caption{Measured and calculated QED differential cross section
         to $ O(\alpha^{3}) $ level (solid line) from VENUS, TOPAZ,
         ALEPH, DELPHI and L3.}
\label{VTADL}
\end{figure}
The measurements of the differential cross section from DELPHI
and an overlay of all available data together also
normalized to $ 91.2 $ GeV is displayed in Fig.\ref{Dall}.
%
\begin{figure}
\begin{center}
 \epsfig{file=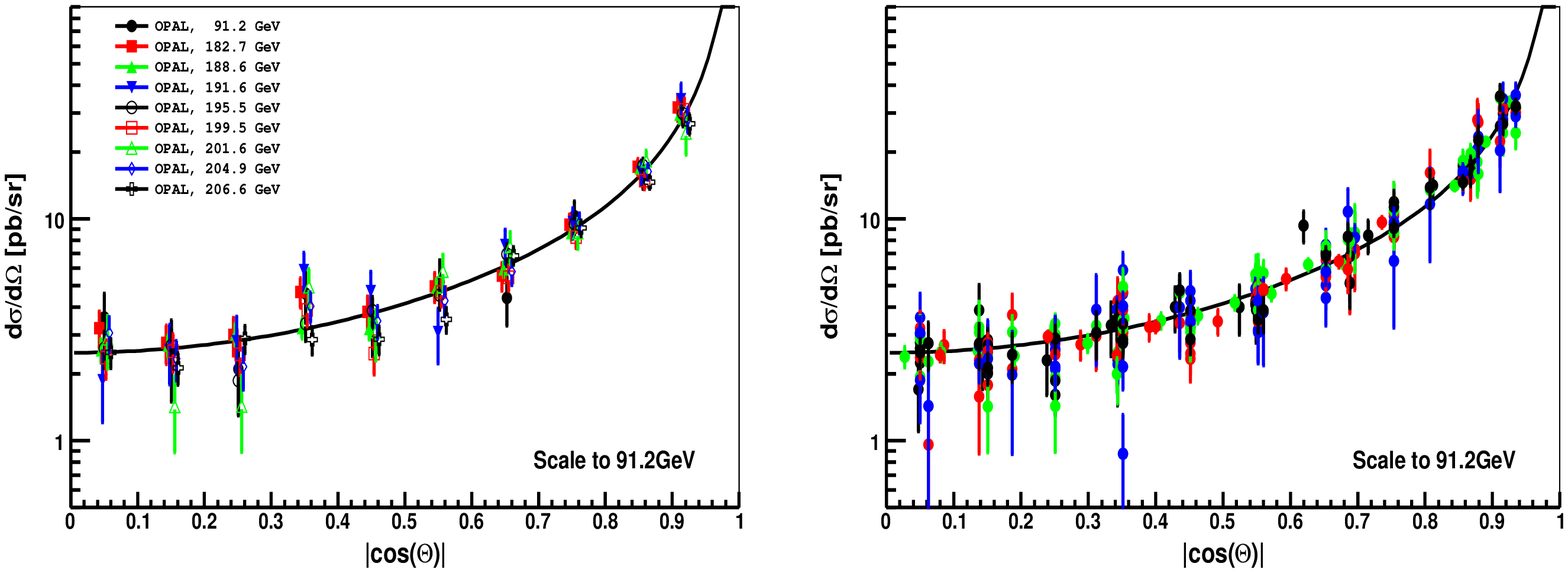,width=17.0cm,height=5.0cm}
\end{center}
\caption{Measured and calculated QED differential cross section
         to $ O(\alpha^{3}) $ level (solid line) from OPAL
         and an overlay of all available data together. }

\label{Dall}
\end{figure}

The agreement between the data and the QED
predictions can be used to constrain models
of the existence of an excited
electron of mass $ m_{e^{*}} $ which
replaces the virtual
electron in the QED process \cite{ExcitedElectron}, or to constrain
a model with deviation from QED arising from an
effective interaction with non-standard
$ e^{+} e^{-} \gamma $ couplings and
$ e^{+} e^{-} \gamma \gamma $ contact terms \cite{DirectContact}.

We first discuss a possible excited electron and
choose the approach from ref. \cite{LITKE}. In this case
the heavy exited electron could couple
to an electron and a photon via magnetic interaction
with an effective Lagrangian of .

\begin{equation}
{\mathcal L}_{\text{excited}}=\frac{e\lambda}{2m_{e^{*}}}
\overline{\psi_{e^{*}}}\sigma_{\mu\nu}\psi_{e}F^{\mu\nu}
\label{LITKElag}
\end{equation}

In this equation $ \lambda $ is the coupling constant,
$ F^{\mu\nu} $ the electromagnetic field tensor,
$ \psi_{e^{*}} $ and $ \psi_{e} $ are the wave function
of the heavy electron and the electron respectively.
The model has $ \lambda $ and $ m_{e^{*}} $ as parameters.

It is possible to write down the corresponding
differential cross-section by adding a deviation term
$ \delta_{new} $ to the QED differential cross-section
including radiative effects up to $ O(\alpha^{3}) $. The modified
equation is

\begin{equation}
(d\sigma/d\Omega)_{theo}=(d\sigma/d\Omega)_{O(\alpha^{3})}
(1+\delta_{new})
\label{LITKEcross}
\end{equation}

If the center-of-mass energy $\sqrt{s}$ satisfies
the condition $ s/m^{2}_{e^{*}} << 1$, then $\delta_{new}$ would
read as

\begin{equation}
\delta_{new}= s^{2}/2 ( 1/\Lambda^{4} )
(1-\cos^{2}\Theta)
\label{eq.4}
\end{equation}

In this approximation, the parameters $\Lambda $ are the
QED cut-off parameters with
$ \Lambda^{2}=m^{2}_{e^{*}}/\lambda $. In the case of arbitrary
$\sqrt{s}$ the full equation of ref.\cite{LITKE} is
used to calculate $ \delta_{new} = f(m_{e^{*}}) $.
The angle $ \Theta $ is the angle of the
two most energetic emitted photons with angle $ \Theta_{1} $ and
$ \Theta_{2} $ with respect to the
beam axis defined in Eq.\ref{COStheta}.

\begin{equation}
 \mid cos ( \Theta ) \mid = 1/2 ( \mid cos ( \Theta_{1} ) \mid +
                     \mid cos (2\pi - \Theta_{2})\mid)
\label{COStheta}
\end{equation}

The third order QED differential cross section
up to $ O(\alpha^{3}) $ is calculated by numerically
generating a high number of Monte Carlo ( MC ) $ \EEGG $ events
\cite{GAMMAgenerator,L3diffCrossSectionB}.
The angular distribution of these
events was fitted with a high order polynomial function to
get an analytical equation for the cross section as function
of the scattering angle defined in Eq.\ref{COStheta}.

We performed about the published differential cross sections an
overall $ \chi^{2} $ test between 55 GeV and 207 GeV.
The used luminosity at the different energies together
with the name of the detector is displayed
in Tab.\ref{Lumi6detectors}.
\begin{table}
\caption{ The luminosity used from the different experiments. }
\begin{center}
\begin{tabular}{||l|l|l|l|l|l|l||}                               \hline
 GeV       & VENUS            & TOPAZ            & ALEPH
           & DELPHI           & L3               & OPAL
                                                              \\ \hline
 55        & \small $ 2.34 pb^{-1} $  \cite{VENUSdiffCrossSection}
           & & & & &
                                                             \\ \hline
 56        & \small $ 5.18 pb^{-1} $  \cite{VENUSdiffCrossSection}
           & & & & &
                                                             \\ \hline
 56.5      & \small $ 0.86 pb^{-1} $  \cite{VENUSdiffCrossSection}
           & & & & &
                                                             \\ \hline
 57        & \small $ 3.70 pb^{-1} $  \cite{VENUSdiffCrossSection}
           & & & & &
                                                             \\ \hline
 57.6      & & \small $ 52.26 pb^{-1} $  \cite{TOPAZdiffCrossSection}
           & & & &
                                                             \\ \hline
 91        & &
           & \small $ 8.5   pb^{-1} $  \cite{ALEPHdiffCrossSection}
           & \small $ 36.9  pb^{-1} $  \cite{DELPHIdiffCrossSection}
           & \small $ 140   pb^{-1} $  \cite{L3diffCrossSectionB}
           & \small $ 7.2   pb^{-1} $  \cite{OPALdiffCrossSection}
                                                              \\ \hline
 133       & & &
           & \small $ 5.92   pb^{-1} $  \cite{DELPHIdiffCrossSection}
           & &
                                                              \\ \hline
 162       & & &
           & \small $ 9.58   pb^{-1} $  \cite{DELPHIdiffCrossSection}
           & &
                                                              \\ \hline
 172       & & &
           & \small $ 9.80   pb^{-1} $  \cite{DELPHIdiffCrossSection}
           & &
                                                              \\ \hline
 183       & & &
           & \small $ 52.9 pb^{-1} $  \cite{DELPHIdiffCrossSection}
           & \small $ 54.8  pb^{-1} $  \cite{L3diffCrossSection}
           & \small $ 55.6  pb^{-1} $  \cite{OPALdiffCrossSection}
                                                              \\ \hline
 189       & & &
           & \small $ 151.9 pb^{-1} $  \cite{DELPHIdiffCrossSection}
           & \small $ 175.3 pb^{-1} $  \cite{L3diffCrossSection}
           & \small $ 181.1 pb^{-1} $  \cite{OPALdiffCrossSection}
                                                              \\ \hline
 192       & & &
           & \small $ 25.1  pb^{-1} $  \cite{DELPHIdiffCrossSection}
           & \small $ 28.8  pb^{-1} $  \cite{L3diffCrossSection}
           & \small $ 29.0  pb^{-1} $  \cite{OPALdiffCrossSection}
                                                              \\ \hline
 196       & & &
           & \small $ 76.1  pb^{-1} $  \cite{DELPHIdiffCrossSection}
           & \small $ 82.4  pb^{-1} $  \cite{L3diffCrossSection}
           & \small $ 75.9  pb^{-1} $  \cite{OPALdiffCrossSection}
                                                              \\ \hline
 200       & & &
           & \small $ 82.6  pb^{-1} $  \cite{DELPHIdiffCrossSection}
           & \small $ 67.5  pb^{-1} $  \cite{L3diffCrossSection}
           & \small $ 87.2  pb^{-1} $  \cite{OPALdiffCrossSection}
                                                              \\ \hline
 202       & & &
           & \small $ 40.1  pb^{-1} $  \cite{DELPHIdiffCrossSection}
           & \small $ 35.9  pb^{-1} $  \cite{L3diffCrossSection}
           & \small $ 36.8  pb^{-1} $  \cite{OPALdiffCrossSection}
                                                              \\ \hline
 205       & & & &
           & \small $ 74.3  pb^{-1} $  \cite{L3diffCrossSection}
           & \small $ 79.2  pb^{-1} $  \cite{OPALdiffCrossSection}
                                                              \\ \hline
 207       & & & &
           & \small $ 138.1 pb^{-1} $  \cite{L3diffCrossSection}
           & \small $ 136.5 pb^{-1} $  \cite{OPALdiffCrossSection}
                                                              \\ \hline
\end{tabular}
\end{center}
\label{Lumi6detectors}
\end{table}
As fit parameter $ ( 1/\Lambda^{4} ) $ is used and the
coupling constant is set to $ \lambda = 1 $. In
Tab.\ref{FITresult} are all
$ 1/\Lambda^{4} [ $GeV$^{-4}] $ minima displayed together with
the quality value of the fit $ \chi^{2}/dof $.
Negative values are depicted in red positive values in black color.
%
\begin{table}
\caption{ The result of the fit parameter $ 1/\Lambda^{4} [ 1/ $GeV$^{4} ] $
          for the different experiments. }
\begin{center}
\begin{tabular}{||l|l|l|l|l|l|l||}                               \hline
           & VENUS            & TOPAZ            & ALEPH
           & DELPHI           & L3               & OPAL          \\
 GeV       & $(1/\Lambda^{4})$& $(1/\Lambda^{4})$&$(1/\Lambda^{4})$
           & $(1/\Lambda^{4})$& $(1/\Lambda^{4})$&$(1/\Lambda^{4})$
                                                                 \\ \hline
\scriptsize 55        & \tiny  \color{red} $-(4.26\pm2.52) 10^{-8}$
                      & & & & &                                  \\
                      & \tiny  \color{red}$ \chi^{2}/dof = 12.9/8$
                      & & & & &
                                                            \\ \hline
\scriptsize 56        & \tiny  $ (3.24\pm1.88) 10^{-8}$
                      & & & & &                             \\
                      & \tiny  $ \chi^{2}/dof = 9.48/8$
                      & & & & &
                                                            \\ \hline
\scriptsize 56.5      & \tiny  \color{red}$ -(2.11\pm3.96) 10^{-8}$
                      & & & & &                             \\
                      & \tiny  \color{red}$ \chi^{2}/dof = 4.93/8$
                      & & & & &
                                                             \\ \hline

\scriptsize 57        & \tiny  \color{red}$ -(1.49\pm2.02) 10^{-8}$
                      & & & & &                             \\
                      & \tiny  \color{red}$ \chi^{2}/dof = 8.82/8$
                      & & & & &
                                                             \\ \hline
\scriptsize 57.6      & & \tiny  \color{red}$ -(1.59\pm5.61) 10^{-9}$
                      & & & &                              \\
                      & & \tiny  \color{red}$ \chi^{2}/dof = 7.32/5$
                      & & & &
                                                             \\ \hline
\scriptsize 91        & &
                      &\tiny $ (0.07\pm2.98) 10^{-9}$
                      &\tiny  \color{red}$ -(2.29\pm1.70) 10^{-9}$
                      &\tiny  \color{red}$ -(6.88\pm8.00) 10^{-10}$
                      &\tiny  \color{red}$ -(0.93\pm3.59) 10^{-9}$
                                                                 \\
                      & &
                      & \tiny  $ \chi^{2}/dof = 9.96/9$
                      & \tiny  \color{red}$ \chi^{2}/dof = 3.54/6$
                      & \tiny  \color{red}$ \chi^{2}/dof = 11.1/15$
                      & \tiny  \color{red}$ \chi^{2}/dof = 6.92/8$
                                                              \\ \hline
\scriptsize 133       & & & & \tiny  \color{red}$ -(0.48\pm1.26) 10^{-9}$
                      & &                                     \\
                      & & & & \tiny  \color{red}$ \chi^{2}/dof = 2.60/3$
                      & &
                                                             \\ \hline
\scriptsize 162       & & & & \tiny  \color{red}$ -(2.35\pm5.40) 10^{-10}$
                      & &                                    \\
                      & & & & \tiny  \color{red}$ \chi^{2}/dof = 4.59/4$
                      & &
                                                             \\ \hline
\scriptsize 172       & & & & \tiny  $ (0.74\pm5.19) 10^{-10}$
                      & &                                    \\
                      & & & & \tiny  $ \chi^{2}/dof = 1.09/4$
                      & &
                                                             \\ \hline
\scriptsize 183       & & &
                      &\tiny  \color{red}$ -(2.54\pm1.60) 10^{-10}$
                      &\tiny  \color{red}$ -(1.48\pm1.37) 10^{-10}$
                      &\tiny   $ (2.05\pm1.43) 10^{-10}$
                                                                 \\
                      & & &
                      & \tiny  \color{red}$ \chi^{2}/dof = 5.27/4$
                      & \tiny  \color{red}$ \chi^{2}/dof = 11.0/9$
                      & \tiny  $ \chi^{2}/dof = 5.86/9$
                                                              \\ \hline

\scriptsize 189       & & &
                      &\tiny  $ (0.14\pm1.01) 10^{-10}$
                      &\tiny  \color{red}$ -(8.58\pm7.16) 10^{-11}$
                      &\tiny  \color{red}$ -(2.05\pm6.89) 10^{-11}$
                                                                 \\
                      & & &
                      & \tiny  $ \chi^{2}/dof = 2.67/4$
                      & \tiny  \color{red}$ \chi^{2}/dof = 17.2/9$
                      & \tiny  \color{red}$ \chi^{2}/dof = 5.13/9$
                                                              \\ \hline
\scriptsize 192       & & &
                      &\tiny  \color{red}$ -(3.59\pm2.07) 10^{-10}$
                      &\tiny  \color{red}$ -(5.79\pm1.41) 10^{-10}$
                      &\tiny   $ (0.13\pm1.63) 10^{-10}$
                                                                 \\
                      & & &
                      & \tiny  \color{red}$ \chi^{2}/dof = 1.03/4$
                      & \tiny  \color{red}$ \chi^{2}/dof = 16.9/9$
                      & \tiny  $ \chi^{2}/dof = 12.6/9$
                                                              \\ \hline
\scriptsize 196       & & &
                      &\tiny  \color{red}$ -(0.43\pm1.19) 10^{-10}$
                      &\tiny  \color{red}$ -(1.93\pm0.89) 10^{-10}$
                      &\tiny  \color{red}$ -(1.62\pm9.37) 10^{-10}$
                                                                 \\
                      & & &
                      & \tiny  \color{red}$ \chi^{2}/dof = 16.4/4$
                      & \tiny  \color{red}$ \chi^{2}/dof = 7.84/9$
                      & \tiny  \color{red}$ \chi^{2}/dof = 7.48/9$
                                                              \\ \hline
\scriptsize 200       & & &
                      &\tiny  \color{red}$ -(0.88\pm1.12) 10^{-10}$
                      &\tiny  \color{red}$ -(2.58\pm0.90) 10^{-10}$
                      &\tiny  \color{red}$ -(1.65\pm0.84) 10^{-10}$
                                                                 \\
                      & & &
                      & \tiny  \color{red}$ \chi^{2}/dof = 8.07/4$
                      & \tiny  \color{red}$ \chi^{2}/dof = 13.8/9$
                      & \tiny  \color{red}$ \chi^{2}/dof = 8.63/9$
                                                              \\ \hline
\scriptsize 202       & & &
                      &\tiny  \color{red}$ -(1.11\pm1.51) 10^{-10}$
                      &\tiny  \color{red}$ -(1.49\pm1.24) 10^{-10}$
                      &\tiny  \color{red}$ -(1.47\pm1.16) 10^{-10}$
                                                                 \\
                      & & &
                      & \tiny  \color{red}$ \chi^{2}/dof = 2.94/4$
                      & \tiny  \color{red}$ \chi^{2}/dof = 15.2/9$
                      & \tiny  \color{red}$ \chi^{2}/dof = 17.8/9$
                                                              \\ \hline
\scriptsize 205       & & & &
                      &\tiny  \color{red}$ -(1.07\pm0.84) 10^{-10}$
                      &\tiny  \color{red}$ -(3.81\pm7.99) 10^{-11}$
                                                                 \\
                      & & & &
                      & \tiny  \color{red}$ \chi^{2}/dof = 12.9/9$
                      & \tiny  \color{red}$ \chi^{2}/dof = 6.26/9$
                                                              \\ \hline
\scriptsize 207       & & & &
                      &\tiny  \color{red}$ -(9.14\pm5.99) 10^{-11}$
                      &\tiny  \color{red}$ -(1.52\pm0.57) 10^{-10}$
                                                                 \\
                      & & & &
                      & \tiny  \color{red}$ \chi^{2}/dof = 23.6/9$
                      & \tiny  \color{red}$ \chi^{2}/dof = 10.7/9$
                                                              \\ \hline
\end{tabular}

\end{center}
\label{FITresult}
\end{table}
The single results of the different $ 1/\Lambda^{4} [ 1/ $GeV$^{4} ] $ minima
get combined in three groups from TRISTAN, LEP 1, LEP 2 and finally an
overall fit result is displayed in Tab.\ref{FITresultcompined}.
\begin{table}
\caption{ The parameter $ 1/\Lambda^{4} [ $GeV$ ^{-4} ] $ for the combined
          $ \chi^{2} $ test. }
\begin{center}
\begin{tabular}{||l|l||}                                       \hline
  TRISTAN       & $ ( 2.49\pm5.05)\times 10^{-9} $           \\
                & $ \chi^{2}/dof = 50.0/41 $
                                                              \\ \hline
  LEP I         & $ -( 9.20\pm6.90)\times 10^{-10} $           \\
                & $ \chi^{2}/dof = 32.3/41 $
                                                              \\ \hline
  LEP II        & $ -( 1.10\pm0.20)\times 10^{-10} $           \\
                & $ \chi^{2}/dof = 267/203 $
                                                              \\ \hline
  All Data      & $ -( 1.11\pm0.20)\times 10^{-10} $           \\
                & $ \chi^{2}/dof = 351/287 $
                                                              \\ \hline
\end{tabular}
\end{center}
\label{FITresultcompined}
\end{table}
The fit results of all the different steps are displayed in
Fig.{\ref{plot_fitRESULTS}}.
%
\begin{figure}[htbp]
\begin{center}
 \includegraphics[width=9.0cm,height=12.0cm]{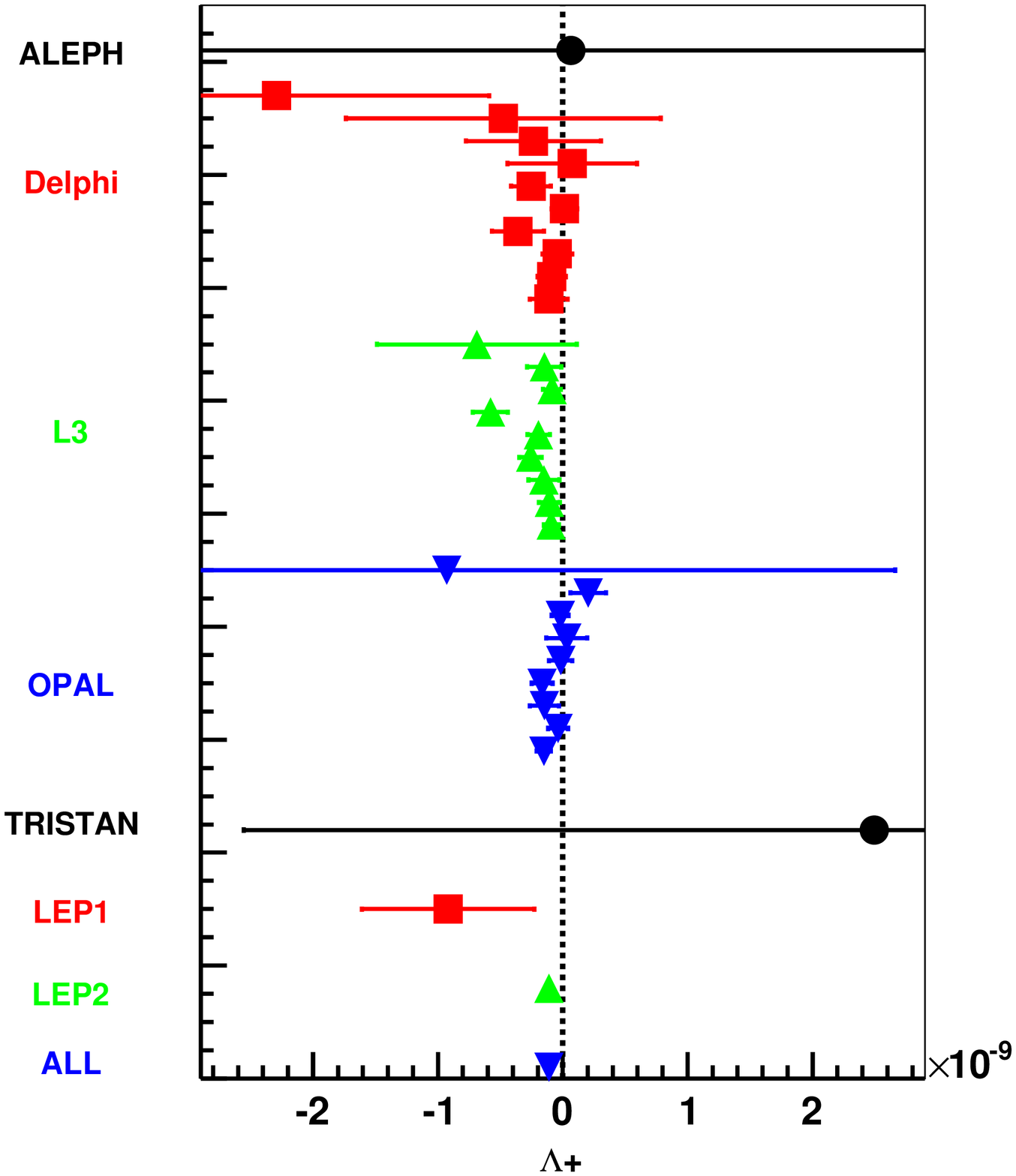}
\end{center}
\caption{The $\chi^{2}$ minima of all $ 1/\Lambda^{4}_{\pm} [ GeV^{-4}] $
               values of the different steps of the fit procedure. }
\label{plot_fitRESULTS}
\end{figure}

Systematic errors arise from the luminosity evaluation, from
the selection efficiency, background evaluations, the choice to use
the Born level or $ \alpha ^{3} $ theoretical QED cross section
as reference cross section,
the choice of the fit procedure, the choice of the fit parameter and
the choice of the scattering angle $ | cos \Theta | $ in particular
in comparison between data and theoretical calculation.

The maximum estimated error for the value of the fit from the luminosity,
selection efficiency and background evaluations is approximately
$ \delta \Lambda/\Lambda = 0.01 $ \cite{SystematicERROR}. The choice of the
of the theoretical QED cross section was studied with 1882 $ \EEGG $ events
from the L3 detector \cite{SystematicERROR,WUthesis}. In 
Fig.{\ref{plot_QEDcrossBornALPHA3}} the measured data points of the $ \EEGG $
reaction together with the QED Born and the $ \alpha ^{3} $ level
are displayed. In part b) the sensitivity of the measured data points
to both QED cross sections is visible.
%
\begin{figure}[htbp]
\vspace{-10.0mm}
\begin{center}
 \includegraphics[width=11.0cm,height=12.0cm]{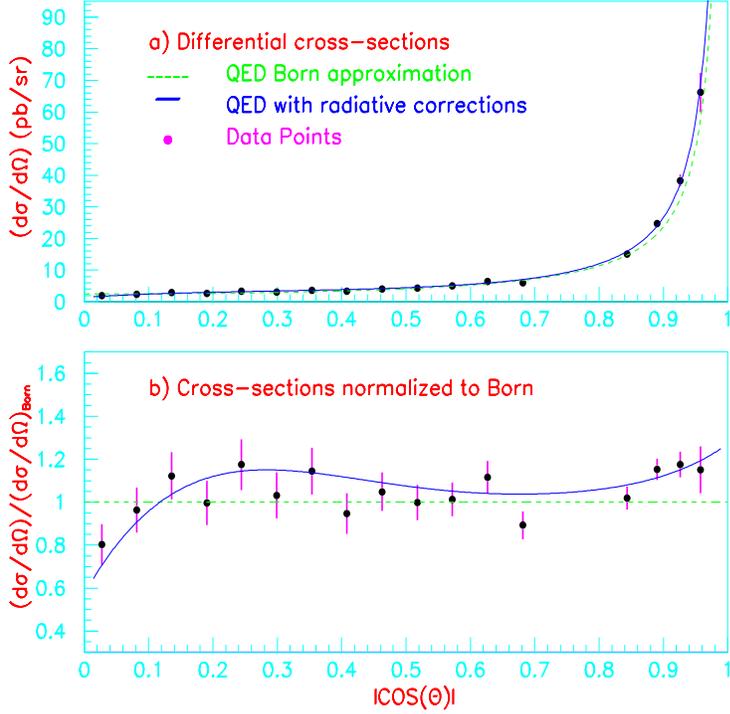}
\end{center}
\caption{Comparison of theoretical QED differential cross section of Born
         and $\alpha^{3}$ level with experimental data. }
\label{plot_QEDcrossBornALPHA3}
\end{figure}
A trop in the $ \chi^{2} $ by a approximately a factor two favors the
QED $ \alpha ^{3} $ level to be used for the fit. At a small sample
of $ \EEGG $ events the fit values $ \Lambda $ are compared
for $ \chi^{2} $, Maximum-Likelihood, Smirnov-Cramer von Misis, Kolmogorov
test all with and without binning \cite{ERRORfitMETHOD}. An approximately
$ \delta \Lambda /\Lambda = 0.005 $ effect is estimated for the overall fit.
A linear fit parameter $ P = ( 1/\Lambda^{4} ) $ prejudices a minimum in the
$ \chi^{2} $ fit whereas the direct fit to $ \Lambda $ ends in a limit
without minimum.
Fig.{\ref{plot-paramterLAMDA}} displays as example on the left side the plot of the
$ \chi^{2} $ as function of $ P = ( 1/\Lambda^{4} ) $ of the data under
discussion and on the right side the $ \chi^{2} $ as function of
$ \Lambda_{6} $ from a L3 data set \cite{SystematicERROR} we will discuss
next in the effective contact interaction.
%
\begin{figure}[htbp]
\begin{center}
 \includegraphics[width=18.0cm,height=6.0cm]{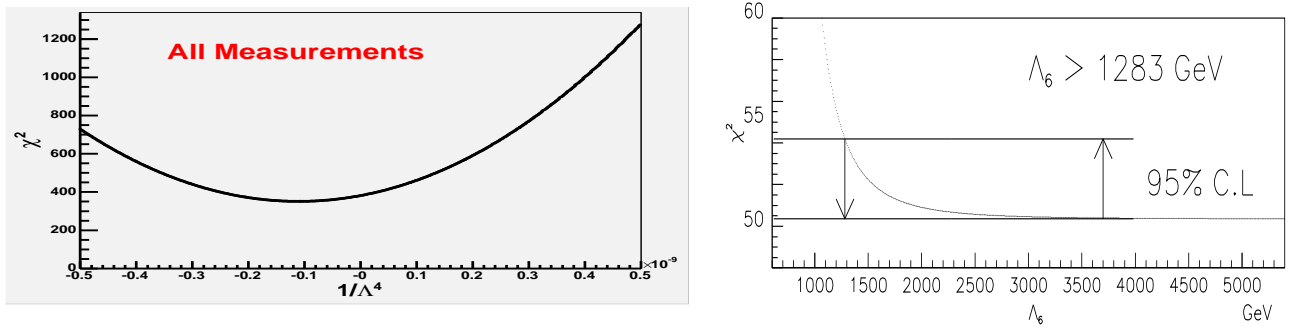}
\end{center}
\caption{Comparison of $ 1/\Lambda^{4} $ and
         $ \Lambda_{6}=\Lambda_{C} $ as fit parameter. }
\label{plot-paramterLAMDA}
\end{figure}
The use of different definitions of scattering angles \cite{TOPAZdiffCrossSection}
introduces in the $ \mid cos ( \Theta ) \mid $ an error of approximately
$ \delta \mid cos ( \Theta ) \mid  = 0.0005 $. In a worst case
scenario for scattering angles close to $ 90^{o} $
the $ | cos ( \Theta ) |_{experiment} \sim 0.05  $. This would
result in $ ( \delta \Lambda /\Lambda )_{\delta |\cos(\Theta)|} = 0.01 $.
Adding all these errors in quadrature give a total systematic error
of $ \delta \Lambda /\Lambda  \approx 0.015 $.
A similar possibility to estimate the systematic error is
to scale up all errors in the
overall fit by a factor $ S = \sqrt ( \chi^{2}/dof ) $ this
changes $ \sigma $ by $ \delta \sigma = 0.5 $.
In Tab. \ref{FITresultscaleerror} first column the
$ ( 1/\Lambda^{4} ) $ values together with the
$ ( \chi^{2}/dof )_{fit} = 351/287 $ and
$ ( \chi^{2}/dof )_{fit} = 287/287=1 $ are displayed. In the
second column the corresponding $ \Lambda $ values and the
from $ \sigma $ estimated uncertainty $\delta \Lambda $ is shown.
\begin{table}
\caption{ The parameter $ 1/\Lambda^{4} $ for the combined
          $ \chi^{2} $ test and including a scale factor. }
\begin{center}
\begin{tabular}{||l|l|l||}                                        \hline
  All Data      & $ -( 1.11\pm0.20)\times 10^{-10}~~$GeV$ ^{-4} $
                & $ \Lambda = 308 \pm 56 ~~$GeV$ $                 \\
                & $ \chi^{2}/dof = 351/287 $
                &

                                                               \\ \hline
  All Data with scale factor
                & $ -( 1.11\pm0.22)\times 10^{-10}~~ $GeV$ ^{-4} $
                & $ \Lambda  = 308 \pm 61~~ $GeV$ $                 \\
                & $ \chi^{2}/dof = 287/287 $
                &
                                                              \\ \hline
\end{tabular}
\end{center}
\label{FITresultscaleerror}
\end{table}

The used hypothesis in Eq.\ref{LITKElag} and Eq.\ref{LITKEcross}
assumes that an exited electron will increase the total QED-$ \alpha^{3} $
cross section and change the angular distribution of the QED cross section.
Opposite to the hypothesis the fit expresses a minimum with a negative
fit parameter $ 1/\Lambda^{4} $ of a significance of
approximately five $ \sigma $ .

In the case of effective contact interaction with non-standard
coupling a cut-off parameter $ \Lambda_{C} $
is introduced to describe the scale of the interaction
with the following Lagrangian \cite{DirectContact}.

\begin{equation}
{\mathcal L}_{\text{contact}}=i\overline{\psi_{e}}
\gamma_{\mu}(D_{\nu}\psi_{e})
\left(\frac{\sqrt{4\pi}}{\Lambda^{2}_{C6}}F^{\mu\nu}+
\frac{\sqrt{4\pi}}{\tilde{\Lambda}^{2}_{C6}}\tilde{F}^{\mu\nu}\right)
\label{LagDirectContactQ}
\end{equation}

The effective Lagrangian chosen in this case has an operator
of dimension 6, the wave function of the electrons
is $ \psi_{e} $, the QED covariant derivative is $ D_{\nu} $,
the tilde on $ \tilde{\Lambda}_{C6} $ and $ \tilde{F}^{\mu\nu} $
stands for dual.
As in the case of excited electron
it is possible to write down the corresponding
differential cross-section by adding a deviation term
$ \delta_{new} $ to the QED differential cross-section
including radiative effects up to $ O(\alpha^{3}) $
see Eq.\ref{LITKEcross} and $\delta_{new}$ reads as

\begin{equation}
\delta_{new}=s^{2}/(2\alpha)(1/\Lambda^{4}_{C6} + 1/\tilde{\Lambda}^{4}_{C6})
(1-\cos^{2}\Theta)
\label{eq.5}
\end{equation}

The angle $ \Theta $ is the angle of the
emitted photons with respect to the
beam axis defined in Eq.\ref{COStheta}.
For all later discussed fit procedures we set
$ \Lambda_{C6}=\tilde{\Lambda}_{C6}=\Lambda_{C} $.

As discussed in Fig.\ref{plot-paramterLAMDA} the choice
of the fit parameter $ ( 1/\Lambda^{4},1/\Lambda^{4}_{C} )$
or $ ( \Lambda,\Lambda_{C} ) $ leads in the
fit in the $ ( 1/\Lambda^{4},1/\Lambda^{4}_{C} ) $ case
to a minimum and in the
$ ( \Lambda,\Lambda_{C} ) $ case to limits. The significance $ \sigma $
of the minimum will finally be
used to interprete the minimum as a real effect with high $ \sigma $
or also a limit with low $ \sigma $.
To demonstrate numerical
the impact of the fit procedure we discuss both possibilities.

We first discuss the case to choose $ ( 1/\Lambda^{4}_{C} )$ as fit parameter.
The for the hypothesis of the exited electron in Eq.\ref{LITKElag}
performed $ \chi^{2} $ fit was repeated for the hypothesis of the
effective contact interaction Eq.\ref{LagDirectContactQ}
using $ ( 1/\Lambda^{4}_{C} ) $
as fit parameter. As in the hypothesis of the exited electron also
for the effective contact interaction an increase of the total
QED-$\alpha^{3} $ cross section and a change of the angular
distribution
Eq.\ref{LITKEcross} and Eq.\ref{eq.5} is expected.
Opposite to both hypothesis also the best fit value of all data
$ ( 1/\Lambda^{4}_{C} )_{best}=-(4.05\pm0.73)\times 10^{-13}~$GeV$^{-4} $ is
negative with significance about $ 5 \times \sigma $. The fit does not allow to
distinguishing between both hypothesis.
Using the best values of
$ (1/\Lambda)^{4}_{C} $ it is possible to calculate
the scale factor
$ (\Lambda_C)_{best} = 1253.2 $ GeV which translates in a finite
size of the interaction area of $ r \approx 15.7 \times 10^{-18} $ cm. The
results of the $ \chi^{2} $ fit for the hypothesis of the exited electron
and the the effective contact interaction are summarized in
Tab. \ref{AllFitvalues}.
\begin{table}
\caption{ The best fit value $ ( 1/\Lambda^{4})_{best} $ and
          $ ( 1/\Lambda^{4}_C)_{best} $ for the combined
          $ \chi^{2} $ test. }
\begin{center}
\begin{tabular}{||l|l|l|l||}                                   \hline
  $ ( 1/\Lambda^{4})_{best} $
                & $ -( 1.11\pm0.20)\times 10^{-10}~$GeV$ ^{-4} $
                & $ \Lambda  = 308~$GeV$ $
                &
                                                             \\ \hline
  $ ( 1/\Lambda^{4}_C)_{best} $
                & $ -(4.05\pm0.73)\times 10^{-13}~$GeV$ ^{-4} $
                & $ \Lambda_{C}  = 1253.2~$GeV$ $
                & $ r \approx 15.7 \times 10^{-18} $ cm
                                                             \\ \hline
\end{tabular}
\end{center}
\label{AllFitvalues}
\end{table}

Second we discuss an example for similar data sets from LEP
if the fit parameter is
direct $ \Lambda $ or $ \Lambda_{C} $ \cite{L3diffCrossSectionB}
and compare it to L3
\cite{L3diffCrossSection} and LEP \cite{Threefamilies} measurements
with $ ( 1/\Lambda )^{4} , ( 1/\Lambda_{C} )^{4} $ as fit parameter.
Including a slightly different data set from the L3 date shown in
Tab. \ref{Lumi6detectors}
we performed a bin-free Likelihood test \cite{L3diffCrossSectionB}.
From the model of the exited electron and contact interaction
defined from Lagrangian Eq.\ref{LITKElag} and Eq.\ref{LagDirectContactQ}
$ \Lambda $ values
can be introduced with either constructive $ ( + ) $ or destructive
$ ( - ) $ interference between the Standard Model process and
the exited electron and contact interaction.
For the hypothesis of the exited electron as parameter was
used $\LAMP=\Lambda$ Eq.\ref{eq.4} and $\LAMM$
\cite{L3diffCrossSectionB}. For the effective contact interaction
only the parameter was $ \LAMP=\Lambda=\Lambda_{C} $ was used.
The limits at $ 95\% $ CL of the fit parameters $ \LAMP$,$\LAMM $ and
$ \Lambda $ are displayed in Tab. \ref{FITresultLAMBDA} first column.
L3 published an analysis of the data set from
Tab. \ref{Lumi6detectors} \cite{L3diffCrossSection}.
As fit parameters $ ( 1/\LAMP^{4}$,$1/\LAMM^{4} ) $ and $ ( 1/\Lambda^{4} ) $
was used. The low significance allows to set limits on these
parameters on $ 95\% $ CL shown in the second column of
Tab. \ref{FITresultLAMBDA}. LEP published a similar result
for all combined LEP data shown in the third column
of Tab. \ref{FITresultLAMBDA}. The limit of the
effective contact interaction $ \Lambda $ sets a limit on the
geometrical size of the interaction area of
$ R_{QED}~<~(\hbar \times c ) / \Lambda $. The numerical
values are displayed in Tab. \ref{FITresultLAMBDA} last row
for all three data sets.

In conclusion to the electromagnetic interaction it is remarkable
that for both hypothesis the excited electron and effective contact
interaction the $ \chi^{2} $ test leads to a best fit value
$ (1/\Lambda^{4})_{best} $ and $ (1/\Lambda^{4}_{C})_{best} $ for
the complete data set of Tab. \ref{Lumi6detectors} what is negative
with a significance of about $ 5 \times \sigma $ shown in
Tab. \ref{AllFitvalues}. This behavior is already visible in the
detailed fit results of Tab. \ref{FITresult},
Tab. \ref{FITresultcompined} and Tab. \ref{FITresultscaleerror}.
The Likelihood test of L3 \cite{L3diffCrossSectionB},
\cite{L3diffCrossSection} and LEP \cite{Threefamilies} uses a
smaller data sample in particular
without the VENUS \cite{VENUSdiffCrossSection} and
TOPAZ \cite{TOPAZdiffCrossSection} data. These tests lead to limits.
It should be noticed that the L3 data \cite{L3diffCrossSection}
and LEP \cite{Threefamilies} date result in a also in negative
$ (1/\Lambda^{4})_{best} $ and $ (1/\Lambda^{4}_{C})_{best} $ values
but in a very shallow minimum with insignificant $ \sigma $-value.

The results indicate that the used
data set prefers to decrease the cross section of $ \EEGG $ with respect
to that predicted by pure QED.
The calculation of the QED-$ \alpha^{3} $ cross section assumes
a scattering center of a point. If the electron would be an extended
object the microstructure of the electron will modify the QED cross
section if the test distances ( CM-scattering energy ) is smaller
as the size of this structure. It is not obvious that only an increase
of the cross section is possible. For example destructive interference effects
originated from the micro structure could also lower the cross section
at high energies.

\begin{table}
\caption{ The limits at $ 95 \% $ CL of the different fit parameters}
\begin{center}
\begin{tabular}{||l|l|l|l||}                               \hline
           & L3 \cite{L3diffCrossSectionB}
           & L3 \cite{L3diffCrossSection}
           & LEP \cite{Threefamilies}
                                                              \\ \hline
 $\LAMP~[TeV]   >$ & 0.433 & 0.4 & 0.392
                                                              \\ \hline
 $\LAMM~[TeV]   >$ & 0.276 & 0.3 & 0.364
                                                              \\ \hline

 $\Lambda~[TeV] >$ & 1.762 & 1.6 & 1.595
                                                              \\ \hline
 $R_{QED}~[cm]<$ & $1.1 \times 10^{-17}$
                 & $1.2 \times 10^{-17}$
                 & $1.2 \times 10^{-17}$
                                                              \\ \hline
\end{tabular}
\end{center}
\label{FITresultLAMBDA}
\end{table}
\subsection { Electroweak Interaction }
\label{sec:Electroweak Interaction}

The $ ep $ accelerator HERA and the $ e^{+}e^{-} $
accelerator LEP  test excited and
non-point-like couplings of quarks and leptons.
In the entrance channel the reaction proceeds
via magnetic and weak interaction and in the exit channel
all three interaction participate.

\subsubsection{Excited and non-point-like quarks}

The electron-proton interaction at high energies
allows to search for excited quarks.
The magnetic transition coupling of quarks
includes  a single production of excited quarks
through $t$-channel gauge boson exchange between
the incoming electrons and quarks.

The H1 data \cite{H1_EXQ} give
for the $ q^{*}\rightarrow qg $ decay channel
a compositeness scale $ \Lambda $.
For a $ q^{*} $ of mass $ 100 $~GeV the
limit on $ \Lambda $ moves from $ 60 $ GeV
to $ 290 $ GeV.
In the $ q^{*}\rightarrow q + \gamma $ decay channel
the $ ep $ data
exclude at 95 \% CL large
 regions of the
cross section times branching ratio
$\sigma(q^{*})\times BR(q^{*}\rightarrow q + \gamma) $
for $ q^{*} $ masses from $ 50 $ GeV to $ 250 $ GeV.
A similar search has been performed by the ZEUS
collaboration~\cite{ZEUS_EXQ}.

At LEP, excited quarks could be
produced via a $ Z^{0}, \gamma $
coupling to fermions. The quarks
are generated singly
or in pairs. In the single production
case it is possible to search for $ q^{*} $
masses up to $ M_{q^{*}} \leq M_{Z} - M_{q} $,
while in the pair production  $ q^{*} $ masses
are constrained by the beam energy.

ALEPH  investigated the
$ q^{*}\rightarrow q + g $ and $ q^{*}\rightarrow q + \gamma $
decay channels \cite{ALEPH_EXQ}.
Substantial areas of the
form factor times branching ratio of
$ q^{*} $ masses between $ 0 $ and $ 50 $ GeV
are excluded.
It is possible to set for the single
production limits on the parameter $\lambda/m_{q^*} $
for masses up to $ 85 $ GeV. For the pair
production the mass limits for Standard Model
coupling are at $ 45 $ GeV.

L3 investigated the $ q^{*}\rightarrow q + \gamma $
decay channel and
gave an upper limits, at $ 95 $\% CL, on the single production
$\sigma(e^{+} e^{-}\rightarrow Z^{o} \rightarrow q^{*} q )
 \times BR(q^{*}\rightarrow q \gamma ) \le 10~$pb$ $
up to $82$ GeV \cite{L3_EXQ}.
For the pair production the limit
is
$\sigma(e^{+} e^{-}\rightarrow Z^{o} \rightarrow q^{*} q^{*} )
 \times BR^{2}(q^{*}\rightarrow q \gamma ) \le 2~$pb$ $
up to $ M_{q^{*}} $ to $ 45 $ GeV. The OPAL
\cite{OPAL_EXQ} has reported similar results.

The search for non-point-like coupling of the quarks has been
also performed  with $ e^{+} e^{-} $
accelerators. As in the case of the QED
contact interaction, an effective Lagrangian
is introduced \cite{BOUNDS_CON1}:

\begin{equation}
{\mathcal L}_{\text{contact}}=\frac{1}{1+\delta_{ef}}
\sum_{i,j=L,R} \eta_{ij}
\frac{g^{2}}{\Lambda^{2}_{ij}}
(\bar{e}_{i}\gamma^{\mu}e_{i})(\bar{f}_{j}\gamma^{\mu}f_{j})
\label{eq.6}
\end{equation}

The four-fermion contact interaction
is characterized by a coupling strength, g,
and by an energy scale  $ \Lambda $.
The Kronecker symbol $ \delta_{ef} $
is zero except for the $ e^{+} e^{-} $ final
state when it is equal to 1. The parameter
$ \eta_{ij} $ defines the contact interaction
model by choosing the helicity amplitudes which
contribute to the reaction
$ e^{+} e^{-}\rightarrow f \bar{f} $. The
wave function $ e_{i} $ and $ f_{j} $
denote the left- and right-handed
initial-state electron and final-state fermion.
The value of $ g/\Lambda $ determines the characteristic scale
of the expected effects. In a general search
the energy scale $ \Lambda $ is chosen by
convention such that $ g^{2}/4\pi = 1 $
and $ | \eta_{ij} |=1 $ or $ | \eta_{ij} |=0 $
is satisfied.

The LEP Collaborations searched for new effects involving
four fermion vertices contact interactions in all exit channels at
center-of-mass energies between $ 132 $ GeV
and $ 207 $ GeV \cite{LEP-DC-2006}.
The LEP Collaborations investigated the pure contact interaction
amplitudes $ e^{+} e^{-}\rightarrow q \bar{q} $
and the deviations from the
Standard Model.

Four helicity amplitudes $ \eta_{LL} $,
$ \eta_{RR} $, $ \eta_{LR} $ and
$ \eta_{RL} $ are investigated for eight
different models each. In accordance with the QED reaction the
corresponding energies scales for the models with constructive
or destructive interference are denoted by $ \LAMP $ and $ \LAMM $
respectively.
Limits for the
$ q \bar{q} $ final state range from
$ \LAMM = 8.1 $ TeV to $ \LAMP = 9.3 $ TeV,
for $ u \bar{u} $ from
$ \LAMM = 9.6 $ TeV to $ \LAMP = 14.3 $ TeV and
for $ d \bar{d} $ from
$ \LAMM = 13.3 $ TeV to $ \LAMP = 7.7 $ TeV,
for $ b \bar{b} $ from
$ \LAMM = 11.5 $ TeV to $ \LAMP = 15.3 $ TeV and
for $ c \bar{c} $ from
$ \LAMM = 8.2 $ TeV to $ \LAMP = 10.3 $ TeV,

These scales allow to estimate an upper limit
for characteristic size
$ r_{q}^{-} $ and $ r_{q}^{+} $
related to strong interaction
of the quarks. Depending on the different
helicity amplitudes and models this scale
ranges from
$ r_{q}^{-} < 2.5 \times 10^{-18} $ cm to
$ r_{q}^{+} < 2.2 \times 10^{-18} $ cm.

\subsubsection{Excited and non-point-like leptons}

The electron-proton interaction at high energies
allows to search, as in the quark case,
for excited leptons. For example excited electrons $ e^{*} $
and neutrinos $ \nu^{*} $ can be probed via the same
magnetic type coupling of the quarks.

The H1 data \cite{H1_EXQ} describe for the
$ e^{*} $ case the electromagnetic and weak decay channel.
For the
$ \nu^{*} $ the channel $ \nu^{*} \rightarrow \nu \gamma  $
is measured. The experiment is able to set limits on the product
$ \sigma \times BR^{*} $ of the production cross section
and the branching ratio for different decay channels.
Big regions of the product
$ \sigma \times BR^{*} $ in the decay channels of
$ e^{*} \rightarrow e \gamma $,
$ e^{*} \rightarrow e Z^{0}  $,
$ e^{*} \rightarrow e W  $
$ \nu^{*} \rightarrow \nu \gamma  $ are excluded by the data in
the  $ e^{*} $ and $ \nu^{*} $ mass range up to $ 250 $ GeV
at $ 95 $\% CL.
Using specific models \cite{HAGIWARA} it is possible for a typical
coupling constant $ c^{2}_{\gamma e^{*} e } = 1/4 $
and an $ e^{*} $ mass of $ 100 $ GeV to restrict
a compositeness scale parameter by
$ \Lambda < 440 $ GeV. In the case of the
$ \nu^{*} $ the same parameter would be
$ \Lambda < 51 $ GeV.

With the $ e^{+} e^{-}$ accelerator LEP excited
leptons can be produced via
$ s- $ and $ t- $ channel for $ Z^{0}, \gamma $
and W
coupling to fermions, in particular to $ e^{*} $
$ \nu^{*} $.

The ALEPH  investigated the
$ 16 $ decay channels from the
states $ l^{*}l $, $ \nu^{*} \nu $,
$ l^{*}\bar{l}^{*} $ and
$ \nu^{*} \bar{\nu^{*}} $ \cite{ALEPH_EXQ}.
No evidence for weak decay of excited
leptons has been found and stringent coupling
limits are set. By combining all radiative
channels under investigation
a  lower limit on the
compositeness maximal scale $ \Lambda > 16 $ TeV
has been set.

The L3 experiment investigated  similar decay channels
at energies ranging from $ \sqrt {s} $ $ 189 $ GeV to
$ 206 $ GeV \cite{L3exidetLEPTON03}.
No evidence for charged and neutral excited leptons
of any flavor was found. Lower mass limits ranging
from $ 91.3 $ GeV to $ 101.5 $ GeV at $ 95 $\% CL
are derived for any value of the excited lepton
couplings. Upper limits according to the excited
lepton flavor and mass, are set in the mass
range from $ 100 $ GeV to $ 200 $ GeV.

The search for non-point-like coupling of the
leptons is also performed with $ e^{+} e^{-} $
accelerators. As in the case of the quarks
discussed in previous subsection the same effective Lagrangian
Eq.\ref{eq.6} has been used to define
characteristic energy scale $\Lambda $.

The LEP Collaborations \cite{LEP-DC-2006} investigated in the same
energy range as in the $ e^{+} e^{-}\rightarrow q \bar{q} $
case also the pure contact interaction
$ e^{+} e^{-}\rightarrow l \bar{l} $
and searched for deviations from the Standard Model.
Four helicity amplitudes $ \eta_{LL} $,
$ \eta_{RR} $, $ \eta_{LR} $ and
$ \eta_{RL} $ are investigated for eight
different models each.
The corresponding energies scales for the models with constructive
or destructive interference are also denoted by $ \LAMP $ and $ \LAMM $
respectively.
Limits in the
$ l^{+} l^{-} $ final state ranging from
$ \LAMM = 16.0 $ TeV to $ \LAMP = 21.7 $ TeV,
for $ e^{+} e^{-} $ from
$ \LAMM = 18.0 $ TeV to $ \LAMP = 15.9 $ TeV,
for $ \mu^{+} \mu^{-} $ from
$ \LAMM = 14.3 $ TeV to $ \LAMP = 19.7 $ TeV and
for $ \tau^{+} \tau^{-} $ from
$ \LAMM = 14.2 $ TeV to $ \LAMP = 14.5 $ TeV.
These scales result in estimates of characteristic size
for weak interaction area $ r_{l}^{-} $ and $ r_{l}^{+} $
of the leptons. Depending on the different
helicity amplitudes and models this scale
ranges from
$ r_{l}^{-} < 1.3 \times 10^{-18} $ cm to
$ r_{l}^{+} < 0.9 \times 10^{-18} $ cm
similar to the quark case.

\subsection {Conclusion for the experimental limits}
\label{sec:Conclusion for the experimental limits}

The investigations of the pure electromagnetic interaction
$ \EEGG $ using the complete set of differential cross
sections available from VENUS, TOPAZ and LEP lead to
a $ 5 \times \sigma $ effect for the hypothesis
of an excited electron and the effective contact interaction.
A $ 2.6 \times \sigma $ effect for axial-vector contact
interaction in the data on $ \EEEEG $ at center of mass
energies $ 192 - 208 $ GeV is reported \cite{Bourilkov}.
All investigations of the strong and electroweak interaction
searching for exited fermions or contact interaction
lead to lower and upper experimental limits. All these
values are summarized in Fig.{\ref{SizeLimits}.
\begin{figure}[htbp]
\begin{center}
 \includegraphics[width=12.0cm,height=8.0cm]{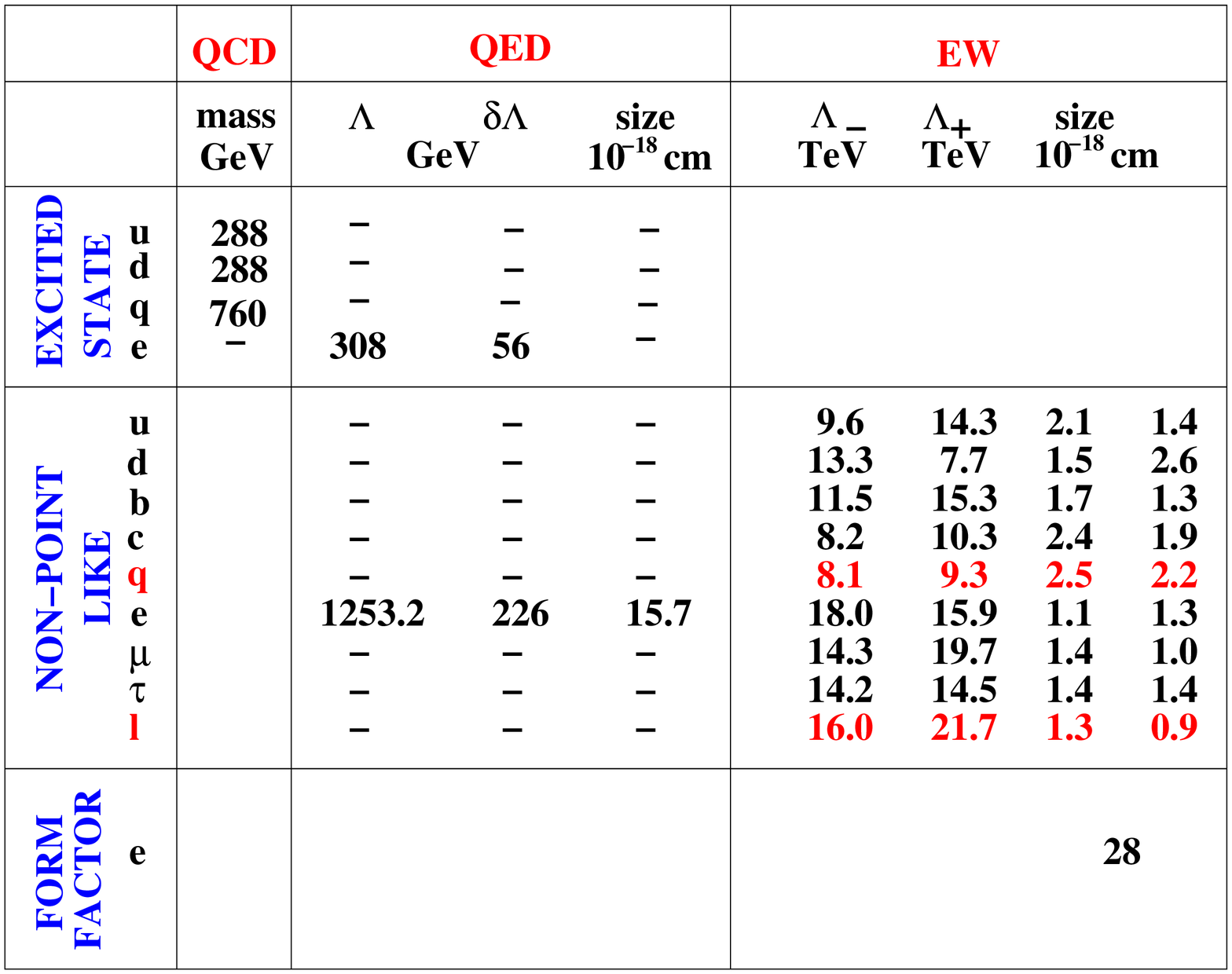}
\end{center}
\caption{ Summary of all experimental values and limits of sizes of fundamental
          particles. }
\label{SizeLimits}
\end{figure}
This figure sorts the maximum values after the
three interactions QCD, QED and EW in the first row and according
the discussed hypothesis exited fermions, contact interaction and
for completeness an investigation concerning form factors
\cite{Bourilkov} in the first column. In the QCD column are
displayed the maximum experimental limits. The QED column
shows the $ \Lambda $ and $ \delta \Lambda  $ values of the
overall $ \chi^{2} $ fit according with the size $ r $ ,
the EW column displays the maximum
possible $ \LAMM~~ $,$ \LAMP~~ $,
size $ r^{-}_{i}<(\hbar \times c ) / \LAMM $
and size $ r^{+}_{i}<(\hbar \times c ) / \LAMP~ $
$ ( i=u,d,b,c,q;e,\mu,\tau,l ) $
values of the eight models under investigation.
The geometrical size of the form factor is shown in the last row.
In Fig.\ref{Datalimts} the data and most stringent limits of
Tab. \ref{SizeLimits} are displayed as function of the mass of the
particles and compared with the Compton wavelength
\( \lambda\!\!\!\!-_c=\hbar/mc\).
\begin{figure}[htbp]
\begin{center}
 \includegraphics[width=10.0cm,height=10.0cm]
                 {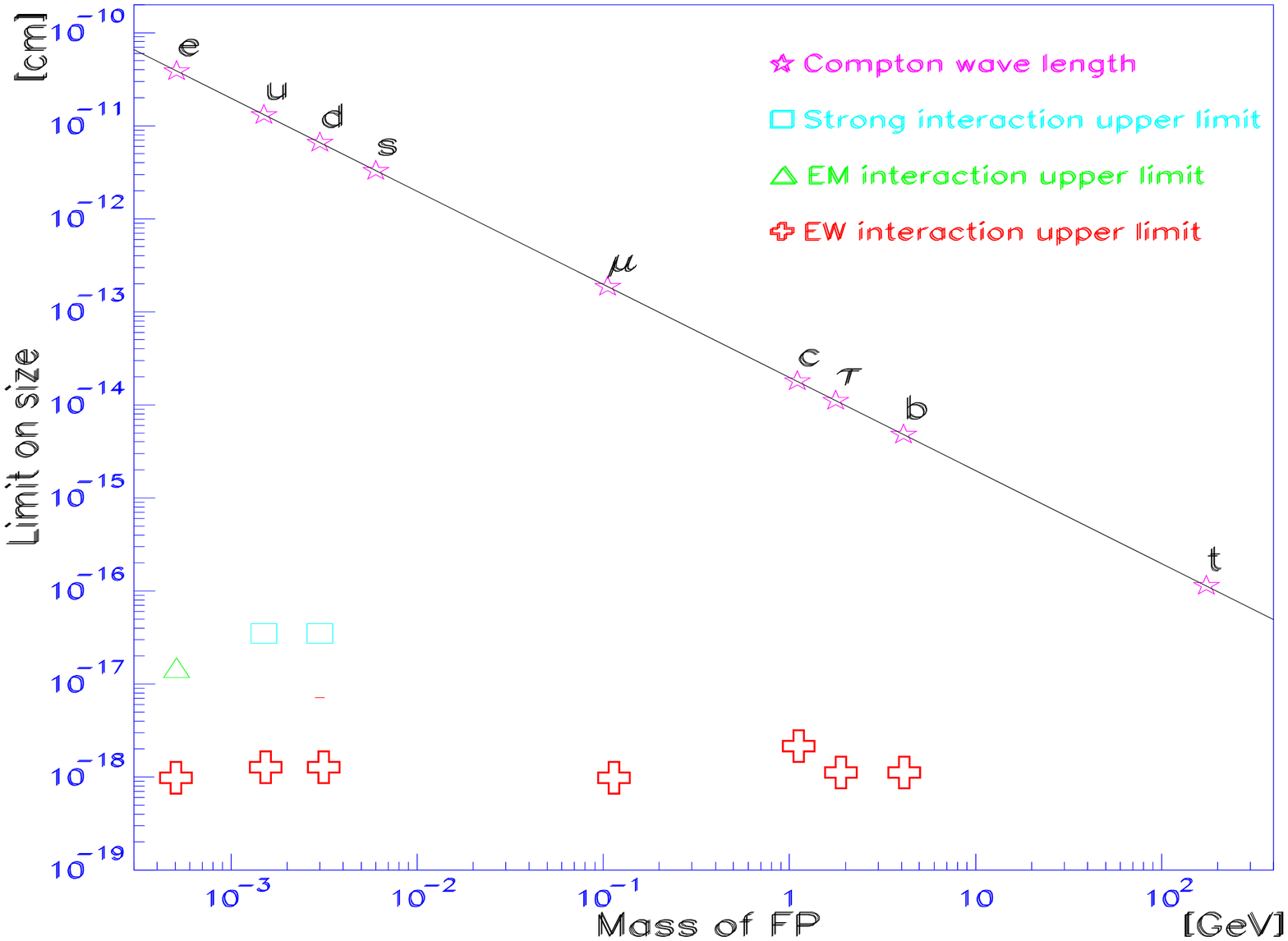}
\end{center}
\caption{ Comparison of Compton wavelength of FP with current
experimental limits
measured according strong, electromagnetic and weak interaction.}
\label{Datalimts}
\end{figure}
All the limits show that the Compton wavelength
\( \lambda\!\!\!\!-_c\)
is bigger as the
characteristic size $r_s$ of the interaction area.
This experimental fact \( r_s \le \lambda\!\!\!\!-_c\) is used in section
particle-like structure related to gravity to estimate
the size of FP and the mass of the Higgs.
\section{Cosmology and the early Universe}
\label{sec:Cosmology and the early Universe}
\subsection{Evolution of the universe}
\label{sec:Evolution of the universe}

Observations suggest that the universe at large scales is homogeneous and
isotropic. To quantify the idea of homogeneity one can calculate for example
the mass of a sphere of radius $R$ at different locations of the universe. In a
strictly uniform Universe with density $\bar\rho$, one will always get the mass
to be $\bar M=\frac{4\pi}{3}\bar\rho R^3$. In reality, we will obtain different
 values when the sphere is centered at different locations so that there will
be a fractional rms fluctuations $\sigma (R)=(\delta M/M)_R=<(M-\bar
M)^2>^{1/2}/\bar M$ around the mean mass $\bar M$. If this quantity decreases
with $R$ and becomes sufficiently small for large R, then it makes sense to
assign the universe as a smooth at large scales with mean density $\bar\rho$. 
In our universe, $\sigma (R)\approx 0.1$ at $R\approx 50\invh\
\Mp$~\footnote{$1\ \Mp\approx 3\times 10^{24}\ {\rm cm}$.} and $\sigma (R)$
decreases roughly as $R^{-2}$ at larger scales.

This suggests that we can model the universe as made up of a smooth background
with an average density $\rho (t)$ being, at best, only a function of time
superposed with fluctuations $\delta\rho (t,{\bf x})$, which are large at small
scales but decreases as scale increased. So, at sufficiently large scales, the
universe may be treated as being homogeneous and isotropic with a uniform
density. 

A simple form of a large scale motion of an object located at point ${\bf r}$
consistent with the idea of isotropy can be given as $\dot {\bf r}(t)={\bf
v}(t)=f(t){\bf r}$. An observer at the origin will see all particles moving with
a speed proportional to the distance from him. Another observer located at
${\bf r_1}$ will see the object at ${\bf r}$ to move with the velocity ${\bf
v'}={\bf v-v_1}=f(t)({\bf r-r_1})$ so that are also moving away with a speed
proportional to the distance from the observer~\footnote{Here and further on in
this chapter we consider a nonrelativistic context of velocity addition.}.
Integrating this relation, $\dot {\bf r}(t)=f(t){\bf r}$, we can describe the
position of any material body in the universe in the form ${\bf r}(t)=a(t){\bf
x}$ where $a(t)$ is a scale factor related to $f(t)$ by  $f(t)=\dot a/a$ and
${\bf x}$ is a constant for any given material body in the universe. It is
convenient to call ${\bf x}$ and ${\bf r}$ the comoving and proper coordinate
of a body and $a(t)$ the expansion (scale) factor. Therefore the dynamics of
the universe is entirely determined by the function $a(t)$. The simplest choice
would be to make $a(t)$ a constant, which would imply no motion in the universe
and all matter would be distributed uniformly in a static configuration.
However, it is clear that such a configuration should be violently unstable
when mutual gravitation forces of the bodies are taken into account. Any such
instability will eventually lead to random motion of particles in localized
region thereby destroying the initial homogeneity. Observationally, however, is
 indicated that the relation ${\bf v}=(\dot a/a){\bf r}$ does hold in our
universe with $\dot a>0$. 

The dynamics of  $a(t)$ can be understood from the application of general
relativity to a homogeneous and isotropic distribution of matter with $\rho $
interpreted as the energy density. Indeed, in the case of a homogeneous and
isotropic spacetime, the interval obeys a particular simple form:
\beq
ds^{2}=dt^{2} - a^{2}(t)\left[\frac{dr^{2}}{1-kr^{2}}+r^{2}d\theta^{2} +
       r^{2}\sin^{2}\theta d\phi^{2}\right],
\label{RMmetrix}
\eeq 
where the parameter $k$ modifies the metric of 3-space and the function $a(t)$
scales the distances between any two points in the 3-dimensional space. One of
the Einstein's equations then reduces to
\beq
\frac{\dot a}{a}+\frac{k}{a^2}=\frac{8\pi G}{3}\rho (t),
\label{e1}
\eeq
while the rest of the equations are satisfied if the equation of motion for
matter are satisfied. 

Observations suggest that our universe today $t=t_0$~\footnote{We will refer
to quantities evaluated at the present time epoch with a subscribe zero.} is 
governed by Eq.\ref{e1} with
\beq
\left(\frac{\dot a}{a}\right)_0=H_0=0.3\times 10^{-17}h\ {\rm s^{-1}}=100\ h\
{\rm km\ s^{-1}}\ \Mp^{-1},
\eeq
where $h$ ranges from $0.6$ to $0.8$. The relation  ${\bf v}=(\dot a/a){\bf
r}=H_0{\bf r}$ is known as Hubble's law and $H_0$ is called Hubble's constant.
Using $H_0$ one can construct the time scale $t_U=H_0^{-1}\approx 10^{10}\invh\
{\rm yr}$ and the length scale $cH_0^{-1}\approx 3000\invh\ \Mp$; $t_U$
characterizes the evolutionary time scale of the universe and $cH_0^{-1}$ is of
the order of the largest length scale currently accessible in cosmological
observations. 
A common value to describe the evolution of the universe
is the deceleration parameter $q(t)$ of Eq.\ref{DecI}

\begin{equation}
q(t)=-\left(\frac{\ddot{a}}{a}\right) \left/ \left(\frac{\dot{a}}{a}\right)^{2}
    = - \frac{1}{H(t)^{2}}\frac{\ddot{a}}{a}
\label{DecI}
\right.
\end{equation}

More importantly, one can also construct out of $H_0$ a quantity
with the dimension of density called critical density
\beq
\rho_c=\frac{3H_0^2}{8\pi G}=1.88h^2\times 10^{-29}\ {\rm gr\
cm^{-3}}\ \ \footnote{Some other useful representations of the
critical density: $\rho_c=2.8\times 10^{11}h^2M_{\odot}\ \Mp^{-3}=1.1\times
10^4h^2\ {\rm eV}\ {\rm cm^{-3}}=1.1\times 10^{-5}h^2\ {\rm photons\
cm^{-3}}$.}.
\label{rho_c}
\eeq 
It is useful to measure all other mass and energy densities in the universe in
terms of critical density defining the ration $\Omega_i=\rho_i/\rho_c$, where
$\rho_i$ is the mass or energy density associated with a particular species of
particles. The parameters used so far to describe the evolution of the universe
are summarized in Tab. \ref{EvolutionUnivers}.

\begin{table}
\caption{ Parameters describing the evolution of the universe. }
\begin{center}
\begin{tabular}{||l|l||}                                      \hline
  parameter     & value
                                                               \\ \hline
  Hubble constant     & $ H_{0}= 100~h~km~s^{-1}~$Mpc$^{-1}$
                                                               \\ \hline
  Evolutionary time   & $ t_{U}=H_{0}^{-1} \approx 10^{10}~h^{-1}~$yr$ $
                                                               \\ \hline
  Length scale        & $ c~H_{0}^{-1} \approx 3000~h^{-1}~$Mpc$ $
                                                               \\ \hline
  critical density    & $ \rho _{c}=1.88~h^{2}\times 10^{-29} $gr~cm$^{-3} $
                                                               \\ \hline
  critical density    & $ \rho _{c}=2.8 \times 10^{11}~h^{2}~M_{\odot}~$Mpc$^{-3}$
                                                               \\ \hline
  critical density    & $ \rho _{c}=1.1\times 10^{4}~h^{2}~$eV~cm$^{-3} $
                                                               \\ \hline
  critical density    & $ \rho _{c}=1.1\times 10^{-5}~h^{2}~$photons~cm$^{-3}$
                                                               \\ \hline
  parameter h         & $ 0.6 < h < 0.8 $
                                                               \\ \hline
\end{tabular}
\end{center}
\label{EvolutionUnivers}
\end{table}

The numerical value of $k$ can absorbed into the definition of $a(t)$ by
rescaling it so that we can treat $k$ as having one of the three values
$(0,-1,+1)$. The choice among these three values for $k$ is decided by
Eq.\ref{e1} depending on whether $\Omega$ greater than, equal or less than
unity. The fact that $k$ is proportional to the total energy of the dynamical
system described by Eq.\ref{e1} shows that $a(t)$ will have a maximum value
followed by a contracting phase to the universe if $k=1$, $\Omega >1$.
The Tab. \ref{ParameterSBPmodel} summarize the three possibilities.

\begin{table}
\caption{ The Big Bang model parameters $ q $, $ k $ and
          $ \Omega $ define three possibilities of an expanding universe. }
\begin{center}

\begin{tabular}{||l|l|l|l||}                                      \hline
  deceleration  & curvature
                & density
                & $ a(t) $
                                                               \\ \hline
  $ 0\leq q(t) < 1/2 $
                & $ k=-1 $
                & $ \Omega < 1 $
                & monotone increasing
                                                              \\ \hline
  $ q(t)=1/2 $
                & $ k=0 $
                & $ \Omega = 1 $
                & monotone increasing
                                                              \\ \hline
  $ q(t) > 1/2 $
                & $ k=1 $
                & $ \Omega > 1 $
                & final cyclic
                                                              \\ \hline
\end{tabular}
\end{center}
\label{ParameterSBPmodel}
\end{table}

If a light was emitted at $a=a_e$ and received today (when $a=a_0$), the
wavelength will change by the factor $(1+z_e)=(a_0/a_e)$, where $z_e$ is called
the redshift corresponding to the epoch of emission, $a_e$. Thus, one can
associate a redshift $z(t)$ to any epoch $a(t)$ by $(1+z)^{-1}=a(t)/a_0$. It is
very common in cosmology to use $z$ as a time coordinate to characterize an
epoch in the past. 

To determine the nature of the cosmological model we need to know the value of
$\Omega$ for the universe, taking into account all forms of energy densities
existing at present. In particular, to determine the form of $a(t)$
from Eq.\ref{e1}  we need to determine how $\rho$ varies with $a$. As a
particular species contributes an energy density $\rho$ and pressure $p$ one
needs to integrate the equation of motion $d(\rho
a^3)=-pd(a^3)$~\footnote{The equation reflects thermodynamical relation
$dE=-pdV$ with $V\propto a^3$.} to determine the behavior of $\rho$ with respect
to $a$. The simplest equation of state which is adequate in describing the 
large scale dynamics of the universe is $p=\omega\rho$. Then the equation
\beq
d(\rho a^3)=-\omega\rho d(a^3)
\label{pv} 
\eeq
can be immediately integrated to give
$\rho\propto a^{-3(1+\omega)}$. Doing that for each component of the energy
density,  Eq.\ref{e1} can be written in the form
\beq
\frac{\dot
a}{a}=H_0^2\sum_i\Omega_i\left(\frac{a_0}{a}\right)^{3(1+\omega_i)}-\frac{k}{a^2
},
\label{e1i}
\eeq
where each of the species is determined by density parameter $\Omega_i$ and the
equation of state characterizing by constant $\omega_i$. The most familiar
forms of energy densities are those due to pressure less matter (dust) or
radiation. For the dust we have $\omega_i=0$, while for radiation
$\omega_i=1/3$. 

Observation situation regarding the composition of our universe can be
summarize as follows.

Our universe has $\Omega_{\rm tot}\approx 1$ or, more precisely, 
$0.98\le\Omega_{\rm tot}\le 1.08$. The value $\Omega_{\rm tot}$ is determined
from the angular anisotropy of CMBR., with reasonable assumption $h>0.5$. This
observations now show that we live in a universe with density close to the
critical density. 

Measurements of primordial; deuterium produced in the BBN as well as the CMBR
observations show that the total amount of baryons in the universe contributes
about $\Omega_B=(0.024\pm 0.0012)h^{-2}$. Given the independent observations on
the Hubble constant which fix $h=0.72\pm 0.07$, we conclude that
$\Omega_B$ ranges between 0.01 and 0.06. Therefore most of the universe is
non-baryonic. 

Host of observations related to the large scale structure and dynamics (such as
rotation curves of galaxies, estimate of cluster masses, gravitational
lensing, galaxy surveys) all suggest that the universe is populated by a
non-luminous component of matter, the dark matter (DM), made of weakly
interacting massive particles which does cluster at galactic scales. This
component contributes about $\Omega_{\rm DM}\simeq 0.20\ -\ 0.35$.

Combining the last observation with the first one described above we conclude 
that there should exist at least one more component to the energy density of the
universe contributing about 70\% of the critical density. Early analysis of
several observations indicated that this component is unclustered and has
negative pressure. This is confirmed dramatically by more resent
supernova observations and analysis of CMBR data obtained by WMAP. The
observations suggest that the missing component, the dark energy (DE), has

$\omega=p/\rho\le -0.78$ and contributes $\Omega_{DE}\approx 0.60\ -\ 0.75$. 

The simples choice for the dark energy is a fluid with $p=-\rho$ such that
$\omega=-1$. In this case, the equation   Eq.\ref{pv} is identically
satisfied with $\rho =-p=\ {\rm constant}$. That is a fluid, which will have
the same energy density and pressure at all the time as the universe
expands~\footnote{This is to be contrasted with the energy density of normal
matter which decreases as the universe expands, because of the work done by
$pdV$ term with $p>0$. When the pressure is negative, the decrease in the
energy density due to expansion can be compensated by the negative work done
thereby maintaining constant energy density.}. It is indeed possible to mimic
such a fluid by adding a term to Eistein's equations called the cosmological
constant. It is also possible that the energy of the vacuum state of the
universe is non zero and exerts a gravitational influence. One should expect
that the vacuum have an energy density and pressure which are constant in space
and time. The equation Eq.\ref{pv} then demands that it should have an
equation of state $p=-\rho=\ {\rm constant}$. Hence one often calls the dark
energy arising from a fluid with equation of state $p=-\rho$ either as
cosmological constant or vacuum energy. 

The universe also contains radiation (with $p=\rho /3$) contributing an energy
density $\Omega_Rh^2=2.56\times 10^{-5}$ today most of which is due to photons
in the CMBR~\footnote{Assuming that most of the energy density is at temperature
$T=2.73\ {\rm K}$ today, we get $\rho_R=(\pi^2/15)(k_B^4T^4/c^3\hbar^3)$.
Dividing this by $\rho_c$ we get the density above.}. 

Taken together we conclude that our universe has (approximately)
$\Omega_{DE}\simeq 0.7$, $\Omega_{DM}\simeq 0.26$,  $\Omega_{B}\simeq 0.04$,
$\Omega_{R}\simeq 5\times 10^{-5}$. All known observations are consistent with
such an admittedly weird composition for the universe.
Tab. \ref{CosSetPar} summerizes the numerical values of the experimental
measured parameters for the Hubble parameter and density composition of
universe.

\begin{table}
\caption{ Basic set of currently used cosmological parameters. }
\begin{center}
\begin{tabular}{||l|l|l||}                                      \hline
  parameter           & Symbol          & value
                                                               \\ \hline
  Hubble parameter    & h               &   $ 0.72 \pm 0.07 $
                                                               \\ \hline
  Total matter density
                      & $\Omega_{tot} $ & $0.98 \le \Omega_{tot} \le 1.08$
                                                               \\ \hline
  Baryon density      & $\Omega_{B} $   & $0.01 \le \Omega_{B} \le 0.06$
                                                               \\ \hline
  Dark matter density
                      & $\Omega_{DM}$   & $0.20 \le \Omega_{DM} \le 0.35$
                                                               \\ \hline
  Dark energy density & $\Omega_{DE}$   & $0.60 \le \Omega_{DE} \le 0.75$
                                                               \\ \hline
\end{tabular}
\end{center}
\label{CosSetPar}
\end{table}

As far as dynamics concerned, we therefore need to consider three form of energy
densities: $\rho_{\rm NR}$, $\rho_{\rm R}$ and $\rho_{\rm V}$. If neither
particles nor photons are created or destroyed during the expansion, then their
number density will decrease as $n\propto a^{-3}$ as $a$ increases. In the case
of photons, the wavelength will also get stretched as $\lambda\propto a$. Thus
energy density of material particles scales as $\rho_{\rm NR}\propto a^{-3}$,
while that of radiation vary as $\rho_{\rm R}\propto a^{-4}$. The latter means
that as we go to smaller $a(t)$ in the past, radiation energy density grows
faster (as $\Omega_{\rm R}a^{-4}$) compared to matter energy density (which
grows as $\Omega_{\rm NR}a^{-3}$). So even though radiation is dynamically
irrelevant today, it would have been the dominant component in the universe at
sufficiently small $a$: when $a<a_{\rm eq}=a_0(\Omega_{\rm NR}/\Omega_R)$. That
is at redshift larger than
\beq
\label{zeq}
z_{\rm eq}\simeq\Omega_{\rm NR}/\Omega_R\simeq 4\times
10^4\Omega_{\rm NR}h^2.
\eeq
 Further, combining $\rho_R\propto a^{-4}$ with the
result $\rho_{\rm R}\propto T^4$ for thermal radiation, it follows that any
thermal spectrum of photons in the universe will have its temperature varying
as $T\propto a^{-1}$. In the past, when the universe was smaller, it would also
have been denser, hotter and, at sufficiently early epochs, was dominated by
radiation energy density since $(\rho_{\rm R}/\rho_{\rm NR})\propto 1/a$. 

Given all this, the total energy density in the universe at any epoch can be
expressed as
\beq
\label{ed1}
\rho_{\rm total}(a)=\rho_{\rm R}(a)+\rho_{\rm NR}(a)+\rho_{\rm V}(a)
=\rho_c\left[\Omega_{\rm R}\left(\frac{a_0}{a}\right)^4+(\Omega_{\rm
B}+\Omega_{\rm DM})\left(\frac{a_0}{a}\right)^3+\Omega_{\rm V}\right],
\eeq
where $\rho_c$ and various $\Omega$'s refer to their values at $a=a_0$. From
Eq.\ref{e1i} we get
\beq
\label{eiomega}
\frac{\dot a^2}{a^2}+\frac{k}{a^2}=H_0^2\left[\Omega_{\rm
R}\left(\frac{a_0}{a}\right)^4+\Omega_{\rm
NR}\left(\frac{a_0}{a}\right)^3+\Omega_{\rm V}\right].
\eeq
This equation can be cast in a more suggestive form. Namely one can introduce a
dimensionless time coordinate $\tau =H_0t$. Then writing $a=a_0q(\tau)$
and $(k/a^2)=(\Omega_{tot}-1)H_0^2(a_0/a)^2$ one obtains
\beq
\label{motion}
\frac{1}{2}\left(\frac{dq}{d\tau}\right)^2+V(q)=E,
\eeq
where
\beq
V(q)=-\frac{1}{2}\left[\frac{\Omega_{\rm R}}{q^2}+\frac{\Omega_{\rm
NR}}{q}+\Omega_{\rm V}q^2\right];\ E=\frac{1}{2}(1-\Omega_{\rm tot}).
\eeq
Eq.\ref{motion} has the structure of the first integral for motion of a
particle with energy $E$ in a potential $V(q)$. For models with
$\Omega_{\rm tot}=1$ we can take $E=0$ so that $(dq/d\tau )=\sqrt{-V(q)}$. Then
the dynamics of the system can be classified as follows: (i) At high redshift
(small q)~\footnote{For the notations chosen $q=(1+z)^{-1}$.} the universe is
radiation dominated and velocity $\dot q$ is independent of the other
cosmological parameters. (ii) At lower redshift (${\cal O}(1)<z<10^4$) the
universe is matter dominated. (iii) At still lower redshift ($0<z<{\cal
O}(1)$) the velocity $\dot q$ changes from being a decreasing function to an
increasing one. In the other words, the presence of a cosmological constant
leads to an accelerating universe at low redshift. 

Based on the above considerations, we can identify three distinct phases in
the evolution of the universe depending on which form of the energy density
dominates the expansion. At very early epoch, the radiation will dominate over
other forms of energy densities and Eq.\ref{motion} can be easily integrated
to give $a(t)\propto t^{1/2}$. As the universe expands, a time will come
when ($z=z_{\rm eq}$, Eq.\ref{zeq}) the matter energy density becomes
comparable to radiation energy density. Similarly, at very late epochs, the
vacuum energy density (cosmological constant) will dominate over non
relativistic matter and the universe will become "vacuum dominated". This
occurs at a redshift of $(1+z_{\rm V})=(\Omega_{\rm V}/\Omega_{\rm
NR})^{1/3}$. For $\Omega_{\rm V}\approx 0.7$, $\Omega_{\rm NR}\approx 0.3$ this
occurs at $z_{\rm V}\approx 0.33$. During $z_{\rm V}<z<z_{\rm eq}$, the
universe is matter dominated with $a\propto t^{2/3}$. 

Basically it is possible to distinguish eight time intervals
summarized in Fig.\ref{BBMtiming-t}.
From t=0 to t=10$^{-43}$ s the Planck Era, at
t=10$^{-35}$ s the GUT Era, at t=10$^{-10}$ s 
the Electroweak Era, at t=0.001 s  the Particle Are,
at t=3 min the Era of nucleon synthesis, at t=300 000  a 
the Era of Nuclei, at 1 billion years  Era of Atoms and until today
the Era of galaxy formation. 

\subsection{Key events in the life of the universe}
\label{sec:Key events in the life of the universe}

\subsubsection{Overview and present status of inflation}
\label{sec:Overview and present status of inflation}

Inflation was introduced to solve several outstanding problems of the
standard Big Bang model \cite{guthsato} and has now become an important 
part of the standard cosmology. It provides a natural mechanism
for the generation of scalar density fluctuations that seed large scale
structure, thus explaining the origin of the temperature anisotropies in
the cosmic microwave background (CMB), and for the generation of tensor
perturbations (primordial gravitational waves) 
\cite{kt,libros,mass,hu,fluc}.

A distinct aspect of inflationary perturbations is that they are 
generated by quantum fluctuations of the scalar field(s) that drive 
inflation. After their wavelength becomes larger than the Hubble radius, 
these fluctuations are amplified and grow, becoming classical and 
decoupling from causal microphysical processes. Upon re-entering the 
horizon, during the radiation and matter dominated eras, these classical 
perturbations seed the inhomogeneities which generate structure upon 
gravitational collapse \cite{libros,hu,fluc}. 
A great diversity of inflationary models predict fairly generic features: 
a
gaussian, nearly scale invariant spectrum of (mostly) adiabatic scalar 
and tensor primordial fluctuations, which provide an
excellent fit to the highly precise wealth of data provided by the
Wilkinson Microwave Anisotropy Probe (WMAP)\cite{WMAP1,WMAP3,WMAP5}
making the inflationary paradigm fairly robust.
Precision CMB data reveal peaks and valleys in the temperature
fluctuations resulting from  acoustic oscillations in the
electron-photon fluid at recombination. These are depicted in 
Fig.\ref{acus} where up to five peaks can be seen.

Baryon acoustic oscillations driven by primordial fluctuations
produce a peak in the galaxy correlations at 
$ \sim 109 \; h^{-1} $ Mpc (comoving
sound horizon) \cite{eis}. This peak is the real-space version of the 
acoustic oscillations in momentum (or $l$) space and are confirmed 
by large scale structure (LSS) data \cite{eis}.

\begin{figure}[h]
\includegraphics[height=9.cm,width=12.cm]{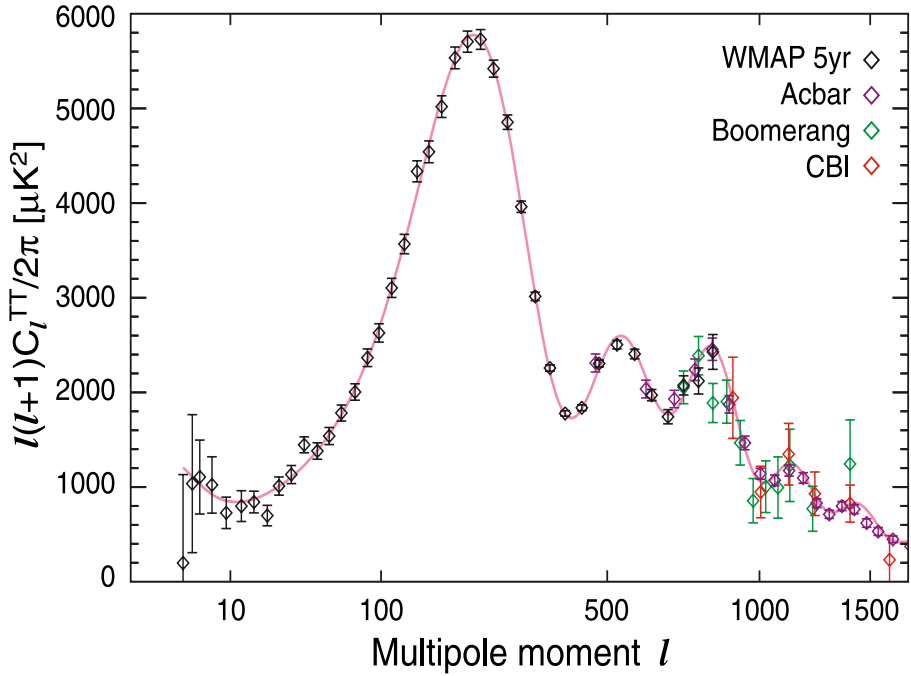}
\caption{Acoustic oscillations from WMAP 5 years data set plus other 
CMB data. Theory and observations nicely agree except for the lowest 
multipoles: the quadrupole CMB suppression.} 
\label{acus}
\end{figure}

Perhaps the most striking validation of inflation
as a mechanism for generating superhorizon fluctuations
is the anti correlation peak in the temperature-polarization (TE) angular
power spectrum at $ l \sim 150 $ corresponding to superhorizon
scales \cite{WMAP1} and depicted in Fig.\ref{TE}. 
The observed TE power spectrum can only 
be generated by fluctuations that exited the horizon during inflation
and re-entered the horizon later, when the expansion of the universe
decelerates.

\begin{figure}[htbp]
\includegraphics[height=9.cm,width=12.cm]{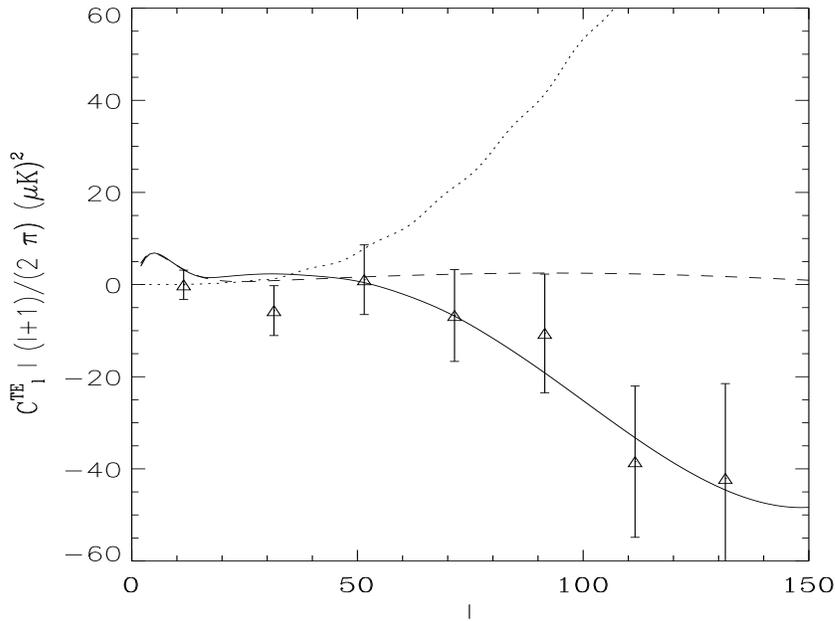}
\caption{Temperature-Polarization angular power spectrum.
The large-angle TE power spectrum predicted in primordial adiabatic
models (solid), primordial isocurvature models (dashed) and by defects
such as cosmic strings (dotted). The WMAP TE data (Kogut et al.
\protect\cite{WMAP1})
are shown for comparison, in bins of $ \Delta l = 10 $. Superhorizon
adiabatic modes from inflation fit the data while subhorizon sources of
fileTE power go in directions opposite to the data.}
\label{TE}
\end{figure}

\begin{figure}[htbp]
\includegraphics[height=9.cm,width=12.cm]{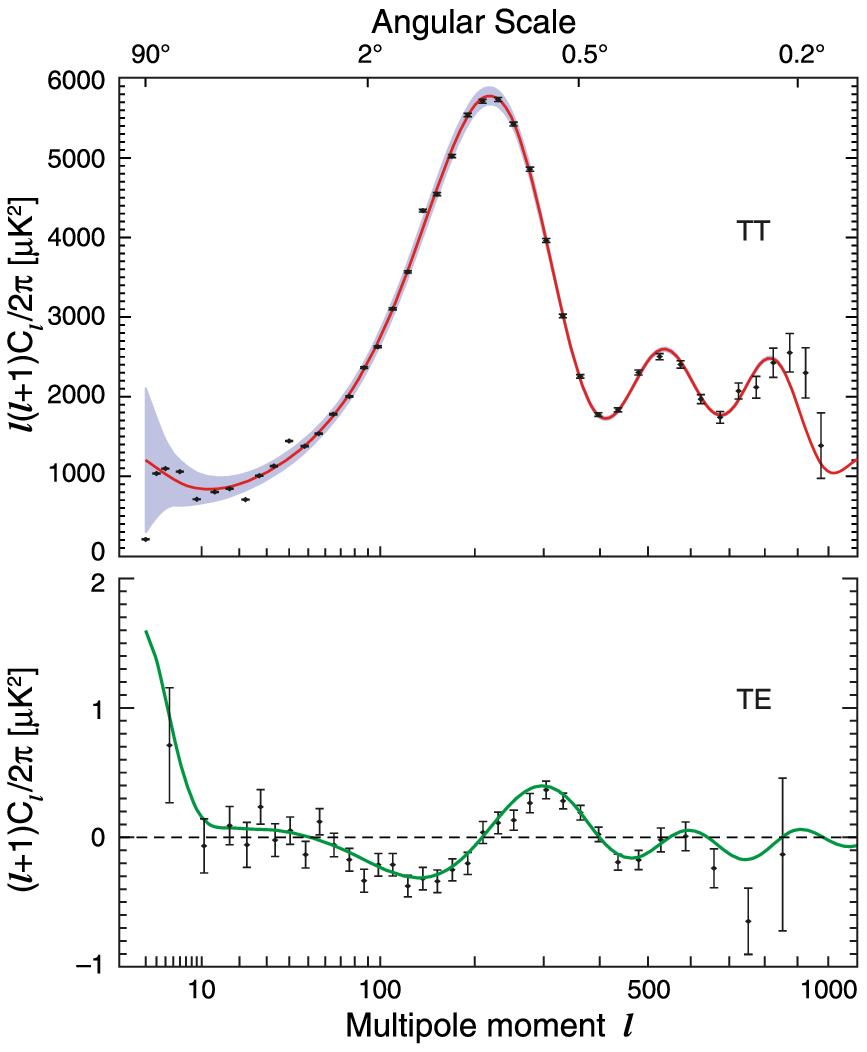}
\caption{The temperature (TT) and temperature-polarization correlation
(TE) power spectra based on the 5 year WMAP data \protect\cite{WMAP5}.}
\label{TEwmap5}
\end{figure}

The confirmation of many of the robust predictions of
inflation by current high precision observations places inflationary
cosmology on solid grounds.

Amongst the wide variety of inflationary scenarios, single field slow-roll
models provide an appealing, simple and fairly generic description of
inflation. Its simplest implementation is based on a scalar field (the
inflaton) whose homogeneous expectation value drives the dynamics of the
scale factor, plus small quantum fluctuations. The inflaton potential is
fairly flat during inflation and it dominates the universe energy during 
inflation. This flatness not only leads to a slowly varying Hubble 
parameter, hence ensuring a sufficient number of efolds of inflation,
but also provides an explanation for the gaussianity of the fluctuations 
as
well as for the (almost) scale invariance of their power spectrum. A flat
potential precludes large non-linearities in the dynamics of the 
fluctuations of
the scalar field.

The current WMAP data are validating the single field slow-roll scenario 
\cite{WMAP1,WMAP3,WMAP5}. Furthermore, because the potential is flat the
scalar field is almost massless, and modes cross the horizon with an
amplitude proportional to the Hubble parameter. This fact combined with a
slowly varying Hubble parameter yields an almost scale invariant 
primordial
power spectrum.  The slow-roll approximation has been recently cast as a 
systematic $ 1/N $ expansion \cite{1sN}, where $ N \sim 60 $ is the number 
of efolds before the end of inflation when modes of cosmological relevance 
today first crossed the Hubble radius.

The observational progress begins to discriminate among 
different inflationary models, placing stringent constraints on them. 
The upper bound on the ratio $ r $ of tensor to scalar fluctuations 
obtained by WMAP convincingly excludes the massless monomial 
$ \varphi^4 $ potential \cite{WMAP1,WMAP3,WMAP5} and hence 
 strongly suggests the presence of a mass
term in the single field inflaton potential \cite{ciri,infwmap}.
Hence, as a minimal single field model, one should consider a sufficiently 
general polynomial, the simplest polynomial potential bounded
from below being the fourth order trinomial potential \cite{mcmc}.

The observed low value of the CMB quadrupole with respect to
the $\Lambda$CDM theoretical value
has been an intriguing feature on large angular scales since 
first observed by COBE/DMR \cite{cobe}, and confirmed by the WMAP data 
\cite{WMAP3,WMAP5}. In the best fit $\Lambda$CDM model
using the WMAP5 data we find that the probability that the quadrupole is 
low or lower than the observed value is just 0.031. Even if one does not as
care about the specific multipole and looks for any multipole as low or 
lower than the observed quadrupole with respect to the $\Lambda$CDM model 
value, then the probability remains smaller than 5\%. Therefore, it is 
relevant to find a cosmological explanation of the quadrupole suppression 
beyond the $\Lambda$CDM model.

\subsubsection{Primordial nucleosynthesis}
\label{sec:Primordial nucleosynthesis}

When the temperature of the universe was higher than the temperature
corresponding to the atomic ionization energy, the matter content in the
universe was a high temperature plasma. For $t<t_{\rm eq}$, in the radiation
dominated phase, $a(t)\propto t^{1/2}$; therefore 
\beq
\label{aT}
\left(\frac{\dot a}{a}\right)^2=H^2(t)=\frac{8\pi
G}{3}g\left(\frac{\pi^2}{30}\right)T^4,
\eeq
where $g$ counts the relativistic degrees of freedom. As the temperature
decreases, more and more particles become non-relativistic; thus $g(T)$ is a
slowly decreasing function of $T$. Numerically,
\beq
\label{tT}
t\simeq 0.3g^{-1/2}\left(\frac{m_{\rm Pl}}{T^2}\right)\simeq 1{\rm
s}\left(\frac{T}{1{\rm MeV}}\right)^{-2}g^{-1/2}.
\eeq

When the temperature of the universe is higher than the binding energy of the
nuclei ($\simeq {\rm MeV}$), none of the heavy elements (helium and metals)
could have existed in the universe. The binding energies of the first four
light elements, $^2H$, $^3H$, $^3He$ and $^4He$ are 2.22~MeV, 6.92~MeV,
7.72~MeV and 28.3~MeV respectively. This would suggest that these nuclei could
be formed when the temperature of the universe was in the range between 1 and
30~MeV. The actual synthesis takes place only at a much lower temperature,
$T_{nuc}=T_n\simeq 0.1{\rm MeV}$. The mean reason for this delay is the 'high
entropy' of our universe, i.e., the high value of the photon-to-baryon ratio,
$\eta^{-1}$, given as:
\beq
\label{eta}
\eta =\frac{n_B}{n_{\gamma}}=5.5\times 10^{-10}\left(\frac{\Omega_{\rm
B}h^2}{0.02}\right).
\eeq 

The nucleosynthesis requires protons and neutrons combining together to form
bound nuclei of heavier elements like deuterium, helium etc.. The abundance of
these elements are going to be determined by the relative abundance of
neutrons and protons in the early universe. Therefore one needs first to worry
about the maintenance of thermal equilibrium between protons and the neutrons.
As long as the inter-conversion between $n$ and $p$ through the weak
interaction processes: $\nu +n\leftrightarrow p+e$, $\bar e +n\leftrightarrow
p+\bar\nu$ and the decay $n\leftrightarrow p+e+\bar\nu$, is rapid compared to
the expansion rate, thermal equilibrium will be maintained. Then the
equilibrium $n/p$ ratio will be
\beq
\label{ntop}
\left(\frac{n_n}{n_p}\right)=\left(\frac{X_n}{X_p}\right)=\exp (-Q/T),
\eeq
where $Q=m_n-m_p=1293\ {\rm MeV}$. Therefore, at high ($T\gg Q$) temperature,
there will be equal number of neutrons and protons but as the temperature drops
below about 1.3~MeV, the neutron fraction will start dropping exponentially
provided thermal equilibrium is still maintained. The expansion rate
Eq.\ref{aT} at $T=Q$ is given by $H\approx1.1\ {\rm s}$. The neutron to
proton conversion rate, for example, is well approximated by
\beq
\label{nprate}
\lambda_{np}\approx 0.29\ {\rm
s}^{-1}\left(\frac{T}{Q}\right)^5\left[\left(\frac{Q}{T}\right)^2+6\left(\frac{Q
}{T}\right)+12\right].
\eeq 
Therefore, at $T=Q$, this gives $\lambda\approx 5\ {\rm s}^{-1}$, slightly more
rapid than the expansion rate. As $T$ drops below $Q$, this decreases rapidly
and the reaction ceases to be fast enough to maintain thermal equilibrium. Using
the rate equation, which is the basis of the general procedure for studying non
equilibrium abundances in an expanding universe, one arrive to the conclusion
that the neutron fraction falls out of equilibrium when temperatures drop below
1~MeV and it freezes to about $X_n=n_n/(n_n+n_p)0.15$ at temperatures below
0.5~MeV. 

As the temperature decreases further, the neutron decay with a half life of
$\tau_n\approx 886.7\ {\rm s}$ becomes important and starts depleting the
neutron number density. The only was the neutrons can survive is through the
synthesis of light elements. As the temperature falls further to
$T=T_{\rm He}\simeq 0.28\ {\rm MeV}$, significant amount of He could have been
produced if the reaction rates were high enough. All possible reactions which
produce $^4{\rm He}$ are based on D, $^3{\rm He}$ and  $^3{\rm H}$ and do not
occur rapidly enough because the mass fractions of D, $^3{\rm He}$ and  $^3{\rm
H}$ are still quite small ($10^{-12}$, $10^{-19}$ and $5\times 10^{-19}$
respectively) at $T\simeq 0.3\ {\rm MeV}$. The reaction $n+p\leftrightarrow
d+\gamma$ will lead to an equilibrium abundance ration of deuterium, which 
passes through unity (for $\Omega_{\rm B}h^2=0.02$) at the temperature of about
0.07~MeV which is when the nucleosynthesis can really begin.  

Therefore, one needs to determine the neutron fraction at $T=0.07\ {\rm MeV}$
given that it was about 0.15 at 0.5~MeV. During this epoch, the
time-temperature relationship is given by $t=130\ {\rm s}(T/0.1\
{\rm MeV})^{-2}$. The neutron decay factor is $\exp (-t/\tau_n)\approx 0.74$ for
$T=0.07\ {\rm MeV}$. This decreases the neutron fraction to $0.15\times
0.74=0.11$ at the time of nucleosynthesis. When the temperature becomes $T\le
0.07\ {\rm MeV}$, the abundance of $D$ and $^3{\rm He}$ builds up and these
elements further react to form $^4{\rm He}$. A good fraction of D and $^3{\rm
He}$ is converted into $^4{\rm He}$. The resultant abundance of  $^4{\rm He}$
can be easily calculated by assuming that almost all neutrons end up in $^4{\rm
He}$. Since each $^4{\rm He}$ nucleus has two neutrons, $(n_n/2)$ per unit
volume of helium nuclei can be formed if the number density of neutrons is
$n_n$. Thus the mass fraction of  $^4{\rm He}$ will be 
\beq
\label{Y}
Y=\frac{4(n_n/2)}{n_n+n_p}=\frac{2(n/p)}{1+(n/p)}=2x_c,
\eeq
where $x_c=n/(n+p)$ is the neutron abundance at the time of production of
deuterium. For $\Omega_Bh^2=0.02$, $x_c\approx 0.11$ giving $Y\approx 0.22$.
Increasing baryon density to $\Omega_Bh^2=1$ will make $Y\approx 0.25$. An
accurate formula for the dependence of helium abundance on various parameters is
given by 
\beq
\label{Yexact}
Y=0.226+0.025\log\eta_{10}+0.0075(g_*-10.75)+0.014(\tau_{1/2}(n)-10.3\ {\rm
min}),
\eeq
where $\eta_{10}$ measures the baryon-photon ratio today via the relation
\beq
\label{eta10}
\Omega_{\rm B}h^2=3.65\times 10^{-3}\left(\frac{T_0}{2.73\ {\rm
K}}\right)^3\eta_{10}
\eeq
and $g_*$ is the effective number of relativistic degrees of freedom
contributing to the energy density and $\tau_{1/2}(n)$ is the neutron half
life. 

As the reactions converting D and $^3{\rm H}$ into $^4{\rm He}$ proceed, the
number density of D and $^3{\rm H}$ is depleted and the reaction rates- which
are proportional to $\Gamma\propto X_A(\eta n_{\gamma})<\sigma v>$- become
small. These reactions soon freeze-out leaving a residual fraction of D and
$^3{\rm H}$ (a fraction of about $10^{-5}$ to $10^{-4}$). Since
$\Gamma\propto\eta$ it is clear that the fraction of (D, $^3{\rm H}$) left left
unreacted will decrease with $\eta$. In contrast, the $^4{\rm He}$ synthesis,
which is not limited by any reaction rate, is fairly independent of $\eta$ and
depends only on the $(n/p)$ ration at $T\simeq 0.1\ {\rm MeV}$. The best fits,
with typical errors, to deuterium abundance calculated from the theory, for the
range $\eta =(10^{-10}\ -\ 10^{-9})$ reads
\beq
\label{Y2}
Y_2\equiv\left(\frac{\rm D}{\rm H}\right)_p=3.6\times 10^{-5\pm
0.06}\left(\frac{\eta}{5\times 10^{-10}}\right)^{-1.6}.
\eeq

The production of still heavier elements, even those like $^{16}{\rm C}$,
$^{16}{\rm O}$ which have higher binding energies than $^4{\rm He}$, is
suppressed in the early universe. Two factors are responsible for this
suppression: (1) For nuclear reactions to proceed, the participating nuclei
must overcome their Coulomb repulsion. The probability to tunnel through the
Coulomb barrier for heavier nuclei (with large $Z$) is suppressed. (2) Reaction
between helium and proton would have led to an element with atomic number 5
while the reaction of two helium nuclei would have led to an element with
atomic mass 8. However, there are no stable elements in the periodic table with

The current observations indicate, with reasonable certainty that: (i) $({\rm
D/H})\ge 1\times 10^{-5}$. (ii) $({\rm D+^3He})/{\rm H}\simeq (1\ -\ 8)\times
10^{-5}$ and (iii) $0.236\le ({\rm ^4He/H})\le 0.254$. These observations are
consistent with the predictions if $10.3\ {\rm min}\le\tau\le 10.7\ {\rm min}$,
and $\eta =(3\ -\ 10)\times 10^{-10}$. Using $\eta =2.68\times
10^{-8}\Omega_{\rm B}h^2$, this leads to the important conclusion:
$0.011\le\Omega_{\rm B}h^2\le 0.037$. When combined with the broad bound on $h$,
$0.6\le h\le 0.8$, say, we can constrain the baryon density of the universe to
be  $0.01\le\Omega_{\rm B}\le 0.06$. It shows that, if $\Omega_{\rm total}\simeq
1$ then most of the matter in the universe must be non baryonic.

Since ${\rm ^4He}$ production depends on $g$, the observed abundance of ${\rm
^4He}$ restricts the total energy density present at the epoch of
nucleosynthesis. In particular, it constraints the number $N_{\nu}$ of light
neutrino~\footnote{That is, neutrinos with $m_{\nu}\le 1\ {\rm MeV}$ which would
have been relativistic at $T\simeq 1\ {\rm MeV}$).}. The observed abundance is
best explained by $N_{\nu}=3$, is barely consistent with  $N_{\nu}=4$ and rules
out $N_{\nu}>4$. As we have mentioned in the previous chapter, the laboratory
bound on the total number of particles including neutrinos, which couples to
$Z^0$ is  $N_{\nu}=2.9841\pm 0.0083$, which is consistent with the cosmological
observations.

The major events since Big Bang are summarized in
Fig.{\ref{BBMtiming-t}}

\begin{figure}[htbp]
\begin{center}
\includegraphics[width=1.0\linewidth]{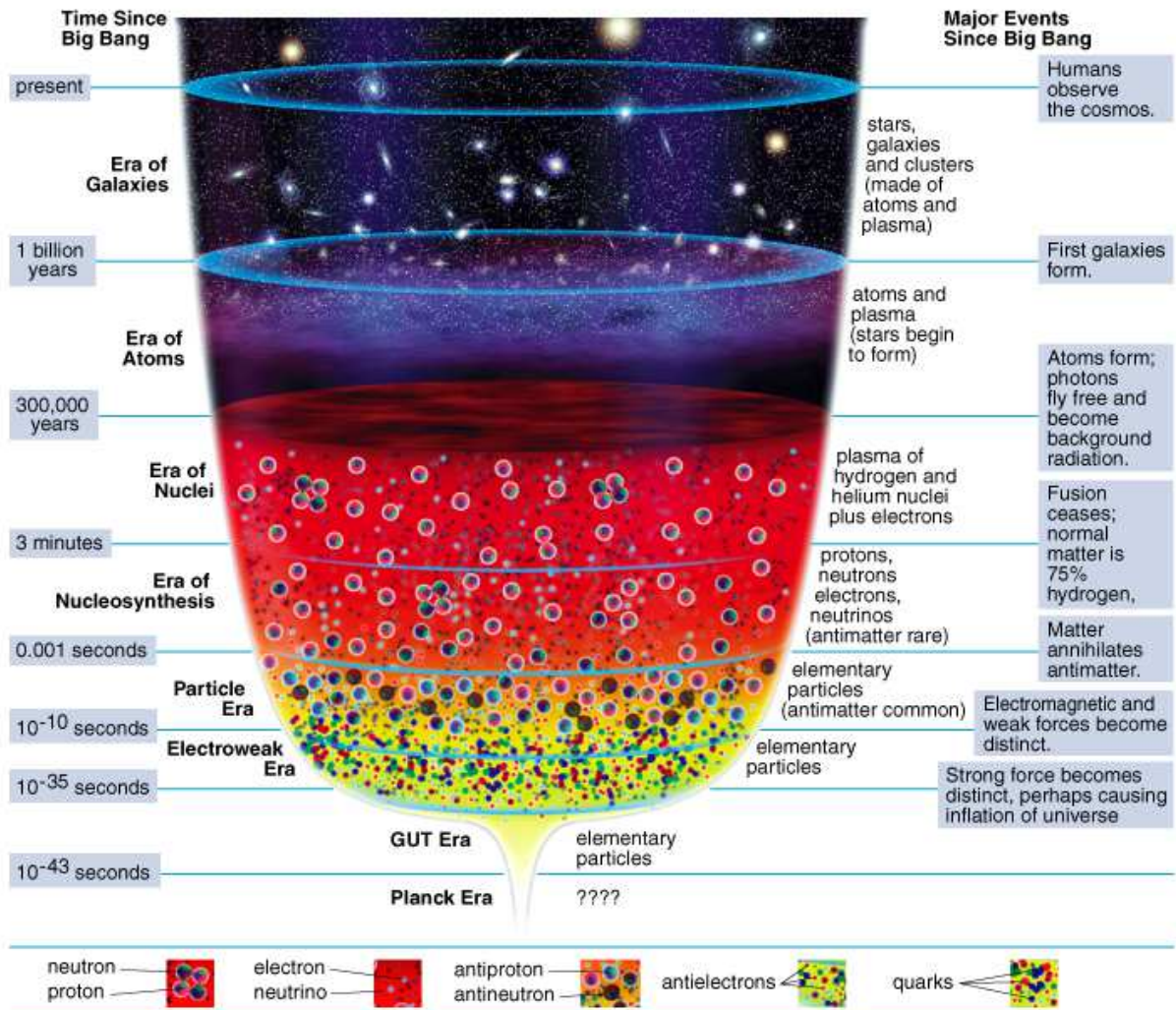}
\end{center}
\caption{Big Bang Model timing.}
\label{BBMtiming-t}
\end{figure}

\section{ Summary of experimental status of Standard
          and Big Bang Model }

The Standard Model is defined by 19 parameters and in the Big Bang models
for e.g. in the $\Lambda$CDM case by 11 basically
cosmological parameters
\cite{WMAP3,COSparameterI,COSparameterII}.
Both theories are
extensively experimental tested and proven to describe our nature.
The quantized SM model is applicable in particular at small
distances describing the fundamental particles and the interaction
between them whereas the Big Bang Model
describes distances from the Planck Scale up to size of the universe
today. We like to stress that both theories can describe only one
nature. The logically consequence about this fact is that between
both theories similarities or links must exist.

In Tab. \ref{linksSMBigBang} we discuss these links in more detail.
The Dirac equation 
and the Einstein equation 
are examples of the principle equations in both theories.
The are embed in the same four dimensional space $ x,y,z,t $
and use the same definition of energy E. To calculate radiative effects
for example in the $\EEGG$ reaction Eq.\ref{LITKEcross} it is
necessary to introduce a vacuum which allows virtual corrections.
The experimental evidence of this corrections is for example
visible in the differential cross section of the $\EEGG$ reaction in
Fig.{\ref{plot_QEDcrossBornALPHA3}} where the measured data prefer
the $ \alpha^{3} $ level. In a microscopic picture the vacuum
produces for example statistical virtual particles and anti-particles
like a flashing volume. Similar statistical properties are needed
in the Big Bang theory to explain the small temperature fluctuations
on the micro wave background of the actual WMAP experiment \cite{WMAP}.
It seems for this reason very likely both vacuum used in the Standard ( SM ) and
Big Bang Model ( BBM ) are the same. The generation of mass is performed in the SM
via the Higgs mechanism. In the BBM produces every stress energy tensor
$ T^{i}_{k} \neq 0 $ an curved space $ R \neq 0 $ what means mass.
As for all four interactions an tensor $ T^{i}_{k} $ exist mass will
be produced accordingly these interactions. Similar as the mass
generation also for the momentum both theories a very different.
The SM momentum follows the quantized regime and the BBM the classical
ansatz. The most remarkable difference is that the SM
uses the wave function $ |\psi|^{2} $ to localize the
probability of a coordinate of a particle whereas such an approach in
the Big Bang model is absent. A mass accumulation in this model
must be described by particle density and the coordinates
$ x,y,z,t $. It seems very unlikely that the size of a particle
in the SM with finite rest mass is zero, because the mass density
of the FP would be $ \rho = \infty $.
A particle in the BBM is a cluster of mass possible including charges.
The validity of the BBM is limited by the quantization of the gravitation
by the Planck scale $ l_{ Planck } $ and density  $ \rho_{ Planck } $.
A discussion about a classical particle in the BBM is for this reason
only sensible if the size $ r $ of such a particle is
$ r > l_{ Planck } $ and the density $ \rho < \rho_{ Planck } $.
The charges ( color, EM, weak ) of particles and the adjacent
interactions are domiciled in both theories, whereas the BBM
includes the most important interaction for this theory the gravity.
The coupling $ \alpha_{1} $, $ \alpha_{2} $ and $ \alpha_{3} $ in the SM
model are depending about the energy scale $\mu $ as displayed in
Fig.{\ref{fig.4}}. An experiment at higher test energy $ Q $ has to
respect the change in the coupling, above the Grand Unification scale
only one interaction is left until the Planck scale. Above this
scale it is assumed that gravitation is dominating. The test energy $ Q $
is not only an energy the uncertainty principle defines
an according size of test distance in an experiment. In other words
a test of the point like behavior of the FPs to a test distance zero
would imply infinite high test energies. For example at energies
$ Q > 10^{15} $ GeV the experiment has to deal with only one interaction.
A similar behavior of scale dependence we find in the BBM in
Fig.{\ref{BBMtiming-t}} in particular at energies from the $ M_{w} $
up to the Planck scale and above. 

The energy scale in this case
corresponds simultaneously to size of the universe and the time
after Big Bang. This similarity between SM and BBM has the
interesting consequence that an increase of our accelerator
energies for example at LHC test the conditions in a very early
universe.

\begin{table}
\caption{ Possible links between Standard and Big Bang model.}
\begin{center}
\begin{tabular}{||l|l|l|l||}                               \hline
           & Standard model
           & LINK
           & Big Bang models
                                                              \\ \hline
 principle equations
           & $ ( i\gamma^{\mu}\partial_{\mu}-m)\psi=0 $
           & --
           & $\mathcal{R}^{i}_{k}-\frac{1}{2}R\delta^{i}_{k}=
             -\frac{8\pi G}{c^{4}}T^{i}_{k}-\Lambda \delta^{i}_{k}$
                                                              \\ \hline
 parameter                       & 19      & --  & 11
                                                              \\ \hline
 4 dimensional space             & x,y,z,t & yes & x,y,z,t
                                                              \\ \hline
 energy ( eV )                   & E       & yes & E
                                                              \\ \hline
 vacuum                          & V       & yes & V
                                                              \\ \hline
 mass                            & Higgs   & --  & $ T^{i}_{k} \neq 0 $
                                                   for $ \Lambda =0 $
                                                              \\ \hline
 momentum                        & $\bar{p}=-i\hbar \bar{\nabla} $
                                 & --
                                 & $ m \bar{v} $
                                                              \\ \hline
 particle location               & $|\psi|^{2}$
                                 & --
                                 & x,y,z,t
                                                              \\ \hline
 particle size $ r $             & $ r = 0 $
                                 & --
                                 & $ r > l_{ Planck } $
                                                              \\ \hline
 particle density $ \rho $       & $ \rho = \infty $
                                 & --
                                 & $ \rho < \rho_{ Planck } $
                                                              \\ \hline
 particle charge                 & color, EM  , weak
                                 & yes
                                 & color, EM  , weak
                                                              \\ \hline
 particle interaction            & strong, EM , weak
                                 & yes
                                 & strong, EM , weak, gravitation
                                                              \\ \hline
 interaction scale $ Q $         & $ \alpha = f ( Q )= f ( size ) $
                                 & yes
                                 & $ Q  = f ( size )
                                        = f ( t_{After Big Bang} ) $
                                                              \\ \hline
\end{tabular}
\end{center}
\label{linksSMBigBang}
\end{table}

\section{ Possible micro structure of Fundamental Particles}
\label{sec:Possible micro structure of Fundamental Particles}

The investigation of physics beyond the SM and BBM model was
performed in the past with tremendous effort.

1783 John Michell introduced the Black Hole a body so massive that even 
light could not escape \cite {BlackHoles1}.
The black hole has a one-way surface, the event horizon, into which
objects can fall, but nothing is able to escape. The Black Holes are 
still the subject of intense investigations like
non static black holes, Einstein-Yang-Mills sphalerons and black 
holes,vacuum black holes, soliton and black holes,
five-dimensional Black Hole, Skyrme black hole, tiny black holes, hairy
black holes,cosmic colored black holes and
non-abelian black holes \cite{BlackHoles2}.

1931 Paul Dirac introduced the magnetic monopole is a hypothetical 
particle that is a magnet with only one pole.
He showed that if magnetic monopoles exist, then it would explain the 
quantization of electric charge in the universe \cite{Monopols1}.
Recently calculations of non-Abilian monopols, monopols in BHI-Higgs 
theory, gravitate  lumps, self-gravitating monopols and
monopole-antimonoplo pair
get published \cite{Monopols2}.

Since the introduction of the Yang-Mills theory 1954  
\cite{YangMills1} the theoretical effort
increased to search for fundamental particle like structures.

1958 T. Skyrme introduced Skyrmions a mathematical model used to form 
baryons a subatomic particle \cite{Skyrmions1}.
Their topological charge has been identified with the baryon number. The 
are not individual
baryons but coherent states of known baryons and higher resonances on a 
compact manifold associated with the spin and flavor symmetry group.
In recent years extensive theoretical studies about the different 
parameters of Skyrmions are performed.
Skyrmions and their stability \cite{Skyrmions2},
vibration energies and excitation \cite{Skyrmions3},
calculating the density , mass and size \cite{Skyrmions4},
barion nuclei production \cite{Skyrmions5},
spin and isospin \cite{Skyrmions6},
Skyriome-Black holes \cite{Skyrmions7},
Gravitating Skyriom \cite{Skyrmions8},
Skyriom stars \cite{Skyrmions9} and
monopole-Skyrioms \cite{Skyrmions10}.
Various theoretical models are studied like Landau-Lifshits ansatz,
Yang-Mills ansatz, Einstein-Skyrme model or condensates,
lattices and fluids. Also a scattering of Skyrmions was
investigated \cite{Skyrmions11}.

1961 Yoichiro Nambu and Giovanni Jona-Lasinio introduce the
Nambu--Jona--Lasinio model \cite{NJLModel1}.
The model is a theory of nucleons and mesons constructed 
from interacting Dirac fermions with chiral symmetry, it
is a dynamical model of elementary particles based on an analogy with 
superconductivity.
The dynamical creation of a condensate from fermion interactions 
inspired many theories of electroweak
symmetry breaking, such as technicolor \cite{NJLModel2}, the top quark
condensate \cite{NJLModel3}, quark condensate \cite{NJLModel4}
and nonlocal quark model \cite{NJLModel5}.

1968 Rosen introduced a configuration of a charged scalar field that 
are classically stable against small perturbations \cite{Q-ball1}.
These stable configurations of multiple scalar fields were studied by 
Friedberg, Lee and Sirlin in 1976 \cite{Q-ball2}.
The name "Q-ball" and the proof of quantum-mechanical stability come 
from Sidney Coleman \cite {Q-ball3}. Recently
investigations of spinning Q-balls are published \cite{Q-ball4}.

1984 F. Klinkhamer and N. Manton describe first time a Sphaleron 
\cite{Sphaleron1} a static solution to the electroweak field
equations of the Standard Model of particle physics.
A sphaleron is a saddle point of the electroweak potential energy
of the surface $ z = x^{2} - y^{2} $ 
in three dimensional analytic geometry
and it is involved in processes that violate baryon and lepton number. 
In some theories, at the higher temperatures of the early universe,
sphalerons convert an imbalance of the number of leptons and antileptons 
formed by the first leptogenesis in
an imbalance in the numbers of baryons and antibaryons
\cite{Sphaleron2}.
The main theoretical investigations are electroweak Sphalerons with spin 
and charge \cite{Sphaleron3},
Sphalerons and Strings \cite{Sphaleron4},
Sphalerons and Black Holes \cite{Sphaleron5},
Sphalerons and vortex rings \cite{Sphaleron6},
Sphalerons and Higgs \cite{Sphaleron7} and
and QCD Sphalerons \cite{Sphaleron8}.

1997 articles of the Dilaton a hypothetical particle in string 
theory appears.
The Dilaton is a particle of a scalar field, connected to gravity.
Investigation are performed about the mass of the Dilaton 
\cite{Dilaton1}, dilatonic Inflation \cite{Dilaton2},
Dilaton and Quantum Cosmology \cite{Dilaton3}, 
Dilaton and Dark Energy \cite{Dilaton4} and
Dilaton in Einstein-Yang-Mills theory \cite{Dilaton5}.

1997-2008 recent researches of Solitons are published.
The soliton phenomenon was first described by 
John Scott Russell (1808 - 1882) \cite{solitons1}.
A soliton is a self-reinforcing solitary wave
that maintains its shape while it travels at constant speed.
Solitons arise as the solutions of a class of weakly nonlinear 
dispersive partial differential equations describing physical systems.
The play an important role in particle physics.
In the last years investigations get performed
of non-Abelian Solitons, gravitating and spinning
Solitons , Solitons, quark-solitons,
sine-Gordon Solitons, anomalous Abelian Solitons and quark-solitons
\cite{solitons2}.

2000 semiclassical instability of De Sitter space and Black hole pair 
creation in De Sitter space get published \cite{DeSitter1}.
De Sitter space is the maximally symmetric, vacuum solution of
Einstein's field equation with a repulsive cosmological 
constant $ \Lambda $. De Sitter space was discovered 1917
by Willem de Sitter \cite{DeSitter2}.

2004 investigations of excitation of physical vacuum are performed 
\cite{Vacuum1}. The vacuum state \cite{Vacuum2} is the quantum state with
the lowest possible energy. The vacuum state contains fleeting
electromagnetic waves and particles that pop into and out of existence.
In perturbation theory the properties of the vacuum are 
analogous to the properties of the ground state of a
harmonic oscillator. In the Standard Model the non-zero vacuum expectation
value of the Higgs field
is the mechanism by which the other fields in the theory acquire mass.
The uncertainty principle in the form 
$\Delta E \Delta t \ge \hbar$ implies that in the vacuum one 
or more particles with energy
$ \Delta E $ above the vacuum may be created for a short time $ \Delta t $.

In 2007 theoretical solutions for Einstein-Yang-Mills strings and 
superconducting electroweak strings are
presented \cite{SuperString}.

Following the commissioning of LEP 1989 initiated all four
LEP detectors a program to search for physics beyond the SM,
in particular via the reaction $ \EEGG $ about
not pointlike behavior of FPs. Induced from the investigation
in chap.~\ref{sec:Status of experimental limits on the sizes of
Fundamental Particles}, chap.~\ref{sec:Electroweak Interaction} and
chap.~\ref{sec:Conclusion for the experimental limits} our
interest get focused to study the possible consequences
of a not pointlike behavior of FPs.

In the quantized SM it is absolute sufficient to describe
our experimental knowledge in physics with 19 parameters.
It is not necessary to use any micro structure of FP.
But it is necessary to accept, that a FP like an electron
is geometrical a point
with a finite mass, charge, spin, magnetic moment and an
electric dipole moment. In a microscopic picture the density
of FP will be infinite, the charge must be a point,
imagining a point with diameter zero what is able to generate
a magnetic moment and an electric dipole moment
is conflicting with a three dimensional logic space structure.
If we trop the condition that the FP must
be a point and allow an extended microscopic structure it is
possible to avoid these objections in understanding.
Such an approach points to the un-quantized BBM.In this model
an object with rest mass must be described by a density mass
accumulation originated from an stress energy tensor
$ T^{i}_{k} \neq 0 $
in a finite space. The particle would be already an
extended object. We like to stress that both very successfully
theories describe the same physical object. For this reason
it seems usefully to search for an ansatz of
a geometrical extended object in the the transition between
SM and BBM.

In the following sections we introduce an
{ \it Empirical Toy Ansatz about a Microstructure of 
Fundamental Particles } { \bf  ( ETAMFP ) }. We follow
first the current knowledge of the running coupling
constants of the Standard Model together with the
time development of the very early Big Bang theory to
develop a microstructure of FPs.
Both theories are coupled by a common 
energy scale $ \Lambda $ which defines according the
Uncertainty Principle a size. We assume the time-scale-size 
development of both theories is stored in the size
of a fundamental particle. This implements that the
part of the fundamental particles which is close to
a diameter zero is liked historical to the highest
possible scales of the Standard Model and Big Bang theory,
the Planck Scale. This logic forms the extent of the
Fundamental Particle from a size zero to the size today.
The size of the Fundamental Particles is in this sense 
the direct consequence of the time development of the
Standard Model and Big Bang theory from the time
$ t = 0 $ up to the time of nucleosynthesis
$ t = t_{nucleosynthesis} $.

To test the validity
of this ansatz we step by step confront this model with
the experimental measured parameters describing the FPs and ask
is the ansatz able to reproduce these parameters.
We first test is the ansatz able to organize all the known fermions
and bosons in a common scheme. Next we study the consequences of
magnetic and electric dipole moment of FPs about the size.
Next we are leaving this empirical ETAMFP ansatz and compare 
it with the microstructure of a neutral particles and
the model for a charged particle 
\cite{SizePaper2003,ChargedFP}, which are developed
from general theory of relativity. Finally we
conclude.

\subsection {Empirical toy ansatz about a microstructure of a fundamental particle}
\label{sec:Empirical toy ansatz about a microstructure of a fundamental particle}

As already discussed to
test in an scattering experiment small distances for example
in the discussed $ \EEGG $ reaction two basic conditions have to be
full filled. First it is necessary to perform the test at the correct
test energy $ Q = E_{CM} $ Eq.\ref{alphadistance}, \ref{alphasol})
because the test size $ \lambda $ is direct inverse
proportional to the test energy $ \lambda \sim f ( 1/E_{CM} ) $.
Second it is necessary to collect a high amount of events
about the reaction under investigation. Including radiative
corrections of the SM it is possible to push the test size
substantial down, as for example demonstrated in the measurement
of the electric dipolmoment of the electron \cite{electricDIPOL}.
As the FPs are point particles in the SM model an experimental prove
of this fact would require an infinite high test $ E_{CM} $ energy or
infinite high statistics. Both conditions would pass
the Grand Unification and Planck scale because it would be
necessary to perform the experiment at such $ E_{CM} $ energies or
calculate radiative corrections up to these scales.
The predictions of the SM and BBM model summarized in Fig.{\ref{fig.4}
and Fig.\ref{BBMtiming-t} demonstrate that the physics conditions
have changed at these scales dramatically. 

It is not feasible to test in an experiment this extreme energy
conditions or perform it with infinite high number of events.
The experimental available energies in planning and performed
of the known accelerators are displayed in
Fig.\ref{BBMtimingexcelerator} reaching about 15 TeV and simulate
a time after Big Bang of about $ 10^{-20} $ s. The leading parameters
which define the physical conditions
in the SM model are the test energy $ Q $ and the scale $ \mu $
well demonstrated in Fig.{\ref{fig.4}. Essential in the BBM model
are the time after Big Bang $ t $ Eq.\ref{tT},
the temperature $ T $ Eq.\ref{aT} and
the scale $ \mu $ shown in Fig.\ref{BBMtiming-t} and 
Fig.\ref{BBMtimingexcelerator}.

To reconstruct for a certain parameter set in the SM and
BBM model the same physical conditions it is essential
that both model fulfill time reversed invariance.
If an experiment is performed at a defined scale $ \mu $
according Fig.\ref{BBMtimingexcelerator}, we measure as a
function of this scale always the same physical facts,
for example at LEP energies $ W^{\pm} $ and $ Z^{0} $.
This result is not a function of the date the experiment
was performed, we measure in spring or autumn always at the
LEP scale $ W^{\pm} $ and $ Z^{0} $. The logical consequence
of this experimental fact is, the universe
remembers how it was created. 

%
\begin{figure}[htbp]
\begin{center}
 \includegraphics[width=14.0cm,height=9.0cm]{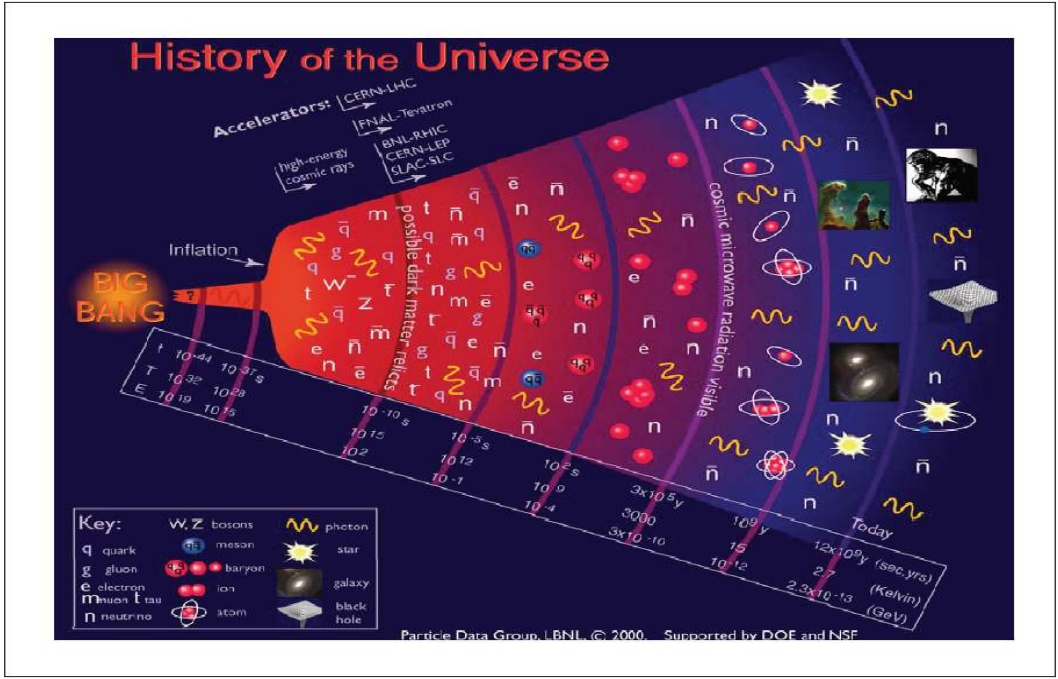}
\end{center}
\caption{Big Bang Model timing including excellerator energies for
         $ E = \mu $. }
\label{BBMtimingexcelerator}
\end{figure}

The real experiment was performed by nature itself at the creation of
the universe, all the scale and time dependence must be stored in the
physical laws of SM and BBM model. The fundamental particles get born
in these process. We assume for this reason the same happens
during the creation of the FPs. Following this philosophy
our empirical toy ansatz about a geometrical
microstructure of fundamental particles assumes the development
about the size of the FPs from size zero to the size today is a result
of the early history of the universe stored in the geometrical
extension of the FPs, or the geometrical size of the
FPs is a result of the time and size development of the history
of the very young universe.

\subsubsection { The experimental paradox to measure small distances }
\label{sec:The experimental paradox to measure small distances }

The experimental paradox to measure small distances is well documented
in the $ \EEGG $ reaction. It is not adequate to increase only the
test energy $ Q = E_{CM } $ of the experiment to measure smaller distances,
parallel the physical parameters what are depending about the
energy scale $ \mu $ change the conditions for the whole experiment.
In this sense the experiment is a race between test size
and the scale $ \mu $ defining the conditions for the whole experiment.

We like to discuss this paradox in more detail 
in a gedanken experiment. Beginning with the
$ \EEGG $ reaction at LEP is the $ E_{CM} $ energy 
approximately 200 GeV. This is above
the production threshold to generate W-boson pairs
$ m^{\pm}_{W} = 91.1876 \pm 0.0021 $ GeV.
In the SM model the $ E_{CM} $ energy is located after
Fig.{\ref{fig.4} and Eq.\ref{alphasol} at a scale of approximately
$ \mu = 200 $ GeV. Including the high statistic of the $ \EEGG $ reaction 
is this a test size of $ 10^{-18} $ cm displayed in Fig.\ref{SizeLimits}.
At this scale the physical conditions are three families
of FPs ( see Tab.~\ref{ParemeterFamily} and Fig.{\ref{fig.1},
three interactions
shown in Fig.{\ref{fig.3} and Fig.{\ref{fig.4}. It is possible to generate
$ Z^{0} $ and $ W^{\pm} $-bosons. According Eq.\ref{alphasol} with increasing
test energy the set of parameters will be not change until the
Grand Unification scale of approximately $ 10^{15} $ GeV.
At this scale the three interactions unify depending about
the Standard Model or SUSY at different energies or in one 
energy. Beyond this scale until the Planck Scale
only one interaction is left. In this era massive super heavy particles
are under discussion. This particles would play an important
role in the proton decay \cite{HalsenMartin}.

Beyond the Planck Scale of $ 10^{19} $ GeV
or $ 10^{-33} $ cm test size the quantized gravitation has to
be taken into account. 
1955 John Wheeler \cite{Spacetime1} devised beyond this scale
the spacetime foam. The foam is supposedly the foundations 
of the fabric of the universe \cite{Spacetime2}.
At this scale the uncertainty principle allows
particles to be created and annihilate without
violating conservation laws, the space shrinks and
the energy of the virtual particles increases.
According Einstein's theory of general
relativity curves energy spacetime,
this suggests the energy of the fluctuations would be large
enough to cause significant departures from the
smooth spacetime seen at larger scales.
These fluctuations could serve as a candidate for the
seeds of the primordial perturbations in the
cosmic inflation because the Large-Scale Structure
of the universe is assumed to form as a result of
the growth of initially small density perturbations
due to gravitational instability \cite{PrimordalPerturbations}.
The character of this foam is still under discussion
\cite{Spacetime3} a possibility the foam could be
a condensate was discussed by B.L.Hu \cite{Spacetime4}.
If this is true the gedanken experiment will detect
between $ 10^{-33} $ cm and zero cm a condensate.

We used up to this stage of
the experiment only the SM model which only implicit defines
a test size via the uncertainty principle. Explicitly shown
is the test size of our experiment in particular at this very
high scales in the BBM model. In this model the horizon distance 
or size of the universe in the vicinity of the Planck Scale
defines unambiguously an upper limit for the test size
of our gedanken experiment, if we request the test size should be
causally connected. If we use the link of the common scale
between SM and BBM as discussed in Tab. \ref{linksSMBigBang}
it is possible to repeat the gedanken experiment in the BBM
model. For the discussion it is sensible to repeat the experiment
starting in the vicinity of the Planck Scale down to the scale $ \mu = 200 $ GeV
and use the timing of the BBM discussed in 
chap.{\ref{sec:Cosmology and the early Universe}
in particular Fig.\ref{BBMtiming-t}. If time reversal invariance
holds at the Planck size down to a test size zero we like to
stress that every experimental physicist unambiguously will
measure at this era a superfluid condensate.
Between Planck Era and GUT Era the horizon distance or
the size of the universe increases, the experiment will be
dominated by one interaction only and mass exist already.
The gedanken experiment accordingly will measure one
interaction and mass or super heavy massive particles.
At the GUT scale the size of the universe is sufficient
that suddenly three interactions appear accompanied by three types
of charges. In the following Electroweak Era the size of the
universe will be big enough to host FPs. 
The gedanken experiment will
measure the three families of FPs including the three interactions
as discussed in the chap.{\ref{sec:Standard Model Parameters}.
We will stop at this scale our gedanken experiment.

If time reversal invariance holds for the SM and
BBM model from the vicinity of the Planck Era to the era today
the universe is able to remember how it was created. We like to
discuss two scenarios beginning with the Planck Era
how this fact could be stored in the universe in more detail,
in particular focusing on the extent of FPs:

\begin{itemize}

 \item
The condensate in the Planck Era exist only in this era until
the Planck scale and disappears after the universe cross this scale. 
In the GUT Era is get replaced by super heavy particles from
the size $ s $ of the universe $ 0 \le s \le s_{GUT} $. After
the GUT Era the regime of the three interaction dominates
the universe $ 0 \le s \le s_{today} $. 
No condensate or super heavy particles
and no phase transitions energies 
from the different eras exist at the scale today.
The parameters defining the different era act precisely
at the energy the universe passes the adjacent scale $ \mu $
of the particular phase transition.

 \item
In the from the gravitation dominated Planck Era
the condensate of this era still exist after the
universe passed the phase transition from Planck to GUT Era.
Primordial inhomogeneities caused by quantum fluctuations
in the Planck Era are able to pass to the GUT Era.
As the universe expands down to the GUT scale the
horizon distance of the condensate increases but stays
inside the GUT size. In the GUT Era the primordial inhomogeneities
form an onion like aggregates with an inner condensate kernel
and outer mass shell. Hypothetical super heave neutral particles
are under discussion in this era. These aggregates would be
a candidate for the seeds in the inflation for the
generation of scalar density fluctuations 
( See chap.\ref{sec:Key events in the life of the universe} ).
The sofar discussed conditions including the 
super heave neutral particles still exist after the universe
passes the GUT scale, inside the size of the universe at this
scale. After the GUT scale these aggregates could serve
as kernels adding outside
three types of charges ( colour, electro magnetic and weak ).
The three adjacent interaction strong, electro magnetic
and weak become distinct. In particular if the strong
force get distinct, it seems this is the cause for inflation.
After the universe undergoes the
inflation the horizon distance is much bigger as every
possible generated elementary particle and every relict size
of Planck and GUT scale. With increasing time $ t $ after Big Bang
and decreasing temperature $ T $ a quark gluon plasma will exist.
Protons, neutrons,  electrons and neutrinos get generated in
a hight temperature plasma. Finally the universe follows the in 
chap.\ref{sec:Primordial nucleosynthesis} described
regime of Eq.\ref{aT}
to Eq.\ref{Y2}. Out of a high temperature plasma, FPs
and finally atoms get generated and stable. 

In this scenario is the geometrical structure of a Fundamental Particle
the direct consequence of the development of the 
primordial inhomogeneities of the Planck Era up to the era today.
Still a relict of the condensate of the Planck Era, the mass
shell of the GUT Era and the following charges together
with remains of the size of the Planck scale and GUT
scale should exist today. After our scenario should this structure
exist inside every Fundamental Particle.

\end{itemize}

In the first scenario no information of the development of the
universe would exist today. The conditions would be linked
direct to the scale energy and appear in the time $ t $ after
Big Bang. 

In the second scenario
comparing the Planck Scale at $ 10^{19} $ GeV or Planck length
$ l_{Pl} = 10^{-33} $ cm with the experimental limits of the
size of FP in Fig. \ref{SizeLimits} of approximately  $ 10^{-18} $ cm
it is obvious that FPs are not created at this scale. The FPs are
$ 15 $ magnitudes bigger. But the this scenario traces
a possibility from the Planck scale until the scale today
how the inner structure of a FP could be described. We use
for this reason the second scenario for the further discussion.

\subsubsection {The geometrical approach }
\label{sec:The geometrical approach}

As just discussed
our empirical toy ansatz about a microstructure of a
fundamental particle is that particle contain every energy
state a cross of its radius which the Universe passed through
during its evolution shown in Fig.\ref{BBMtiming-t}.
For such a case it is possible to read this microstructure
direct from the SM model Fig.{\ref{fig.4} and the BBM model
Fig.\ref{BBMtiming-t}. Using these both figures a
schematic development of a geometrical
extended FP is shown in Fig.{\ref{SchematicDevFP}}.

%
\begin{figure}[htbp]
\vspace{8.0mm}
\begin{center}
 \includegraphics[width=10.0cm,height=7.0cm]{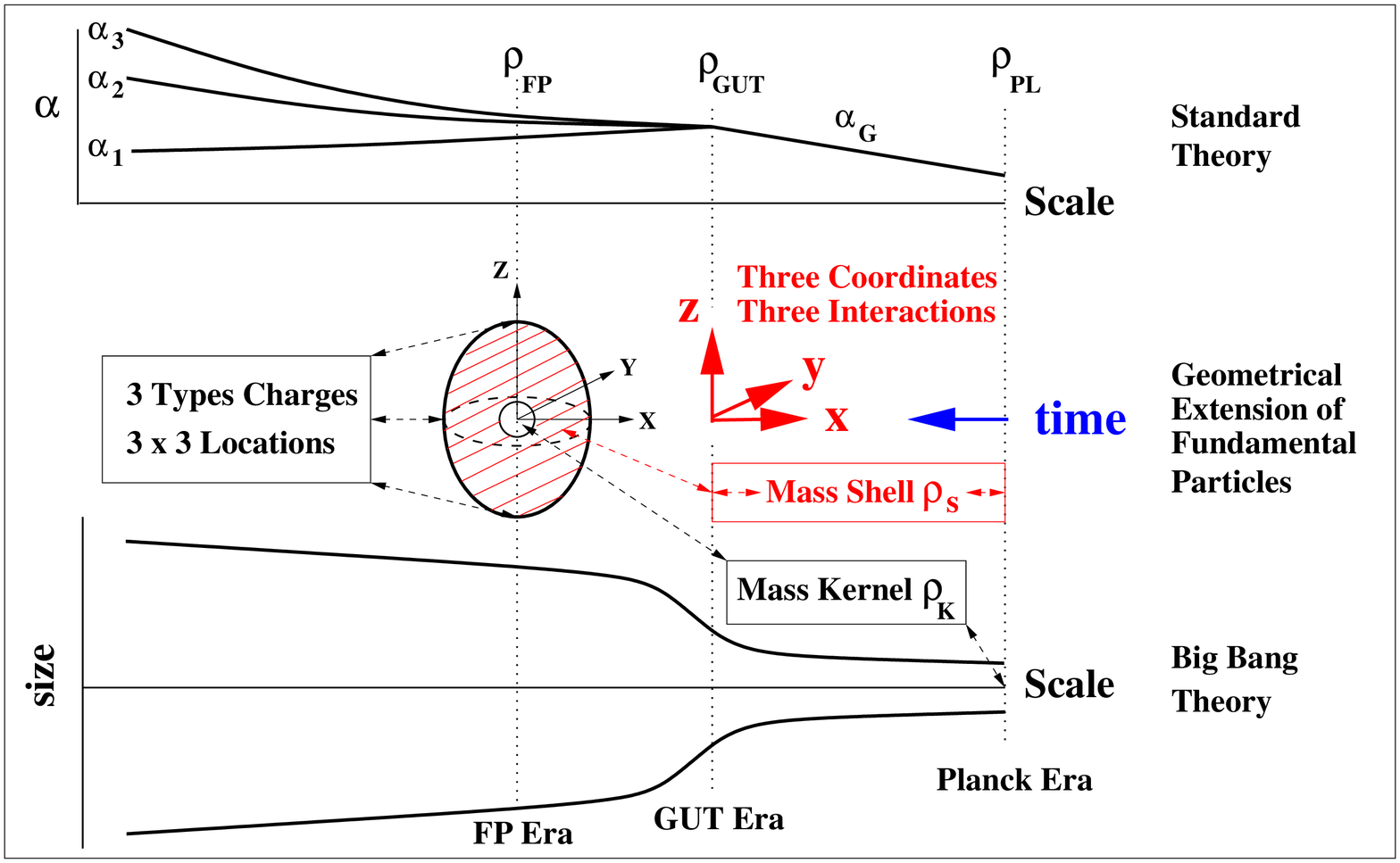}
\end{center}
\caption{ Schematic development of geometrical extended FP. }
\label{SchematicDevFP}
\end{figure}

In the middle part of Fig.{\ref{SchematicDevFP}} we
describe the historical procedure initiated from $ t = 0 $
of Big Bang time about mass density, size,
charge distribution a point-like observer would measure following
the creation of a fundamental particle. The mass of the universe
known today is a composite of $ \Omega_{B} = 0.044 \pm 0.004 $
baryonic matter, $ \Omega_{DM} = 0.22 \pm 0.04 $ dark matter
and $ \Omega_{DE} = 0.7 $ dark energy. As all these contributions
are created from one origin the discussed procedure should be
generic for all three contributions. We concentrated in the
following discussion only on the baryonic matter because the
FPs are located in this part.

In the upper part schematically displayed is the scale
dependence of the coupling of $ \alpha_{3} $, $ \alpha_{2} $
and $ \alpha_{1} $ as function of the scale of the
SM model Fig.{\ref{fig.4}. In the lower part we
relate the time development of the Big Bang theory of
Fig.\ref{BBMtiming-t} to the SM model with the according
energy states Planck Era, GUT Era and the FP Era.

If the spectator initiates the measurement from the Big Bang
time  t = 0 to the Planck scale he enters an era our
knowledge is very poor, all assumptions have to be taken
as highly speculative. With this in mind 
we introduce some hypothesis.
In accordance with our previous discussion the spectator 
will measure a quantized
energy mass condensate, originated from a volume explosion with
a density of $ \rho_{PL} $. In this mass-
time dominated condensate the charges color, electric and weak are
already existing. The constituents of this condensate at a distance
below $ r \le 10^{-33} $ are in direct contact not able to form
an interaction similar the field theory in the SM model.
The statistical distributed
volume explosion does not introduce a total quantized  spin different
from zero.
The consequences of this assumption will be discussed in the
next 
chap.\ref{sec:Links between continuous symmetries gravitation and spin}

Next the spectator will cross the domain boundary of the
Planck scale at $ Q = 10^{19} $ GeV
at a distance from the center of $ r = 10^{-33} $ cm and enters the
era to the GUT scale.
Usually nature performs smooth transitions from one era to next. We
expect for this reason that the gravitational mass dominated regime
changes slowly to the GUT scale.
It would be likely that a decreasing
mass density get generated forming a mass shell enclosing the mass
kernel of the Planck Era. This era is dominated by only one interaction.
If it is possible to describe this era already by a field 
theory of high group \cite{HalsenMartin}
a geometrical structure of two particles interacting via a gauge
particle would be possible. If the total spin of universe would be zero
these particles describing the mass shell could carry a spin which
adding to a total spin of zero. All well known spins $ 1/2 $,
$ 1 $ and zero which are able to adding vectorial to the total spin zero
would be possible.
As a further consequence the mass shell with
a mass density of $ \rho_{s} $ could in a geometrical microscopic
picture rotate around the Planck kernel. 

Next the spectator on its time space journey will cross the
the domain wall of the GUT Era at $ Q = 10^{15} $ GeV at
a mass density $ \rho_{GUT} $.
At the Grand unification scale in the time development of the universe
the three interaction strong, electro magnetic and weak accompanied
by the adjacent three types of charges strong, electro magnetic and
weak get dominant. This is  shown in the upper
part of Fig.{\ref{SchematicDevFP}}.
The interactions are at this scale much stronger
as the gravitation and the size of the universe is big enough
to host three intepented interactions
in the three dimensional space.
For the spectator who is concentrated on the particle like
aggregates with spin axis, which are carried on from the quantum fluctuations
of the Planck Era to this scale,
three possible geometrical locations
appear to place charges on one coordinate axis. One position
at the center and two
respectively at the surface of the object.
In total $ 3 \times 3 = 9 $ positions are possible.
As shown in a more detailed view in
Fig.{\ref{BasicSCHEMEFP2006}} of the fundamental particle
it would be possible to place on the x-coordinate
three electric charges, the z-coordinate three colors and on the
y-coordinate three weak charges.

%
\begin{figure}[htbp]
\vspace{8.0mm}
\begin{center}
 \includegraphics[width=10.0cm,height=7.0cm]{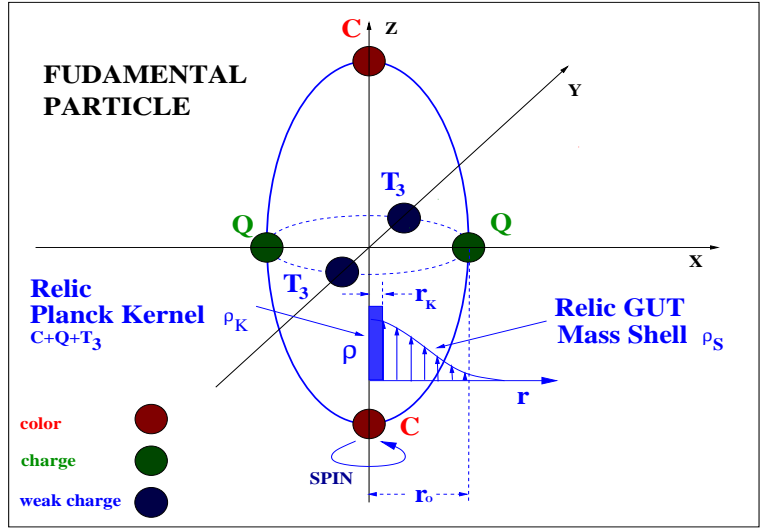}
\end{center}
\caption{Basic scheme of extended fundamental particle in the ETAMFP model.}
\label{BasicSCHEMEFP2006}
\end{figure}

The assumed spin axis in Fig.{\ref{BasicSCHEMEFP2006}} allows
to distinguish between left and right on one coordinate axis
with lead to three position if we assume one position exist
as discussed at the Planck era on the Planck kernel.
The appearance of the three interactions in the vicinity of
the GUT scale occurs in the SM
model at three or in the SUSY model at one scale
energy. Both theories sharing the fact out of one interaction three
interaction get born. 
It is likely that in the first moment after the GUT scale
all three possible charges are in direct contact.
With increasing
time after the GUT scale the full picture of quantum field theory
with FPs and gauge bosons would be formed up.
In the direct contact regime the displacement between the charges is
tiny, repulsive forces could be superimposed by attractive forces.
A superposition of similar strong repulsive and attractive forces
opens the possibility at a certain distance from the center of the FP 
stable geometrical positions could exist, where the different
forces balance each other. Such positions would be ideal
to place charges. For example the positive electric charge 
on the right side of the x-axis in such a position would allow
via these stability conditions only a second charge opposite
on the left side of the x-axis. In such a regime with increasing
number of charges the discussed $ 3 \times 3 = 9 $ 
positions are possible. This regime would form an energetic minimum.
We will discuss this hypothesis in more detail in the 
chap.\ref{sec:Forces and stability}.

With increasing time after the Big Bang the universe
follows in the SM model the conditions of the running coupling 
constants discussed in the 
chap.\ref{sec:Standard Model Parameters} and
undergoes in the BBM model the inflation with all consequences discussed in the
discussed in the chap.\ref{sec:Cosmology and the early Universe}.
Concerning about the FPs, the universe is big enough to host
all FPs, gauge particles and hypothetical particles already
exist before the GUT scale. All these particles would be in 
our ETAMPF model the outcome of primordial quantum fluctuations
already in the Planck Era. In particular important would be
scalar density fluctuations that seed large scale structures.
The size of the universe will decouple from the size
of the FPs. The geometrical size of the universe will increase
according the comoving scale but the size of the FPs which
is stabilized by the four fundamental forces will be the same.
According our ETAMPF model also remains of the Planck Era
including Planck size and GUT Era including GUT size will
get stable and remain inside the FPs. 
The character of the condensate transferred
from the Planck scale until the scale today including
also the size of this condensate today and the conditions of 
mass shell together with the GUT size will be discussed
in more detail in the chap.\ref{sec:Classical approach}.

After the universe passed the plasma conditions it will follow
the discussed scheme of primordial nucleonsynthesis
( See chap.\ref{sec:Primordial nucleosynthesis} ).
The spectator will cross the energy of the  $ \EEGG $ experiments.
We determine at this energy the journey of our spectator.

The status of the FP at a scale of about $ 200 $ GeV
is displayed
in the middle of Fig.{\ref{SchematicDevFP}} and
Fig.{\ref{BasicSCHEMEFP2006}}. It is an object with a non
rotating mass kernel with a density of $ \rho_{k} $
what is the heritage of the heavily attenuated
Planck density $ \rho_{PL} $. The kernel carries mass
and the three charges color $ C $, electric $ Q $
and weak $ T_{3} $. The
size of the kernel radius $ r_{k} $ is the relict of the
Planck radius by the time development of the universe.
The kernel ends at a domain wall
separating the kernel from the rotating mass shell ( Spin )
of GUT Ea with a density of
$ \rho_{s} $ and an outer radius of $ r_{0} $ where again
a domain wall exist. The mass density $ \rho_{s} $ is a heritage of the
heavily attenuated GUT scale density $ \rho_{GUT} $ and the size $ r_{0} $
is the by the time development of the universe expanded
radius at the GUT scale. The locations of the three
charges at nine positions
are the same as discussed at the object existing at
the GUT scale. We could place on the x-coordinate three
electric charges $ Q $, on the y-coordinate three
weak charges $ T_{3} $ and on the z-coordinate
three colors $ C $.

\subsubsection {Links between continuous symmetries gravitation and spin }
\label{sec:Links between continuous symmetries gravitation and spin}

To associate the distributions of the charges and mass of the FPs
in Fig.{\ref{BasicSCHEMEFP2006}}
with the SM and BBM model we used the running coupling constant of
the SM shown in Fig.{\ref{fig.4} and the space-time evolution
of the BBM in Fig.\ref{BBMtiming-t}. Embedded in the continuous
symmetries of the four dimensional space-time is more information
about the two so fare missing connections to the gravitation and spin
of the discussed microstructure of of FPs. We like to discuss,
four possible links to associate the gravitational interaction
to the time of the four dimensional space-time and discuss later
the consequences for the FPs about the local isotropy of the 
space against rotation ( Spin ).

First at scales below the Grand unification the three interactions
strong, electro magnetic and weak appear well separated from
each other, because the geometrical size of the universe is
sufficient to make them distinguishably from each other. These
three interaction are embedded in the space for example described
by the Cartesian coordinates x, y and z.
But in total exist four coordinates
and four interactions. The last possible 
coordinate is the time. In accordance with our
discussion above it would be sensible to connect the time to
the gravitational interaction. This would of particular importance in
the regime above the GUT scale where the gravitational interaction
is dominating but it should also hold at the scale today.

Second we like to stress that an experiment which investigates the
center of a SM model point like FP will, meet unambiguously
the conditions at the Grand unification scale where
the gravitation is not negligible any more and next the
the Planck Scale where the gravitation get dominant.
At the Planck Scale
the Planck conditions of mass, length, time and density of
Eq.\ref{PlMass}, Eq.\ref{Pllenght}, Eq.\ref{Pltime} and
Eq.\ref{Pltensity} of the gravitation are absolute crucial.
\begin{equation}
m_{Pl}=\sqrt{\hbar c / G } \simeq 2 \times 10^{-5} g
\label{PlMass}
\end{equation}

\begin{equation}
l_{Pl}=\sqrt{\hbar G / c^{3} } \simeq 1.6 \times 10^{-33} cm
\label{Pllenght}
\end{equation}

\begin{equation}
t_{Pl}=l_{Pl}/c\simeq 5 \times 10^{-44} s
\label{Pltime}
\end{equation}

\begin{equation}
\rho_{Pl}=c^{5}/(\hbar G^{2}) \simeq 5 \times 10^{93} g/cm^{3}
\label{Pltensity}
\end{equation}
At these conditions a complete geometrical extended three
dimensional FP does not exist
because the scale is to high or the test distance to small. But
if our assumption is true that the FP has accumulated the history
of the SM and BBM in his geometrical structure this
implements that the inner part of the FP is dominated
by the gravitation and the structure is prevailing one dimensional.
As the three dimensional test distance is extreme small below $ 10^{-33} $
cm again the time would be a good candidate for this dimension.

Third the mass-energy equivalence $ E = m \times c^{2} $ proposed
and interpreted 1905 by Einstein, as a general principle which 
follows from the relativistic symmetries of space and time,
points also to the time as connection the gravitation.

Fourth the uncertainty principle $ \Delta E \times \Delta t \sim \hbar $
connects again the time to the energy.The energy mass equivalent
point again to a connection between time and the gravitation.

We use this four indications to connect in the further discussion
of this paper the time to the gravitational interaction.

The local isotropy of the space-time leads
in the SM via the Noether- Theorem
to the angular momentum conservation and finally to the spin of
the FPs existing at the scale today.
In the SM is the spin a quantum number to describe
one feature of the particle. For the point particle of the SM
has the spin no microscopic meaning, because a rotating point with
diameter zero is meaningless. If we trop the condition of a point
particle and allow an geometrical extension of the FPs like in
Fig.{\ref{BasicSCHEMEFP2006}} a rotation of the object is
possible and the spin, magnetic moment and electric dipole moment
is natural included in the object. From experiments we know
that for example the spin axis of an electron is anti-parallel to the
axis of the magnetic moment of the electron. For our extended object
in Fig.{\ref{BasicSCHEMEFP2006}} implies this that
the charge is rotating together with the mass center in the same
rotation plane. The geometrical extension of the mass $ \rho(r) $ in the
center, the position of the charges at $ r_{0} $ generate a classical
angular momentum, magnetic moment and electric dipole moment.
As we discussed in the geometrical approach, remains the question
at which time in the development on the universe the spin appears?
The spin is associated with the creation of the mass of the FP
because in a microscopic picture only a rotating mass like a
gyroscope will lead to a stable spin axis. The mass of the FP
is in our ETAMFP model a relic  of the time
between $ t = 0 $ and the Planck time as
Planck kernel and between Plank time and GUT time as mass shell.
It seems unlikely that the volume explosion before the Planck time
generates a spin because this process is statistical equal
distributed over the whole volume of explosion. This would
implement that the universe has a total spin zero. But after
the Planck scale the universe could be already geometrical
extended enough to form a particles with a fined size which
direct interact with each other like in the field theory of a
high group \cite{HalsenMartin}. If in this case a 
quantized spin get created
two particle with spin interact via a field particle with
each other. The spin of the particles involved would be
well defined. If we assume like in the SM model the particles
carry spin $ 1/2 $ the field particles must carry spin $ 1 $
or zero, because only
$ 2 \times \vec{\frac{1}{2}} + \vec{1} = \vec{0} $ or
$ 2 \times \vec{\frac{1}{2}} + \vec{0} = \vec{0} $ add to zero. In
a self interacting mode with two field particles also higher
spins like $ 2 $ are possible because $ \vec{2} + \vec{2} = \vec{0} $.
It seems after this discussion
reasonable to assume the spin appears in the
universe in the time window between Planck and GUT scale.

\subsubsection { Forces and stability }
\label{sec:Forces and stability}

As discussed in the geometrical approach the dependence of the
coupling constants $ \alpha_{3} $, $ \alpha_{2} $ and $ \alpha_{1} $
as function from the scale $ Q ( GeV ) $ in the SM shown in
Fig.{\ref{fig.4} and the time-size dependence of the
same scale in the BBM displayed in Fig.\ref{BBMtiming-t}
contains information about which forces exist at which scale and
what is the dependence of these forces about the scale of
particle size under investigation.

At the LEP scale at approximately 200 GeV a straight conception
about forces and FPs exist. The SM contains the three forces
strong, electro magnetic and weak shown in Fig.{\ref{fig.4}
which interact with three families of FPs as shown in Fig.{\ref{fig.1}.
Up to the Grand unification scale the forces are sorted after
there strength $ \alpha_{3} > \alpha_{2} >\alpha_{1} $. At this
scale at about $ 10^{15} $ GeV the three forces get unified
in one force. The associated test distance in our discussed
gedanken experiment is at this scale so small that the
concept of the SM of two geometrical separated FPs interacting
via an also geometrical separated gauge boson is questionable.
In this
case it seems reasonable to assume the interaction is similar
to a direct contact term interaction for example used in the
Lagrangian Eq.\ref{LagDirectContactQ}.
Between the Grand unification scale
and Planck scale only one force exist above the Planck scale
the gravitation is dominating. From the
experiment the known FPs are in particular in the case of
the electron with a mean life time $ \tau > 4.6 \times 10^{26}$ yr 
extreme stable. This experimental fact require that the distance
dependence of the forces acting in the FPs must lead
to highly stable conditions.

The target of the following discussion is to develop a general
scenario guided from the experimental facts and the scale-time-radius
dependence of the SM and BBM model to imagine a possible radius
dependence of the known four forces what could lead to such highly
stable conditions.

If we start at a radius $ r=0 $ in the center
of the FP it is necessary to discuss first the radius dependence
of the gravitation at distances close to the Planck scale $ r_{P} $
and the Grand Unification scale $ r_{ GU } $. If we follow the classical
force of the gravitation $ G_{SS} $ as function of the distance
between to masses in a Schwarzschild \cite{Schwarzschild}
vacuum as displayed in
Fig.\ref{ForcesSM-BBM} ( blue broken line ) the force $ f1 $
will be infinite at $ r = 0 $. Such a behavior will not lead to
a stable mass core needed from our experimental knowledge for a
FP for example like an electron. The mass core will only be stable if
the attractive Schwarzschild force $ f1 $ in the environment of $ r_{P} $
will change to a repulsive force $ f2 $ like a De Sitter \cite{DeSitter}
behavior $ G_{DS}$ ( Red solid line in Fig.\ref{ForcesSM-BBM} ).
\begin{figure}
\begin{center}
 \epsfig{file=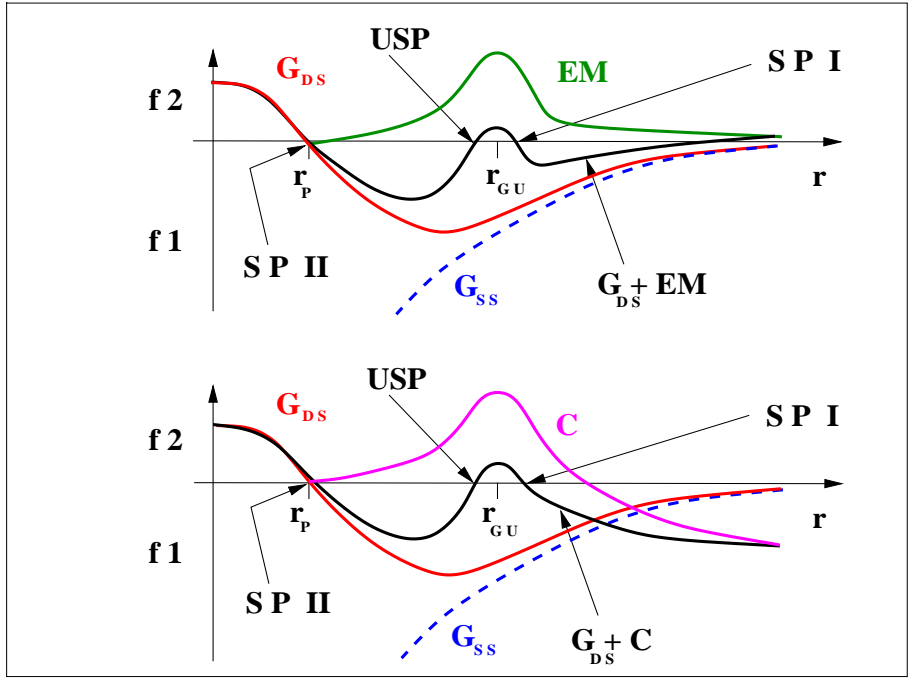,width=12.0cm,height=9.0cm}
\end{center}
\caption{ Scenario of the radius dependence of the forces of SM
          and BBM model close to the Planck and Grand unification scale.
          Upper part unification of electromagnetic ( EM ) force with
          gravitational force ( G$_{DS}$ ) and lower part
          unification of color force ( C ) with G$_{DS}$.}
\label{ForcesSM-BBM}
\end{figure}
In such a case in our direct contact term scenario a repulsive force
will lead to a repulsive pressure which get balanced by an
attractive pressure generated by the Schwarzschild force. Point SPII
in Fig.\ref{ForcesSM-BBM} describes such a situation. If the stable
radius $ r $ shown as black circle in Fig.\ref{ForcesSTABLE} get smaller
a repulsive force $ \Delta f = f2 - f1 > 0 $ will push the mass core back
in the stable position 
( left middle red circle broken line in Fig.\ref{ForcesSTABLE}).
\begin{figure}
\begin{center}
 \epsfig{file=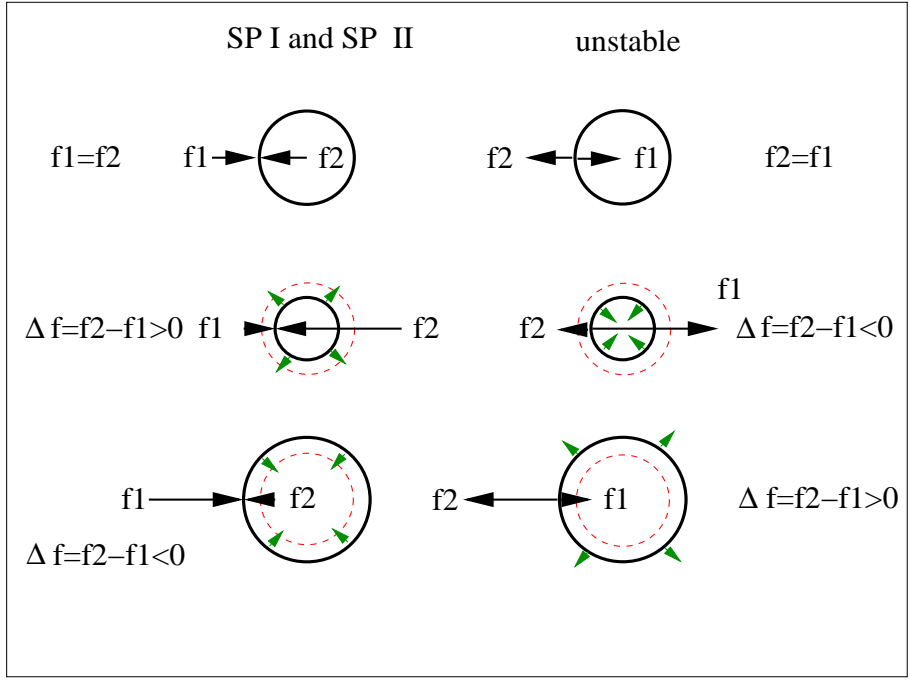,width=12.0cm,height=9.0cm}
\end{center}
\caption{ Stability conditions for a mass core of FPs. Stable 
          conditions left side and unstable conditions right side. }
\label{ForcesSTABLE}
\end{figure}
In the contrary if the stable radius increases the radius get pushed
back from a attractive force $ \Delta f = f2 - f1 < 0 $ to the 
left lower broken line in Fig.\ref{ForcesSTABLE}.
In conclusion only a De Sitter like
vacuum would lead to a stable mass core of the FPs.

Second it is necessary to discuss the unification of the electromagnetic
force EM, the color force C and the weak force at the Grand
unification radius $ r_{GU} $ and the involvement of the gravitation
between Planck and Grand unification scale.

Between the radius
$ r_{P}<r<r_{GU} $ the SM and BBM predicts one force including
one coupling constant and for $ r>r_{GU} $ the three discussed
forces appear including the three charges color, electromagnetic
and weak. The consequence of this prediction is that the charges
should be geometrical located at the vicinity of 
the radius $ r_{GU} $. For simplicity
we assume in the further discussion the charge themselves have a point-like
geometrical character.

The attractive-repulsive character of the gravitational force
in the environment of $ r_{GU} $ was investigated in
ref.\cite{DeSitter} and ref.\cite{IRINA2}.
Following the concept to find a possible
stable scenario for FPs we assume that all three forces change
the attractive or repulsive character like the assumed
De Sitter behavior of the gravitation at the radius $ r_{P} $.
Numerous combinations
to unify the four discussed forces between $ r_{P}<r<r_{GU} $ are
possible. We pick one possible scenario put like to point out that
also other scenarios will lead to the same result.

The electromagnetic
interaction EM is at distances of $ r >> r_{GU} $ repulsive as shown
in Fig.\ref{ForcesSM-BBM} ( Green solid line ) and we assume between
$ r_{P}<r<r_{GU} $ the force will decrease to zero or is very small
at $ r_{P} $. If we unify the EM force with the gravitation
$ G_{DS} + EM $ as shown in the black solid line upper part in
Fig.\ref{ForcesSM-BBM} a stable point $ SP I $ appears like
$ SP II $ of the gravitation the repulsive inner core forces
balance the attractive forces $ r > r_{GU} $.

An electromagnetic
charge located at $ SP I $ would be stabilized geometrical very
similar as the
gravitational mass core at $ r<r_{p} $ shown in Fig.\ref{ForcesSTABLE}.
If the stable
radius $ r $ shown as black circle in Fig.\ref{ForcesSTABLE} get smaller
a repulsive force $ \Delta f = f2 - f1 > 0 $ will push the EM charge back
in the stable position ( Middle left red circle broken line in Fig.\ref{ForcesSTABLE}).
In the contrary if the stable radius increases the radius get pushed
back from a attractive force $ \Delta f = f2 - f1 < 0 $ 
( Lower broken red circle in Fig.\ref{ForcesSTABLE} ).
In conclusion a De Sitter like behavior
of the EM force would lead to a stable geometrical position
of an EM charge at point $ SP I $.

If we choose Cartesian coordinates
for the FPs , put the center of the FPs at $ r=0 $ and locate
on the +x-axis at $ SP I $ one $ EM $ charge opposite at -x-axis
a second stable point will exist because this charge will
detect for the distance more as $ 2 \times r_{GU} $ an repulsive force to the
first charge because the force $ G_{DS} + EM $ will be repulsive
for $ r>>r_{GU} $. As the
$ EM $ charge appears explicitly at the $ r_{GU} $ radius it is likely
that the charge already exist at $ r < r_{GU} $. At approximately
$ r_{P} $ the prediction of the SM and BBM is that the gravitation
dominates all other forces. Even a very small attractive coupling between
mass and the three charges would locate one more stable position
for these charges at the geometrical position with maxim mass
concentration. According Fig.\ref{ForcesSM-BBM} would this be
direct in the center of the FPs.

The crossing point $ USP $ in Fig.\ref{ForcesSM-BBM} leads
to a geometrical not stable regime as shown in Fig.\ref{ForcesSTABLE}
right side.
The force $ \Delta f = f2 - f1 < 0 $ on the right side middle part
will collapse to position $ SP II $ and the force $ \Delta f = f2 - f1 > 0 $
will expand $ USP $ the position to $ SP I $ at the right side lower part.

The color force C and weak force is attractive at $ r >> r_{GU} $ if
we assume again a change of character close to $ r_{GU} $ to a repulsive
force and put it at $ r=r_{P} $ to zero or very small the unification
with the gravitation $ G_{DS}+C $ shown as black line in the lower part
of Fig.\ref{ForcesSM-BBM} will generate a stable point $ SP I $.
The scenario for the week force would be very similar and will be
not further discussed. It would be possible to place at point
$ SP I $ color charges. After the previous discussion of the
stability in Fig.\ref{ForcesSTABLE} would also the color charge
be stabilized close to $ SP I $. If we put the color charge on
the + z-axis at an opposite stable condition would exist on the
z-axis because the color force would be for $ r>r_{GU} $ for some
distance repulsive to stabilize the second color charge opposite
to the first charge. The scenario for the week force would be
similar on the last free y-axis.

In conclusion to the discussion of the behavior of the four
forces it turns out, that the above scenario generates six stable
geometrical positions to place charges at approximately $ r=r_{GUT} $
plus one position for each charge in the center of the FPs
including a stable mass core. The connection between the three
axis of the Cartesian coordinates and the three interactions
would be explained be simple geometrical reasons.

\subsubsection { The flashing vacuum }
\label{sec:The flashing vacuum}

The ETAMFP model Fig.{\ref{BasicSCHEMEFP2006}} includes a
charge circling a kernel. An accelerated charge will
emit electromagnetic radiation. Such an object is losing
energy, is not stable with a finite life time. If we
consider for example an electron with a life time
of $ \tau > 4.6^{~26}$ yr it is obvious a radiation free
path as postulated in the Bohr model \cite{BorModel} must exist.
A well known example for a stable geometrical extended object
is the Hydrogen atom. In the SM model the probability
distribution $ | \psi |^2 = f ( r ) $ of an electron in $ s $ state
peaks in the center of the proton but not in a delta function at
$ r = 0 $, similar the half stabel $ p $ state peaks at a
a distance $ r > 0 $ outside the center of the proton. The electron
circles the center of the hydrogen atom on a radiation free path,
statistical appears and disappears in a certain distance $ r $ from
centers of the hydrogen atom and at the same time generate with
these flashes a radiation free path around the center of the
hydrogen. To interprete this statistical appearing and disappearing
a second state would be necessary to give the electron the
possibility to disappear. An approbiate candidate for this
state would be the vacuum. The vacuum state \cite{Vacuum2} is
the quantum state with the lowest possible energy. This state
is characterized by distinct properties. 
In perturbation theory for example in the $ \EEGG $ reaction
the QED differential cross-section Eq.\ref{LITKEcross} is calculated
including radiative effects up to $ O(\alpha^{3}) $.
It contains fleeting
electromagnetic waves and particles that pop into and out of existence.
The uncertainty principle in the form 
$\Delta E \Delta t \ge \hbar$ implies that in the vacuum one 
or more particles with energy
$ \Delta E $ may be created for a time $ t < \Delta t $.
In the environment of a black hole the vacuum get unstable and
loses energy via Hawking radiation \cite{Hawkinradiation}.
Vacuum fluctuations cause a particle-antiparticle pair to appear
close to the event horizon of a black hole. One of the pair falls into 
the black hole whilst the other escapes. In this quantum tunneling effect
a particle-antiparticle pair get screened and one will tunnel outside 
the event horizon. If the total energy is conserved and this particle
escapes from the black hole it will cause and energy or mass loss.
The Hawking radiation appears near a black hole with all
properties of a black hole in particular a Schwarzschild horizon
\cite{Schwarzschild} and a singular gravitational potential.
In this case the particle pair get screened. 
A further scenario would be that not a particle anti-particle pair
get created but only one particle oscillates statistical between
real state and vacuum state for and backward.
In all discussed scenarios the vacuum
has the possibility to flash statistical
on every geometrical location from time to time. 
It will also be possible close to a fundamental particle. A
FP is after our knowledge not a black hole but discussions
on the way a FP could be described by a De Sitter-Schwarzschild 
geometry a black hole whose singularity is replaced with de Sitter 
core of some fundamental scale \cite{werner,IRINA1,particle}.
( See chap.~\ref{sec:Particle-like structure related to gravity})
If the gravitational potential is not
singular any more the possibility appears after the stability discussion
in Fig.\ref{ForcesSM-BBM} that one more vacuum property is possible.
The vacuum is fluctuating close to a FP similar to the Hawking radiation
put with the difference that one pair partner is not escaping
from the FP it just appearing statistical at a certain location
whereas his counter-partner stays in the vacuum, similar to
a mirror particle in a Dirac sea \cite{Diracsea} or it
oscillates between real and vacuum state.
As in all models unter discussion ( See Tab. \ref{linksSMBigBang} )
a vacuum exist, we introduce for this reason as working
hypothesis the flashing vacuum \cite{flashingvac}.

Translated in our ETAMFP model Fig.{\ref{BasicSCHEMEFP2006}}
would the previous discussion implement for the classical
charge rotating on radius $ r_{o} $ an aggregation of statistical
flashing charges located with a certain distribution function
in the vicinity of the radius $ r_{o} $. The quantum of
flashes per time together with the distribution function
would define the radius $ r_{o} $, if we assume the charge is geometrical
substantial smaller as the FP. The integral summed over all
flashes would define the total charge of the FP.
If the flashes are collective
overlaid be a rotational speed would this be the deeper
reason for the magnetic moment and the electric dipole moment.
The discussion so far we focused on the electric charge, but
in particular in perturbation theory radiative effects
including strong, electro magnetic and weak interactions.
These implements also the attached charges color, electric
and weak in the fluctuations. The vacuum properties
are also very important for the Higgs mechanism
to create mass for the FPs. For this reason it is very
likely that the vacuum is also able to flash in a
quantum matter mode. In our ETAMFP model
the statistical quantum flashes per time
together with the a distribution function would define
the density distribution $ \rho (r) $ as function of
the radius $ r $. The total integral would allow to calculate
the total mass of the FP and similar as the charges would an
overlaid rotational speed of the flashes generate an
angular momentum and the integral the total spin. Again the
geometrical extension of such a mass quantum must be smaller
as the size of a FP. We assume as an ad hoc hypothesis
the frequency of these
quantum of flashes is high enough that it can be calculated
in a classical ansatz. This approach we will follow
in the next sections of this paper.

\subsubsection { Conclusion of the empirical toy ansatz about
                 a microstructure of a fundamental particle }

We first discussed the experimental paradox to measure very
small distances. In the case of the point like Standard Model FPs
it would be necessary to measure distances down to zero.
This unambiguously request infinite high test energies,
the CM energy of the experiment will cross the
GUT scale and Planck scale. The logic consequence is,
the whole physics conditions according to the energy scale
of the SM- or BBM-model change to the physics laws existing
at this scales. The experiment must be performed below
the Planck time of $ t = 10^{-44} $ and the Planck length
of $ 10^{-33} $ cm. 
To reconstruct for a certain scale $ \mu $ in the SM and
BBM model the same physical conditions it is essential
that both model fulfill time reversed invariance.
This is confirmed from experiment up to the LEP
energies. For our discussion we assume this is correct
up to Planck Era.
The logical consequence of this fact is, the universe
remembers how it was created.
In our empirical toy ansatz ETAMFP
about a microstructure of a
fundamental particle we propose that the FPs contain every energy
state a cross of its radius which the Universe passed through
during its evolution.
In this scenario is the geometrical structure of a Fundamental Particle
the direct consequence of the development of the 
primordial inhomogeneities of the Planck Era up to the era today.
Still a relict of the condensate of the Planck Era, the mass
shell of the GUT Era and the following charges together
with remains of the size of the Planck scale and GUT
scale exist today
inside every Fundamental Particle.
The ETAMFP model changes the point like FPs of the SM
to geometrical extended objects. It follows
the evolution of the
universe in the BBM model in particular the primordial
nucleosynthesis, but the FPs are not points any more.

Second we developed an approach of a geometrical extended
fundamental particle. We use the obvious proposal that
the SM and BBM describe the same nature and use the
common link of the scale $ \mu $ between both models.
This links the scale dependence of the couplings
of the SM model $ \alpha_{1} $, $ \alpha_{2} $ and $ \alpha_{3} $
together with the time $ t $ after Big Bang. In this case it
is possible to read the microstructure of the FPs,  
as a remain of the time development of the early 
universe direct from the SM and BBM model
in three major steps.
At the time window $ 0 < t < t_{ Planck } $ a condensate of
mass and the three charges color, electromagnetic and weak
get created including the geometrical size of the phase transition
Planck to GUT Era,
between $ t_{ Planck } < t < t_{ GUT } $ mass
creation dominates the geometrical size of the phase transition
GUT to era today get stored
and finally at the window $ t_{ GUT } < t < t_{ FP } $
the primordial inhomogeneities carried over from the Planck Era
get equipped with the three charges color,
electromagnetic and weak according the three
accordingly interactions strong, electro magnetic and weak.
The whole structure follows the inflation and ends at the 
FP Era as fundamental particles.

Third we discussed links between continuous symmetries
gravitation and spin. After the discussion of
four possible links to associate the gravitational interaction
to the time of the four dimensional space-time we
conclude that the three interactions
strong, electro magnetic and weak are dominant in the three
dimensional space and the fourth interaction
the gravitation is linked the time.
Following the timing of the the SM- or BBM-model we
assume the spin appears in the universe in the time
window between Planck and GUT scale.

Fourth we search for a link between forces acting inside a FP
and its stability. As for example the electron is a highly stable
particle the radius dependence of the forces inside the electron
must lead to a very highly stable condition. It turns out that
only stable conditions are possible if the known forces strong,
electro magnetic, weak and gravitation change from repulsive
to attractive or vice versa close or inside the FP where the
strength of the gravitational force is similar or even dominating.
The discussion about the radius behavior of the four
forces shows that the above scenario generates six stable
geometrical positions to place charges at approximately $ r=r_{GUT} $
plus one position for each charge in the center of the FPs
including a stable mass core. 
The geometrical locations of the different charges
would be explained be simple geometrical reasons and a
energy minimum of the FP mass.

Finally fifth we investigate the question, why the rotating
highly accelerated charge of the ETAMFP model does not radiate
electro magnetic energy? We conclude from the electron with a
life time of $ \tau > 4.6^{~26}$ yr a radiation free path similar
postulated in the Bohr model must exist. We use the stable Hydrogen atom
to search for a microscopic picture how an electron in a p-state
is able to circle the center of the Atom without radiation.
We introduce the flashing vacuum as candidate to give the electron
the possibility to appear and disappear statistical to form
the well known probability distribution $ | \psi |^2 = f ( r ) $.
We extend the in the SM and BBM model defined vacuum properties
by the possibility
the vacuum is fluctuating close to a FP similar to the Hawking radiation
put with the difference that one pair partner is not escaping
from the FP it just appearing statistical at a certain 
location, whereas his counter-partner stays in the vacuum, similar to
a mirror particle in a Dirac sea or it
oscillates between real and vacuum state.
For a working hypothesis we
assume the frequency of quantum flashes is big
enough that we can use it for a classical ansatz of a radiation free
path of the charges circling around the center of the FP.

The geometrical structure in our ETAMFP model is generic every 
extended stable or semi stable structure originated from the
primordial inhomogeneities or quantum fluctuations in the
Planck Era are fundamental. In the introduction of
chap.{\ref{sec:Possible micro structure of Fundamental Particles}}
we summarized the main various structures what are under
discussion today. All the discussed particle like structures
should be indicated as fundamental particles. To simplify
the discussion in the followings chapters we like to
follow the common habit to restrict the name of fundamental
particles to fermions, bosons and higgs.

\subsection{ Scheme of geometrical extended fundamental
             particles and anti-particles }

To confront the ETAMFP model of an geometrical extended FP
described in Fig.{\ref{BasicSCHEMEFP2006}} we test first
is the model able to sort all known fundamental particle ( Fermions ),
fundamental anti-particle ( Anti-fermions )  and interaction particles
( Bosons ) in one common scheme.

\subsubsection { The charges of fundamental particles }
\label{sec:The charges of fundamental particles}

After the experiment of R. A. Millikan (1911) \cite { Millikan }
we learned from experiment most of fundamental particle carries charges.
The charges of the FPs of the four interactions assigned in Fig.{\ref{fig.3}
localize geometrical the location of the source of the interaction
and the strength of the strong, electromagnetic, weak and
gravitational interaction.

\subparagraph { A) The charge of the strong, electromagnetic and
                weak interaction. }

To test the scheme it is first necessary to
summerize the charges of the
standard theory at a low scale at about $ M_{W} $
as displayed for example in Fig.{\ref{fig.4}.
The strong interaction is described by the three
charges red ( R ), green ( G ) and blue ( B ). The most fundamental charge of
the electromagnetic interaction is $ Q= \pm 1/3 $.
The charge of the weak interaction is the weak isospin $ T $ and $ T_{3}$.
The hypercharge quantum numbers $ Y $ is a function of the electric charge
and the weak isospin $ T_{3}$ according Eq.\ref{WeakQN}.

\begin{equation}
Q = T_{3} + \frac {Y}{2}
\label{WeakQN}
\end{equation}

Besides the colors all the charges of the fermions are summarized
in Tab. \ref{Hypercharge} of ref. \cite{HalsenMartin}.

%
\begin{table}
\caption{Hypercharge quantum numbers of Lepton and Quarks.}
\label{Hypercharge}
\begin{center}
\begin{tabular}{||l|l|l|l|l|l|l|l|l|l||}                         \hline
$Lepton$   &$T$&$T_{3}$&$Q$&$Y$&$Quark$&$T$&$T_{3}$&$Q$&$Y$
                                                            \\  \hline
 $\nu_{e}$ &$1/2$&$1/2$&$0$&$-1$&$u_{L}$&$1/2$&$1/2$&$2/3$&$1/3$
                                                            \\  \hline
$e^{-}_{L}$&$1/2$&$-1/2$&$-1$&$-1$&$d_{L}$&$1/2$&$-1/2$&$-1/3$&$1/3$
                                                            \\  \hline
           &     &      &    &    &$u_{R}$&$0  $&$0   $&$2/3 $&$4/3$
                                                            \\  \hline
$e^{-}_{R}$&$0  $&$0   $&$-1$&$-2$&$d_{R}$&$0  $&$0   $&$-1/3$&$-2/3$
                                                            \\  \hline
\end{tabular}
\end{center}
\end{table}

\subparagraph { B) The pseudo charge of the gravitational interaction.}

The model in Fig.{\ref{BasicSCHEMEFP2006}} describes so far in three
dimensions how to place on the x-axis the electromagnetic charge
$ Q $, on the y-axis the weak charge $ T_{3} $ and on the z-axis
the color charge $ C $. As we know that the SM and BBM model
uses four dimensions it is essential to study which type
of charge belongs to the fourth dimension the time.
After we discussed in the chapter about links between continuous
symmetries gravitation and spin we
reached the conclusion to link in this paper the time to the
gravitational interaction. Two indications point to the mass
as pseudo charge of the gravitational interaction.

First we consider the definition of a charge as source of a
force field in field theory. The field in the environment
of a charge contains energy. If we investigate the Einstein
equation in Tab. \ref{linksSMBigBang}
and ignore for the discussion the
cosmological term $\Lambda $ energy will implement energy stress
tensors $ T_{k}^{i} \not= 0 $. It means also the curvature
will be not zero. As consequence in the environment
of the charge after the Einstein equation mass must exist.
If we assume a charge is the source of a force field and the Einstein
equations are valid for our case a charge can not exist without mass.
The fact is unique for all known charges color, electric and weak.
As consequence fundamental particles ( fermions and bosons ) 
which carry on type of this charge must also carry mass.

A special case is the mass them self in the Einstein equations. If we
extent the definition of a charge as source of a gravitational field
on the mass, the mass get a double significance. The mass would behave
just as mass in ( gr ) in our usually definition but would be at the
same time a charge like the color, electric and weak charge.
This self interacting phenomenon will equip every spatial
energy accumulation of a field with mass, in particular
the gamma and Higgs. The gamma is a hybrid because it is a
spatial energy accumulation of electromagnetic character, it
carries via $ E = m \times c^{2} $ induced mass and is
electromagnetic. It is sensitive via the induced mass to gravitational
interaction and the gauge particle for the electromagnetic interaction.

These considerations unify the known four interactions strong,
electro magnetic, weak and gravitation with four charges color,
electric, weak and mass to one scheme and at the same time connect
the four interactions and charges to the four coordinates as discussed
in the chapter geometrical approach.

Second is the mathematical
structure of the forces $ F_{Q} $ between two charges $ Q_{1} $
and $ Q_{2} $ in the Coulomb law Eq.\ref{Coulombforce},

\begin{equation}
F_{Q} = \frac{1}{4\pi \varepsilon }\frac{Q_{1}Q_{2}}{r^{2}}
\label{Coulombforce}
\end{equation}

the force $ F_{M} $  between two magnetic poles $p_{1}$ and $p_{2}$
Eq.\ref{MagnetPOL}

\begin{equation}
F_{M} = \frac{1}{4\pi \mu }\frac{p_{1}p_{2}}{r^{2}}
\label{MagnetPOL}
\end{equation}

and the the gravitational force $ F_{G} $ between two masses
$ m_{1} $ and $ m_{2} $ Eq.\ref{GravitationalFORCE}

\begin{equation}
F_{G} = G \frac{m_{1}m_{2}}{r^{2}}
\label{GravitationalFORCE}
\end{equation}

absolute analog. But the analogy between the charges $ C $, $ Q $, $ T_{3} $
and the mass is not general valid. For example is the amount of the
electric charge $ Q $ not depending about the velocity if the FPs
in contrary to the rest mass of the particle which follows the
Lorentz-Transformation, which pushes the mass to infinite if
the velocity of the particle approaches the velocity of light.

For a working hypothesis we use in the following discussion of
this paper the mass as pseudo-charge
of the gravitational force. The extention of the FP can be originated
from the size of the mass kernel and/or the size of the color,
electric and weak charge. Without limiting the general discussion
we assume that the color, electric and weak charge is point-like
and the whole extention of the FP is originated from the density
distribution of the mass $ \rho = f ( t,r ) $ as function
of time $ t $, radius $ r $ and the geometrical position $ r_{0} $
the charges $ C $ , $ Q $ and $ T_{3} $ of the FPs get placed.

\subparagraph { C) Effect of charges for a geometrical extended object. }

An electric charge located at the border of an extended
object with spin will generate a magnetic moment and under
certain conditions an electric dipole moment. Inspecting the
basic scheme of a FP in Fig.{\ref{BasicSCHEMEFP2006}} it is
possible to assume the principle for the electric charge is also
correct for the color, weak and mass charge. This assumption
would have the consequence that the four charges generate
four moments and for dipole moments. Originated from the
color charge a color moment, from the electric charge
a magnetic moment, from the weak charge a weak moment
and from the mass a mass moment. The same charges
are able to generate the equivalent color-, electric-,weak-
and mass-dipole moment. The FPs carry usually not only
one charge for example in the quark case color, electric,
weak and mass charge. This would lead to the the possibility
of mixed moments and dipole moments.

So fare experimental evidence exists for the magnetic moment
of the electron and muon \cite{MagneticMomentElecMoun} and a hint
about the electric dipole-moment of the electron and muon
\cite{MagneticMomentElecMoun}. In addition experimental
limits and
discussions about quark color magnetic moments and weak
magnetic moments are on the way \cite{SpeculationMoments}.
The introduced scheme in Fig.{\ref{BasicSCHEMEFP2006}} respects
the experimental knowledge we have today. The color charges
on the spin axis ( z-axis ) would generate a small or zero
color moment, the electric charges on the x-axis would generate
a magnetic moment and the weak charges on the y-axis would generate
a weak moment. According Fig.{\ref{BasicSCHEMEFP2006}} we assume
for the mass a geometrical distribution of a mass density
$ \rho ( t,r ) $ together with the spin-rotation will this
generate an angular momentum the spin. It will not generate
a mass-dipole moment unless the mass distribution is geometrical
somehow clustering. In the Einstein equations a mass-dipole moment
is forbidden, which implements the assumption the mass density
distributions is not clustering.

\subsubsection { Extension of the geometrical
                 approach to four dimensions }

To extend the three-dimensional scheme of Fig.\ref{BasicSCHEMEFP2006}
to four dimensions, we introduce the scheme for an electron in
Fig.\ref{Electron2006}. The three Cartesian coordinates are assigned
at the left side of the figure by black solid lines. Left of the line
are written down the three coordinates x, y and z and on the
right the attached charges $ Q $, $ T_{3} $ and $ C $. The fourth
coordinate is shown on the right side. Left the time and right the
pseudo charge mass.
The geometrical center of the FP is a green dot.
The geometrical position $ r_{0} $ of the charges
$ Q $, $ C $ and $ T_{3} $ is symbolized by the length of the black
line. Above the black line for a coordinate is shown the
amount of the charge at the position left, center and right in
blue numbers. The integral of the charge measured far away
from the FP is shown below the black line in red color.
The common used symbol for the FP is displayed in the center
of the figure above a big red dot. The time coordinate
on the right side of the scheme has only a center ( gree dot ) and
a black line to right, symbolizing the fact that negative absolute
time does not exist. Above the black line a mass is assigned
in blue color if the particle is stable. Below a mass is
assigned if the particle has a rest mass. Mass assigned
$ + m $ means matter and $ -m $ means antimatter. The length
of the black time line symbolizes that the FP has a mass.
We do not distinguish between rest mass and energy.

\begin{figure}
\begin{center}
 \epsfig{file=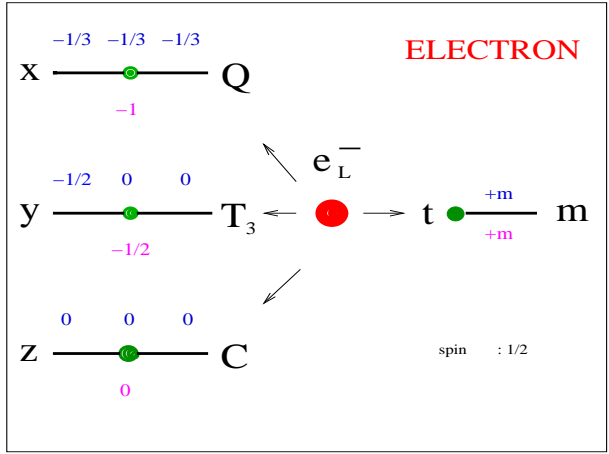,width=8.0cm,height=6.0cm}
\end{center}
\caption{ Scheme of electron. }
\label{Electron2006}
\end{figure}

Before we apply the scheme on all FPs we demonstrate
how to use the scheme to describe for example
an electron in detail as shown in Fig.\ref{Electron2006}.
At the $ M_{W} $ scale four
independent interactions exist and four orthogonal 
coordinates are available. Following the discussion of section,
empirical toy ansatz about a microstructure
of a fundamental particle,
the mass kernel of the FP will be time like and the strong,
electromagnetic and weak interaction space like. The spatial
extended mass kernel together with the spin of the FP assigns
for every axis three distinguished positions to place a point like
charge, as shown in Fig.\ref{BasicSCHEMEFP2006}.
Two at the border of the mass kernel and one in the kernel.
The spin axis of the FP we coincide as distinguished axis with
the z direction. As discussed we keep the z-axis free for the
color charge, place the weak charge on the y-axis and place the
electric charge on the x-axis. The electron carries no color charge
which sets the charge $ C $ on the z-axis to zero.
The weak
hypercharge $ Y $ ranges from $ 0 $ to $ 4/3 $ and is according
Eq.\ref{WeakQN} a function of the charge $ Q $ and the
weak isospin $ T_{3}$ which ranges from $ 0 $ to $ \pm 1 $
in units of $ 1/2 $.
These number is more fundamental and for this reason used
as quantum number for the discussed scheme.
We have the possibility to place the weak charges $ T_{3} $
what could be $ 0 $ , $ \pm 1/2 $ and $\pm 1 $ on the y-axis.
In case of the electron $ T_{3} = - 1/2 $. Three possible
positions left center and right are open. If the electron would
carry a weak moment it would be necessary to put
the charge $ T_{3} = - 1/2 $ left or right of the y-axis. As this is
not excluded so far we use the position y-axis left. The total
charge measured from far outside the electron would be
also $ T_{3} = - 1/2 $.
The electric charge $ 1/3 $ must be placed on the x-axis or y-axis because
only in this position it is possible to generate together with
the spin a magnetic moment and electric dipole moment. We choose
the x-axis. On the x-axis all the three open positions will be
occupied with $ Q = - 1/3 $.
The total charge measured from a spectator far outside
the mass kernel would be the integral of all three charges
$ Q = - 1 $. 

The real geometrical position of the charges will be
usually not
precisely on the x-axis or y-axis because the
forces between the charges will place the charges
in a position of an energetic minimum discussed in
chap.\ref{sec:Forces and stability}. 
For example in the case of the electron
would only a second charge $ T_{3} $ on the free
position of the y-axis, lead to a total balanced
symmetric system which would place the different charges
precisely on the different axis. Our interest is focused
in the following chapters on a schematic diagram. We
ignore for this reason this detail.

The electron is a very stable particle for this reason
the mass is assigned with $ + m $ on the time axis above in blue
and below in red because this mass is also the rest mass of
the electron. The electron is a fermion with spin $ 1/2 $ as
shown in Fig.\ref{Electron2006} right low corner.

\subsubsection { Scheme of the lightest left handed
                 fundamental particles. }
\label{sec:Scheme of the lightest left handed FP}

In Fig.\ref{AllleftFP2006} are displayed the scheme of all
the four left handed lightest FPs.
\begin{figure}
\begin{center}
 \epsfig{file=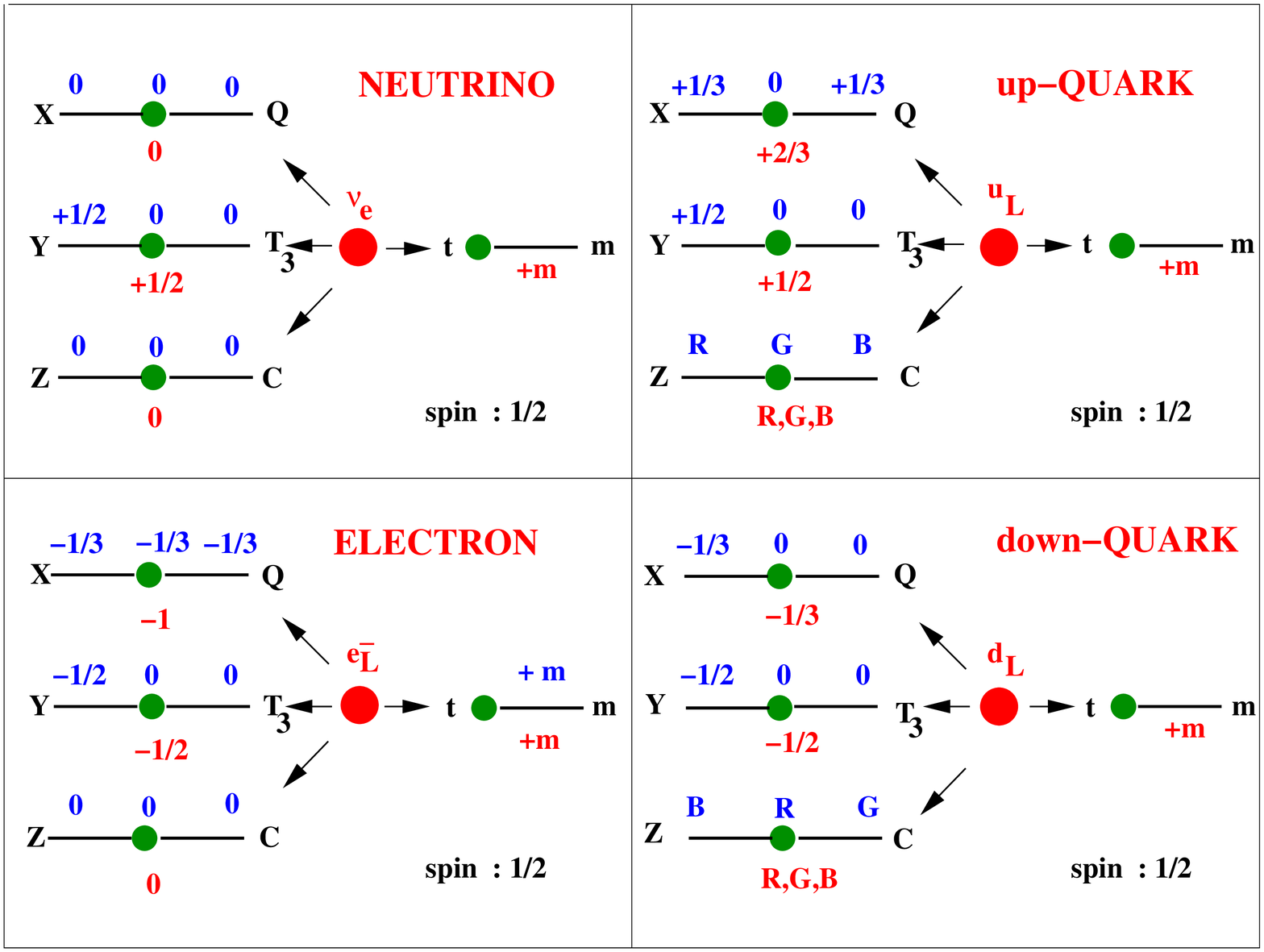,width=13.0cm,height=8.0cm}
\end{center}
\caption{ Scheme of the lightest left handed fundamental particles. }
\label{AllleftFP2006}
\end{figure}
The experimental results for the electron neutrino $ \nu_{e} $
are $ Q = C = 0 $, the mass is $ m < 3~ eV $ \cite{Magmomentneutrino}
and $ T_{3} = + 1/2 $.
The spin is $ 1/2 $ and the particle is left-handed. We choose
for this reason all the electric charges $ Q $ zero in all possible
positions above the x-axis and accordingly also the integral charge zero
below the x-axis.
A  limit for the magnetic moment of
$ \mu < 1.0 \times 10^{-10} \mu_{B} $ \cite{Magmomentneutrino}
for the electron neutrino $ \nu_{e} $ exist,
what opens the possibility the $ \nu_{e} $ carries electric
charges. In this case the $\nu_{e}$ could be composed
in a charge $ Q=-1/3 $ and an anti-charge $Q=+1/3 $ placed
above at two of the three free positions of the x - axis. The scheme
would be still in agreement with the experiment because the
integrated charge below the x-axis $ Q $ would be zero.
For the weak charge $ T_{3} = + 1/2 $ three possible positions
above the y-axis: left, center and right are open.
If the $ \nu_{e} $ would
carry a weak moment it would be necessary to put
the charge $ T_{3} = + 1/2 $ left or right above the y-axis. As this is
not excluded so far we use the position y-axis left above. If the
$ \nu_{e} $ would carry no weak moment the charge
$ T_{3} = + 1/2 $ must be placed
in the center position above the y-axis. The total
weak charge of the $ \nu_{e} $ measured from far outside
is also $ T_{3} = + 1/2 $ shown below the y-axis. The $ \nu_{e} $ carries no
color what sets the color charge $ C = 0 $ above and below the z-axis.
The experiments about life time and rest mass of the
$ \nu_{e} $ are not definite for the time being
\cite{Magmomentneutrino} we choose a non stable particle with
rest mass $+m $ below the t-axis.

The left handed electron $ e^{-}_{L} $ we discussed just in very
much detail in chapter 5.2.2 for this reason we repeat
only the main points of these discussion. The particle
carries three electric charges $ Q = - 1/3 $ we placed on the
three free position above the x-axis, the integrated charge is
$ Q = -1 $ below the x-axis , $ T_{3} = - 1/2 $ placed left above
the y-axis and the integrated charge is also
$ T_{3} = - 1/2 $ placed below the y-axis and the
color is zero consequently are all color above and below the
z-axis zero.
The particle has spin $ 1/2 $, with a mass of
$ m= 0.51 $ MeV and is very stable \cite{MagneticMomentElecMoun},
for this reason is the mass assigned $ + m $ above and below the
t-axis.
The scheme request in agreement with the experiment
\cite{MagneticMomentElecMoun} that the electron generates
a magnetic moment, electric dipole moments and weak moments are
possible.

The left-handed up-Quark $ u_{L} $ is composed
out of two anti charges $ Q=+1/3 $
which could be placed on two of the three free positions above the
x-axis. The integrated charge is $ Q = + 2/3 $ shown below the x-axis.
All the possible positions of this two charges would to lead
a magnetic moment of the  $ u_{L} $.
Similar like the for the $ \nu_{e} $, for the weak charge of the $ u_{L} $
$ T_{3} = + 1/2 $ three possible positions
above the y-axis left, center and right are open.
If the $ u_{L} $ would
carry a weak moment it would be necessary to put
the charge $ T_{3} = + 1/2 $ left or right above the y-axis. As this is
not excluded so far we use the position y-axis left above
\cite{SpeculationMoments}. If the
$ u_{L} $ would carry no weak moment the charge
$ T_{3} = + 1/2 $ must be placed
in the center position above the y-axis. The total
weak charge of the $ u_{L} $ measured from far outside
is also $ T_{3} = + 1/2 $ shown below the y-axis.
The color could be places above the z-axis
in total 18 combinations ( $ RRR $, $ GGG $, $ BBB $, $ RBG $,
$ RGB $, $ BRG $, $ BGR $, $ GRB $, $ GBR $, $ RGG $, $ GRG $,
$ GGR $, $ RBB $, $ BRB $, $ BBR $, $ BRR $, $ RBR $ and $ RRB $ ).
The color is assigned symbolic $ R $, $ G $, $ B $ above the
z-axis. The integrated total possible 18 charge combinations
are shown symbolic $ ( R,G,B ) $ below the z-axis. The
color charges could generate color moments.
The particle has spin $ 1/2 $, with a rest mass between
$ m= 1.5~ to~4 $ MeV \cite{Quarkmass} and is stable only in a nucleus.
We assigned it non stable with rest mass $ + m $ below the t-axis.

The left-handed down-Quark $ d_{L} $ is
composed out of one charges $ Q=-1/3 $
which could be placed at one of the three free positions above the
x-axis. The choice of the position will generate on the mass kernel
no magnetic moment and at the border a magnetic moment with opposite
sign to the up-Quark $ u_{L} $. If the position of the two charges
of the up-Quark $ u_{L} $ is like in Fig.\ref{AllleftFP2006} the magnetic
moment of the up-Quark $ u_{L} $ must be $ |\mu_{u}|~>|\mu_{d}| $.
Similar like for the $ \nu_{e} $, for the weak charge of the $ d_{L} $
is $ T_{3} = - 1/2 $. Three possible positions
above the y-axis ( left, center and right )  are open.
If the $ d_{L} $ would
carry a weak moment it would be necessary to put
the charge $ T_{3} = - 1/2 $ left or right above the y-axis.
Depending about further experiments
\cite{SpeculationMoments} it would be possible to discover
a weak moment or exclude it. In case of the exclusion the position
of the weak charge must change accordingly to the center.
The total
weak charge of the $ d_{L} $ measured from far outside
is also $ T_{3} = - 1/2 $ shown below the y-axis.
Like for the $ u_{L} $ quark the color could be places above the z-axis
in total 18 combinations ( $ RRR $, $ GGG $, $ BBB $, $ RBG $,
$ RGB $, $ BRG $, $ BGR $, $ GRB $, $ GBR $, $ RGG $, $ GRG $,
$ GGR $, $ RBB $, $ BRB $, $ BBR $, $ BRR $, $ RBR $ and $ RRB $ ).
The color is assigned symbolic $ R $, $ G $, $ B $ above the
z-axis. The integrated total possible 18 charge combinations
are shown symbolic $ ( R,G,B ) $ below the z-axis. The
color charges could generate color moments.
The particle has spin $ 1/2 $, with a rest mass between
$ m= 4~ to~8 $ MeV \cite{Quarkmass} and is stable only in a nucleus.
We assigned it non stable with rest mass $ + m $ below the t-axis.

\subsubsection { Scheme of the lightest right handed
                 fundamental particles }
\label{sec:Scheme of the lightest right handed FP}

In Fig.\ref{AllrightFP2006} are displayed the scheme of all
the right handed lightest FPs. The change from left handed
FPs to right handed FPs is performed by the change of the
weak charge $ T_{3} $ from $ T_{3}=|1/2| $ to $ T_{3}=0 $.
The rest of the quantum numbers stays the same. For this reason
we discuss the scheme of the right handed FPs only brief.

\begin{figure}
\begin{center}
 \epsfig{file=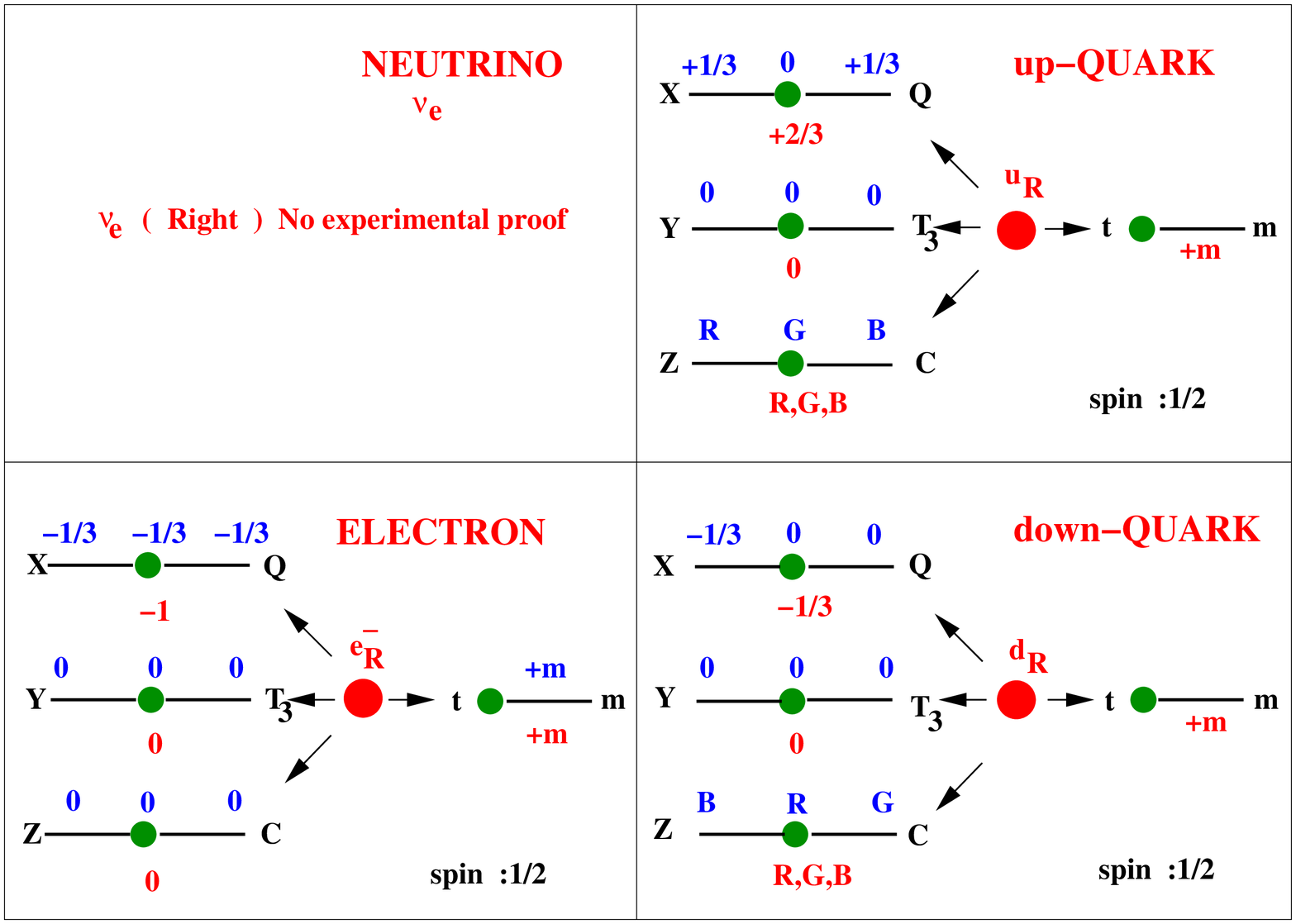,width=13.0cm,height=8.0cm}
\end{center}
\caption{ Scheme of the lightest right handed fundamental particles. }
\label{AllrightFP2006}
\end{figure}

All experiments proofed that no right handed neutrino exist.
For this reason is the scheme for the $ \nu_{e}(right) $
empty.

The right handed electron $ e^{-}_{R} $
carries three electric charges $ Q = - 1/3 $ we placed on the
three free position above the x-axis, the integrated charge is
$ Q = -1 $ shown below the x-axis.
The weak charge $ T_{3} = 0 $ and accordingly the integrated charge
$ T_{3} = 0 $ and the color is zero. For this reason are all
weak and color charges above and below the y-axis and z-axis
zero. The particle has spin $ 1/2 $, with a mass of
$ m= 0.51 $ MeV and is very stable \cite{MagneticMomentElecMoun}
for this reason we assign the mass $ + m $ above and below
the t-axis.
The scheme request in agreement with the experiment
\cite{MagneticMomentElecMoun} that the electron generates
a magnetic moment, electric dipole are possible.

The right-handed up-Quark $ u_{R} $ is composed
out of two anti charges $ Q=+1/3 $
which could be placed on two of the three free positions above the
x-axis. The integrated charge is $ Q = + 2/3 $ shown below the x-axis.
All the possible positions of this two charges would to lead
a magnetic moment of the  $ u_{R} $.
The weak charge of the $ u_{R} $ is $ T_{3} = 0 $. Accordingly we set
all weak charges above and the integrated weak charge below the
y-axis to zero.
The color could be places above the z-axis
in total 18 combinations ( $ RRR $, $ GGG $, $ BBB $, $ RBG $,
$ RGB $, $ BRG $, $ BGR $, $ GRB $, $ GBR $, $ RGG $, $ GRG $,
$ GGR $, $ RBB $, $ BRB $, $ BBR $, $ BRR $, $ RBR $ and $ RRB $ ).
The color is assigned symbolic $ R $, $ G $, $ B $ above the
z-axis. The integrated total possible 18 charge combinations
are shown symbolic $ ( R,G,B ) $ below the z-axis. The
color charges could generate color moments.
The particle has spin $ 1/2 $, with a rest mass between
$ m= 1.5~ to~4 $ MeV \cite{Quarkmass} and is stable only in a nucleus.
We assigned it non stable with rest mass $ + m $ below the t-axis.

The right-handed down-Quark $ d_{R} $ is
composed out of one charges $ Q=-1/3 $
which could be placed at one of the three free positions above the
x-axis. The choice of the position will generate on the mass kernel
no magnetic moment and at the border a magnetic moment with opposite
sign to the up-Quark $ u_{R} $. If the position of the two charges
of the up-Quark $ u_{R} $ is like in Fig.\ref{AllrightFP2006} the magnetic
moment of the up-Quark $ u_{R} $ must be $ |\mu_{u}|~>|\mu_{d}| $.
The weak charge of the $ d_{R} $ is $ T_{3} = 0 $. Accordingly we set
all weak charges above and the integrated weak charge below the
y-axis to zero.
Like for the $ u_{R} $ quark the color could be places above the z-axis
in total 18 combinations ( $ RRR $, $ GGG $, $ BBB $, $ RBG $,
$ RGB $, $ BRG $, $ BGR $, $ GRB $, $ GBR $, $ RGG $, $ GRG $,
$ GGR $, $ RBB $, $ BRB $, $ BBR $, $ BRR $, $ RBR $ and $ RRB $ ).
The color is assigned symbolic $ R $, $ G $, $ B $ above the
z-axis. The integrated total possible 18 charge combinations
are shown symbolic $ ( R,G,B ) $ below the z-axis. The
color charges could generate color moments.
The particle has spin $ 1/2 $, with a rest mass between
$ m= 4~ to~8 $ MeV \cite{Quarkmass} and is stable only in a nucleus.
We assigned it non stable with rest mass $ + m $ below the t-axis.

\subsubsection { Scheme of the lightest right handed
                 anti fundamental particles }
\label{sec:Scheme of the lightest right handed anti FP}

In Fig.\ref{AllrightantiFP2006} are displayed the scheme of all
the right handed lightest anti-FPs.
\begin{figure}
\begin{center}
 \epsfig{file=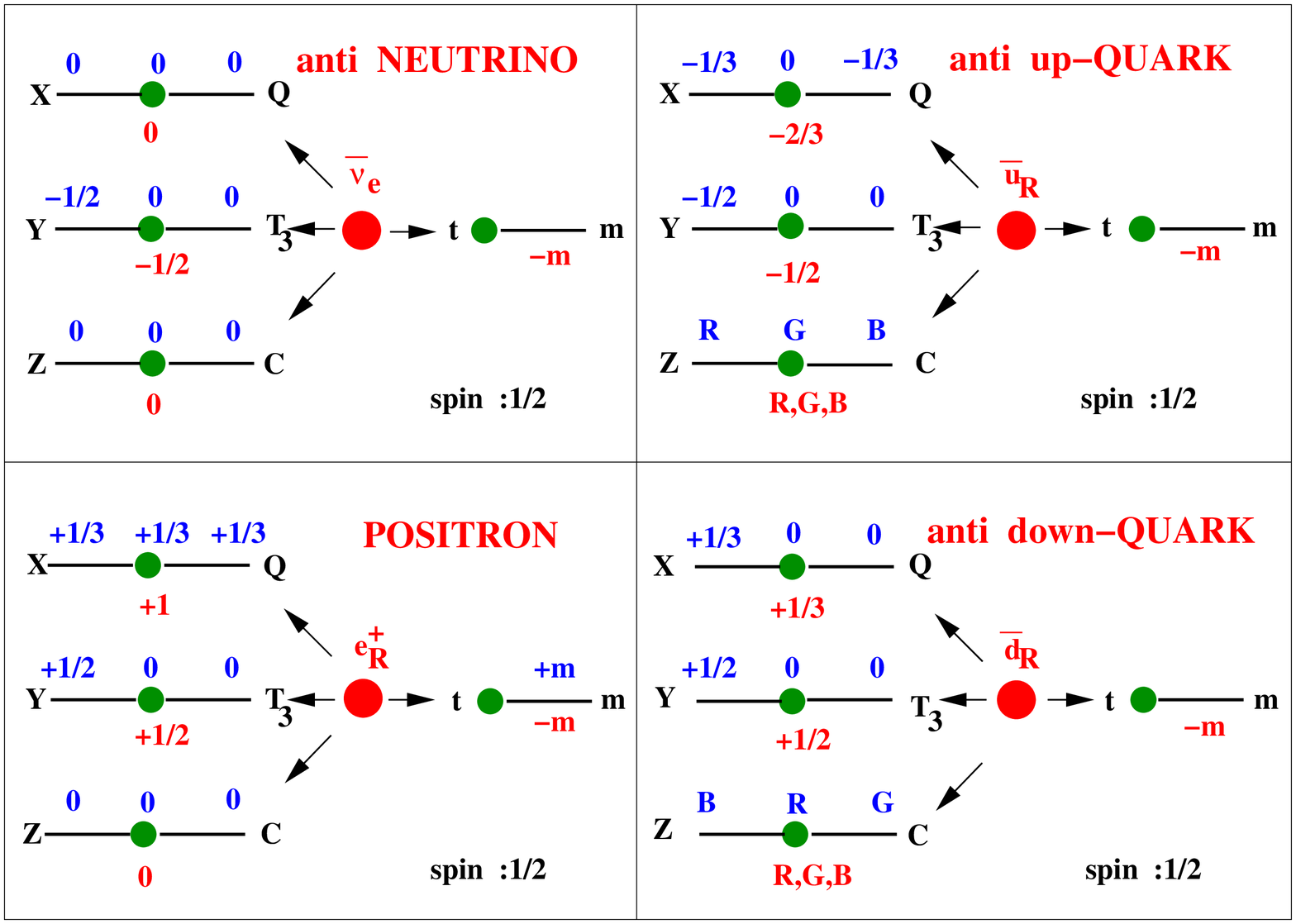,width=13.0cm,height=8.0cm}
\end{center}
\caption{ Scheme of the lightest right handed anti fundamental particles. }
\label{AllrightantiFP2006}
\end{figure}
The experimental results for the right handed anti electron neutrino
$ \bar{\nu}_{e} $ are $ Q = C = 0 $, the mass is small
and $ T_{3} = - 1/2 $. We place the electric charge $ Q $ above the
x-axis to $ Q = 0 $ and the integral charge below also to $ Q = 0 $.
It is possible like for the left handed neutrino $\nu_{e} $
that the particle carries electric
charges. In this case the $\bar{\nu}_{e}$ could be composed
in a charge $ Q=-1/3 $ and an anti-charge $Q=+1/3 $ placed
on two of the three free positions of the x - axis. The scheme
would be still in agreement with the experiment because the
integrated charge $ Q $ would be zero.
The weak charge $ T_{3} = -1/2 $ we place above the left side of the y-axis.
As we have no experimental evidence of a weak moment is this
a random decision.
In this position the $ \bar{\nu}_{e} $ could generate a weak
moment, the same would happen in the right position. The center
position would generate a zero weak moment. The integrated
weak charge below the y-axis is $ T_{3} = -1/2 $.
The color charge $ C $ above the z-axis is zero also the integrated
color charge below the z-axis. The stability and mass of the particle
is not experimentally not settled. We assume an unstable particle
with a finite anti-rest mass and set the anti-mass $ -m $ below the t -axis.
Experimental are the neutrino-antineutrino intrinsic properties not
settled. It is not known is each neutrino mass eigenstate
$ \nu_{i} $ identical to its antiparticle $ \bar{\nu}_{i} $ or
distinct from it, for $ \nu_{i}=\bar{\nu}_{i} $ we would
have Majorana particles and for $ \nu_{i} \neq \bar{\nu}_{i}$
we would call it Dirac particles \cite{Neutrino-Antineutrino}.

The right-handed positron $ e^{+}_{R} $ is composed of
three anti charges $ +1/3 $
assigned above the x-axis and the integrated charge $ Q=+1 $ below
x-axis. Similar to the left-handed electron $ e^{-}_{L} $
it carries a weak charge $ | T_{3} | = 1/2 $ but with positive
sign $ T_{3}  = + 1/2 $. Three possible
positions ( left, center and right ) are open. If the positron would
carry a weak moment it would be necessary to put
the charge $ T_{3} = + 1/2 $ left or right above the y-axis. As this is
not excluded so far we use the position of he y-axis left above. The total
weak charge measured from far outside the positron would be
also $ T_{3} = + 1/2 $ shown below the y-axis. The positron carries no
color what sets the color charge $ C = 0 $ above and below the z-axis.
The particle has spin $ 1/2 $, with a anti-mass of
$ m= 0.51 $ MeV and is very stable ,
for this reason is the mass assigned $ + m $ above and the anti-mass $ -m $
below the t-axis. The scheme request that the positron generates
a magnetic moment like the electron $ e^{-}_{L} $ an
electric dipole moment is possible.

The right-handed anti-up-Quark $ \bar{u}_{R} $ is composed
out of two charges $ Q=-1/3 $
which could be placed on two of the three free positions above the
x-axis. The integrated charge is $ Q = - 2/3 $ shown below the x-axis.
All the possible positions of this two charges would to lead
a magnetic moment of the  $ \bar{u}_{R} $.
The weak charge of the $ \bar{u}_{R} $
$ T_{3} = - 1/2 $. Three possible positions
above the y-axis ( left, center and right ) are open.
If the $ \bar{u}_{R} $ would
carry a weak moment it would be necessary to put
the charge $ T_{3} = - 1/2 $ left or right above the y-axis. As this is
not excluded so far we use the position y-axis left above
\cite{SpeculationMoments}. If the
$ \bar{u}_{R} $ would carry no weak moment the charge
$ T_{3} = - 1/2 $ must be placed
in the center position above the y-axis. The total
weak charge of the $ \bar{u}_{R} $ measured from far outside
is also $ T_{3} = - 1/2 $ shown below the y-axis.
The anti-color could be places above the z-axis
in total 18 combinations ( $ RRR $, $ GGG $, $ BBB $, $ RBG $,
$ RGB $, $ BRG $, $ BGR $, $ GRB $, $ GBR $, $ RGG $, $ GRG $,
$ GGR $, $ RBB $, $ BRB $, $ BBR $, $ BRR $, $ RBR $ and $ RRB $ ).
The anti-color is assigned symbolic $ R $, $ G $, $ B $ above the
z-axis. The integrated total possible 18 charge combinations
are shown symbolic $ ( R,G,B ) $ below the z-axis. For simplicity we
use in Fig.\ref{AllrightantiFP2006} for color and anti-color the
same symbol. The anti-color charges could generate anti-color moments.
The particle has spin $ 1/2 $, with an anti- rest mass between
$ m= 1.5~ to~4 $ MeV \cite{Quarkmass}.
We assigned it non stable with anti-rest mass $ - m $ below the t-axis.

The right-handed anti-down-Quark $ \bar{d}_{R} $ is
composed out of one anti-charge $ Q=+1/3 $
which could be placed at one of the three free positions above the
x-axis. The choice of the position will generate on the mass kernel
no magnetic moment and at the border a magnetic moment with opposite
sign to the anti-up Quark $ \bar{u}_{R} $. If the position of the two charges
of the anti-up-Quark $ \bar{u}_{R} $ is like in
Fig.\ref{AllrightantiFP2006} the magnetic
moment of the anti-up Quark $ \bar{u}_{R} $ must be
$ |\mu_{\bar{u}}|~>|\mu_{\bar{d}}| $.
Similar like for the $ \nu_{e} $, the weak charge of the $ \bar{d}_{R} $
is $ T_{3} = + 1/2 $. Three possible positions
above the y-axis ( left, center and right )  are open.
If the $ \bar{d}_{R} $ would
carry a weak moment it would be necessary to put
the charge $ T_{3} = + 1/2 $ left or right above the y-axis.
Depending about further experiments
\cite{SpeculationMoments} it would be possible to discover
a weak moment or exclude it. In case of the exclusion the position
of the weak charge must change accordingly to the center.
The total
weak charge of the $ \bar{d}_{R} $  measured from far outside
is also $ T_{3} = + 1/2 $ shown below the y-axis.
Like for the $ \bar{u}_{R} $ quark the anti-color could be places above the z-axis
in total 18 combinations ( $ RRR $, $ GGG $, $ BBB $, $ RBG $,
$ RGB $, $ BRG $, $ BGR $, $ GRB $, $ GBR $, $ RGG $, $ GRG $,
$ GGR $, $ RBB $, $ BRB $, $ BBR $, $ BRR $, $ RBR $ and $ RRB $ ).
The anti-color is assigned symbolic $ R $, $ G $, $ B $ above the
z-axis. The integrated total possible 18 charge combinations
are shown symbolic $ ( R,G,B ) $ below the z-axis.
For simplicity we
use in Fig.\ref{AllrightantiFP2006} for color and anti-color the
same symbol. The anti-color charges could generate anti-color moments.
The particle has spin $ 1/2 $, with an anti- rest mass between
$ m= 4~ to~8~$ MeV \cite{Quarkmass}.
We assigned it non stable with anti-rest mass $ - m $ below the t-axis.

\subsubsection { Scheme of the lightest left handed
                 anti fundamental particles }
\label{sec:Scheme of the lightest left handed anti FP}

In Fig.\ref{AllleftantiFP2006} are displayed the scheme of all
the left handed lightest anti-FPs. The change from right handed
anti-FPs to left-handed anti-FPs is performed by the change of the
weak charge $ T_{3} $ from $ T_{3}=|1/2| $ to $ T_{3}=0 $.
The rest of the quantum numbers stays the same. For this reason
we discuss the scheme of the right handed FPs only brief.

\begin{figure}
\begin{center}
 \epsfig{file=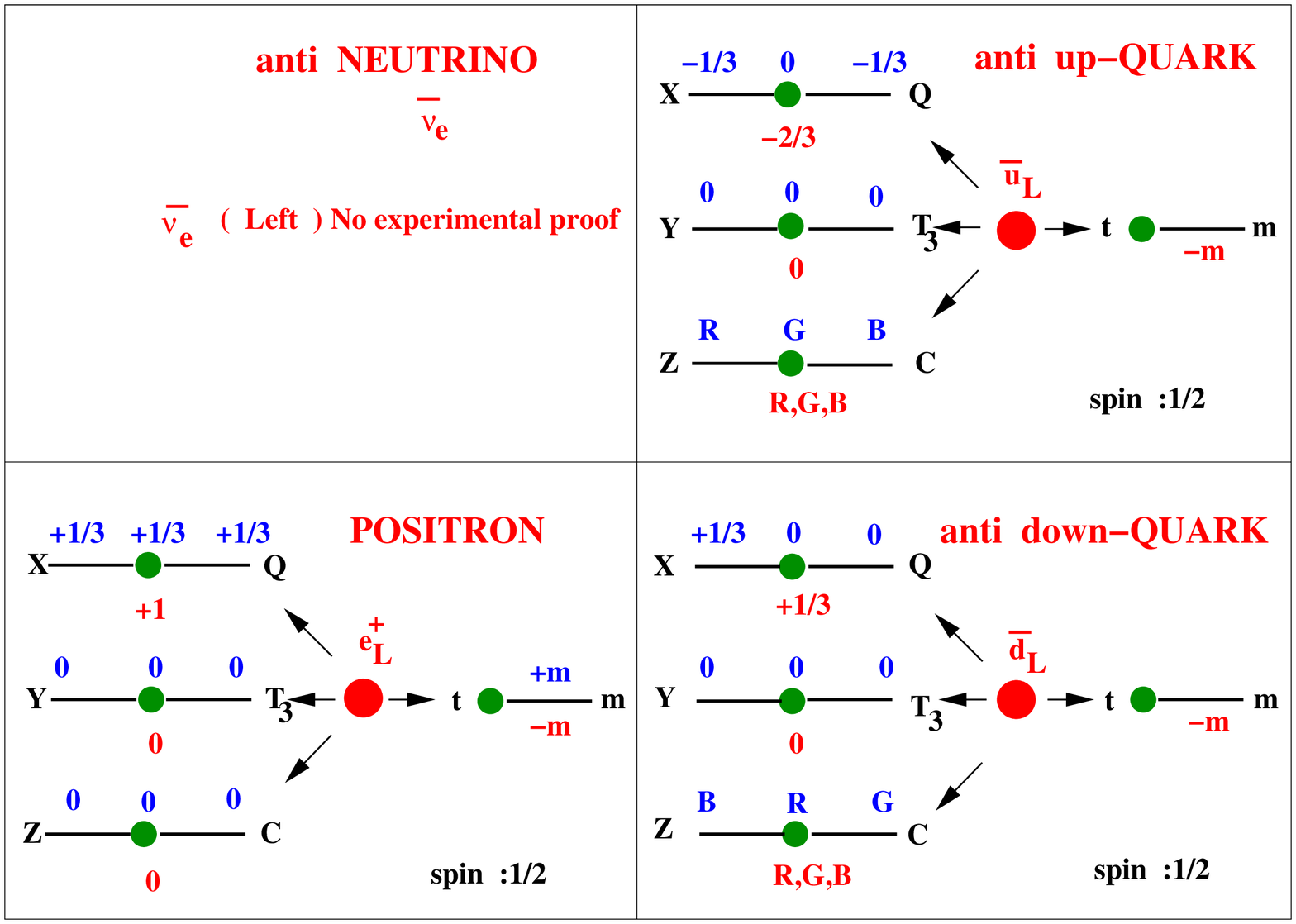,width=13.0cm,height=8.0cm}
\end{center}
\caption{ Scheme of the lightest left handed fundamental anti-particles. }
\label{AllleftantiFP2006}
\end{figure}

No experimental proof of a left handed anti-neutrino exist.
For this reason is the scheme for the $\bar{\nu}_{e}(left) $
empty.

The left handed positron $ e^{+}_{L} $
carries three electric anti-charges $ Q = + 1/3 $ we placed on the
three free position above the x-axis, the integrated charge is
$ Q = + 1 $ shown below the x-axis.
The weak charge $ T_{3} = 0 $ and accordingly the integrated charge
$ T_{3} = 0 $ and the color is zero. For this reason are all
weak and color charges above and below the y-axis and z-axis
zero. The particle has spin $ 1/2 $, with a mass of
$ m= 0.51 $ MeV and is very stable \cite{MagneticMomentElecMoun}
for this reason we assigned the mass $ + m $ above the t-axis and
with a finite anti-rest mass $ - m $ below the t-axis.
The scheme request that the electron generates
a magnetic moment an electric dipole would be possible.

The left-handed anti-up-Quark $ \bar{u}_{L} $ is composed
out of two charges $ Q=-1/3 $
which could be placed on two of the three free positions above the
x-axis. The integrated charge is $ Q = - 2/3 $ shown below the x-axis.
All the possible positions of this two charges would to lead
a magnetic moment of the  $ \bar{u}_{L} $ .
The weak charge of the $ \bar{u}_{L} $ is $ T_{3} = 0 $. Accordingly we set
all weak charges above and the integrated weak charge below the
y-axis to zero.
The anti-color could be places above the z-axis
in total 18 combinations ( $ RRR $, $ GGG $, $ BBB $, $ RBG $,
$ RGB $, $ BRG $, $ BGR $, $ GRB $, $ GBR $, $ RGG $, $ GRG $,
$ GGR $, $ RBB $, $ BRB $, $ BBR $, $ BRR $, $ RBR $ and $ RRB $ ).
The anti-color is assigned symbolic $ R $, $ G $, $ B $ above the
z-axis. The integrated total possible 18 charge combinations
are shown symbolic $ ( R,G,B ) $ below the z-axis.
For simplicity we
use in Fig.\ref{AllleftantiFP2006} for color and anti-color the
same symbol. The anti-color charges could generate anti-color moments.
The particle has spin $ 1/2 $, with an anti- rest mass between
$ m= 1.5~ to~4 $ MeV  \cite{Quarkmass}.
We assigned it non stable with anti-rest mass $ - m $ below the t-axis.

The left-handed anti-down-Quark $ \bar{d}_{L} $ is
composed out of one anti-charges $ Q=+1/3 $
which could be placed at one of the three free positions above the
x-axis. The choice of the position will generate on the mass kernel
no magnetic moment and at the border a magnetic moment with opposite
sign to the anti-up-Quark $ \bar{u}_{L} $. If the position of the two charges
of the anti-up-Quark $ \bar{u}_{L} $ is like in
Fig.\ref{AllleftantiFP2006} the magnetic
moment of the anti-up-Quark $ \bar{u}_{L} $ must be $ |\mu_{u}|~>|\mu_{d}| $.
The weak charge of the $ \bar{d}_{L} $ is $ T_{3} = 0 $. Accordingly we set
all weak charges above and the integrated weak charge below the
y-axis to zero.
Like for the $ \bar{u}_{L} $ quark the anti-color could be places above the z-axis
in total 18 combinations ( $ RRR $, $ GGG $, $ BBB $, $ RBG $,
$ RGB $, $ BRG $, $ BGR $, $ GRB $, $ GBR $, $ RGG $, $ GRG $,
$ GGR $, $ RBB $, $ BRB $, $ BBR $, $ BRR $, $ RBR $ and $ RRB $ ).
The anti-color is assigned symbolic $ R $, $ G $, $ B $ above the
z-axis. The integrated total possible 18 charge combinations
are shown symbolic $ ( R,G,B ) $ below the z-axis.
For simplicity we
use in Fig.\ref{AllleftantiFP2006} for color and anti-color the
same symbol. The anti-color charges could generate anti-color moments.
The particle has spin $ 1/2 $, with an anti- rest mass between
$ m= 4~ to~8~$ MeV \cite{Quarkmass}.
We assigned it non stable with anti-rest mass $ - m $ below the t-axis.

\subsubsection { Scheme of the gauge bosons }
\label{sec:Scheme of the gauge bosons}

In the simplest structure of the SM two fundamental spin $1/2$
particles interacting
with each other through spin $1$ point particles for the strong, electro magnetic
and weak interaction as shown in Fig.{\ref{fig.3}.
If our scheme for geometrical extended fundamental particles
introduced for example in Fig.{\ref{BasicSCHEMEFP2006}}
is general valid for FPs
it is important to test the scheme is also able to describe the
the gluons, $ \gamma $ and $ Z^{0}, W^{\pm }$.

\subparagraph { Scheme of the eight gluons of the strong interaction }

In the SM are eight gluons necessary for the strong interaction between
the six quarks. All the gluons are combinations of colors
and anti colors located at one single point. The gluons carry no
electromagnetic or weak charge $ Q = T_{3} = 0 $. Also the
rest mass is zero but the carry induced mass of $ m=E/c^{2} $.
In a scheme where the gluon is a geometrical extended object
has this the consequence that the x-axis and y-axis is not
populated. The object look from the point of view of the three charges
of the SM like a string. Only the induced mass in the kernel
will lead to an geometrical extension. To respect these experimental
findings we populate only the z-axis in agreement with
Fig.{\ref{BasicSCHEMEFP2006} and do not show the x-axis and y-axis in the
scheme. The t-axis we keep as before. We follow
in Fig.{\ref{EightGluons} this pattern
and locate the nine possible color anti color
combinations ( $ R\bar{G} , R\bar{B} , G\bar{R} , G\bar{B} ,
B\bar{R} ,  B\bar{G} , R\bar{R} , G\bar{G} , B\bar{B} $ )
like for the fermions at three positions above the z-axis.
The choice of the position left, center or right would depend
about a color moment of the gluons. As no experimental evidence
about such a moment exist we took a random choice.
The eight integrated gluon combinations seen from far
outside, the gluon ( $ R\bar{G} , R\bar{B} , G\bar{R} , G\bar{B} ,
B\bar{R} ,  B\bar{G} , R\bar{R}-G\bar{G},
R\bar{R}+G\bar{G}-2 B\bar{B} $ )  are placed below the z-axis.
The gluons carry only induced mass of $ m=E/c^{2} $ we placed it below the
t-axis.
The spin of the gluons assigns on specific direction
of the object, we choose in accordance with Fig.{\ref{BasicSCHEMEFP2006}
the z-axis.
To locate the carrier of the color force color and anti-color in
one location of the three geometrical axis is not the only possibility
in the scheme under discussion. It would be possible to separate
the color and anti-color at two geometrical separated positions.
For the combination $ R\bar{R}+G\bar{G}-2B\bar{B} $ maximal six
free positions would be necessary. These position would exist at the
three axis x, y and z surrounding the porter of the mass kernel as shown for
example in Fig.{\ref{BasicSCHEMEFP2006}. Such a
configuration would generate a spherical closed structure but
would need three coordinates.
We follow in Fig.{\ref{EightGluons} the assumption that every free
space coordinate is acting like a free parameter for the three
interactions strong, electromagnetic and weak and use only
one axis the z-axis.

\begin{figure}
\begin{center}
 \epsfig{file=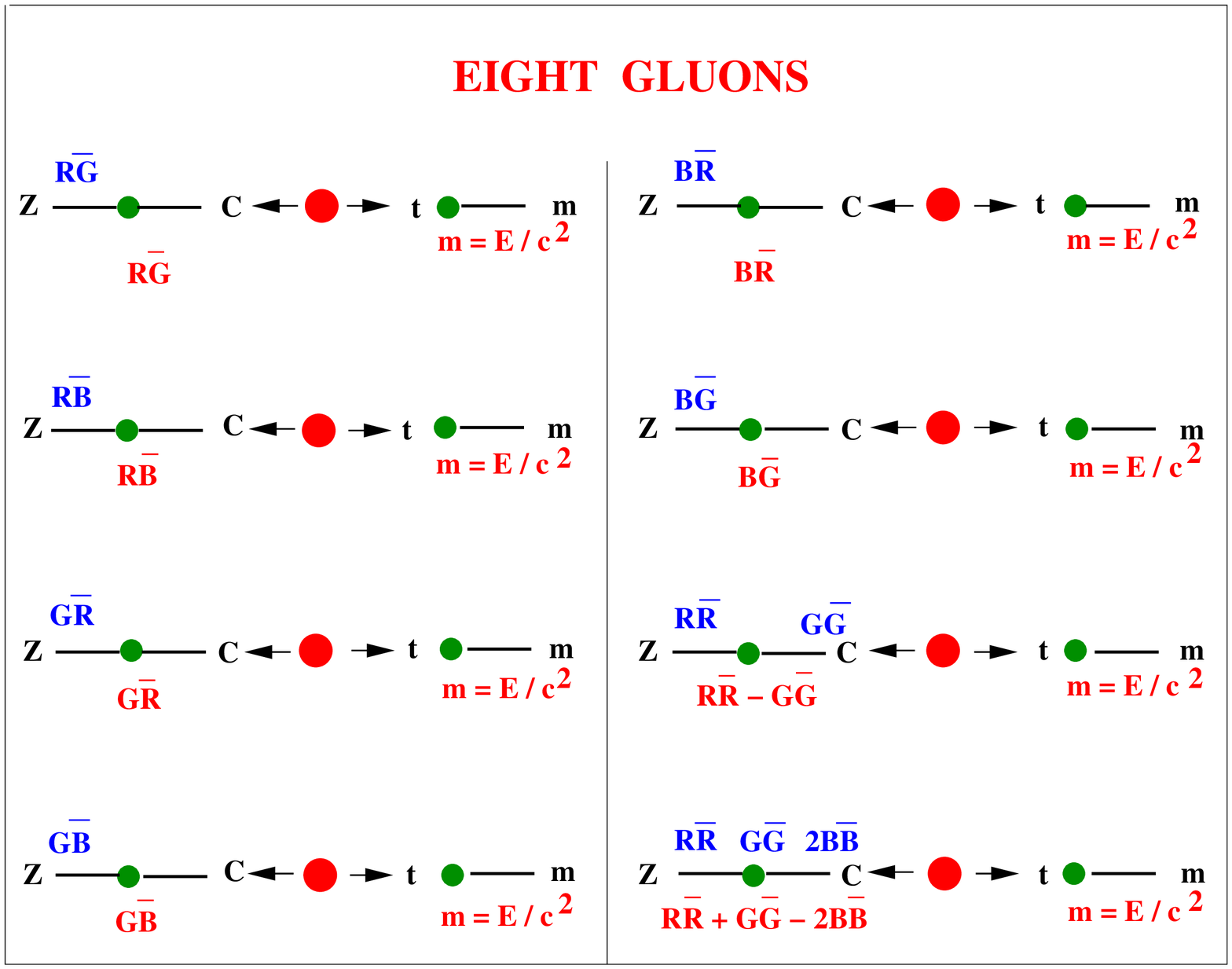,width=13.0cm,height=8.0cm}
\end{center}
\caption{ Scheme of the eight gluons of the strong interaction. }
\label{EightGluons}
\end{figure}

\subparagraph { Scheme of the gauge bosons $ \gamma ,
                Z^{0}$  and $ W^{\pm}$ }
\label{sec:Scheme of the gauge bosons}

The most interesting particle is the $ \gamma $, it carries no charges
$ Q=T_{3}=C=0 $ , no rest mass only induced mass or energy
and $ spin=1 $. The spin assigns a distinct axis and polarization
features as longitudinal or circular polarization.
From experiment we know it is a wave or a particle
depending about the experimental question we ask. It is his own anti-
particle. An geometrical extension could be originated from a
induced mass kernel surrounded by an charge $ Q=-1/3 $ and
anti-charge $ Q=+1/3 $ or self interacting electromagnetic field.
In the scheme shown in Fig.{\ref{GaugeBosons} we set all charges
above and below the three geometrical axis to zero. The
induced mass or energy $ m=E/c^{2} $ we place below the
t-axis. Our scheme would define the $ \gamma $ as a poor time
like object with an induced mass kernel.

\begin{figure}
\begin{center}
 \epsfig{file=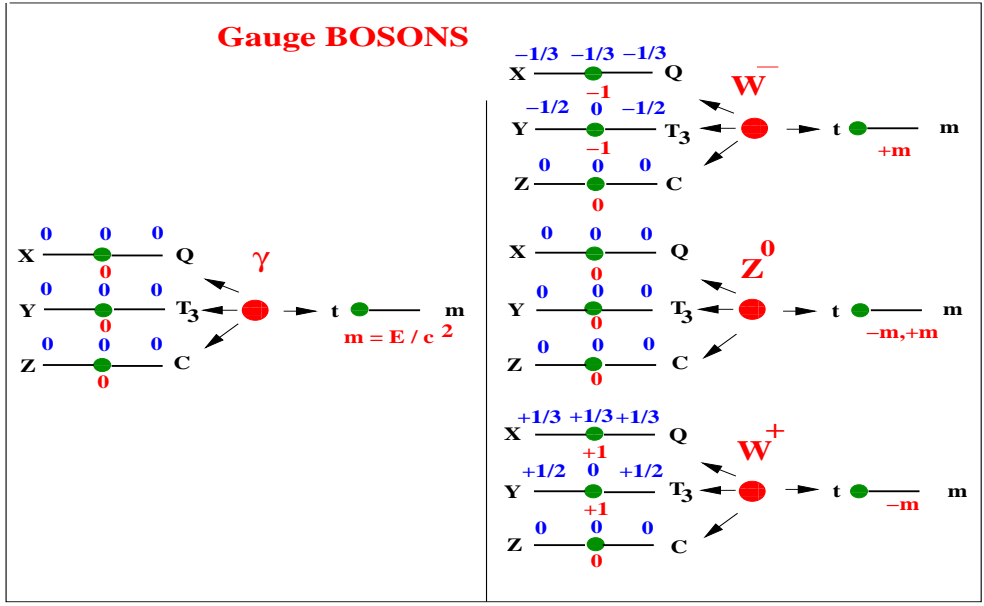,width=13.0cm,height=8.0cm}
\end{center}
\caption{ Scheme of the gauge bosons $ \gamma $, $ Z^{0} $ and $ W^{\pm} $. }
\label{GaugeBosons}
\end{figure}

The carrier of the weak interaction in the SM model
are the weak bosons $ Z^{0} $ and $ W^{\pm} $.
The $ Z^{0} $ is the partner of the $ \gamma $
with charges $ Q=T_{3}=C=0 $ , with $ spin = 1 $ but with a
rest mass of $ m=91.1876 \pm 0.0021 $ GeV \cite{massZandW}.
It is his own anti-particle.
A geometrical extension could be very
similar as for the $ \gamma $ but the
mass kernel would be generated by a high rest mass as displayed in
Fig.{\ref{GaugeBosons} below the t-axis in $ -m $ and $ + m $.
As for the $ Z^{0} $ are the charges  $  Q=T_{3}=C=0 $, we set
in Fig.{\ref{GaugeBosons} all the charges above and
below the three geometrical axis to zero.

The weak boson $ W^{-} $ carries three electric
charges $ Q=-1/3 $ , two weak charges $ T_{3} = -1/2 $
no color charge $ C = 0 $ and
a rest mass of $ m= 80.425 \pm 0.038 $ GeV \cite{massZandW}.
In Fig.{\ref{GaugeBosons} we place the three electric charges
$ Q=-1/3 $ above the x-axis on the three free positions.
The scheme request in accordance with the SM,
that the $ W^{-} $ carries a magnetic moment \cite{MagMomentW}.
The integrated charge is $ Q= -1 $ shown
below the x-axis. The two weak charges $ T_{3} = -1/2 $
we put above the y-axis left and right. The integrated weak charge
is $ T_{3} = -1 $ placed below the y-axis. The weak bosons
$ W^{\pm} $ are the FPs with an integrated weak charge $ T_{3} = |1| $
together with the fermions with an integrated weak charge
$ T_{3} = 0 $ and $ T_{3} = |1/2| $ support this fact the
assumption to cluster the weak charge in units of
$ 1/2 $. This leads to the sequence of the weak charge
already discussed in Fig.\ref{Electron2006} of
$ T_{3} = 0, \pm 1/2 , \pm 1 $. The geometrical position
of the charge $ T_{3} $ ( left,right ) will generate a weak moment
what will
be bigger as the moment generated by the position ( center,right )
or ( center,left ) like
$ \mu_{weak}^{left,right} > \mu_{weak}^{centre,right} $. As no
experimental evidence of a weak moment of the $ W^{-} $ exist
we took a random choice about the position.
The color charge above and below the z-axis we set to zero because
the $ W^{-} $ carries no color charge.
The particle has $ spin = 1 $ with a rest mass $ + m $
we assigned below the t-axis.

The weak boson $ W^{+} $ carries three electric
anti-charges $ Q=+1/3 $ , two weak charges $ T_{3} = +1/2 $,
no color charge $ C = 0 $ and
a rest mass of $ m= 80.425 \pm 0.038 $ GeV \cite{massZandW}.
In Fig.{\ref{GaugeBosons} we place the three electric anti-charges
$ Q=+1/3 $ above the x-axis on the three free positions.
The scheme request in accordance with the SM,
that the $ W^{+} $ carries a magnetic moment with opposite sign
to the $ W^{-} $ moment  \cite{MagMomentW}.
The integrated anti-charge is $ Q= +1 $ shown
below the x-axis. The two weak charges $ T_{3} = +1/2 $
we put above the y-axis left and right. The integrated weak charge
is $ T_{3} = +1 $ placed below the y-axis.
Similar like for the $ W^{-} $ will the geometrical position
of the charge $ T_{3} $ ( left,right ) generate a weak moment
what is
bigger as the moment generated by the position ( center,right )
or ( center,left ) like
$ \mu_{weak}^{left,right} > \mu_{weak}^{centre,right} $. As no
experimental evidence of a weak moment of the $ W^{+} $ exist
we took a random choice about the position.
The color charge above and below the z-axis we set to zero because
the $ W^{+} $ carries no color charge.
The particle has $ spin = 1 $ with a very similar rest mass like the
$ W^{-} $ it is $ ( m_{W^{+}} - m_{W^{-}} ) = -0.2 \pm 0.6 $ GeV 
\cite{massZandW}. We assigned the anti-rest mass of the $ W^{+} $
as $ - m $ below the t-axis.

\subsubsection { Excited states of fermions }
\label{sec:Excited states of fermions}

The discussed scheme for geometrical extended fundamental particles
includes so far all lightest left and right handed FPs,
lightest left and right handed anti-FPs of the
first particle family and the gauge bosons. The majority
of the FPs belong to the second and third family. To test
the reliability of the scheme under discussion it is essential
to proof is the scheme able to describe also the
second and third family.

\subparagraph { Coincidence between three space coordinates
                and three families.}

It would be
possible to include also these particles in the described scheme, but
all these following scheme structures would be very similar like in
Fig.\ref{AllleftFP2006}, Fig.\ref{AllrightFP2006},
Fig.\ref{AllrightantiFP2006} and Fig.\ref{AllleftantiFP2006}.
It is remarkable that the charges, position of charges and
the left/right handed structure of the heavier FPs is inside
the families the same. The leading parameter to characterize
from the experiment the difference between the three families
is the rest mass of the fermions. The rest mass from family
one, two and three is exponential increasing with the family
number $ m_{\mu}/m_{e} \sim 207 $ and $ m_{\tau}/m_{e} \sim 3477 $
\cite{MagneticMomentElecMoun}.
Concerning our ETAMFP model in Fig.{\ref{BasicSCHEMEFP2006}}
is the mass of the geometrical extended FPs located in the
center of the particle. It is for this reason sensible to
search for a possibility how the mass sphere could increase
its mass or energy. As the spin of all three particle families
is the same a vibration of the mass sphere of the fermions
would be a possibility to increase the rest mass or energy.
The mass sphere has in accordance with the three Cartesian
coordinates three basic degrees of freedom to vibrate.
This fact coincide with the three fermion
families. 
For example the
$ \mu $ and $ \tau $ would be the excited states of the
$ e $. A harmonic oscillator with the vibration energy
$ E=\hbar \omega (n_{x}+n_{y}+n_{z}) $ ( neglecting the zero energy )
would allow to use the three space coordinates $x$, $y$ and $z$
of our scheme to install at the x-axis $ n_{x}=0, 1, 2 ..$, on the
y-axis $ n_{y}=0, 1, 2 ..$ and on the z-axis $ n_{z}=0, 1, 2 ..$
vibration states. The ground state would be $ n_{x}=1 $ and
$ n_{y}=n_{z}=0 $, the first excited state $ n_{x}=n_{y}=1 $
and $n_{z}=0 $ and the third excited state $ n_{x}=n_{y}=n_{z}=1 $.
The mass sphere of the FPs would fulfill a rotational
movement superimposed be a vibration in three axis as
shown in Fig.\ref{FPvibration}. The oscillation is limited by
the essential condition not to change the spin of the FPs.

\begin{figure}
\begin{center}
 \epsfig{file=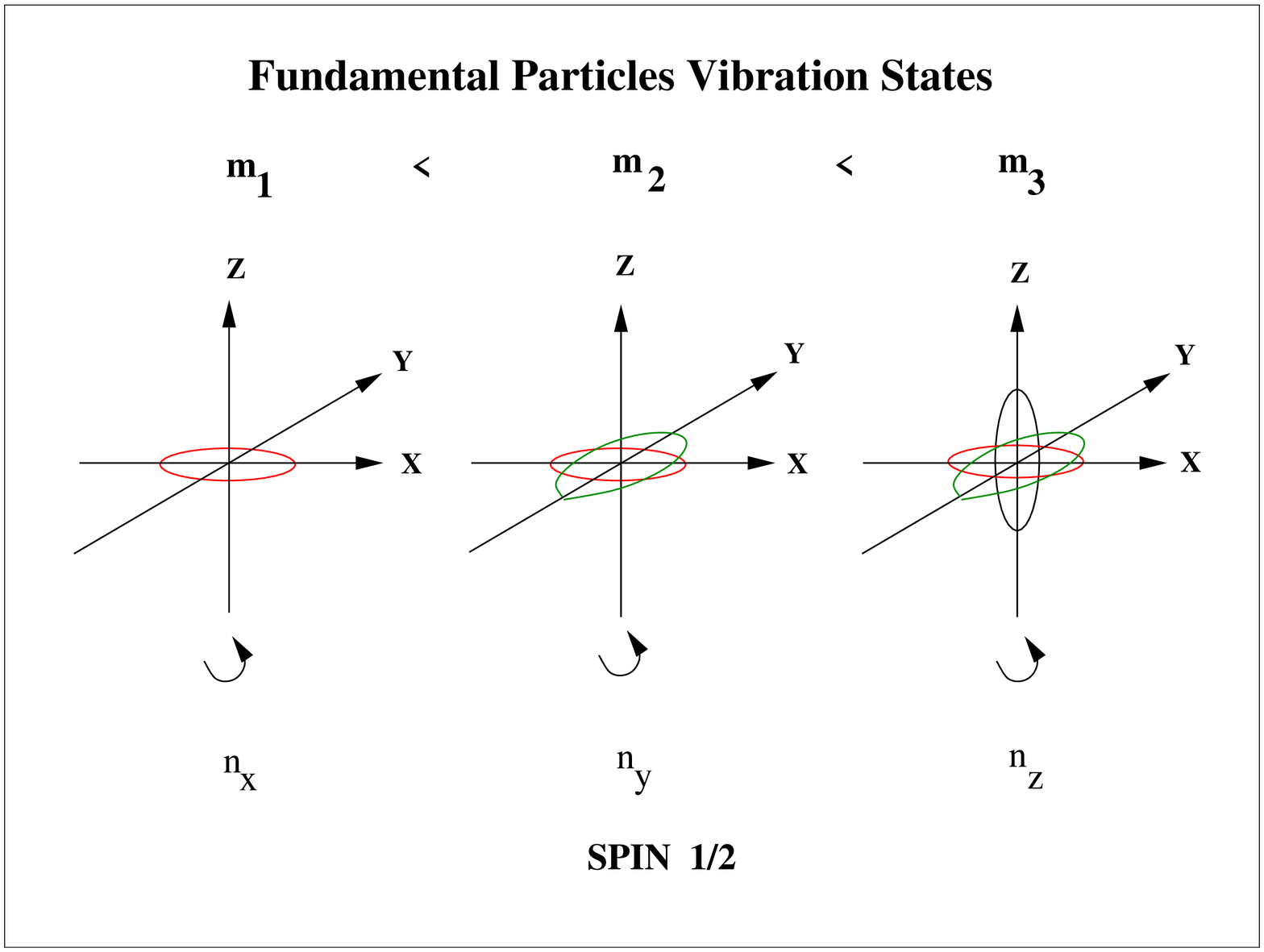,width=13.0cm,height=8.0cm}
\end{center}
\caption{ Excited states of fundamental particles. }
\label{FPvibration}
\end{figure}

In the center of Fig.\ref{FPvibration}
the three vibration states are symbolized by three coordinate systems.
The mass of the three
families $ m_{1} <  m_{2} <  m_{3} $ is displayed on top of z-axis
of these coordinates. The spine $ 1/2 $ is symbolized by a twisting
arrow below the z-axis. The spin axis is chosen to be parallel
the z-axis. The ground state on the x-axis $ n_{x} $ is shown
left, the first excited state on the y-axis $ n_{y} $
in the middle and the second exited state on the z-axis
$ n_{z} $ on the right in Fig.\ref{FPvibration}.
The ground state on the x-axis would be a rotating spherical mass sphere
with a distinct rotating axis parallel the spin symbolized
as red circle in the x-y plane.
The first excited state on the y-axis would be a rotating
ellipsoid mass sphere around the z-axis and vibrating at the same time
with his mean axis in x and y direction, symbolized
by an red and green ellipsis in the x-y plane.
The first excited state on the z-axis would be a rotating
ellipsoid mass sphere around the z-axis and vibrating at the same time
with his mean axis in x,y and z direction, symbolized by
a red, green and black ellipsis in the x-y plane and x-z plane.

\subparagraph { Empirical attempt to sort all three particle
                families in one scheme.}

The discussed conditions open the possibility to sort all the masses
of fermions in an empirical attempt of a 
rotator superimposed with a highly degenerated oscillator.
Depending about the number
of parameters and constant factors what will be used
it is possible to calculate the
masses of all fermions. Using 2 parameters
and 7 constants for 12 fermion masses
it is possible to reach a moderate agreement between calculation
and experimental date \cite{QEDtest}. We introduce an example
with 2 parameters and 10 constants for 12 fermion masses
which allows to calculate precise all masses of the fermion and anti-fermions
including a prediction about the mass limits for the
$\nu_{\mu}$ and $\nu_{\tau}$ neutrino in Eq.\ref{FPmass}.

\begin{eqnarray}
\label{FPmass}
E(k_{i};Q) & = & (A+B|Q|+CQ^{2}+D|Q|^{3})(k_{i})^{f(Q,k_{i})}  \\
f(Q,k_{i}) & = & (R+|Q|V(k_{i}-1)+|Q|(|Q|-1)        \nonumber  \\
           &   & (S(|Q|-1/3)+W(|Q|-1/3)(k_{i}-1)    \nonumber  \\
           &   & +T(|Q|-2/3)+Z(|Q|-2/3)(k_{i}-1)))  \nonumber
\end{eqnarray}


Eq.\ref{FPmass} contains the two parameters family number $ k_{i} $
and the electric charge $ Q $ of the fermions. For simplicity the
weak charge and color charge is not used, but is implicit hidden
in the 10 constants.
The first parameter is the family number
$ k_{i}=(n_{x}+n_{y}+n_{z}) $ according
Fig.\ref{FPvibration} with $ n_{x,y,z} $ as fibration state for the
x,y and z-axis. The first family is $ n_{x}=1 $ and $ n_{y}=n_{z}=0 $
which leads to $ k_{1}=(n_{x}+n_{y}+n_{z}) = 1 + 0 + 0 = 1 $.
The second family is $ n_{x}=n_{y}=1 $ and $ n_{z}=0 $
which leads to $ k_{2}=(n_{x}+n_{y}+n_{z}) = 1 + 1 + 0 = 2 $.
The third family is $ n_{x}=n_{y}=n_{z}=1 $
which leads to $ k_{3}=(n_{x}+n_{y}+n_{z}) = 1 + 1 + 1 = 3 $.
More exited states are mathematical possible. Experimental
evidence and limits are discussed in
chap.\ref{sec:Status of experimental limits on the sizes of
Fundamental Particles} and  chap.\ref{sec:Electroweak Interaction}.
The second parameter is the
charge $ Q = 0, \pm 1, \pm 2/3, \pm 1/3 $ for the leptons/anti-leptons and
quarks/anti-quarks. Included in Eq.\ref{FPmass} are the 10 constant
factors $ A<3\times 10^{-6}$ MeV, $ B= 42.1358$ MeV, $ C = -87.7995$ MeV 
and $ D = 46.1747$ MeV to define the masses of the first family. The
constant factors for the leptons/anti-leptons  are $ R = 7.9617 $
and $ V = -0.2698 $ ,for the quark/anti-quark family $ Q = \pm 2/3 $
the factors $ S = 5.2528 $ and $ W = -19.3806 $ and for the
for the quark/anti-quark family $ Q = \pm 1/3 $
the factors $ T = -77.3429 $ and $ Z = 26.8191 $.

\subparagraph { Comparison of calculation and experimental data. }

The comparison of calculation and experimental data is shown in
Fig.\ref{Fermionmass}. The plot displays on the x-axis the
family number $ k_{i} $ and on the y-axis the rest mass of
the fermions in MeV. The experimental data are shown in
black rectangles. The calculation for the rest mass of the
$ u, c, t $ quark is displayed as black dotted line, for the
$ d, s, b $ quark displayed as black dot dashed line, for the
$ e, \mu, \tau $ leptons displayed as black dashed line and
for the
$ \nu_{e}, \nu_{\mu}, \nu_{\tau} $ neutrinos
displayed as black solid line.

\begin{figure}
\begin{center}
 \epsfig{file=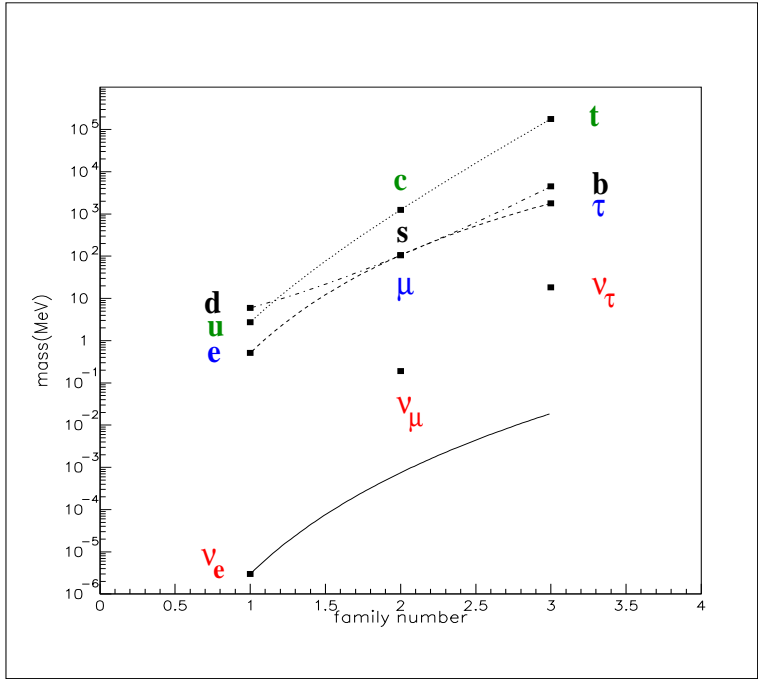,width=10.0cm,height=9.0cm}
\end{center}
\caption{ Excited states of fermions. }
\label{Fermionmass}
\end{figure}

The experimental data are taken from ref. \cite{massTOPLEP}.
For the quarks with the charge $ Q=2/3 $, $u$,$c$ and $t$,
we used as data the geometrical middle value of the experimental
mass limits from ref. \cite{massTOPLEP}. The same middle
values for the experimental rest mass
we used for the quarks with the charge $ Q=-1/3 $ $d$,$s$ and
$b$. For the charge leptons $ e $, $ \mu $ and $ \tau $ we used
the well measured rest masses from ref.\cite{massTOPLEP}.
For the
neutrinos exist mass limits only. For this reason we use the
mass limit for the $ \nu_{e} $ as start value and calculate
the mass limits for the $ \nu_{\mu} $ and $ \nu_{\tau} $ ( black solid line).
We assume for the calculation the same exponent as for the charged
leptons.
The experimental limits ( black rectangles ) are much higher
as the calculation. Further experiments are necessary to proof
is this a fact of the accuracy of the experiment or an effect
of the calculation.
The perfect agreement between experimental measured rest mass
and calculation of Eq.\ref{FPmass} for the quarks and leptons
is obvious, because total 12 parameters allow to
calculate 12 rest masses.

The mass increase for the first family from
$ m_{\nu_{e}} << m_{e} <  m_{u}  <  m_{d} $
follows approximately the strength of the weak, electro magnetic
and strong coupling constants of the three interactions.
It is visible in Fig.\ref{Fermionmass} that the gradient of the
mass increase per family is very similar for the charged leptons
and the quarks with charge $ Q=2/3 $. The quarks with charge  $ Q=-1/3 $ in
particular the mass of the $d$ quark does not follow this trend.
The upper experimental mass limit for the $u$ quark with 4 MeV just
touches the lower mass limit of the $d$ quark with also
4 MeV. With improved experimental date in the next years
this disagreement will may be disappear.

\subsubsection { Conclusion about the scheme of geometrical
                 extended fundamental particles and anti-particles }

In conclusion it was possible to demonstrate, the hypotheses,
the time and size development of the history of the very young
universe forms the geometrical extension of the FPs, allows to sort
all FPs in a common scheme.
The mass increase for the first family 
follows approximately the strength of the weak, electro magnetic
and strong coupling constants of the three interactions.
The scheme predict a general similar tendency of the 
gradient of the mass increase per family for the leptons and quarks.
The scheme shows that the link between geometrical space and time
is of essential importance. Four interactions coincide with
the four coordinates of our universe.
The three families are in this philosophy the possibility of the
mass kernel to oscillate ( link time ) in
these three space coordinates.
The scheme also shows the most simple charged highly symmetrical
geometrical object is the electron. It serves as ground state
for all leptons with a extremely high stability.

\subsection {Classical approach to estimate the size of an electron
             determined from its experimental measured parameters}
\label{sec:Classical approach}

In the discussed scheme in Fig.\ref{BasicSCHEMEFP2006},
Fig.\ref{Electron2006} and Fig.\ref{AllleftFP2006}
is the electron the most fundamental
particle. It has a spherical geometrical structure, is the ground state
of the leptons and is very stable with a life time of
$ \tau_{e} > 4.6\times 10^{26}$ yr.
It is for this reason a very important object to study the
consequences of the rest mass, charge, spin, magnetic
and possible electric dipole in accordance with our particle scheme
of a geometrical extended object. Inspecting our basic scheme of
an geometrical extended FP in Fig.\ref{BasicSCHEMEFP2006} the
mass center and the charges $ Q $ and $ T_{3} $ will fulfill
a rotation orbiting the coordinate center $ x = y = z = 0 $.
Immediately three questions concerning about a classical approach
are coming up. Is the size of the electron bigger as the
Planck size of $ 10^{-33} $ cm ? Latest at the Planck
scale gravity must be quantized and a classical picture would
get meaningless. Does a scenario about the forces in the electron
exist to form such a stable object? Why are the accelerated
charges $ Q $ and $ T_{3} $ are not emitting energy in form of radiation
for example light?

The experimental limits about the size
of fermion shown in Fig.\ref{Datalimts} are $ 10^{-18} $cm and for the
electric dipole moment
$ d_{e} < ( 0.07 \pm 0.07 ) 10^{-26} $ e cm. These limits  are fare above
the Planck Scale. The difference is with $ 15 $ respectively $ 7 $
magnitudes so far away from the Planck scale that we believe a classical
approach has still some meaning.

In the chapter forces and stability ( chap.\ref{sec:Forces and stability} )
we discussed a possible scenario
in Fig.\ref{ForcesSM-BBM} and  Fig.\ref{ForcesSTABLE}
how such a stable object schematically could exist. In case of the
electron point $ SP II $ in Fig.\ref{ForcesSM-BBM} would be the
not rotating inner mass kernel, the rotating mass  center would extent
to point $ SP I $ on where the charges $ Q $ and $ T_{3} $
would be placed. This scenario would assume at the radius $ r_{m} $
the mass density goes to zero and the radius $ r_{e} $ where the charges
are placed are about the same
$ r_{m} \sim r_{e} \sim r_{SP I } $. The same stability scenario also explains
that two electric charges $ Q = - 1/3 $ are place opposite to each
other on the x-axis shown in Fig.\ref{AllleftFP2006} and one
charge $ Q = - 1/3 $ at $ x = y = z =0 $. As consequence the magnetic
moment and possible electric dipole moment of the electron get
only generated from a total charge $ Q = - 2/3 $ and not $ Q = - 1 $.
For simplicity of the further discussion we will neglect the weak
charge $ T_{3} $.

The extreme high experimental proved stability of the electron ask
for a radiation free path of the charges around the center of the
electron. This experimental fact request the charges do not
undergo any acceleration because charge rotating on a circle is
accelerated. In the chapter about a flashing vacuum 
( chap.~\ref{sec:The flashing vacuum} )
we discussed
a microscopic scenario about such a radiation free path.
In this picture the vacuum has the important property to
move a charge geometrical over a certain curvaceous path
without acceleration. In this manner the curvaceous path
for a charge get radiation free. An important consequence
of such a type of path is that the geometrical length of the path
will be contracted by the Lorentz transformation if the velocity
of the charges approach the speed of light.

For the following calculations we need a numerical basis
about the parameter mass, charge, spin, magnetic and possible
electric dipole moment of the electron together with the
Planck Constant, speed of light and Bohr Magneton. These
parameters are summarized in Tab. \ref{Parameterelectron} and taken from
ref.\cite{MagneticMomentElecMoun}.

\begin{table}
\caption{ Parameters of electron. }
\label{Parameterelectron}
\begin{center}
\begin{tabular}{||l|l|l||}                                      \hline
  parameter           & Symbol        & value
                                                               \\ \hline
  Total Mass         & $ E_{tot} $   & $ 0.51099892 \pm 0.00000004 $ MeV
                                                               \\ \hline
  Charge
                      & $ e  $        & $ 1.60217653(14) \times 10^{-19} $ A s
                                                               \\ \hline
  Weak Charge
                      & $ T_{3}  $    & $ - 1/2 $
                                                               \\ \hline
  Spin                & $ S=\frac{1}{2}\cdot\hbar $ &
                        $ \frac{1}{2} \cdot 6.58211915(56)\times 10^{-22} $ MeV s
                                                               \\ \hline
  Magnetic Moment
                      & $ \mu_{e} $   &
                        $ 1.001159652187 \pm 0.000000000004 \cdot \mu_{B} $
                                                               \\ \hline
  Electric Dipole Moment
                      & $ d_{e} $     & $( 0.07 \pm 0.07 ) 10^{-26} $ e cm
                                                               \\ \hline
  Planck Constant     & $\hbar$       & $ 6.58211915(56) \times 10^{-22} $ MeV s
                                                               \\ \hline
  Speed of Light
                      & $ c $         & $ 299792458 $ m/s
                                                               \\ \hline
  Bohr Magneton
                      & $ \mu_{B}=\frac{e\hbar}{2m_{e}} $ &
                        $ 5.788381804(39) \times 10^{-11} $ MeV/T
                                                               \\ \hline
\end{tabular}
\end{center}
\end{table}

In Tab. \ref{Parameterelectron}
the electron charges are the source of a electric, weak and gravitational
force and should be able to polarize and curve the vacuum
in the finiteness of the mass core. Following Einstein
equation in Tab. \ref{linksSMBigBang}
the mass of the electron would be originated
from these curvatures.
The spin of an geometrical extended electron is
classically described by a gyroscope with a defined spin
axis and a moment of inertia. The spin axis is perpendicular to
a surface of rotation. The same counts for the classical
magnetic moment of a charge. The charge rotates in our scheme
like a point around the mass core of the object. As we
know from experiment that the axis of the spin is anti-parallel
to the axis of the magnetic moment it is likely that the
charge rotates in the same surface of rotation as the mass
core. If the charge in Fig.{\ref{BasicSCHEMEFP2006}}
is considered as point it must
cause an electric dipole moment if an spectator measures
the dipole moment in the x-y plane and the electron is
polarized parallel to the z-axis. The degree of polarization
and the conditions of the experimental apparatus will affect
the measured value of the dipole moment. The experiment is
extreme difficile and is not settled so far.

In the following sections in this chapter we study
with an increasing refinement of the classical
electron model the impact of the different
parameters of the electron on the size of the electron.

\subsubsection { Mass and charge radius of an electron
                 as gyroscope }
\label{sec:Mass and charge radius of an electron as gyroscope}

To study the order of magnitude in a first simple approach we
describe the microscopic structure of the electron as a
rotating sphere like a gyroscope filled with mass as shown
in Fig.{\ref{BasicSCHEMEFP2006}}. The charges, spin and
spin axis are in accordance also with
Fig.{\ref{BasicSCHEMEFP2006}}. In our scheme the electron as
ground state of the leptons carries no fibration energy.
The total energy $ E_{tot} $ of a gyroscope is under this
conditions the sum of the energy of the electron
$ m_{0}c^{2} $ and the rotational energy of the
gyroscope $ E_{rot} = \frac {1}{2} \theta \omega^{2}_{m} $
as displayed in Eq.\ref{Etot}.

\begin{equation}
E_{tot} = m_{0} c^{2} + \frac {1}{2} \theta \omega^{2}_{m}
\label{Etot}
\end{equation}

The angular momentum $ L $ is shown in Eq.\ref{Ltot}

\begin{equation}
L = \theta \omega_{m} = \frac{1}{2} \hbar
\label{Ltot}
\end{equation}

and the magnetic moment of the electron $ \mu_{e} $ in
Eq.\ref{MagMoment}.

\begin{equation}
\mu_{e} = \frac{1}{2} e \omega_{e} r_{e}^{2}
\label{MagMoment}
\end{equation}

$ \theta $ is the moment of inertia of the mass core,
$\omega_{m}$ and $\omega_{e}$ the angular velocity of the mass core
and the charge and $ r_{e} $ the radius the point like charge
is located from the center.

The intention of this first approach is to study only the
order of magnitude of the size of the electron.
For this reason we calculate possible
limits on the angular velocity $ \omega_{m} $
from the experimental values. A lower limit for $ \omega_{m}(min) $
can be deduced from Eq.\ref{Ltot} which allows a small
but not zero $ \omega_{m}(min) > 0 $ for $ \theta \rightarrow \infty $.
An upper limit for $ \omega_{m} $
can be calculated if we set in Eq.\ref{Etot} the rest mass
$ m_{0} c^{2} = 0 $.
Inserting Eq.\ref{Etot} in Eq.\ref{Ltot}
and set $ m_{0} c^{2} = 0  $ it is possible
to calculate with values from Tab. \ref{Parameterelectron} the
angular velocity $\omega_{m} ( max ) $ in Eq.\ref{AgularVelocite}

\begin{equation}
\omega_{m}(max) = 4 \frac{E_{tot}}{\hbar} = 3.1 \cdot 10^{21}~~1/s
\label{AgularVelocite}
\end{equation}

As discussed it is known from the experiment
that the plane of rotation from
the charge and the mass of the electron is the same no experimental
information exist about the angular velocity $\omega_{m}$
and $\omega_{e}$ of the mass and charge. It is also not known
in which distance from the center $ r_{m} $
the mass density goes to zero and on
which radius $ r_{e} $ the point-like charge is located.
If we follow
our discussion about forces and stability in the introduction
of this chapter it is likely to assume between
$\omega_{m}$ and $\omega_{e}$ a certain functional interaction
like $\omega_{m} = f ( \omega_{e}) $ exist.

For a working hypothesis we follow this philosophy to use first the
simplest possible assumptions set $ r_{m} = r_{e} = r $ and
$\omega_{m} = \omega_{e} = \omega $. Both hypotheses are supported
by the discussion in chapter the pseudo charge of the gravitational
interaction ( chap.~\ref{sec:The charges of fundamental particles} ),
because the mass of the electron will be with high
likelihood generated by the electric charge of the electron. This
charge is mostly located after our ETAMFP model at the radius $ r_{e} $
and the charge is rotating around the center what supports the
assumption that the angular velocity of the charge and mass is
the same.
Using Eq.\ref{MagMoment}, the angular velocity $ \omega $
and for simplicity set the charge $ Q = | 1 | $ it is possible
to calculate with $ \omega ( min ) > 0 $ the radius $ r(max) $
in Eq.\ref{RaduisAmax} and with Eq.\ref{AgularVelocite}
$ r(min) $ in Eq.\ref{RaduisAmin}.

\begin{equation}
r (max) = \sqrt \frac {2 \mu_{e}}{ e \omega(min) } < \infty
\label{RaduisAmax}
\end{equation}

\begin{equation}
r (min) = \sqrt \frac {2 \mu_{e}}{ e \omega(max) } > 1.93 \cdot 10^{-13}~~m
\label{RaduisAmin}
\end{equation}

The in this simple approach estimated limits for the radius
are very huge. In particular the radius $ r (max) $ is totally
in conflict with our experimental data discussed for example in
Fig.\ref{Datalimts}. We will for this reason ignore this limit
for the further discussion. For the guidance of the next
approach more interesting is the limit for $ r(min) $.

The radius $ r(min) $ is fare away from the Planck
or Grand unification scale $ r (min) >> r_{Planck} $, what supports our
assumption to use for this calculations a classical ansatz.
The velocity of the outer mass radius and the charge
$ v = r (min) \times \omega = 6.0 \cdot 10^{8}~~m/s $
( $ \beta = 2.0 $ ) is bigger
as the velocity of light $ v > c $.

To include this fact for the next approach it would be possible
to use the special relativity or the general relativity. We
discussed already the charge of the electron orbits on a radiation
free path. This is the same condition as the charge would
move with constant velocity along a straight-line. We use
for this reason in the following section
the special relativity in form of the Lorentz
transformations to avoid velocities $ \beta > 1 $.

\subsubsection { Charge radius of electron
                 as Lorentz contracted gyroscope }

To study the next order of magnitude in a second approach we
describe the microscopic structure of the electron as before as
rotating sphere like a gyroscope filled with mass as shown
in Fig.{\ref{BasicSCHEMEFP2006}}. The charges, spin and
spin axis are in accordance also with
Fig.{\ref{BasicSCHEMEFP2006}}. We keep the condition
of the radius $ r_{m} = r_{e} = r $.
But we include in this approach
the result of the last calculation and take into account 
the velocity of the outer mass radius and the charge is
bigger as the velocity of light and we include
the mass $ m_{0} c^{2} $.

We just discussed the radiation free
path length $ p = 2 \pi r_{m} = 2 \pi r_{e} = 2 \pi r $
of the mass radius $ r_{m} $ and charge $ r_{e} $ must be
corrected be the special relativity.
It is necessary for this correction to distinguish
between the radius $ r_{0} $ an observer placed on
the rotating system will measure and the radius $ r_{e} $
an observer far away in rest will measure.
This implements the observer in rest from fare
outside the electron will measure
a Lorentz contracted $ p $ if the velocity of the charge
$ v=\omega_{e} r_{0} $ approaches $ c $. The charge is not sensitive
to the special relativity and the angular velocity $ \omega_{e} $
is visible only for an observer in rest. The moving charge
does not behave like a clock on the plane of rotation. For
this reason exist only the angular velocity $ \omega_{e} $.
The measured magnetic moment is originated from a charge
moving along the path $ p = 2 \pi r $ what is really contracted
in this rotating system is the radius $ r_{0} $
shown in Eq.\ref{Rcontracted}.

\begin{equation}
r_{0}=\frac{r_{e}}{\sqrt {1-\beta^{2}}}
\label{Rcontracted}
\end{equation}

\subparagraph{ Lorentz contraction affects magnetic and
               electric dipole moment of electron }

Using the data from Tab. \ref{Parameterelectron} for the magnetic
and the electric dipolmoment of the electron it is remarkable
the $ \mu_{e} $ is a finite well measured number and
$ d_{e} $ is very small in the frame of the error still
zero. Our just discussed simple approach points in a direction
the speed of the charge is close to the speed of light.
Concerning the Lorentz contraction of the radius $ r_{0} $
in Eq.\ref{Rcontracted} to $ r_{e} $ the question rises
which radius $ r_{0} $ or $ r_{e} $ together with the charge
is generating the magnetic and electric
dipolmoment of the electron measured from an observer in rest
in the laboratory. In other words to which radius $ r_{0} $
or $ r_{e} $ is the experiment sensitive to measure $ \mu_{e} $
or $ d_{e} $. We study for this reason the limits for
$ \mu_{e} $ and $ d_{e} $ given from the experimental values
and our ETAMFP model to use $ r_{0} $ or $ r_{e} $.

To set a limit on $ \mu_{e} $ for the case $ r_{0} $
generates the magnetic moment of the electron measured from an
outside observer in rest, we assume after the discussion above
the most likely case $ \beta \rightarrow 1 $. The radius
$ r_{0} $ get contracted from Eq.\ref{Rcontracted} to $ r_{e} $.
The magnetic moment reads under this circumstances like
in Eq.\ref{Magmomentr0}.

\begin{equation}
\mu_{e}=\frac{1}{2} e \omega_{e} r_{0}^{2}
       =\frac{1}{2} e \omega_{e} \frac {r_{e}^{2}}{(1-\beta^{2})}
\label{Magmomentr0}
\end{equation}

The velocity of the charge $ v $ is displayed in Eq.\ref{Qvelocityr0}

\begin{equation}
\beta=\frac{v}{c}=\frac{\omega_{e} r_{0}}{c}
                 =\frac{\omega_{e}}{c}\frac{r_{e}}{\sqrt{1-\beta^{2}}}
\label{Qvelocityr0}
\end{equation}

Replacing $ r_{e} $ in Eq.\ref{Magmomentr0} with Eq.\ref{Qvelocityr0}
allows to calculate the discussed limit in Eq.\ref{Limitmur0} if we
insert the numerical values from $ e $, $ c $ from
Tab. \ref{Parameterelectron} and $ \omega_{e} $ from
Eq.\ref{AgularVelocite}.

\begin{equation}
\lim_{\beta\to~ 1}\mu_{e} =
\lim_{\beta\to~ 1} \frac{1}{2} e \frac{\beta^{2} c^{2}}{\omega_{e}}
= e \frac{1}{2} \frac{c^{2}}{\omega_{e}} = 2.3\times 10^{-24 }~~Am^{2}
\label{Limitmur0}
\end{equation}

This limit is close to the experimental measured value shown in
Eq.\ref{Comparisomuexpmulim}.

\begin{equation}
\mu_{e} ( experiment ) = 9.3 \times 10^{-24} ~~Am^{2} \sim
\mu_{e} ( limit ) = 2.3 \times 10^{-24}~~Am^{2}
\label{Comparisomuexpmulim}
\end{equation}

To set a limit on $ \mu_{e} $ for the case $ r_{e} $
generates the magnetic moment of the electron measured from an
outside observer in rest, we assume after the discussion above
again the most likely case $ \beta \rightarrow 1 $. The radius in the
rotating system $ r_{0} $ will be finite and contracted
by equation Eq.\ref{Rcontracted} to $ r_{e} \rightarrow 0 $.
The angular velocity $ \omega_{e} $ will by finite.
The limit is zero as shown in Eq.\ref{MagMomentlowere}.

\begin{equation}
\lim_{r_{e}\to~ 0}\mu_{e} =
\lim_{r_{e}\to~ 0} \frac{1}{2} e \omega_{e} r_{e}^{2} = 0
\label{MagMomentlowere}
\end{equation}

This limit is absolute in disagreement with the
experimental measured value shown in
Eq.\ref{Comparisomuexpmulimre}.

\begin{equation}
\mu_{e} ( experiment ) = 9.3 \times 10^{-24} ~~Am^{2} >>
\mu_{e} ( limit ) = 0 ~~Am^{2}
\label{Comparisomuexpmulimre}
\end{equation}

To set a limit on $ d_{e} $ for the case $ r_{0} $
generates the electric dipole moment of the electron measured from an
outside observer in rest, we assume again
the most likely case $ \beta \rightarrow 1 $. The radius
$ r_{0} $ get contracted from Eq.\ref{Rcontracted} to $ r_{e} $.
The electric dipole moment reads under this circumstances like
in Eq.\ref{Elecmomentr0} if we replace $ r_{0} $ by $ r_{e} $
using Eq.\ref{Rcontracted}.

\begin{equation}
d_{e} = e r_{0} = e \frac{ r_{e} }{\sqrt{1-\beta^{2}}}
\label{Elecmomentr0}
\end{equation}

Replacing $ r_{0} $ in Eq.\ref{Elecmomentr0} with Eq.\ref{Qvelocityr0}
leads to the limit for $ d_{e} $ in Eq.\ref{Limitder0} if we use again
$ c $ from Tab. \ref{Parameterelectron} and $ \omega_{e} $ from
Eq.\ref{AgularVelocite}.

\begin{equation}
\lim_{\beta\to~ 1} d_{e} =
\lim_{\beta\to~ 1} e \frac{c}{\omega_{e}}\beta = e \frac{c}{\omega_{e}}
= 9.6 \times 10^{-14}~em
\label{Limitder0}
\end{equation}

This limit is about 17 magnitudes away from the experimental
measured value shown in Eq.\ref{Comparisondeexpr0}.

\begin{equation}
d_{e} ( experiment ) = 0.07\times 10^{-28} ~~e m <<
d_{e} ( limit ) = 9.6 \times 10^{-14}~~e m
\label{Comparisondeexpr0}
\end{equation}

To set a limit on $ d_{e} $ for the case $ r_{e} $
generates the electric dipole moment of the electron measured from an
outside observer in rest, we assume
again the most likely case $ \beta \rightarrow 1 $. The radius in the
rotating system $ r_{0} $ will be finite and contracted
by equation Eq.\ref{Rcontracted} to $ r_{e} \rightarrow 0 $.
The angular velocity $ \omega_{e} $ will by finite.
The limit is zero as shown in Eq.\ref{LimitDipolre}

\begin{equation}
\lim_{r_{e}\to~ 0}d_{e} =
\lim_{r_{e}\to~ 0} e r_{e} = 0
\label{LimitDipolre}
\end{equation}

This limit is approximately in agreement with the experimental
measured limit shown in Eq.\ref{Comparisondeexpre}.

\begin{equation}
d_{e} ( experiment ) = ( 0.07 \pm 0.07 ) \times 10^{-28} ~~e m \sim
d_{e} ( limit ) = 0 ~~e m
\label{Comparisondeexpre}
\end{equation}

Inspecting the just discussed limits,
the answer from the comparison between the experimental values
and the calculated limits for the magnetic and electric dipole
moment of the electron concerning which radius $ r_{0} $ or
$ r_{e} $ has to be used from the Lorentz contraction is rather
clear. The measurement of the magnetic moment of the electron
is sensitive to the radius $ r_{0} $. This is the radius an
observer located in the center of the electron would
measure from the center to the charge if he rotates
with $ \omega_{e} $.
The  measurement of the electric dipole moment $ d_{e} $ is
sensitive to the Lorentz contracted radius $ r_{e} $ an
observer will measure if he is located fare outside the electron
in rest. In this sense the Lorentz contraction in our
discussed model Fig.{\ref{BasicSCHEMEFP2006}} would explain
why the magnetic moment
$ \mu_{e} $ is a finite well measured number and
$ d_{e} $ is very small in the frame of the error still
zero.

\subparagraph{ Limits of the parameters of the electron after
               Lorentz contraction. }

Our approach about an geometrical extended electron is at this stage
so far developed to estimate the parameters $ \omega $, rest mass
$ m_{0} c^{2} $, rotational energy $ E_{rot} $, $ r_{0} $ and $ r_{e} $
under the condition the speed of the rotating charge is close to
the speed of light $ c $. Further following this stage the
magnetic moment of the electron $ \mu_{e} $
is described by the Eq.\ref{Magmomentr0} and the angular velocity
$ \omega_{e} $ from Eq.\ref{Qvelocityr0} where $ \mu_{e} $ and
$ \omega_{e} $ is depending about $ r_{0} $.

To calculate the angular velocity $ \omega_{e} $ we insert
Eq.\ref{Qvelocityr0} in Eq.\ref{Magmomentr0} and find
the Eq.\ref{AngularVelocityr0}.

\begin{equation}
\omega_{e}=\frac{1}{2} \frac{e(c\beta)^{2}}{\mu_{e}}
\label{AngularVelocityr0}
\end{equation}

To calculate the rest mass $ m_{0} c^{2} $ we modify
slightly Eq.\ref{Etot} to Eq.\ref{EtotErot}

\begin{equation}
E_{tot} = m_{0} c^{2} + E_{rot}
\label{EtotErot}
\end{equation}

We assumed as in the
first approach the angular velocity of the mass center
$ \omega_{m} $ and the charge is the same
$ \omega_{m} = \omega_{e} = \omega $. This allows to write
down together with Eq.\ref{Ltot} the rotational
energy $ E_{rot} $ in the form of Eq.\ref{ErotOmegaLh}

\begin{equation}
E_{rot} = \frac{1}{2} \theta \omega^{2} = \frac{1}{2} L \omega
        = \frac{1}{4}\hbar \omega
\label{ErotOmegaLh}
\end{equation}

Inserting Eq.\ref{ErotOmegaLh} in Eq.\ref{EtotErot}
allows to calculate the rest mass $ m_{0} c^{2} $
in Eq.\ref{Restmass}.

\begin{equation}
m_{0}c^{2} = E_{tot} - \frac{1}{4} \hbar \omega
\label{Restmass}
\end{equation}

The last missing parameter the radius $ r_{0} $
can be easily calculated rewriting Eq.\ref{Qvelocityr0} to
Eq.\ref{Radiusr0}

\begin{equation}
r_{0} = \frac{\beta c }{\omega}
\label{Radiusr0}
\end{equation}

If we set $ \beta = 1 $ and take for the charge $ e $ ,
the speed of light $ c $ and the magnetic moment
$ \mu_{e} $ the numerical values from
Tab. \ref{Parameterelectron} is is possible to
calculate $ \omega $. We calculated $ \omega $
for the discussed option $ \mu_{e} $ is generated from
a total charge $ 2/3 \times e $ as discussed for our ETAMFP model
in Fig.{\ref{BasicSCHEMEFP2006}} and for the possibility
this charge is $ 1 \times e $. The numerical values are
calculated with Eq.\ref{AngularVelocityr0} and shown
together with the assumption $ \beta \leq 1 $
in Tab. \ref{modelbeta1} first three lines. If the angular
velocity $ \omega $ is known for the case the charge responsible
for $ \mu_{e} $ is $ 2/3 \times e $ or $ 1 \times e $ it is possible to
calculate with Eq.\ref{Restmass} the rest mass for both
options $ m_{0} c^{2} (2/3~e ) $ and $ m_{0} c^{2} (1~e ) $
displayed in Tab. \ref{modelbeta1} in the fourth and
fifth line.
For the same both options it is possible to calculate
with Eq.\ref{ErotOmegaLh} the rotational energy
$ E_{rot}~(2/3~e) $ and $ E_{rot}~(1~e) $ taking into
account $ \omega (2/3~e ) $ and $ \omega (1~e ) $ and
the total energy $ E_{tot} $ with $ \hbar $ from
Tab. \ref{Parameterelectron}. The numerical values are shown
in Tab. \ref{modelbeta1} in the sixth and seventh line.
Also for both options it is possible to calculate
with Eq.\ref{Radiusr0} the radius $ r_{0} (2/3~e)$
and $ r_{0}(1~e)$ taking into
account $ \omega (2/3~e ) $ and $ \omega (1~e ) $ and
the $ c $ from Tab. \ref{Parameterelectron} including
$ \beta = 1 $. The numerical values are shown
in Tab. \ref{modelbeta1} in the eighth and ninth line.
In the last line of Tab. \ref{modelbeta1} the numerical
value for the Lorentz contracted radius
$ r_{e} = r_{0} \sqrt{( 1- \beta^{2})} \ge 0 $ is shown.
This value must be of course close to zero.

\begin{table}
\caption{Size of electron and charge radius for $\beta = 1$.}
\label{modelbeta1}
\begin{center}
\begin{tabular}{||l|l|l||}                                      \hline
  parameter           & Symbol        & value
                                                               \\ \hline
  velocity charge     & $ \beta  $    & $ \le 1 $
                                                               \\ \hline
  angular velocity $ 2/3~e $
                      & $ \omega (2/3~e ) $
                                      & $ \le 5.1696325 \times 10^{20} $ 1/s
                                                               \\ \hline
  angular velocity $ 1~e $
                      & $ \omega (1~e ) $
                                      & $ \le 7.7544488  \times 10^{20} $ 1/s
                                                               \\ \hline
  rest mass $ 2/3~e $ & $ m_{0} c^{2} (2/3~e ) $
                      & $ \ge 0.425931077 $ MeV $ ( 83.3526374 \% ) $
                                                               \\ \hline
  rest mass $ 1~e $   & $ m_{0} c^{2} (1~e ) $
                      & $ \ge 0.383397155 $ MeV $ ( 75.028956 \% ) $
                                                               \\ \hline
  rotational energy $ 2/3~e $
                      & $ E_{rot}~(2/3~e) $
                      & $ \le 0.085067843 $ MeV $  ( 16.6473626  \% ) $
                                                               \\ \hline
  rotational energy $ 1~e $
                      & $ E_{rot}~(1~e) $
                      & $ \le 0.127601765 $ MeV $  ( 24.971044   \% ) $
                                                               \\ \hline
  charge radius $ 2/3~e $
                      & $ r_{0} (2/3~e)$ & $ \ge 5.7991058 \times 10^{-13} $ m
                                                               \\ \hline
  charge radius $ 1~e $
                      & $ r_{0}(1~e)$    & $ \ge 3.8660705 \times 10^{-13} $ m
                                                               \\ \hline
  charge radius rest system
                      & $ r_{e} = r_{0} \sqrt{( 1- \beta^{2})} $
                      & $ \ge 0 $ m
                                                               \\ \hline
\end{tabular}
\end{center}
\end{table}

The status of the just discussed estimation of this approach is
the speed of the rotating charge is close to the speed of light
$ c $. Including the Lorentz contraction of the charge circling
the center of the electron it explains why the magnetic moment
is rather big and the electric dipole moment very small. Certainly
after the discussion about the scale dependence of the SM and BBM
the radius $ r_{e} $ will be not zero. It is also not possible in this
approach to distinguish between the $ (2/3~e) $ and $ r_{0}(1~e)$
for the calculated parameters $ \omega $,  $ m_{0} c^{2} $, $ E_{rot} $ and
$ r_{0} $.

\subsubsection { Mass and charge radius of electron
                 as Lorentz contracted gyroscope }
\label{sec:Mass and charge radius of electron as Lorentz 
                contracted gyroscope}

So far our discussed model did not include a specific assumption
about the mass core of the electron. In the discussion about the
geometrical approach of the FPs summarized in
Fig.{\ref{BasicSCHEMEFP2006}} we assumed the mass  of the
of the FPs contains two parts. An inner non rotating mass kernel,
the seeds of a quantized energy mass condensate of the Planck Era 
and a rotating
mass density distribution the seeds of the mass between Planck and GUT Era.
In the chapter about exited states of fermions 
( chap.\ref{sec:Excited states of fermions} ) we
discussed in Fig.\ref{Fermionmass}
the possibility to describe the electron as the vibration ground
state of the leptons. 
An energetic favorable object of such a mass core would be a
classical rotating sphere like a gyroscope with a
rotating axis as spin axis.

\subparagraph{ Mass core as Lorentz contracted rigid sphere }

Following the philosophy to use
a simple as possible approach the most easily accessible
assumption is to use a rigid sphere, with constant mass density,
circling the spin axis with the angular velocity $ \omega_{m} $
and a border at radius $ r_{0} $. The mass moment of inertia $ \theta $ is
in Eq.\ref{Imoment} a function of the mass $ m_{0} $ and the
radius $ r_{0} $.

\begin{equation}
\theta = \frac{2}{5} m_{0} r_{0}^{2}
\label{Imoment}
\end{equation}

The according rotational energy $ E_{rot} $ is displayed in
Eq.\ref{IErot}

\begin{equation}
E_{rot} = \frac{1}{2} \theta \omega_{m}^{2} = \frac{1}{2} L \omega_{m}
        = \frac{1}{4}\hbar \omega_{m}
\label{IErot}
\end{equation}

We assumed in Eq.\ref{Imoment} the radius $ r_{0}(mass)$ where the
mass density ends coincides like in  Fig.{\ref{BasicSCHEMEFP2006}}
with the radius $r_{0}(charge)$ where the charge is located
$ r_{0}(mass) = r_{0}(charge) = r_{0} $.
If we assume again $\omega_{m}=\omega_{e}=\omega $ like in the
previous chapter about the charge radius, the speed of the
surface of the mass sphere at the position $ r_{0} $ perpendicular
to the spin axis will exceed the speed of light. The conditions
will be very similar as in the case of the charge. In accordance
to our discussion in chap.\ref{sec:The flashing vacuum} we
assume the mass $ m_{0} $ and the radius $ r_{0} $ in Eq.\ref{Imoment}
have to be corrected with the Lorentz transformations. In our simple
model of a rigid sphere will only be the mass ring with a half moon
shaped cross section
in the vicinity of $ r_{0} $ sensitive to $ \beta $.
The mass content of this ring decreases fast to zero if the
inner radius $ r $ of this ring approaches $ r_{0} $.
This mass decrease is damping the increasing sensitivity to $ \beta $,
if $ \beta $ is approaching $ \beta = 1 $ at $ r = r_{0} $.
In contrast is the mass less surface ring of the sphere in the vicinity of $ r_{0} $
fully exposed to the sensitivity to $ \beta $ and will be substantial
contracted because $ \beta $ will be very close to $ \beta = 1 $. 
In the limit of $ \beta = 1 $ the surface ring will be in the
vicinity of $ r_{0} $ contracted to a point in the center of the
sphere.

To apply the Lorentz transformations on Eq.\ref{Imoment} we first
estimate the magnitude of the sensitivity of $ m_{0} $ and $ r_{0} $
to $ \beta $.

We begin with the $ \beta $ sensitivity of $ m_{0} $.
The mass distribution of a rotating sphere is shown in
Fig.\ref{Spheremassbeta}. On the left side in Fig.\ref{Spheremassbeta}
is drawn
the mass sphere in two dimensions  ( z-axis and x-axis )  rotating
parallel with the spin to the  z-axis. On the right side middle upper part
is displayed the mass distribution of the sphere in arbitrary units
$ m_{0}[au] $
as function of the of $ x=X~[au] $ for a constant mass density.
The mass distribution ( solid black line )
decreases from 100 [au]  at $ x = 0 $
to zero at $ x = r_{0} ( \vdash~\Uparrow ) $.
The graph below on the right side of Fig.\ref{Spheremassbeta}
shows the linear increase if $ \beta $ as function of $ x $ according
Eq.\ref{Qvelocityr0}. The graph total right side in
Fig.\ref{Spheremassbeta} displays the Lorentz factor
$ R = 1 / \sqrt { ( 1-\beta^{2} ) } $ as function of $ x $ as
black solid line. 
If we fold the $ \beta $, $ R =  1 / \sqrt { ( 1-\beta^{2} ) } $
and $ m_{0}~[au] $ dependence from $ X~[au] $ it is possible
to calculate the correction for the mass
$ \Delta m ( x,\Delta x ) = \Delta m_{0} ( x,\Delta x ) /
\sqrt { (1 - \beta^{2}) } $. The
correction $ \Delta m ( x,\Delta x ) = 0 $ at $ X[au]=r=0 $,
increases to a maximum at approximately $ X[au]=r=80 $
and approaches $ \Delta m ( x,\Delta x ) = 0 $ at
$ X[au]=r=100 $ or $ r=r_{0} $.
This is displayed
in the mass plot middle upper part ( red broken line ) of
Fig.\ref{Spheremassbeta}. Comparing the black line 
with the red broken line in this plot it is visible that
the Lorentz correction is active only for $ X[au]=r > 40 $
shown in the black hatched half moon shaped area
of Fig.\ref{Spheremassbeta} left side.
The total mass increase originated by the
Lorentz transformation is approximately $ 11 \% $ if $ \beta = 1 $
at $ x = r_{0} ( \vdash~\Uparrow ) $. We used so far in this simple
model a constant mass density distribution from $ r = 0 $ to
$ r = r_{0} $. Following the discussion in
chap.~\ref{sec:The geometrical approach} and
Fig.{\ref{BasicSCHEMEFP2006} we expect a mass distribution
which deceases from $ r = 0 $ to $ r = r_{0} $ sharply
to $ \rho = 0 $ in the vicinity of $ r = r_{0} $. This
will suppress the above discussed edge effect of the
rotating sphere to be negligible at
$ x = r_{0} ( \vdash~\Uparrow ) $. We ignore for this reason
the Lorentz correction of $ m_{0} $ in the following
calculations.

\begin{figure}
\begin{center}
 \epsfig{file=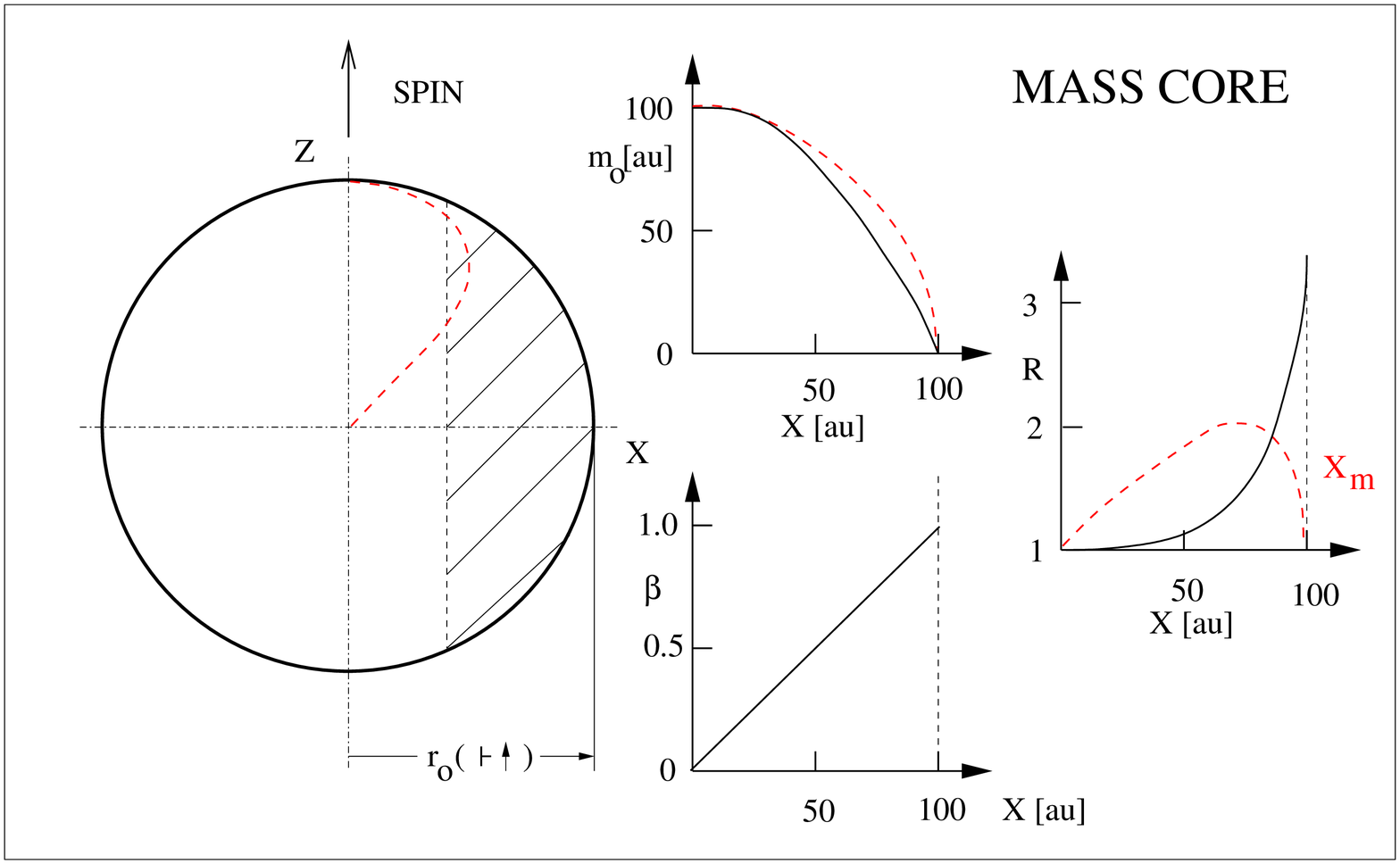,width=14.0cm,height=9.0cm}
\end{center}
\caption{ The spherical electron mass rotating $ \beta \rightarrow 1 $. }
\label{Spheremassbeta}
\end{figure}

Second we estimate the $ \beta $ dependence of $ r_{0} $.
The border of the rotating sphere is a mass less shell rotating
at $ x = r_{0} ( \vdash~\Uparrow ) $ with maximal $ \beta = 1 $.
The surface velocity $ \beta $ decreases if the distance $ x $
from the spin axis get smaller $ x < r_{0} ( \vdash~\Uparrow ) $
and finally will be $ \beta = 0 $ at the spin axis as
shown in the lower plot on the middle side ( black solid line )
in Fig.\ref{Spheremassbeta}. The distance from the spin axis $ x $
in the rotating
system will undergo a Lorentz contraction of Eq.\ref{shellcontaction}
to the distance $ x_{m} $ measured from an experiment
outside the electron.

\begin{equation}
X_{m}=x_{m} = x \sqrt{1-\beta^{2}}
\label{shellcontaction}
\end{equation}

The application of Eq.\ref{shellcontaction} on $ x $ in arbitrary
units is displayed in the plot total right side of Fig.\ref{Spheremassbeta}
( red broken line ). The contraction put $ x_{m} $ for $ \beta = 1 $ at
$ x = r_{0} ( \vdash~\Uparrow ) $ to zero and the radius at
$ x = 0 $ get not affected. A quarter of this contracted infinite thin
shell is shown
in the red broken line of Fig.\ref{Spheremassbeta} left side.
By inspection of Fig.\ref{Spheremassbeta} it is obvious,
the $ \beta $ dependence of $ r_{0} $ must by
included to calculate the moment of inertia in
Eq.\ref{Imoment}.

\subparagraph{ Model factor respecting not rotating inner mass kernel }

After the discussion about
the basic scheme in Fig.{\ref{BasicSCHEMEFP2006}} we expect 
not the whole mass of the electron rotates. This would imply
the inner Planck Kernel does not contribute to the
rotational energy.
To respect this possible mass kernel
we introduce a model
factor $ B $. We expect this factor is $ 0 < B < 1 $ because
the decreasing influence of the inner non rotating mass kernel
dominates the maximum possible increase of the mass at
border of the mass sphere we investigated in the preceding
paragraph.
We modify for this reason the rest mass $ m_{0} $ to model rest mass
$ m_{0}^{model} = B m_{0} $.

The experimental accessible radius from the electron discussed
in  chap.\ref{sec:Status of experimental limits on the sizes of
Fundamental Particles} is sensitive in $ \EEGG $ reaction to the
mass radius of the electron and  the radius the charges
are loaded. The radius limit deduced of the electric dipolmoment
of the electron is sensitive only to the radius the charges
are loaded.
We assume the charge of the electron is placed
at the radius $ x = r_{0} ( \vdash~\Uparrow ) $. 
The limits for this radius we discussed in
Eq.\ref{LimitDipolre} and Eq.\ref{Comparisondeexpre}.
Following this discussion and Eq.\ref{shellcontaction}
we set at $ z = 0 $ the
minimal possible radius $ x_{m} ( min ) = r_{e} $.
If we collect our assumption about the different angular velocities,
radius and model factor we end at
$\omega_{m}=\omega_{e}=\omega $ , $ x_{m} ( min ) = r_{m}=r_{e}=r_{L} $
and $ m_{0}^{model} = B m_{0} $. Including these assumptions we
replace in  Eq.\ref{Imoment}
the radius $ r_{0} $ by the  Eq.\ref{r0mass}.

\begin{equation}
r_{0} = \frac{r_{m}}{\sqrt{1-\beta^{2}}}
\label{r0mass}
\end{equation}

Inserting the model factor and Eq.\ref{r0mass} in Eq.\ref{Imoment}
we get the Lorentz corrected mass moment of inertia $ \theta $
including all our assumptions in in Eq.\ref{LImoment}.

\begin{equation}
\theta = \frac {2}{5} B m_{0} \frac{r_{m}^{2}}{(1-\beta^{2})}
\label{LImoment}
\end{equation}

At this stage of the development of our  ETAMFP model
we investigate the impact of the so far introduced
assumptions, about the mass and charge radius of a electron
considered as Lorentz contracted gyroscope, on the model.
We develop for this reason in a first step
equations for the velocity $ \beta $ ,
the angular velocity $ \omega $,
the energy of $ m_{0} c^{2} $ and $ E_{rot} $, the radius $ r_{0} $
the Lorentz contracted radius $ r_{L} $ as function of the
model factor $ B $. In a second step we study the from
the numerical measured parameters in Tab. \ref{Parameterelectron}
of the electron given numerical limits for the
model factor $ B $ in five plots $ \beta^{2} = f ( B ) $,
$ \omega = f ( B ) $ , $ E_{rot} = f ( B ) $,
$ r_{0} = f ( B ) $ and $ r_{L} = f ( B ) $ .


\subparagraph{ Equations including model factor for rotating mass core }

If we insert Eq.\ref{Qvelocityr0} in Eq.\ref{LImoment} and use
Eq.\ref{IErot} it is possible to calculate $ \beta $ as
function from $ \omega $ and $ B m_{0} $ in Eq.\ref{betaI}.

\begin{equation}
\beta^{2}=\frac{5}{4}\frac{\hbar \omega }{ B m_{0} c^{2}}
\label{betaI}
\end{equation}

Next we replace in Eq.\ref{Magmomentr0} $ \mu_{e} $
with $ \mu_{e} = A \mu_{B} $ ( With $ A $ as anomal magnetic moment ),
use $ r_{L} $ from
Eq.\ref{Qvelocityr0} and find Eq.\ref{AngularVelocityI}
the angular velocity $ \omega_{e} $ as function of
$ \beta^{2} $.

\begin{equation}
\omega_{e}=\frac{1}{2} \frac{e(c\beta)^{2}}{A\mu_{B}}
\label{AngularVelocityI}
\end{equation}

The experimental measured parameters $ e $, $ c $,
$ A=1.001159652187 $,
and $ \mu_{B} $ are known and shown in
Tab. \ref{Parameterelectron}.

Inserting Eq.\ref{AngularVelocityI} in Eq.\ref{betaI} and
using Eq.\ref{IErot} with the definition of
$ \mu_{B}=(e\hbar)/(2m_{e}) $ we find
the angular velocity $ \omega $ as function of $ B $ in
Eq.\ref{omegaMODEL}.

\begin{equation}
\omega = \frac{E_{tot}}{\hbar}(4-\frac{5~\tilde{e}}{B A })
\label{omegaMODEL}
\end{equation}

According our chapter of the schemes of fermions we use in
Eq.\ref{omegaMODEL} the model depending charge $ \tilde{e} $
where $ \tilde{e} $ could be $ \tilde{e} = 1 $ or $ 2/3 $.


To find a similar expression for $ \beta $ we insert
$ m_{0}c^{2} $ from  Eq.\ref{Restmass}
in Eq.\ref{betaI} and replacing $ \omega $
with Eq.\ref{omegaMODEL}. This allows to calculate
the velocity $ \beta^{2} $ in Eq.\ref{betaMODEL}.

\begin{equation}
\beta^{2} = \frac {4~A~B-5~\tilde{e}}{B~\tilde{e}}
\label{betaMODEL}
\end{equation}

This equation depends only from the anomal magnetic moment $ A $,
the reduced charge $ \tilde{e} $ and the model factor $ B $.

A certain values of the  model parameter $ B $ defines
via Eq.\ref{betaMODEL} and  Eq.\ref{omegaMODEL} a
velocity $ \beta $ and the angular velocity $ \omega $.
Inserting $ \omega $ for the calculation of
the energy $ m_{0} c^{2} = f( B ) $
and $ E_{rot} = f( B ) $ we use Eq.\ref{Restmass}
and Eq.\ref{IErot}. For the calculation of the
radius $ r_{0} $ we insert $ \beta $ and $ \omega $
in Eq.\ref{Radiusr0}.

For the last missing parameter $ r_{L} = f(B) $ we
use Eq.\ref{Qvelocityr0} resolved after $ r_{L} $ and
insert the angular velocity $ \omega $ from
Eq.\ref{omegaMODEL}
and the velocity $ \beta $ from Eq.\ref{betaMODEL}
to calculate Eq.\ref{RlLorentz}

\begin{equation}
r_{L} = \frac {c}{\omega}\beta\sqrt{1-\beta^{2}} =
        \frac {c}{\frac{E_{tot}}{\hbar}(4 - \frac{5~\tilde{e}}{B A})}
        \sqrt{\frac{4 A B - 5 ~ \tilde{e}}{B ~\tilde{e}}}
        \sqrt{1-\frac{4 A B - 5 ~ \tilde{e}}{B ~\tilde{e}}}
\label{RlLorentz}
\end{equation}

The status of our calculations permits now to calculate
the velocity $ \beta $, the angular velocity $ \omega $,
the energy of $ m_{0} c^{2} $ and $ E_{rot} $, the radius $ r_{0} $
the Lorentz contracted radius $ r_{L} $ as function of the
model factor $ B $.

\subparagraph{ Numerical limits for model factor set by
               experiment }

As announced in a second step we study the from
the numerical measured parameters in Tab. \ref{Parameterelectron}
of the electron dictated numerical limits for the
model factor $ B $ in five plots $ \beta^{2} = f ( B ) $,
$ \omega = f ( B ) $ , $ E_{rot} = f ( B ) $,
$ r_{0} = f ( B ) $ and $ r_{L} = f ( B ) $ .

Important for the whole following discussions is the test which
charge $ \tilde{e} = 1 $ or $ \tilde{e} = 2/3 $ is in agreement
with our so far discussed model. In Tab. \ref{modelbeta1} it was
not possible to distinguish between the two possibilities.

If we insert $ B=1 $ and $ \tilde{e} = 1 $ in Eq.\ref{betaMODEL}
we calculate an negative $ \beta^{2}=-0.995361392 $ which will result
in an imaginary $ \beta $. If we use the physical allowed limits for
$ 0 \le \beta^{2} \le 1 $ and $ \tilde{e} = 1 $
we find the according limit for
the model factor $ B $ is ranging from $ 1.24 \le B \le 1.66 $.
The imaginary $ \beta $ for $ B=1 $ is in disagreement with
the Lorentz transformation. The model factor $ B $ ranging
from $ 1.24 \le B \le 1.66 $ is in disagreement with our
expectation of $ B \le 1 $.

If we insert in $ B=1 $ and $ \tilde{e} = 2/3 $ in
Eq.\ref{betaMODEL} we calculate a $ \beta $ slightly bigger as one
of $ \beta^{2}=1.006957912 $. Inserting in Eq.\ref{betaMODEL}
the physical allowed limits of $ 0 \le \beta^{2} \le 1 $
and $ \tilde{e} = 2/3 $
we calculate a model factor ranging from
$ 0.832368076 \le B \le 0.998610351 $. It is interesting to
notice that the anomal magnetic moment A and
$ \tilde{e} = 2/3 $ are the leading
parameters in Eq.\ref{betaMODEL} to decide
about the numerical value of $ \beta^{2} $.
According our ETAMFP model we expect a real $ \beta \le 1 $ and
a model factor $ B \le 1 $. This is in agreement
with our geometrical approach of Fig.{\ref{BasicSCHEMEFP2006}}
and the whole scheme of fermions where we assume the
charge for the electron is placed in units $ 1/3 $ at
three possible locations.
For this reasons we will
in the following discussion only consider the $ \tilde{e} = 2/3 $ case.

To calculate the plot $ \beta^{2} = f ( B ) $ we use
Eq.\ref{betaMODEL} and calculate the
the model factor $ B $ in the limit of $ 0 \le \beta^{2} \le 1 $.
\begin{figure}
\begin{center}
 \epsfig{file=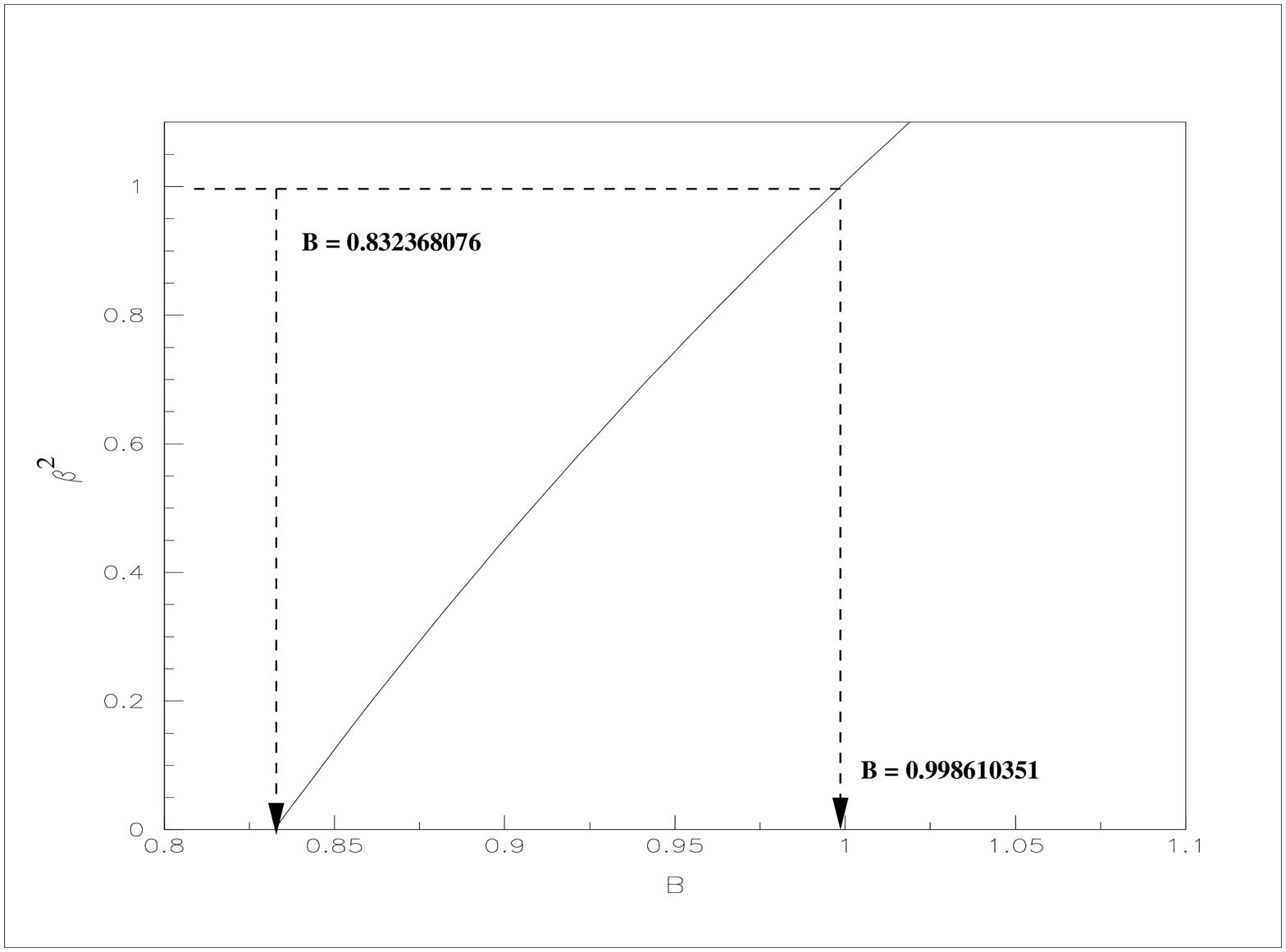,width=12.0cm,height=9.0cm}
\end{center}
\caption{ Possible range of $ B $ as function from $ \beta^{2} $.}
\label{BETAfunctionB}
\end{figure}
The solid black line in Fig.\ref{BETAfunctionB} displays
the discussed function. We find for $ \beta^{2} = 1 $ the
upper limit for $ B=0.99861035 $ and for a not rotating
sphere accordingly $ \beta^{2} = 0 $ a lower limit for
$ B = 0.832368076 $ shown in broken solid lines with
arrow in Fig.\ref{BETAfunctionB}. The limits of $ B $
are used in the following plots
$ \omega = f ( B ) $ , $ E_{rot} = f ( B ) $,
$ r_{0} = f ( B ) $ and $ r_{L} = f ( B ) $
to study the behavior of $ \omega $, $ E_{rot} $ and
$ r_{L} $ in the frame of our ETAMFP model.

To calculate the plot $ \omega = f ( B ) $ we use
Eq.\ref{omegaMODEL} and vary $ B $ from $ B = 0.7 $
to $ B = \infty  $ to display in Fig.\ref{OMEGAfunctionB}
the whole possible range of $ \omega $.
The solid convoluted black line in Fig.\ref{OMEGAfunctionB}
displays the discussed function.
\begin{figure}
\begin{center}
 \epsfig{file=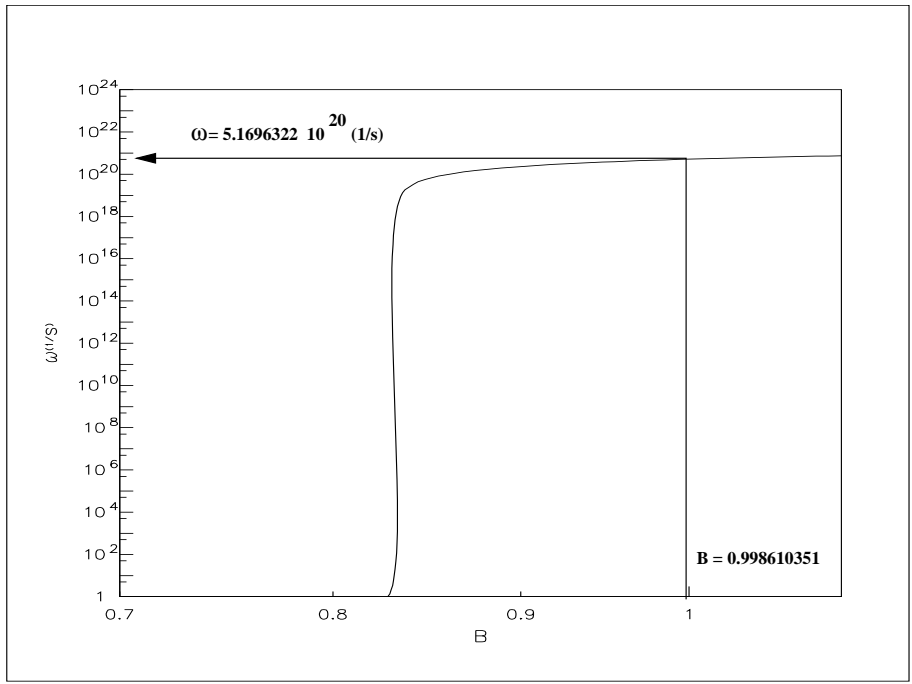,width=12.0cm,height=9.0cm}
\end{center}
\caption{ Possible range of $ \omega $ as function from $ B $. }
\label{OMEGAfunctionB}
\end{figure}
If we insert in Eq.\ref{omegaMODEL} $ B=0.998610351 $ which
is in accordance with $ \beta=1 $ we calculate a
maximum value of angular velocity
$ \omega_{max} = 5.1696322\times10^{20}~(1/s)$. This value
is in agreement with the value in Tab. \ref{modelbeta1} for
$ \omega ( 2/3 e ) $ and shown in Fig.\ref{OMEGAfunctionB}
as solid straight  black line with arrow.
If we insert in Eq.\ref{omegaMODEL} $ B=0.832368076 $ which
is in accordance with $ \beta=0 $ we calculate of course
a minimum value of angular velocity of
$ \omega_{min} = 0.0~(1/s)$ important only for the
completeness of the plot. A similar result we find
if we insert in Eq.\ref{omegaMODEL} $ B = \infty $ we find the
maximum possible $ \omega_{max}=3.1053763 \times 10^{21} (1/s) $.
This value is only of mathematical interest because it would
exceed $ \beta = 1 $.

To calculate the plot $ E_{rot} = f ( B ) $
and $ m_{0} c^{2} = f ( B ) $ we vary the model parameter
$ B $ from $ B = 0.8 $ to $ B = 1.0 $. Next we calculate with
Eq.\ref{omegaMODEL} the according angular velocity
$ \omega $ and subsequent inserting $ \omega $ in
Eq.\ref{Restmass} to get $ m_{0} c^{2} = f ( B ) $ and
Eq.\ref{IErot} to find $ E_{rot} = f ( B ) $.
The black broken line in Fig.\ref{ROTmassfunctionB}
displays $ m_{0} c^{2} = f ( B ) $ and the black
solid line $ E_{rot} = f ( B ) $.

\begin{figure}
\begin{center}
 \epsfig{file=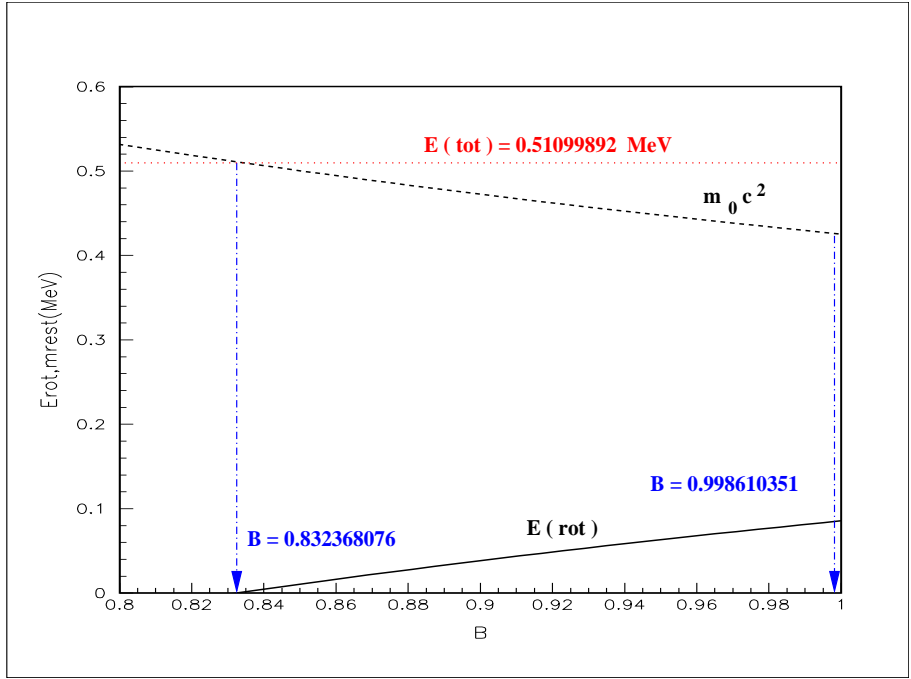,width=12.0cm,height=9.0cm}
\end{center}
\caption{ Possible range of $ m_{0} c^{2}$ and $ E_{rot} $
          as function from $ B $. }
\label{ROTmassfunctionB}
\end{figure}

The maximum possible share for $ B = 0.998610351 $ between
$ m_{0} c^{2} $ and $ E_{rot} $ ist $ m_{0} c^{2} = 0.425931082 $ MeV 
or $ (83.35 \% )$ and $ E_{rot} = 0.085067838 $ MeV or
$ (16.65 \% )$. These values are in agreement with the
numbers for $ m_{0}c^{2}( 2/3 e ) $ and $ E_{rot} ( 2/3 e ) $
in Tab. \ref{modelbeta1} and marked by a blue dot dashed line
with arrow on the
right side in Fig.\ref{ROTmassfunctionB}.
The lower limit for $ B = 0.832368076 $ results in
$ E_{rot} = 0.00 $ MeV and $ m_{0} c^{2} = E_{tot} $ and is
marked as blue dot dashed line with arrow at the left side in
Fig.\ref{ROTmassfunctionB} together with the numerical value
for $ E_{tot} $ as dotted red line.

To calculate the plot $ r_{0} = f ( B ) $
we vary the model parameter
$ B $ from $ B = 0.8 $ to $ B = 1.0 $, calculate with
Eq.\ref{betaMODEL} velocity $ \beta^{2} $ with Eq.\ref{omegaMODEL}
the angular velocity $ \omega $ and insert these values in
Eq.\ref{Radiusr0} to find $ r_{0} = f ( B ) $
shown as black solid line in Fig.\ref{ROfunctionB}.

\begin{figure}
\begin{center}
 \epsfig{file=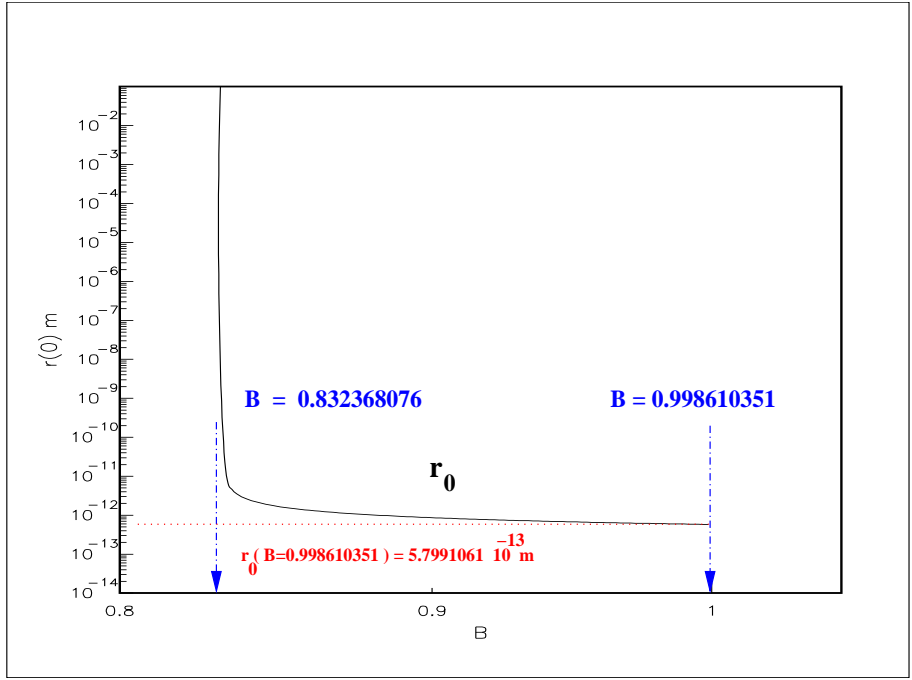,width=12.0cm,height=9.0cm}
\end{center}
\caption{ Possible range of $ r_{0} $ as function from $ B $. }
\label{ROfunctionB}
\end{figure}

An observer located in the middle point of the electron
rotating with $ \omega $ would measure a minimum possible
$ r_{0} $ allowed by the experimental date of
$ r_{0}=5.7991061 \times 10^{-13}~m $ for the condition
$ \beta = 1 $ and $ B=0.998610351 $. This value is in agreement
with the value for $ r_{0} ( 2/3 e ) $ in
Tab. \ref{modelbeta1} and marked by the crossing  of the
red dotted line with the blue dashed dotted line on the right side
in Fig.\ref{ROfunctionB}. The maximum of this radius will increase
with decreasing $ B $ slowly to $ r_{0} \rightarrow \infty $ at
$ B = 0.832368076 $ shown from the blue dashed dotted line
left side in Fig.\ref{ROfunctionB}. This singularity is certain
not supported by the data because the electron has a finite spin,
is only a mathematical option.

To calculate the plot for the Lorentz contracted radius
$ r_{L} = r_{exp} = f ( B ) $ of the electron measured from an
observer outside the electron we vary the model parameter from
$ B = 0.832368076 $ to $ B = 1.0 $ in Eq.\ref{RlLorentz}
and plot the black solid line in Fig.\ref{RexpfunctionB}.

\begin{figure}
\begin{center}
 \epsfig{file=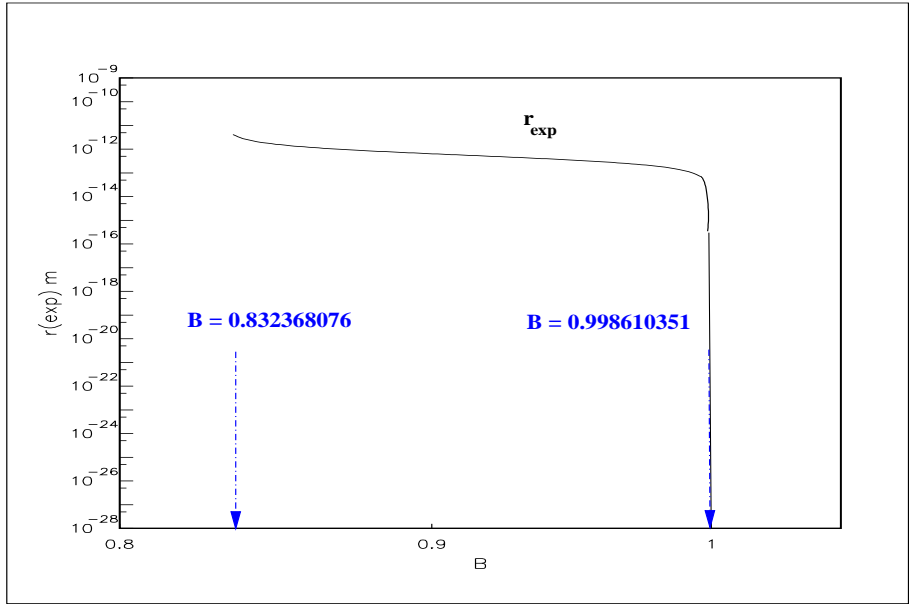,width=12.0cm,height=8.0cm}
\end{center}
\caption{ Possible range of $ r_{L} $ as function from $ B $. }
\label{RexpfunctionB}
\end{figure}

The square root structure of Eq.\ref{RlLorentz} leads to
a steep gradient singularity if $ B \rightarrow 0.998610351 $
what includes that also $ \beta \rightarrow 1 $. This is shown
by a blue dot dashed line with arrow in the right side of
Fig.\ref{RexpfunctionB}. For
$ B \rightarrow 0.832 368 076 $ $ r_{L} $ reaches a maximum
marked by a blue dashed dotted line with arrow left side in
Fig.\ref{RexpfunctionB}. The possible
from the experiment and model allowed values ranging from
is $ 8.6 \times 10^{-12}~(m) \ge r_{L} \ge 0.0 ~ (m) $.

\subsubsection { Implication of the electric dipole moment of the
                 electron on its geometrical extension }

In all calculations so far we did not use the electric dipole moment
$ d_{e} $ of the electron because the experimental value is very small.
The excellent , very important and ambitious experiments \cite{electricDIPOL}
measured so far for $ d_{e} $ a value
( See for example shown in Tab. \ref{Parameterelectron} )
comparable with zero, what has to be considered as a limit. 

In all our considerations of the classical approach the
charges of the electron ( For example in Fig.{\ref{BasicSCHEMEFP2006}} )
are separated by a distance $ r_{0} $. A classical dipole is
constructed from two electric charges of different sign separated by a
distance $ x $. Our ETAMFP model contains two electric charges $ - 1/3 \times e $
from the same sign separated by a distance $ r_{0} $. In this sense is our
model not a classical dipole but certainly very similar.
Our considerations about the sensitivity of the experiment to measure
the electric dipole moment in Eq.\ref{Comparisondeexpr0} to
Eq.\ref{Comparisondeexpre} come to the conclusion that the
experiments measure the distance between the charges in the LAB-system.
If we inspect the value for $ r_{e} $ in Tab. \ref{modelbeta1}
and the function of $ r_{e} = f(B) $ in Fig.\ref{RexpfunctionB}
it is obvious, if the speed of the
charges approach $ c $ the Lorentz contraction will push 
the radius $ r_{e} $ against zero. This is in agreement
with the minute measurement of $ d_{e} $.
But the limit
is still seven magnitudes away from the Planck scale, which is still a save
distance to the quantization of the mass.
We identify for these reasons $ r_{e} $ with the measured electric
dipole moment $ d_{e} $.
In this sense is $ d_{e} $ a direct measurement
of the distance between the two charges $ -1/3 $ circulating
around the mass kernel of the electron measured from an observer
outside the electron in the LAB-system.

Respecting the experimental challenge that a discovery potential of about
$ 5 \times \sigma $ for the measurement of the electric dipole moment
of the electron is not reached so far and take into account the just
discussed conditions, we introduce the dipole moment in our ETAMFP model and use
the for the dipole moment deduced radius $ r_{e} $.
In a first stage we take the from the observer in the LAB-system measured
radius $ r_{e} $ to calculate $ \beta $ , the angular velocity
$ \omega $, rest mass $ m_{0}c^{2} $, rotational energy $ E_{rot} $ , the
radius of the $ r_{0} $ of the electron without any specific assumption
about the mass kernel of the electron. It means we ignore the above discussed
model factor $ B $. In a second stage we include this model factor $ B $.

For the first stage we calculate the radius $ r_{e} $ from the electric
dipole moment.
As displayed in Fig.{\ref{BasicSCHEMEFP2006}} and Fig.\ref{Electron2006} the
three charges $ -1/3 $ are generating in our scheme the electric dipole moment.
It would be possible to group the two outer charges $ -1/3 \times e \times 2 r_{e} $
or the inner charge two times with the outer charges
$ 2 \times ( -1/3 \times e \times r_{e} ) $ what leads to the same result of the
dipole moment shown in Eq.\ref{ElectronDipol}.

\begin{equation}
d_{e} = e~l = (\frac{1}{3} e ) ( 2 r_{e} )
            =  2((\frac{1}{3} e ) ( r_{e} ))
            =  ( 0.07 \pm 0.07 ) \times 10^{-26} e~cm
\label{ElectronDipol}
\end{equation}

The numerical value is taken from Tab. \ref{Parameterelectron}.
The according experimental measured radius $ r_{e} $
is given in Eq.\ref{ElectronDipolRadius}.

\begin{equation}
      r_{e} = \frac{3}{2}~l = 0.105 \times 10^{-28}~m
\label{ElectronDipolRadius}
\end{equation}


To calculate an equation for $ \beta^{2} $ we
insert $ \omega _{e} $ from Eq.\ref{Qvelocityr0} in
Eq.\ref{Magmomentr0} and find
Eq.\ref{BETAsquer}

\begin{equation}
      \beta^{2} = \frac{(\frac{2\mu_{e}}{c~e~r_{e}})^{2}}
                  {1 + {(\frac{2\mu_{e}}{c~e~r_{e}})^{2}}}
\label{BETAsquer}
\end{equation}

To simplify Eq.\ref{BETAsquer} we introduce the
constant factor $ K $ of Eq.\ref{Kfactor}

\begin{equation}
      K = (\frac{2\mu_{e}}{c~e~r_{e}})^{2}
\label{Kfactor}
\end{equation}

The equation for the speed of the charge $ 2/3 \times e $
of the electron $ v = \beta \cdot c $ we express in the
very compact form of $ \beta^{2} $ in
Eq.\ref{BETAKfactor}.

\begin{equation}
      \beta^{2} = \frac{K}{1+K}
\label{BETAKfactor}
\end{equation}

As the quadratic form of $ K \ge 0 $ the value of $ \beta $ is between
$ 0 \le \beta \le 1 $.
To calculate the angular velocity $ \omega_{e} $ of the charges
we resolving Eq.\ref{Qvelocityr0} after $ \omega _{e} $
and insert the K factor from Eq.\ref{Kfactor} in this equation
what is shown in Eq.\ref{omegare}.

\begin{equation}
      \omega_{e} = \frac{c}{r_{e}} \frac{1}{1+K} \sqrt{K}
\label{omegare}
\end{equation}

Inserting Eq.\ref{omegare} in Eq.\ref{Restmass} defines the rest mass
$ m_{0} c^{2} $ in Eq.\ref{restmassK}

\begin{equation}
       m_{0} c^{2} = E_{tot} -\frac{1}{4} \hbar \frac{c}{r_{e}}
                     \frac{1}{1+K}\sqrt{K}
\label{restmassK}
\end{equation}

and inserting Eq.\ref{omegare} in Eq.\ref{IErot} allows to calculate
the rotational energy $ E_{rot} $ in Eq.\ref{Erotre}.

\begin{equation}
       E_{rot} = \frac{1}{4} \hbar \frac{c}{r_{e}} \frac{1}{1+K}\sqrt{K}
\label{Erotre}
\end{equation}

Inserting Eq.\ref{BETAKfactor} in Eq.\ref{r0mass}
allows to calculate the charge radius of the electron in the rotating
system in Eq.\ref{r0fuctionK}.

\begin{equation}
       r_{0} = r_{e} \sqrt { 1 + K }
\label{r0fuctionK}
\end{equation}

We dispense to recalculate
the different parameters of Tab. \ref{modelbeta1},
with Eq.\ref{ElectronDipol} to Eq.\ref{r0fuctionK} because
the change of the parameters is negligible if we assume
$ \beta \approx 1 $.

As announced before in a second stage we include the  model factor $ B $
in our calculations. We first calculate the factor $ B $ themselves as
function of the magnetic moment $ \mu_{e} $, velocity of light $ c $,
charge $ e $ and radius $ r_{e} $ summarized in the factor $ K $
of Eq.\ref{Kfactor}.
Using Eq.\ref{BETAKfactor} and Eq.\ref{betaMODEL} it is possible
to calculate the model factor $ B $ in Eq.\ref{BfactorfunctionK}

\begin{equation}
       B = \frac{(1+K)5\tilde{e}}{4 A ( 1+K ) - K \tilde{e} }
\label{BfactorfunctionK}
\end{equation}

The factor $ B $ is written in the simple form as function of $ K $,
the anomal magnetic moment $ A $ and the reduced charge $ \tilde{e} $.
Inserting the factor $ B $ in they equations from Eq.\ref{betaI} to
Eq.\ref{RlLorentz} it is possible to study 
the influence of
the electric dipole moment of the electron on our Lorentz contracted
mass core.

It is for the further discussion of our ETAMFP model important to study the
numerical impact of the measured value of the electric dipole moment
on our ETAMFP model. We inserting for this reason the electric dipole moment
from Tab. \ref{Parameterelectron} in Eq.\ref{ElectronDipol} to
Eq.\ref{r0fuctionK} to calculate the numerical values of the parameters
$ \beta $, the angular velocity $ \omega $, rest mass $ m_{0}c^{2} $,
rotational energy $ E_{rot} $ and the radius of the $ r_{0} $.
The numerical value are shown in  Tab. \ref{electronPARAMETERSdipol}.

\begin{table}
\caption{Size of electron and charge radius using the
         electric dipole moment.}
\label{electronPARAMETERSdipol}
\begin{center}
\begin{tabular}{||l|l|l||}                                      \hline
  parameter           & Symbol        & value
                                                               \\ \hline
  velocity charge     & $ \beta  $    & $ < 1 $
                                                               \\ \hline
  angular velocity $ 2/3~e $
                      & $ \omega  $
                                      & $ = 5.1696321 \times 10^{20} $ 1/s
                                                               \\ \hline
  rest mass $ 2/3~e $ & $ m_{0} c^{2}  $
                      & $ =  0.425931083 $ MeV $ ( 83.3526386 \% ) $
                                                               \\ \hline
  rotational energy $ 2/3~e $
                      & $ E_{rot} $
                      & $ = 0.085067837 $ MeV $  ( 16.6473613  \% ) $
                                                               \\ \hline
  charge radius $ 2/3~e $
                      & $ r_{0} $ & $ =   5.7991062 \times 10^{-13} $ m
                                                               \\ \hline
  charge radius rest system
                      & $ r_{e} = r_{0} \sqrt{( 1- \beta^{2})} $
                      & $0.105 \times 10^{-28} $ m
                                                               \\ \hline
  model factor
                      & $ B $
                      & $0.998610351$
                                                               \\ \hline
\end{tabular}
\end{center}
\end{table}

The first line in Tab. \ref{electronPARAMETERSdipol} displays the
numerical value of the speed of the outer charges $ 2/3 \times e $ in
form of $ \beta $ calculated with Eq.\ref{BETAKfactor}.
This speed get pushed from the very small value
of the electric dipole moment extreme close to $ \beta = 1 $.
But the numerical value is not one. Inspecting Eq.\ref{BETAKfactor}
it is obvious $ \beta $ must be smaller as one depending about the
numerical value of the factor $ K $ of Eq.\ref{Kfactor} which
is as always positive. The factor $ K $ is a function of the
magnetic dipole moment $ \mu_{e} $, speed of light $ c $, charge $ e $
and the contracted radius $ r_{e} $ derived from the electric dipole
moment of the electron. The numerical extreme small measured value of
the electric dipole moment pushes $ K $ to a very high number
shown Tab. \ref{PrecissionKbetaB} in the first line.
As consequence $ \beta $ calculated
from Eq.\ref{BETAKfactor} is very close to one as displayed in
Tab. \ref{PrecissionKbetaB} second line.

\begin{table}
\caption{ High precession numerical values of the parameter $ K $,
          $ \beta $ and $ B $. }
\begin{center}
\begin{tabular}{||l|l||}                                      \hline
  parameter     & value
                                                               \\ \hline
  K                   & $ 3.0503068128782230372348481033722070616426133911863 \times 10^{33} $
                                                               \\ \hline
  $ \beta $           & 0.99999999999999999999999999999999983608206299476884
                                                               \\ \hline
  B                   & 0.99861035118674254715174743056685316037852026506779
                                                               \\ \hline
\end{tabular}
\end{center}
\label{PrecissionKbetaB}
\end{table}

The second line in Tab. \ref{electronPARAMETERSdipol} displays
the angular velocity $ \omega $ the whole electron considered
as a gyroscope circuits around an axis which coincides with the
spin axis of the electron. The numerical values are the same
as discussed in Tab. \ref{modelbeta1} for the case
$ \omega (2/3 e ) $, what is after the just discussed high
$ \beta $ value a logical consequence.

The third line in Tab. \ref{electronPARAMETERSdipol} shows
the rest mass $ m_{0} c^{2} $ of the electron calculated
with Eq.\ref{restmassK}. Also this numerical value agrees
with the numbers in Tab. \ref{modelbeta1} for the case
$ m_{0} c^{2} ( 2/3 e ) $.

The fourth line in Tab. \ref{electronPARAMETERSdipol} displays
the rotational energy $ E_{rot} $ of the electron calculated
with Eq.\ref{Erotre}. The numerical value agrees
with the numbers in Tab. \ref{modelbeta1} for the case
$ E_{rot} (2/3 e ) $.

The fifth line in Tab. \ref{electronPARAMETERSdipol} displays
the uncontracted radius $ r_{0} $ of the mass boundary of the
electron which coincides with the radius where the charges
$ 1/3 e $ in  Fig.{\ref{BasicSCHEMEFP2006}} are located
calculated with Eq.\ref{r0fuctionK}.
This radius an observer would measure if he is placed in
the center of the electron rotating with $ \omega $.
The numerical value agrees with the Tab.
 \ref{modelbeta1} for the case
$ r_{0} (2/3 e ) $.

The sixth line in Tab. \ref{electronPARAMETERSdipol} displays
the contracted radius $ r_{e} $ of the mass boundary of the
electron. This coincides with the radius where the charges
$ 1/3 e $ in  Fig.{\ref{BasicSCHEMEFP2006}} are located.
This radius an observer would measure if he is placed in
in the LAB-system outside the electron in rest.
The numerical values are taken from Eq.\ref{ElectronDipolRadius} or
recalculated for test of the high precision numeric as
function of $ r_{0} $ and $ \beta $ as shown in
Tab. \ref{electronPARAMETERSdipol}.

Finally we calculated
with Eq.\ref{BfactorfunctionK} the numerical
value of the model factor $ B $ which is shown in
Tab. \ref{electronPARAMETERSdipol} last line. Numerical
is this factor in Tab. \ref{electronPARAMETERSdipol}
in agreement with the factor used in Fig.\ref{BETAfunctionB}
Fig.\ref{RexpfunctionB}
for the upper limit of $ B $ for $ \beta = 1 $ up
to very high precision as shown in Tab. \ref{PrecissionKbetaB}
in the last line. The measurement of the electric dipole moment
of the electron locates the solution for $ \beta $, $ \omega $,
the energies $ m_{0} c^{2} $, $ E_{rot} $, the radius $ r_{0} $ and
$ r_{e} $ as function of $ B $ displayed in
in Fig.\ref{BETAfunctionB} to Fig.\ref{RexpfunctionB} as function of
$ B $ at the upper limit close to $ \beta = 1 $ as shown
in Tab. \ref{PrecissionKbetaB}. 

\subsubsection { Estimation of the size of the mass kernel of
                 the electron }

After we introduced a model factor $ B $ in the calculation of the
rotating mass of the electron the calculation shows that the factor
taken from Tab. \ref{electronPARAMETERSdipol} is $ B = 0.998610351 $,
this is less as one. We conclude from this fact that the electron
contains a non rotating mass kernel what would be a heritage of the
Planck scale as discussed in Fig.{\ref{SchematicDevFP}}. If the
density of the electron mass kernel $ \rho_{K} $ and
the density of the mass shell of the electron $ \rho_{S} $
would be known
and we use the simple relation of Eq.\ref{MassKernel}
\begin{equation}
       m_{K} = m_{0} ( 1-B )
\label{MassKernel}
\end{equation}
it is possible to write down a relation for the radius 
$ r_{K} $ of the mass kernel of the electron for our
ETAMFP like Eq.\ref{KernelRadius}.

\begin{equation}
      r_{K} = r_{0} \sqrt[3]{\frac{(1-B)}
                             {(1+B(\frac{\rho_{K}}{\rho_{S}}-1))}}
\label{KernelRadius}
\end{equation}

If we recall the discussion in 
chap.\ref{sec:Empirical toy ansatz about a
microstructure of a fundamental particle} it is only possible
in the ETAMFP model to predict a general tendency about
$ \rho_{K} $ and $ \rho_{S} $. But if we follow the in 
Fig.{\ref{SchematicDevFP} discussed heritage ansatz it is possible
to estimate the order of magnitude of $ r_{K} $.

To estimate the numerical magnitude of the radius $ r_{K} $
we assume the ratio of the density at the GUT scale to the total
mass density of the electron $ \rho_{GUT} / \rho_{e} $ is the same
as the ratio of the density of the Planck scale to the mass
kernel of the electron $ \rho_{PL} / \rho_{K} $. The density of the
mass kernel reads than like Eq.\ref{KernelDensity}.

\begin{equation}
      \rho_{K} = \frac{\rho_{PL}}{\rho_{GUT} / \rho_{e}}
\label{KernelDensity}
\end{equation}

The density of $ \rho_{GUT} $ depents about $ \Lambda_{GUT} $ like
Eq.\ref{GUTdensity}.

\begin{equation}
      \rho_{GUT} = \Lambda_{GUT}^{4}/(\hbar c )^{3}
\label{GUTdensity}
\end{equation}

For the size of the GUT scale exist only approximate numbers. The
current values in the literature are varying in a big range as
shown in Eq.\ref{GUTscale} \cite{GUTscale}.

\begin{equation}
      10^{13}~GeV < \Lambda_{GUT} < 10^{17}~GeV
\label{GUTscale}
\end{equation}

Implementing five different GUT scales $ \Lambda_{GUT} $ in the
range of Eq.\ref{GUTscale} in Eq.\ref{GUTdensity} allows
to calculate the according GUT densities $ \rho_{GUT} $.
For $ \hbar $ and $ c $ we used the values from
Tab. \ref{Parameterelectron}. Inserting $ \rho_{GUT} $,
$ \rho_{PL} = 2,15 \times 10^{76} $ GeV$^{4} $ and
$ \rho_{e} = 4.7977 \times 10^{-15} $ GeV$^{4} $ in
Eq.\ref{KernelDensity} opens the possibility to calculate the
density of the kernel of the electron $ \rho_{K} $
using Eq.\ref{KernelDensity}. Assuming
the shell density of the electron is approximately
the middle density of the electron
$ \rho_{S} \sim \rho_{e} $ and taking
the numerical values for $ B $ and $ r_{0} $ from
Tab. \ref{electronPARAMETERSdipol} we calculate for
five different GUT scales the kernel radius of the
electron $ r_{K} $ with Eq.\ref{KernelRadius}.

\begin{table}
\caption{ The density $ \rho_{K} $ of the electron kernel and its
          radius $ r_{K} $ as function of five possible GUT scales $ \Lambda_{GUT} $. }
\begin{center}

\begin{tabular}{||l|l|l|l|l|l||}                                \hline
 $ \Lambda_{GUT} $[GeV]& $ 10^{13} $
                       & $ 10^{14} $
                       & $ 10^{15} $
                       & $ 10^{16} $
                       & $ 10^{17} $
                                                               \\ \hline
 $ \rho_{GUT}~[GeV^{4}] $ & $ 9,95 \times 10^{51} $
                       & $ 9,95 \times 10^{55} $
                       & $ 9,95 \times 10^{59} $
                       & $ 9,95 \times 10^{63} $
                       & $ 9,95 \times 10^{67} $
                                                              \\ \hline
 $ r_{K}~[m] $         & $ 5.02 \times 10^{-22} $
                       & $ 1.08 \times 10^{-20} $
                       & $ 2.33 \times 10^{-19} $
                       & $ 5.02 \times 10^{-18} $
                       & $ 1.08 \times 10^{-16} $
                                                              \\ \hline
\end{tabular}
\end{center}
\label{ElectronKernelRadius}
\end{table}

The numerical values are shown in Tab. \ref{ElectronKernelRadius}. In the
first line five different GUT scales are shown. The scales cover the
actual scales used in the literature and allows in this sense to study
the possible range of the size of the electron kernel. The second
line displays the according density $ \rho_{GUT} $. The last line
displays the radius of the electron kernel of the five different
GUT scales. It is interesting to notice that with increasing
$ \Lambda_{GUT} $ also $ \rho_{GUT} $ increases but originated
from the also increasing ratio $ \rho_{GUT} / \rho_{e} $ is
the electron kernel density via Eq.\ref{KernelDensity}
decreasing which causes via Eq.\ref{KernelRadius}
an increase of the kernel radius with increasing GUT scale.

The study of the size of the electron mass kernel demonstrates
in the frame work of our ETAMFP model the biggest radius is still fare
inside the electron and the smallest value
is not interfering with the Planck scale.

\subsubsection { Comparison of the size of electron with
                 the experimental data }

To compare the size of electron with experimental data we
follow two complete different alternative paths.

The experimental measured parameters in
Tab. \ref{Parameterelectron} implemented in the ETAMFP model
result in a extended FP. In particular for the electron
it was possible as summarized in Tab. \ref{electronPARAMETERSdipol},
Tab. \ref{PrecissionKbetaB} and Tab. \ref{ElectronKernelRadius}
to estimate eight parameters describing the electron as
extended object. The most interesting parameter is the
radius of electron where the mass shell end and the outer charges are
located of $ r_{0} = 5.8 \times 10^{-13} $ m. This number is the
outcome of numerous low energy experiment parameters from
Tab. \ref{Parameterelectron} inserted in the ETAMFP model.

As discussed in the beginning of this paper in 
chap.\ref{sec:Status of experimental limits on the sizes of
Fundamental Particles},
an entire other ansatz is to measure
the size of an electron via a $ \EEGG $ or $ \EEff $ reaction at GeV
energies.
The angular distribution of the cross section was fitted in
both reactions with the well known SM theory cross section and
included a non standard term of direct contact interaction shown in the
Lagrangian functions of Eq.\ref{LagDirectContactQ}
and Eq.\ref{eq.6}. The fit
searches for a deviation from the SM model. The results
of these fits are shown in Fig.\ref{SizeLimits}.
In both reactions was measured a scale which
get translated in a geometrical extension via the uncertainty principle.
In case of the pure QED reaction $ \EEGG $ we find a minimum in the fit
with a significance of about $ 5 \times \sigma $ at a distance of
$ r \approx 1.57 \times 10^{-17} $ cm. In the electro weak
$ \EEff $ reaction only a limit was set for all possible exit channels
of $ r <  0.9 \times 10^{-18} $ cm. In the SM model the FPs are
point particles. To describe the direct contact interaction in a
microscopic method with point particles is meaningless, even taking into
account that fact that the square of the wave function in the
Lagrangian functions of Eq.\ref{LagDirectContactQ}
and Eq.\ref{eq.6} include a distance. If we
assume a geometrical extended object like discussed in
Fig.{\ref{BasicSCHEMEFP2006}} it is possible to find a
scenario for a finite scattering distance. In the pure QED
reaction $ \EEGG $ both electrons approaching each other
at a speed close to the velocity of light and annihilate at
scale of $ \Lambda_{C} = 1253.2 $ GeV which is in accordance
with a distance from the point of gravity of the reaction of
$ r \approx 1.57 \times 10^{-17} $ cm. 

The radius $ r_{0} $
in Fig.{\ref{BasicSCHEMEFP2006}} for both electrons will
be under these circumstances Lorentz contracted in
flight direction by a factor
\begin{equation}
      \gamma = \Lambda_{C} / E_{tot} = 2.5 \times 10^{6}
\label{Gammafactor}
\end{equation}
taking the total mass of the electron $ E_{tot} $
from Tab. \ref{Parameterelectron}. The radius $ r_{0} $
is the radius where the mass of the electron starts it would be
the first possibility to initiate annihilate in a direct contact
term. This radius of the electron is also the heritage of the GUT
scale where the three interactions strong, electro magnetic and
weak get independent and a domain wall could exist. For these
reasons is the radius $ r_{0} $ a distinguished geometrical size
for the annihilation reaction $ \EEGG $.
As we discussed in Fig.\ref{Spheremassbeta} the electron spin
induces a Lorentz contraction on the geometrical position
of the other charges and on the
extension of the mass shell of the electron. If we
inspect Fig.\ref{Spheremassbeta} and Fig.\ref{PrecissonradiusElectron}
( See chap.~\ref{sec:Conclusion of the classical approach} )
the contraction of the electron shell disappears at about
$ 0.8 \times r_{0} $ it is for this reason not necessary to consider
different polarization degrees of the electron-positron beam.
The Lorentz contracted radius from the radius in $ r_{0} $ in
Fig.{\ref{BasicSCHEMEFP2006}} can be estimated as a function
of $ \gamma $ and $ 0.8 \times r_{0} $ like
$ r ( model ) \approx ( 0.8 \times r_{0} ) / \gamma $.
The numerical values of both approximations,
the radius of electron via QED $ \EEGG $ and the
radius of electron via ETAMFP model almost agree and
are summarized in
Tab. \ref{CoparisonElectronExperiment}.
The agreement of both numbers is a strong support
for the ETAMFP model because this numbers
are caused from two absolute different paths
to measure and calculate the radius of the
electron.

\begin{table}
\caption{ Comparison of the size of electron with the experimental data. }
\begin{center}
\begin{tabular}{||l|l||}                                      \hline
  radius of electron via QED $ \EEGG $  & radius of electron via ETAMFP model
                                                               \\ \hline
  $ 1.57 \times 10^{-17} $ cm & $ 1.86 \times 10^{-17} $ cm
                                                               \\ \hline
\end{tabular}
\end{center}
\label{CoparisonElectronExperiment}
\end{table}

As discussed the analysis of the reaction $ \EEff $ set only a limit on the
size of the interaction scale $ \Lambda $ in contrast to the
$ 5 \times \sigma $ effect in the pure QED reaction $ \EEGG $ .
The question rises, why so far was in the $ \EEff $ reaction only a
limit detected ?

The majority of the QED the reaction $ \EEGG $ proceeds via the exchange
of a virtual electron in the t and u channel. The non standard direct
contact term only induce a very weak difference from the dominating
QED reaction. I a direct contact scheme
in particular important is the annihilation of the
$ e^{+} $ and $ e^{-} $ because this reaction is sensitive to
the radius $ r_{0} $ in our ETAMFP model.
In the $ \EEff $ process the weak interaction is via s-channel
involved. The majority of the interaction is very well described
by the SM model and the non standard direct contact interaction
is also only a very small contribution to the overall reaction.
In a microscopic picture of the non standard part it would be possible the
$ e^{+} $ and $ e^{-} $ first annihilate, create a small energy
sphere and finally condense to the in the Lagrangian function
Eq.\ref{eq.6} shown final states of fermions. An other
possibility would be that $ e^{+} $ and $ e^{-} $ does not
annihilate and get transferred direct in the different
possible excited states if the fermions following our
discussion in Fig.\ref{Fermionmass}. The sensisivity to the
radius $ r_{0} $ is connected to the annihilation process.
The total
cross section of the $ \EEff $ reaction is much bigger as in the annihilation
reaction $ \EEGG $. In particular the total cross section of the elastic
Bhabha scattering $ \EEEEG $ is about a factor 60 bigger as the QED
cross section at the energies under investigation.
It seems likely that in the $ \EEff $ reaction the much bigger
SM model process is masking heavily the annihilation reaction.
Under this circumstances only a very high statistic
would allow to detect a deviation originated from the
direct contact term in the standard $ \EEff $ reaction.

\subsubsection { Conclusion of the classical approach to estimate
                 the size of an electron determined from its
                 experimental measured parameters }
\label{sec:Conclusion of the classical approach}

The classical approach to estimate the size of an electron
determined from its experimental measured parameters is an
important benchmark test for the ETAMFP model. The model
developed in chap.\ref{sec:Empirical toy ansatz
about a microstructure of a fundamental particle} uses
as input parts from the SM theory and the 
timing of the Big Bang model. In particular the
running coupling constants of the SM theory shown in Fig.{\ref{fig.4}
and the very early timing of the Big Bang model displayed
in Fig.\ref{BBMtiming-t} is leading to the geometrical approach in
in Fig.\ref{SchematicDevFP}. The numerical input in the ETAMFP 
model for the parameters of the electron are measured
in the framework of the SM theory. The test indicates that the
geometrical structure of the electron in agreement
with the discussed schemes of the electron from
chap.\ref{sec:Scheme of the lightest left handed FP}
to chap.\ref{sec:Scheme of the lightest left handed anti FP}.
Important is the comparison of the size of the electron
calculated via the ETAMFP model and the direct measurement
discussed in 
chap.\ref{sec:Status of experimental limits on the sizes of
Fundamental Particles}. This we performed in appropriate action
following the philosophy to use as simple as possible
solutions for the problems in question and refine them step by step.

We assume in the first step, the electron what carries a spin
behaves like a classical gyroscope. The charge attached to electron
at distance $ r_{e} $ from the center will have under this
circumstances a speed which exceeds the speed of light.

In the next step we assume the velocity of light $ c $
must be respected.
As path of the charges circling the electron center
we use the in  chap.\ref{sec:The flashing vacuum} discussed
ansatz about the flashing vacuum. 
We assume for this reason the
path of the charges circling the electron center
is free of radiation loss. The path of the charge follows the
circumference according the distance from the electron center.
The speet of the charge is close to the speed of light.
The length of the circumference will be for this reason
contracted according the Lorentz equations.
Comparing the measurement of the magnetic moment of the
electron and the electric dipole moment with this contraction
shows, that the experiment of the magnetic moment is
sensitive to the un-contracted radius $ r_{0} $ where the
charges $ 1/3 \times e $ are located, whereas in the case of the
electric dipole moment the experiment is sensitive to the
contracted radius $ r_{e} $. This leads to the explanation
why the magnetic moment is a precise measured numerical
value and the measurement of the electric dipole moment is
so challenging small. The Lorentz contraction pushes the measurable
dipole moment in the LAB-system to zero, if the charge velocity
approaches the speed of light. Assuming $ \beta = 1 $ for
the speet of the charge it is possible to estimate the important
parameters of the the electron,the angular velocity of charge, the rest mass,
the rotational energy and the un-contracted radius $ r_{0} $ where the charge
is located.

Including in the next step in the ETAMFP model an ansatz about the shape
of the mass core of the electron, as a rigid sphere and taking
into account the possibility of a non rotating inner mass kernel
implements an important limit in the model. It is only possible
to find a solution for the parameters of the electron if
only $ 2/3 \times e $ of the charge generate the
magnetic moment. The solution
for $ 1 \times e $ get excluded. It also turns out that a non
rotating inner mass kernel could exist.

In the last step of the ETAMFP model development so far, we include
the limit of the electric dipole moment in the ansatz. This confirms
a possible inner mass kernel and the parameters, angular velocity,
rest mass, rotational energy and the un-contracted charge radius of the
electron, already calculated for the case $ \beta = 1 $.
In particular it highlights the fact that the numerical value
of $ \beta $ get pushed from the measurement of the electric dipole
moment of the electron extreme close to $ \beta = 1 $.
\begin{figure}
\begin{center}
 \epsfig{file=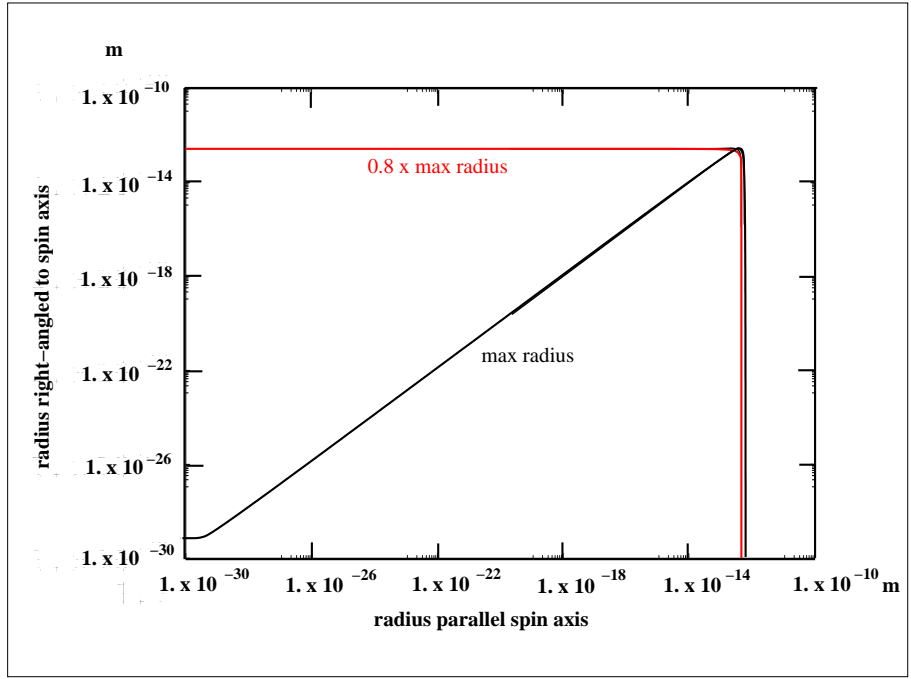,width=12.0cm,height=9.0cm}
\end{center}
\caption{ Contracted electron radius $ r_{e} $ as function of the
          un-contracted radius $ r_{0} $. }
\label{PrecissonradiusElectron}
\end{figure}
We highlight this problem in Fig.\ref{PrecissonradiusElectron}.
The plot shows the contraction of the infinite thin outer spherical
shell rotating with $ \beta $ from Tab. \ref{PrecissionKbetaB}
orbiting the spin axis.
The black solid line is the contracted radius
right-angled to the spin axis of the electron
( y-axis in Fig.{\ref{PrecissonradiusElectron}} or
x-axis in Fig.{\ref{BasicSCHEMEFP2006}} )
as function of the un-contracted radius
parallel to the spin axis ( x-axis in Fig.{\ref{PrecissonradiusElectron}}
or z-axis in Fig.{\ref{BasicSCHEMEFP2006}} )
of the electron. The charges are
located in the extreme minimum at the coordinate
$ x = 0 $ ( or $ z = 0 $ in Fig.{\ref{BasicSCHEMEFP2006}} ) 
and $ y = r_{e} = 0.105 \times 10^{-28} $ m 
( or $ x = r_{e} $ in Fig.{\ref{BasicSCHEMEFP2006}} )
in the lower left corner of the plot.
The contraction is focused only on the extreme outer spherical
shell of the mass core of the electron. If we calculate the
same behavior for a shell $ 20 \% $ deeper inside the core,
where the velocity $ \beta $ is smaller accordingly the smaller radius,
the contraction
is nearly disappeared as shown in the red solid line in
Fig.\ref{PrecissonradiusElectron}.

To visualize the outer shape of the electron from the point of
view of an observer in the CM-system rotating with the angular
velocity $ \omega $ compared to an observer in the LAB-system
in rest, we show the electron in an artistic view
from Fig.\ref{PrecissonradiusElectron}
in Fig.\ref{SketchElectronContracted}.
\begin{figure}
\begin{center}
 \epsfig{file=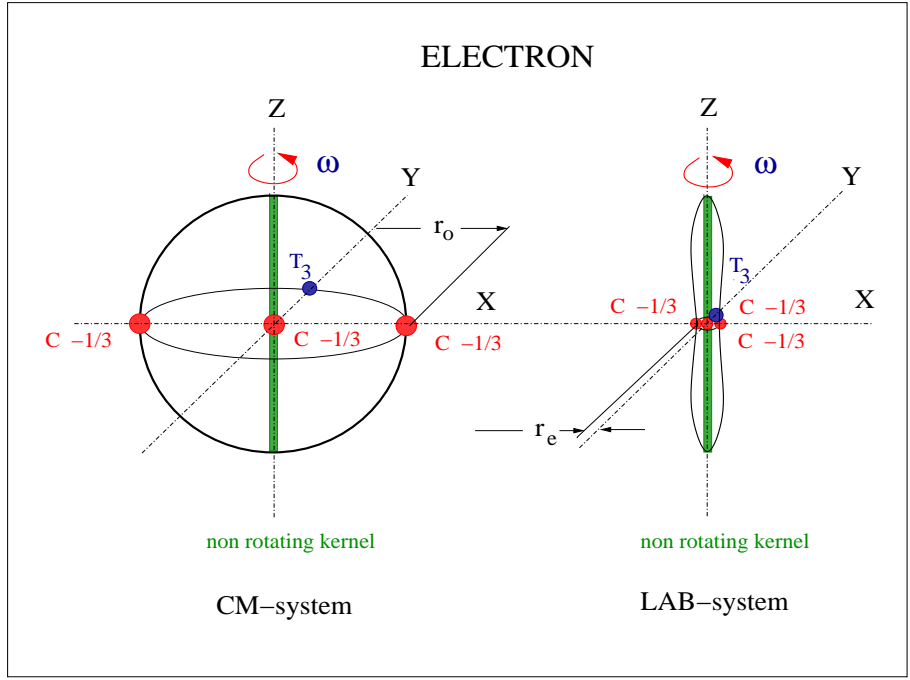,width=12.0cm,height=9.0cm}
\end{center}
\caption{ Sketch of electron including the electric dipole moment.}
\label{SketchElectronContracted}
\end{figure}
Left side in
Fig.\ref{SketchElectronContracted} shows the chape of the electron
mass core ( black circle ) including the location of the three
electric charges $ 1/3 \times e $ on the x-axis ( red dots )
in the CM-system.
The maximum radius of the sphere is assigned as $ r_{0} $.
The weak charge $ T_{3} $ is located on the y-axis ( blue dot )
The spin axis is chosen parallel the z-axis. The possible
non rotating mass kernel is symbolized as a green solid strip
on the z-axis. We expect that this kernel is spherical
shaped but show for artistic reasons a green column.
Right side in
Fig.\ref{SketchElectronContracted} is displayed
exact the same electron measured from an observer in rest in the
LAB-system. Shown is only the contracted outer shell including
the electric and weak charges located at the radius $ r_{e} $.

Comparing our first assumption of an geometrical extended
electron in Fig.{\ref{BasicSCHEMEFP2006}} with
the shape of the electron after we imposed all the discussed conditions
on it, we notice a remarkable agreement. The agreement is in particular
visible by comparing Fig.{\ref{BasicSCHEMEFP2006}} with
Fig.\ref{PrecissonradiusElectron} and
Fig.\ref{SketchElectronContracted}.

An important benchmark test about our whole model discussion is
the comparison of the size of electron with experimental data
generated from complete different experiments. 

First we insert the parameters of the electron measured
at low energies from Tab. \ref{Parameterelectron} in the
ETAMFP model and calculate the parameters of an extended
electron in
Tab. \ref{electronPARAMETERSdipol} 
to Tab. \ref{ElectronKernelRadius}
in particular the radius $ r_{0} $ and $ r_{e} $.

Second we compare these radii with the radius
measured in the direct contact term reaction $ \EEGG $
discussed in 
chap.\ref{sec:Status of experimental limits on the sizes of
Fundamental Particles}
These measurements are complete
independent from the low energy data from the electron in
Tab. \ref{Parameterelectron}.
As shown in Tab. \ref{CoparisonElectronExperiment} is
the radius measured from the high energy $ \EEGG $ reaction
numerical very close the radius calculated via the ETAMFP
model. The agreement of both numbers are not a proof of the
ETAMFP model, but it encouraged us to study this model more carefully
from different points of views. We follow this philosophy in the next
chap.\ref{sec:Particle-like structure related to gravity}
and
chap.\ref{sec:Electrical charged particle-like structure
           coupled to General Relativity}.

\section{Particle-like structure related to gravity}
\label{sec:Particle-like structure related to gravity}


In our paper \cite{SizePaper2003}  we summarized the results of
our studies on the non point-like behavior of fundamental
particles. In this paper it was possible to set a limit on the
size of the electron $ r $ of $ r < 6 \times 10^{-17} $ cm. In the
paper under discussion we analyzed a $ 5 \times \sigma $ effect of
the best fit of all cross sections shown in Fig.\ref{VTADL} and
Fig.\ref{Dall} to the direct contact term of the Lagrangian
Eq.\ref{LagDirectContactQ} scale factor $ \Lambda_{C6} $. The
best fit value sets a radius of the electron to $ r = 1.57 \times
10^{-17} $ cm at a test scale of $ 1.25 $ TeV. Under these
circumstances it is very interesting to study the consequences of
this fact on our in Fig.{\ref{BasicSCHEMEFP2006}} discussed basic
scheme of the model of fundamental particles. We repeat for this
reason in the following chapter the main arguments about the
particle-like structure related to gravity of paper
\cite{SizePaper2003} and analyze the impact of the new data on our
ETAMFP model using the self gravitating particle-like structure
with de Sitter vacuum core.

The ansatz in this paper was to model an extended particle-like
object by a de Sitter-Schwarzschild geometry, which describes a
smooth transition from a de Sitter vacuum in the origin to the
Minkowski vacuum at infinity \cite{IRINA1}. The issue is inspired
by the in the actual paper discussed effort to bound a
gravitational object like in Fig.\ref{ForcesSM-BBM} which have
masses comparable with the masses of FPs. The size of a FP cannot be
defined by the Schwarzschild gravitational radius $r_g=2G m
c^{-2}$ ($m$ is the gravitational mass as measured by a distant
observer). A size is constrained from below by the Planck length
$l_{Pl} \sim 10 ^ {-33}$ cm, and for any elementary particle its
Schwarzschild radius $r_g$ is many orders of magnitude smaller
than $l_{Pl}$. The Schwarzschild gravitational radius comes from
the Schwarzschild solution which implies point-like mass and is
singular at $r=0$. The Schwarzschild metric can be modified by
replacing a singularity with  de Sitter regular core
\cite{IRINA1,DeSitter,particle,lambda}. This modified solution, de
Sitter-Schwarzschild geometry, depends on the limiting vacuum
density $\rho_{vac}$ at $r=0$, satisfying there the equation of
state $p=-\rho_{vac}$. The idea goes back to the mid-60s papers by
Sakharov who suggested that $p=-\rho$ can arise at superhigh
densities \cite{sakharov}, by Gliner who interpreted it as the
vacuum equation of state and suggested that it can be achieved as
a result of a gravitational collapse \cite{gliner}, and by
Zeldovich who connected $\rho_{vac}$ with gravitational
interaction of virtual particles \cite{zeldovich}.

In the context of spontaneous symmetry breaking $\rho_{vac}$ is
related to the potential of a scalar field  in its symmetric
(false vacuum) phase. In this context, as well as in the context
of Einstein-Yang-Mills-Higgs (EYMH) self-gravitating non-Abelian
structures including black holes, $\rho_{vac}$ is related to
symmetry restoration in the origin \cite{dynamics}. In a neutral
branch of EYMH black hole solutions, a non-Abelian structure can
be approximated as a sphere of a uniform vacuum density
$\rho_{vac}$ whose radius is the Compton wavelength of the massive
non-Abelian field (see \cite{maeda} and references therein). The
basic fact of de Sitter-Schwarzschild geometry is that a mass of
an object is generically related to an interior de Sitter vacuum
and smooth breaking of space-time symmetry from the de Sitter
group in the origin to the Poincar\'e group at infinity
\cite{Irina10}.

In de Sitter-Schwarzschild geometry (the particular exact analytic
solution was found in the Ref.\cite{IRINA1}), there exists a
certain critical value of  the mass $m_{cr}$ which selects two
types of objects: a neutral non-singular black hole for $m\geq
m_{cr}$, and for $m <m_{cr}$ a neutral self-gravitating
particle-like structure with de Sitter vacuum core related to its
gravitational mass \cite{particle}. This fact is generic for de
Sitter-Schwarzschild geometry. In the case when stress energy
tensor satisfies the dominant energy condition (density
non-negative for any observer, speed of sounds does not exceed
speed of light), the geometry has two characteristic surfaces. The
surface of zero scalar curvature $r=r_s$ defines the gravitational
size $r_s$ of a particle-like structure. The solution
\cite{IRINA1} belongs to this class, and for an object described
by it most of the mass is within $r_s$. The surface of zero
gravity, $r=r_c<r_s$, there exists also in the case when only weak
energy condition is satisfied (density no-negative for any
observer) and defines a size of an inner vacuum core. Beyond
$r=r_c$ gravitational attraction becomes gravitational repulsion.
Both these surfaces are at the characteristic scale
$\sim{(m/\rho_{vac})^{1/3}}$.

\subsection{ Self gravitating particle-like structure with de Sitter vacuum core }
\label{sec:Self gravitating particle-like structure with de Sitter vacuum core}

De Sitter-Schwarzschild geometry has appeared as describing a
black hole whose singularity is replaced with de Sitter core of
some fundamental scale \cite{werner,IRINA1,particle}. Several
solutions have been obtained by direct matching de Sitter metric
inside to Schwarzschild metric outside of a junction surface
\cite{DeSitter}. Typical for matched solutions is that they have a
jump at the junction surface since the O'Brien-Synge junction
condition $T^{\mu\nu}n_{\nu}=0$ is violated there \cite{werner}.
Poisson and Israel analyzed de Sitter-Schwarzschild transition and
came to the conclusion that a layer of "non-inflationary" material
should be introduced at the interface \cite{werner}. In the Ref.
\cite{IRINA1} this material was specified as a spherically
symmetric anisotropic vacuum (inflationary in the radial
direction, $p_r=-\rho$), with the continuous density and pressure,
responsible for a class of regular metrics asymptotically de
Sitter at the center.
The exact analytical
solution for a neutral spherically symmetric black hole with a
regular de Sitter interior was found in \cite{IRINA1}.

The main steps to find this solution are to insert the spherically
symmetric metric
\begin{equation}
ds^{2}=e^{\nu }c^{2}dt^{2}-e^{\mu }dr^{2}-r^{2}
(d\theta^{2}+sin^{2}\theta d\phi ^{2})
\label{eq.7a}
\end{equation}
into the Einstein equations
\( R _{\mu \nu }-\frac{1}{2}Rg_{\mu \nu }=\frac{8\pi G}{c^{4}}T_{\mu \nu }
\)
which then take the form
\begin{equation}
\frac{-e^{\mu }}{r^{2}}+\frac{\mu ^{\prime }
e^{-\mu}}{r}+\frac{1}{r^{2}}=\frac{8\pi G}{c^{4}}T_{t}^{t}
\label{eq.7b}
\end{equation}
\begin{equation}
\frac{-e^{\mu }}{r^{2}}-\frac{\nu ^{\prime }
e^{-\mu}}{r}+\frac{1}{r^{2}}=\frac{8\pi G}{c^{4}}T_{r}^{r}
\label{eq.7c}
\end{equation}
\begin{equation}
\frac{1}{2}e^{-\mu }(\nu ^{\prime \prime }+\frac{\nu ^{\prime 2}}{2}+\frac
{\nu ^{\prime }-\mu ^{\prime }}{r}-\frac{\nu ^{\prime }\mu ^
{\prime}}{2})=\frac{8
\pi G}{c^{4}}T_{\theta}^{\theta}=\frac{8\pi G}{c^{4}}T_{\phi}^{\phi}
\label{eq.7d}
\end{equation}
The requirement of regularity and weak energy condition leads to
the existence of a family of spherically symmetric solutions
which includes the class of solutions which
connect smoothly the de Sitter metric inside to the Schwarzschild
metric outside. In this class asymptotical behavior of a
stress-energy tensor is  \( T_{\mu \nu }\rightarrow 0 \) as \(
r\rightarrow \infty \) and \( T_{\mu \nu }\rightarrow
\rho_{vac}g_{\mu\nu} \) as
 \( r\rightarrow 0  \), with \( \rho _{vac} \) as de Sitter vacuum density
at \( r = 0 \). The algebraic structure  of the stress-energy
tensor \( T_{\mu \nu } \) is \cite{IRINA1}.

\begin{equation}
T_{t}^{t}=T_{r}^{r}\,\, \textrm{and} \,\, T_{\theta }^{\theta }=T_{\phi }^{\phi }
\label{eq.7e}
\end{equation}

The stress-energy tensor of this structure describes a spherically
symmetric (anisotropic) vacuum, invariant under the boosts in the
radial direction (Lorentz rotations in $(r,t)$ plane)
\cite{IRINA1}. It can be associated with a time-dependent
spatially inhomogeneous cosmological term \cite{lambda}.

It smoothly
connects the de Sitter vacuum at the origin with the Minkowski
vacuum at infinity, and satisfies
the equation of state \cite{IRINA1,werner}

\label{alleqs8}
\begin{equation}
p_r=-\rho;~~p_{\perp}=p_r+{r\over 2} {dp_r\over{dr}}
\label{eq.8}
\end{equation}

where $p_r=-T_r^r$ is the radial pressure and
$p_{\perp}=-T_{\theta}^{\theta} = - T_{\phi}^{\phi}$
is the tangential pressure. In this class of solutions
the metric Eq.\ref{eq.7a} takes the  form

\label{alleqs9}
\begin{equation}
ds^2=\biggl(1-\frac{R_g(r)}{r}\biggr) dt^2 -
\biggl(1-\frac{R_g(r)}{r}\biggr)^{-1} dr^2 - r^2 d{\Omega}^2,
\label{eq.9}
\end{equation}

where
 $d{\Omega}^2$ is the metric on the unit two-sphere,
and

\label{alleqs10}
\begin{equation}
R_g(r)=\frac{2GM(r)}{c^2};~ ~ M(r)= \frac{4\pi}{c^2}\int_0^r{\rho(r)r^2 dr}
\label{eq.10}
\end{equation}

In the model of Ref. \cite{IRINA1}
the density profile $T^t_t(r)= \rho(r) c^2$ has been chosen as
\label{alleqs12}
\begin{equation}
\rho = \rho_{vac} e^{ -4\pi\rho_{vac} r^3/3 m }
\label{eq.12}
\end{equation}
which describes, in the semiclassical limit, vacuum polarization in the gravitational field
\cite{particle}.
Inserting Eq.\ref{eq.12} into Eq.\ref{eq.10} shows that \( R_g(r) \) takes
the form
\label{alleqs13}
\begin{equation}
 R_{g}(r)=r_{g}(1-e^{-4\pi \rho_{vac}r^{3}/3m})
=r_g(1-e^{-r^3/r_0^2r_g})
\label{eq.13}
\end{equation}
where
\label{alleqs14}
\begin{equation}
r_0^2 = \frac{3 c^2}{ 8 \pi G \rho_{vac}}
\label{eq.14}
\end{equation}
is the de Sitter radius,
$ r_g = 2 G m / c^2 $ is the Schwarzschild radius, and \( m \) is
the gravitational mass of an object.
In the limit of \( r <\!\!< (r_{0}^{2}r_{g})^{1/3} \)
Eq.\ref{eq.13}
shows that \( R_g \rightarrow r^3 / r^2_0 \), and the metric
of Eq.\ref{eq.9} takes the de Sitter form with
\begin{equation}
g_{tt}=1- \frac{R_g(r)}{r}=1 - \frac{r^2}{r_0^2}
\label{eq.15}
\end{equation}
In the de Sitter geometry the horizon $r_0$ bounds the causally connected region.
An observer at $r=0$ cannot get information from the region beyond the surface $r=r_0$.

For $r\gg (r_0^2 r_g)^{1/3}$ the metric takes the Schwarzschild form
\label{alleqs16}
\begin{equation}
g_{tt}= 1 - \frac{R_g(r)}{r} = 1 - \frac{r_g}{r}
\label{eq.16}
\end{equation}
in agreement with boundary conditions.

The metric \( g_{tt}\left( r\right)  \) is shown
in Fig.{\ref{fig.5}. The fundamental difference from the Schwarzschild case
is that de Sitter-Schwarzschild black hole has two horizons, the black hole
horizon $r_{+}$ and the internal Cauchy horizon $r_{-}$.
%
\begin{figure}[htbp]
\vspace{-8.0mm}
\begin{center}
 \includegraphics[width=11.0cm,height=7.0cm]{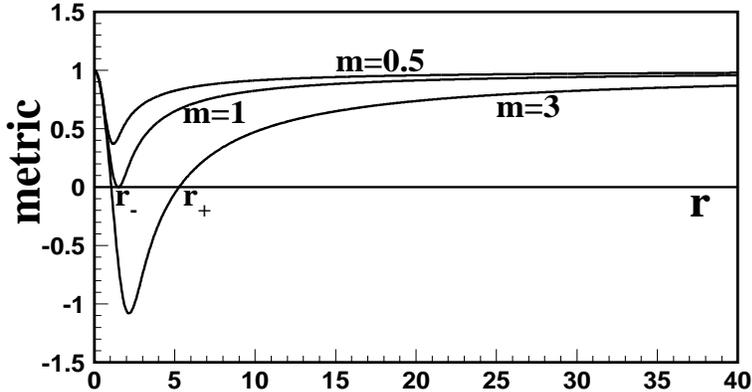}
\end{center}
\caption{De Sitter-Schwarzschild metric (\ref{eq.9})
         $g_{tt}(r)=1-R_g(r)/r$. The mass  $m$
         is  normalized to $m_{cr}$.
         The radius is  normalized to $r_{o}$.
         For $m>1$ we have a black hole,
         $m=1$ corresponds to the extreme
         black hole, and configuration with $m<1$ represents
         a vacuum self-gravitating particle-like structure without horizons.}
\label{fig.5}
\end{figure}
The object is a black hole for $m\geq m_{cr}\simeq{0.3
m_{Pl}\sqrt{\rho_{Pl}/\rho_{vac}}}$, which loses its mass via
Hawking radiation until a critical mass \( m_{cr} \) is reached
where the Hawking temperature drops to zero \cite{particle}. At
this point the horizons come together. The critical value $m_{cr}$
puts the lower limit on a black hole mass. Below $m_{cr}$ de
Sitter-Schwarzschild geometry Eq.\ref{eq.9} describes a neutral
self-gravitating particle-like structure made up of a vacuum-like
material Eq.\ref{eq.7e} with $T_{\mu\nu}\rightarrow
\rho_{vac}g_{\mu\nu}$ at the origin \cite{particle}. This fact
does not depend on particular form of a density profile
\cite{particle,dynamics} which must only satisfy requirement of
regularity at the origin and guarantee the finiteness of the mass
as measured by a distant observer
\label{alleqs17}
\begin{equation}
m = 4\pi \int_0^{\infty} {\rho(r) r^2 dr}
\label{eq.17}
\end{equation}
The interest of this paper is focused on the particle-like structure.
The case of $ m \geq m_{cr} $ is discussed in
\cite{particle,dynamics,Irina10,Irina11}.

De Sitter-Schwarzschild geometry with the density profile
Eq.\ref{eq.12}
has two characteristic surfaces
at the characteristic
scale $ r \sim(r_{0}^2 r_{g})^{1/3}
 $\cite{particle}. The first is the surface of zero scalar curvature.
The scalar curvature
\( R = 8 \pi GT \) changes sign at the surface

\label{alleqs18}
\begin{equation}
r = r_s =
 \biggl ( \frac{m}{\pi \rho_{vac}}\biggr)^{1/3} =
 \frac{1}{\pi^{1/3}}   \biggl(\frac{m}{m_{Pl}}\biggr)^{1/3}
 \biggl(\frac{\rho_{Pl}}{\rho_{vac}}\biggr)^{1/3}l_{Pl}
\label{eq.18}
\end{equation}

which contains the most of the mass $m$. Characteristic
gravitational size of a self-gravitating particle-like structure
can be defined by the radius $r_s$. The second is related to the
strong energy condition of the singularity theorems. It reads $
(T_{\mu\nu} - g_{\mu\nu}T/2)u^{\mu}u^{\nu}\geq 0$, where $ u^{\nu}
$ is any time-like vector. The strong energy condition is
violated, i.e., gravitational acceleration changes sign, at the
surface of zero gravity

\label{alleqs19}
\begin{equation}
 r = r_c =
 \biggl ( \frac{m}{2\pi \rho_{vac}}\biggr)^{1/3} = \frac{1}
{(2\pi)^{1/3}} \biggl(\frac{m}{m_{Pl}}\biggr)^{1/3}
  \biggl(\frac{\rho_{Pl}}{\rho_{vac}}\biggr)^{1/3}l_{Pl}
\label{eq.19}
\end{equation}

The globally regular configuration with de Sitter
core instead of a singularity arises as a result
of the balance between attractive gravity
outside and repulsion inside of the
surface $r = r_c$.
This surface defines the characteristic
size of an inner vacuum core.
For a particle-like structure with $ m <\!\!< m_{Pl} $,
both these sizes are much bigger than
the Schwarzschild radius $r_g$. The ratio of a size of a  vacuum core
to the Schwarzschild radius $r_g$ is given by
\label{alleqs20}
\begin{equation}
\frac{r_c}{r_g}=\frac{1}{2}\frac{1}{(2\pi)^{1/3}}\biggl(\frac{m_{Pl}}{m}\biggr)^{2/3}
\biggl(\frac{\rho_{Pl}}{\rho_{vac}}\biggr)^{1/3}
\label{eq.20}
\end{equation}
 The horizons and characteristic surfaces
of de Sitter-Schwarzschild geometry are shown in the Fig.{\ref{fig.6}
where they are normalized to $r_0$.
\begin{figure}[htbp]
\vspace{-10.0mm}
\begin{center}
 \includegraphics[width=12.0cm,height=8.50cm]{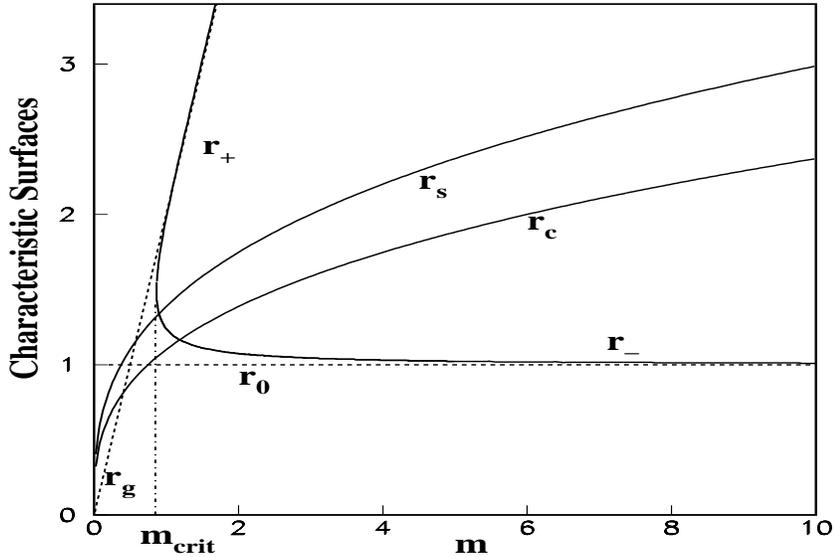}
\end{center}
\caption{Horizons $r_{\pm}$ and surfaces of zero curvature $r_s$ and zero
         gravity $r_c$ of de Sitter-Schwarzschild geometry. Schwarzschild
 radius $r_g$ and de Sitter radius $r_0$ are also shown.}
\label{fig.6}
\end{figure}
As we see from above, de Sitter-Schwarzschild geometry gives rise
to a self-gravitating particle-like structure, kind of a
gravitational vacuum soliton \cite{Irina10}, which is stable for a
wide range of density profiles including the profile
Eq.\ref{eq.12} \cite{Irina8}. The size of such a vacuum soliton
is defined by its mass \( m \) and the vacuum density $\rho_{vac}$
in the vicinity of \( r = 0 \).

\subsection { Modeling of FPs with extended structure
                 by an object with a vacuum interior }

Let us consider a toy model of FP with extended inner mass kernel
like discussed in Fig.\ref{BasicSCHEMEFP2006}. The extended
structure of the kernel should be represented by a size of the De
Sitter-Schwarzschild stable configuration described in the last
paragraph. It is natural to assume ( See Tab.
\ref{linksSMBigBang} ) that in the simplest realization of the De
Sitter-Schwarzschild particle-like object the energy density of a
vacuum-like material inside it can be attributed to the energy
density of some scalar field. We use Higgs ansatz to specify the
potential of a scalar field governing the vacuum interior

\begin{equation}
{V(\phi)}=V(0)-\frac{1}{2}\mu^{2}\phi^{2}+\frac{1}{4}\lambda\phi^{4}
           +\tilde{V}
\label{SCALAR}
\end{equation}

The energy density inside the object is given by
$\rho_{vac}=V(\phi)$ while the term $\tilde{V}$ is added just to
normalize the vacuum energy density outside the object to the
total density the of the Universe, which is fixed by observations.

The density profile of the vacuum core Fig.\ref{fig.5}. can be
approximated by

\begin{equation}
 \rho(r) = \rho_{0} + \rho^{'}(0) r + \rho^{''}(0) r^{2} + \cdot
           \cdot \cdot + \tilde{V}
\label{Self-Scalar1}
\end{equation}

where  $r$ is the distance from the center of
the object, for $ r \rightarrow 0 $
and $\rho_o=\rho_{vac}$ assigns the energy density of the scalar field in the
vicinity of the center of the vacuum core.

Taking into account the integral in Eq.\ref{eq.10}
we get.

\begin{equation}
 \int_0^{r} {\rho(R) R^{2} dR } = \frac{m(r)}{4\pi} + \frac{1}{3} \tilde{V} r^{3} ;
 ~~~g_{tt} = 1 - \frac{(\Lambda + \lambda)}{3} r^{2}
\label{eq.15c}
\end{equation}

Where $ \Lambda = 8 \pi G \rho_{vac} $  and $ \lambda = 8 \pi G \tilde{V} $. We
finally arrive to the following expression of the energy density in the
vicinity of the center.

\begin{equation}
\rho (r \rightarrow 0) = \rho_{vac} + \tilde{V}
\label{eq.15d}
\end{equation}

where $ \tilde{V} $ is the vacuum density related to nonzero today
value of the cosmological constant $ \lambda $.

In the context of spontaneous symmetry breaking the vacuum density
$ \rho_{vac} $ can be related to the vacuum expectation value $ v $
of the Higgs field which supplies a particle with a mass
$ m = g v $. For a Higgs particle $ g = \sqrt { 2 \lambda } $
where $ \lambda $ is its self-coupling.

It is neutral and spinless, and we can approximate it
by a de Sitter-Schwarzschild
self-gravitating vacuum soliton, identifying the central vacuum
density $ \rho_{vac} $ with the self-interaction of the Higgs
scalar in the standard theory:

\begin{equation}
\rho_{vac} = \frac{\lambda v^{4}}{4}
\label{rhovaclambda}
\end{equation}

The gravitational radius of a particle, $ r_{s} $
is after Eq.\ref{eq.18}

\begin{equation}
 r_s =
 \biggl ( \frac{m}{\pi \rho_{vac}}\biggr)^{1/3},
\label{eq.18new}
\end{equation}

confining most of its mass is restricted from above by its Compton
wave length, $ r_{s} \leq \lambda \!\!\!\!-_{C} $, since
$ \lambda \!\!\!\!-_{C} $ constrains the region of localization
for any quantum object. This condition gives us, with taking into
account Eq.\ref{eq.18new}, the constraint for a coupling constant
$ \lambda $:

\begin{equation}
 \frac{r_{s}}{\lambda \!\!\!\!-_{C}}
 = \biggl(\frac{16 \lambda}{\pi}\biggr)^{1/3} \leq 1
 ~~~~~\Rightarrow ~~~~~\lambda \leq \frac { \pi}{16}
\label{limitlambda}
\end{equation}

Than the mass, related to the particle Eq.\ref{SCALAR}

\begin{equation}
 m = \sqrt{(2 \lambda v^{2} )}
\label{massHiggs}
\end{equation}

is constrained from above due to Eq.\ref{limitlambda}.
If a particle gets its mass from the electroweak vacuum
$ v = 246 $ GeV, Eq.\ref{limitlambda} leads to an upper
boundary of a Higgs particle \cite{HiggsLimitIRINA}.

\begin{equation}
 m \leq 154 ~ GeV
\label{LimitHiggsMass}
\end{equation}

Actually in the case of a self-gravitating vacuum soliton with de
Sitter center, a mass is generically related to a smooth breaking
of space-time symmetry from the de Sitter group in the origin to
the Poincar\'e group at infinity in asymptotically flat space-time
\cite{Irina10}. This result is easily extended to the case of
asymptotically de Sitter space-time with another (much smaller)
value of $ \rho_{vac} $ at infinity \cite{Irina11}.

The Minkowski geometry allows the existence of an inertial mass as
the Casimir invariant $ m^{2} = p_{\mu} p^{\mu} $ of a test body.
The high symmetry of this geometry allows the existence of both
inertial frames and quantity $ m $ as the measure of inertia, but
the geometry tells nothing about origin of this quantity.

In the Schwarzschild geometry the parameter $ m $ is responsible
for the geometry and is identified as a gravitational mass of a
source by the asymptotic behavior of the metric at infinity (the
Kepler law ). By the equivalence principle, gravitational mass is
equal to inertial mass which is thus represented by a purely
geometrical quantity, the Schwarzschild radius $ r_{g} = 2 G m $.
But this geometrical fact still does not say anything about the
origin of a mass.

In the de Sitter-Schwarzschild geometry, the parameter $ m $ is
identified as a mass by Schwarzschild asymptotic at infinity. The
geometrical fact of this geometry is that a mass is related to de
Sitter vacuum in the origin where the symmetry is the full Lorentz
group for stress-energy tensor (De Sitter group for a space-time
). This high symmetry is then reduced to the Lorentz boosts in the
radial direction only which allows us to introduce a distinct
point as the center of an object whose mass is defined by
Eq.\ref{eq.17} and related to both de Sitter vacuum trapped
inside and smooth breaking of space-time symmetry.

This picture perfectly agrees with the basic idea of the Higgs mechanism.
In both cases, the de Sitter vacuum is involved and vacuum symmetry
is broken. But in the context of mass originated from breaking
of space-time symmetry, a scalar field is not needed anymore.

\subsection{ Comparison of the self gravitating particle-like structure
                with our ETAMFP model of FPs }

The FPs in our ETAMFP model in Fig.{\ref{BasicSCHEMEFP2006}} are composed as
a function of increasing particle radius out of gravitational energy of
the not rotating inner mass kernel, the rest mass of the shell, the rotating
energy of this shell, possible vibration energy of this shell and the
field energies of the strong, electromagnetic and weak field.
If we recall the discussion about forces and stability of FPs in
Fig.\ref{ForcesSM-BBM} the inner mass kernel stabilizes the whole
aggregate of FP with a repulsive force ( pressure ). The radial
pressure $ p_{r} $ of Eq.\ref{eq.8} fulfills just this
condition. The gravitational force at the radius $ r_{P} $ at point
SP II in Fig.\ref{ForcesSM-BBM} is zero what coincides with the
condition of zero gravity of Eq.\ref{eq.19} for the
self gravitating particle-like structure. We assume for this reason
the mass kernel of our ETAMFP model coincides with the self gravitating
particle-like structure with de Sitter vacuum core.
This ansatz is generic for all FPs in our ETAMFP model.

As comprehensive discussed in the experimental part of this paper
and summarized in Fig.\ref{SizeLimits} we analyzed in the
$ \EEGG $ reaction a radius depended five $ \sigma $ effect
including the direct contact term of Lagrangian Eq.\ref{LagDirectContactQ}.
As we discussed in the section comparison of the size of electron
with the experimental data ( See Tab. \ref{CoparisonElectronExperiment} )
we assumed that the measured radius $ r $ is the distance from the point of
gravity where the heavily Lorentz contracted electrons meet and  annihilation
starts. To calculate the size of the electron in rest it is necessary
to expand this radius with the Lorentz factor
$ \gamma $ of Eq.\ref{Gammafactor}
as in Eq.\ref{Gammafactorexpandet}

\begin{equation}
    r_{s} = r \cdot \gamma
\label{Gammafactorexpandet}
\end{equation}

The quantum constraint on $ r_{s} $ is

\begin{equation}
  r_s \leq \lambda \!\!\!\!-_c \ = \hbar /mc
\label{compton}
\end{equation}
\begin{table}
\caption{ Summary of comparison of self gravitating particle-like
          structure with our ETAMFP model in Fig.{\ref{BasicSCHEMEFP2006}}. }
\begin{center}
\begin{tabular}{||l|l||}                                      \hline
  measured size r ( cm ) from $ \EEGG $
                      & $ 1.57 \times 10^{-17} $
                                                               \\ \hline
  Lorentz Factor $ \gamma $
                      & $ 2.5 \times 10^{6} $
                                                               \\ \hline
  Lorentz expanded size of FP ( cm )
                      & $ 3.93 \times 10^{-11} $
                                                               \\ \hline
  Compton wave length $ \lambda \!\!\!\!-_c \ $ ( cm )
                      & $ 3.85 \times 10^{-11} $
                                                               \\ \hline
  Coupling constant   $ \lambda $ ( FP gravity )
                      & $ \approx \pi / 16  $
                                                               \\ \hline
  Scalar mass ( GeV ) ( FP gravity )
                      & $ \approx 154 $
                                                               \\ \hline
  Size of self gravitating mass kernel $ r_{c}^{e} $ ( cm )
                      & $ \approx 1.7 \times 10^{-19} $
                                                               \\ \hline
  Range of ETAMFP model electron mass kernel $ r_{K} $ ( cm ) 
                      & $ 5.0 \times 10^{-20} < r_{K} < 1.1 \times 10^{-18} $
                                                               \\ \hline
  Range of adjacent GUT scale ( GeV )
                      & $ 1.0 \times 10^{13} < \lambda_{GUT} < 1.0 \times 10^{14} $
                                                               \\ \hline
\end{tabular}
\end{center}
\label{SummaryComparisonGravModel}
\end{table}

As displayed in Tab. \ref{SummaryComparisonGravModel} fourth
and third
line is the Compton wavelength $ \lambda \!\!\!\!-_c \ $ approximately
the size of the measured electron $ r_{s} $ of
Eq.\ref{Gammafactorexpandet}

\begin{equation}
  r_s \approx \lambda \!\!\!\!-_c \ = \hbar /mc
\label{ApproxLamdaI}
\end{equation}

if we use for the calculation the total mass of the electron and
the measured radius $ r $ of the electron with the $ \gamma $ factor shown
in the first and second line of Tab. \ref{SummaryComparisonGravModel}.
Inserting Eq.\ref{eq.18new}
and Eq.\ref{rhovaclambda}
into the condition Eq.\ref{ApproxLamdaI} one can estimate an
approximate value of
on self-interaction constant $ \lambda $ to
( See Tab. \ref{SummaryComparisonGravModel} fifth line )

\label{alleqs21}
\begin{equation}
 1 \approx \frac{r_s}{\lambda \!\!\!\!-_c}=
\biggl(\frac{16\lambda}{\pi}\biggr)^{1/3}
\,  \, \textrm{;} \, \, \lambda \approx \frac{\pi }{16}
\label{ApproxLamda}
\end{equation}

As discussed in the beginning of this chapter the repulsive vacuum
pressure of the inner mass kernel stabilize the whole aggregate
of the FPs. The leading parameter of this vacuum pressure is
$ \rho_{vac} $ of Eq.\ref{eq.19}.
The self gravitating particle-like structure approximates
only the inner not rotating mass kernel of the FPs.
The discussed experiment is sensitive to the
radius $ r_{0} $ in Fig.\ref{BasicSCHEMEFP2006}
of the electron and the its total mass. As $ \rho_{vac} $
stabilize the whole electron we use in Eq.\ref{ApproxLamda}
the total mass of the electron $ m $ to approximate the coupling
constant $ \lambda $. It is also necessary to keep in mind
that in the SM model the Higgs gives via the coupling constant
$ \lambda $ mass to the whole electron.
Inserting the coupling constant $ \lambda $ from
Tab. \ref{SummaryComparisonGravModel} and $ v = 246 $ GeV in
Eq.\ref{massHiggs} allows to approximate the scalar mass
( Higgs mass ) shown in Tab. \ref{SummaryComparisonGravModel}
sixth line.

In the framework of our assumption the masses of  FPs are
related to its gravitationally induced
core with de Sitter vacuum $\rho_{vac}$ at $r=0$.
This allows us to estimate the size of the mass kernel of FPs
as defined by de Sitter-Schwarzschild geometry,
a size of its vacuum core $r_c$, if we
know $\rho_{vac}$ and $m$.
If a particle gets its mass from the electroweak
scale $v$, then its inner core is determined by this scale.
Putting Eq.\ref{rhovaclambda} into the Eq.\ref{eq.19}
we get for a size of a vacuum core of a lepton with the mass $m_l$

\label{alleqs25}
\begin{equation}
{r_c}= \left(\frac{2m_l}{\pi\lambda v^4}\right)^{1/3}
\label{eq.22a}
\end{equation}

Then the approximation on $\lambda$ Eq.\ref{ApproxLamda}
and the mass of the mass kernel of the electron estimated
from Eq.\ref{MassKernel} approximates with Eq.\ref{eq.22a}
the size of the mass kernel of the electron $r_c^{(e)} $ as
shown in Tab. \ref{SummaryComparisonGravModel} seventh
line. We like to stress that in the approximation for the
coupling constant $\lambda$ the whole system of the electron
must be taken into account in contrary to the estimation of the
mass kernel where only the mass of the kernel has to be
considered as realistic choice.

The magnitude of the numerical value of the size
of the self gravitating mass kernel $ r^{e}_{c} $ agrees with the
range of the kernel radius $ r_{K} $ of the ETAMFP model.
We include in Tab. \ref{SummaryComparisonGravModel}
in the last two
lines the corresponding numerical values of the range
of the kernel radius $ r_{K} $ of the electron together with adjacent
range of the GUT scale used for this estimation. The values are
taken from Tab. \ref{ElectronKernelRadius}.


\subsection { Conclusion to the particle-like structure related
                 to gravity }

To study the possible micro structure of FPs we 
start first from the origin of physics and use the pure classical ETAMFP ansatz.
In the just  presented improved approach we approximated the inner kernel
of the FP from the point of view of a self gravitating particle-like
structure with a de Sitter vacuum core and compare this ansatz
with our findings out of the experiment and the pure classical approach
about a possible micro structure of FPs.

For the case of pure QED interaction
the characteristic size for electrons $R_e$ is approximated by
the $ \EEGG $ reaction to $ R_{e} \approx 1.57 \times 10^{-17} $ cm
with a significance of $ 5 \times \sigma $. 
In the framework modeling of FPs by de Sitter-Schwarzschild
geometry with vacuum interior governed by a Higgs scalar field the condition
\( \lambda \!\!\!\!-_c \approx R \) estimates a self-coupling
of corresponding potential to \( \lambda \approx \pi / 16 \).
If the scale of the generation of the FP masses
is the electroweak scale and if we respect that only a small part
of the total mass of the electron belong to the mass kernel
the size of the kernel is approximately
$ r_{c} \approx 1.7 \times 10^{-19} $ cm. The magnitude
of these numbers are
in agreement with the estimation of the kernel radius $ r_{K} $
in Tab. \ref{SummaryComparisonGravModel} of our ETAMFP model
in Fig.{\ref{BasicSCHEMEFP2006}}.

It is important to notice that the
mechanism just under discussion can not only generate the
the inner mass kernel of a FP. It would be also possible to
generate a particle like structure what carries no charge, no spin
only mass. In the SM-model is this a scalar ( Higgs particle ).
According \( \lambda \approx \pi / 16 \) would be
the mass of the corresponding
scalar $ m_{scalar} \approx 154 $ GeV. Following the discussion
in  chap.\ref{sec:Status of experimental limits on the sizes of
Fundamental Particles} is this approximation not a limit any more,
it is the outcome of the $ 5 \times \sigma $ significance
effect of the global fit of the  $ \EEGG $ reaction. 

Self-gravitating particle-like structure with de Sitter core is
generic. It is obtained from the class of solutions to the Einstein
equations with the asymptotic behavior of the de Sitter vacuum at
$r=0$ and Minkowski vacuum at the infinity. The timing and the
process to generate mass for the FPs could follow the spontaneous
symmetry braking ansatz of the SM model but would also allow a
direct process to get mass to the FPs via any process what is able
to install energy in vacuum and lifting a virtual particle on the
mass shell.

Considering our discussion about the geometrical approach in
Fig.{\ref{SchematicDevFP}} belongs the inner mass kernel of the FPs in the
time development of the universe to the time interval between
$ t = 0 < t < t_{ Planck Era } $. The kernel of the FP is in the
ETAMFP model the residue from the Planck Era.
In the chapter of the scheme of FPs
from Fig.\ref{AllleftFP2006} to Fig.{\ref{GaugeBosons} we
tested successfully to sort all known FPs
after the ETAMFP model. Following these experimental guided findings
contains the kernel a mixture of mass and charges. A
candidate of the content would be an energy condensate of the
four charges strong, EM, weak and pseudo charge mass. In the
de Sitter-Schwarzschild geometry would this condensate
contain only the pseudo charge mass.

\section {Electrical charged particle-like structure in Nonlinear
             Electrodynamics coupled to General Relativity }
\label{sec:Electrical charged particle-like structure
           coupled to General Relativity}

In chap.\ref{sec:Particle-like structure related to gravity}
we discussed a scenario of the inner kernel of the FPs.
To further improve the ETAMFP model it is very desirably to search for
a charged particle like structure with geometrical separated
charges which exist according the Big Bang model after the
GUT-time $ t_{GUT Era} $ as displayed in Fig.{\ref{SchematicDevFP}}.

The regular electrically charged vacuum soliton with de Sitter
center in Nonlinear Electrodynamics ( NED ) coupled to General
Relativity is a good candidate for a particle-like structure with
geometrical separated charges \cite{ChargedFP}, because the
geometrical distribution of the electric field and the mass
density distribution are similar to the ETAMFP model.

The Nonlinear Electrodynamics coupled to General Relativity and
satisfying the Weak Energy Condition guarantees the existence of
electrically charged regular structures and provides a cutoff on
self-energy which diverges for a point charge. One should only
discard the requirement of Maxwell weak field limit at the center,
on which non-existence theorem are based, because a field must not
be weak to be regular, then electro vacuum soliton has the regular
center, in which geometry, field and stress-energy tensor are
regular without Maxwell limit as $ r \rightarrow 0 $ which is
replaced with the de Sitter limit representing regular cut-off by
the energy density of (self-interacting) electromagnetic vacuum
\cite{ChargedFP}.

The aim of this section is to present numerical calculations 
of the electric field and mass distributions in
the NED model for the case of an electron.
We compare this distributions with the in 
chap.\ref{sec:Possible micro structure of Fundamental Particles}
extensive discussed ETAMFP model.
Here we consider the NED vacuum
soliton which does not include a spin. For this reason we only
compare the NED soliton with the non-rotating ETAMFP model. 
For the discussed calculations it is sensible
to repeat here the part of the paper of I. G. Dymnikova
in ref. \cite{ChargedFP}, which lead
to the equations of the electric field and mass distribution.

\subsection { Energy Conditions }

The Weak Energy Conditions ( WEC ),
$ T_{\mu,\nu} \xi^{\mu} \xi^{\nu} \ge 0 $ for any time like vector
$ \xi^{\mu} $, which is satisfied if and only if \cite{Hawkin1}

\begin{equation}
       \rho \ge 0 ;~~~ \rho+p_{k} \ge 0,~~~ k = 1,2,3
\label{WECcondition}
\end{equation}

guarantees that the energy density as measured by any local
observer is non-negative.

The Dominant Energy Condition ( DEC ) ,
$ T^{00} \ge | T^{ik} | $ for each $ i,k = 1,2,3 $ which holds if
and only if \cite{Hawkin1}

\begin{equation}
       \rho \ge 0 ;~~~ \rho+p_{k} \ge 0,~~~
        \rho-p_{k} \ge 0,~~~ k = 1,2,3
\label{DECcondition}
\end{equation}

includes WEC and requires each principal pressure
$ p_{k} = - T^{k}_{k} $ never exceed the energy density which
guarantees that speed of sound cannot exceed the speed of light.

The Strong Energy Condition ( SEC ) requires \cite{Hawkin1}

\begin{equation}
       \rho + \sum p_{k} \ge 0
\label{SECcondition}
\end{equation}

and defines the sign of the gravitational acceleration.

\subsection { Symmetry of a source term}

Spherically symmetric electromagnetic field with an arbitrary
gauge invariant Lagrangian $ \mathcal{L}(F)$, $F =
F_{\mu\nu}F^{\mu\nu} $ has stress energy tensor with the algebraic
structure

\begin{equation}
       T^{t}_{t} = T^{r}_{r}
\label{sourceterm1}
\end{equation}

It is invariant under rotation in the $ ( r,t ) $ plane, which
enables to identify it as a vacuum defined by the symmetry
of its stress-energy tensor \cite{Irina8,Irina9}. An observer
moving through a medium with stress-energy tensor of structure
Eq.\ref{sourceterm1}, cannot measure his velocity with respect
to it which is typical for motion in a vacuum \cite{Gliner1,Landau1}.

For the class of regular spherical symmetric geometries with the symmetry
of a source term given by Eq.\ref{sourceterm1}, the Weak Energy Condition
leads inevitable to de Sitter asymptotic at approaching a regular center
\cite{Irina10}.

The basic fact of any geometry with de Sitter center generated by
a source term of type Eq.\ref{sourceterm1},  is that the  mass
of an object is related to both interior de Sitter vacuum  and
smooth breaking of space-time symmetry from the de Sitter group in
the origin to the Poincar\' group at infinity \cite{Irina10}.

For the spherical symmetric stress-energy tensor
with the algebraic structure Eq.\ref{sourceterm1} the
equation of state relating density
$ \rho = T^{t}_{t} $ with the radial pressure
$ p_{r} = -T^{r}_{r} $ and tangential pressure
$ p_{\bot} = -T^{\theta}_{\theta}= -T^{\phi}_{\phi} $ reads
\cite{Irina8,Irina10}

\begin{equation}
       p_{r} = - \rho;~~~p_{\bot} = -\rho - \frac{r}{2} \rho'
\label{sourceterm2}
\end{equation}

where prime denotes differentiation with respect to $ r $.

\subsection { Basic equations }

In nonlinear electrodynamics minimally coupled to gravity, the
action is given by ( in geometrical units $ G = c = 1 $ )

\begin{equation}
       S = \frac{1}{16\pi} \int d^{4}x \sqrt{-g}
           (R - \mathcal{L}(F);~~
           F = F_{\mu\nu} F^{\mu\nu}
\label{Basic1}
\end{equation}

Here $ R $ is the scalar curvature, and
$ F_{\mu\nu} = \partial_{\mu}A_{\nu} - \partial_{\nu}A_{\mu} $
is the electromagnetic field. The gauge-invariant
electromagnetic Lagrangian $ \mathcal{L}(F) $ is an
arbitrary function of $ F $ which should have the Maxwell
limit, $ \mathcal{L} \rightarrow F $,
$ \mathcal{L}_{F} \rightarrow 1 $ in the weak field regime.

The action Eq.\ref{sourceterm1} gives the dynamical field
equations

\begin{equation}
       ( \mathcal{L}_{F} F^{\mu\nu} )_{;\mu} ;~~~
        ^{*} F^{\mu\nu}_{~;\mu} = 0
\label{Basic2}
\end{equation}

where $ \mathcal{L}_{F} = d \mathcal{L}/d F $. In the
spherically symmetric case the only essential components of
$ F_{\mu\nu} $ are a radial electric field
$ F_{01} = - F_{10} = E(r) $ and a radial magnetic field
$ F_{23} = - F_{32} $.

The Einstein equation take the form \cite{Einstein1}

\begin{equation}
       G^{\mu}_{\nu} = -T^{\mu}_{\nu} = 2\mathcal{L}_{F}
                       F_{\nu\alpha}F^{\mu\alpha}
                       -\frac{1}{2}\delta^{\mu}_{\nu}\mathcal{L}
\label{Basic3}
\end{equation}

Definition of $ T_{\mu\nu} $ here differs from standard
definition ( see, e.g., \cite{Landau1} ) by $ 8 \pi $, so
that $ T^{0}_{0(here)} = 8 \pi \rho $, etc.

The density and pressure for electrically charged structures
are given by

\begin{equation}
       \rho = - p_{r} = \frac{1}{2}\mathcal{L} - F \mathcal{L}_{F};~~~
            p_\bot = - \frac{1}{2}\mathcal{L}
\label{Basic4}
\end{equation}

and scalar curvature is

\begin{equation}
       R = 2 ( \mathcal{L} - F \mathcal{L}_{F} ) =
           2 ( \rho - p_\bot )
\label{Basic5}
\end{equation}

Symmetry of a source term Eq.\ref{sourceterm1}
leads to the metric

\begin{equation}
       ds^{2} = g(r) dt^{2} - \frac{dr^{2}}{g(r)} - r^{2} d\Omega^{2}
\label{Basic6}
\end{equation}

where $ d\Omega^{2} $ is the line element on a unit sphere.
The metric function and mass function are given by

\begin{equation}
       g(r) = 1 - \frac{2\mathcal{M}(r)}{r}:~~~
       \mathcal{M}(r) = \frac{1}{2}\int\limits_{0}^{r} \rho(x)x^{2} dx
\label{Basic7}
\end{equation}

Dynamical equation Eq.\ref{Basic2} yield

\begin{equation}
       r^{2} \mathcal{L}_{F} F^{01} = q
\label{Basic8}
\end{equation}

where q is constant of integration identified as an electric
charge by asymptotic behavior in the weak field limit (the Coulomb
Law).

As follows from Eq.\ref{Basic7},

\begin{equation}
       F = 2 F_{01} F^{01} = - \frac{2q^{2}}{\mathcal{L}_{F}^{2} r^{4}}
\label{Basic9}
\end{equation}

Theorems of non-existence require the Maxwell behavior at the
regular center, $ \mathcal{L} \rightarrow 0, \mathcal{L}_{F}
\rightarrow 1, $ as $ F \rightarrow 0 $ . The proof is that
regularity of stress-energy tensor requires $ |F \mathcal{L}_{F} |
< \infty $ as $ r \rightarrow 0 $ while $ F \mathcal{L}_{F}^{2}
\rightarrow -\infty $ by virtue of Eq.\ref{Basic9}, it follows
that $ \mathcal{L}_{F} \rightarrow \infty $
and $ F \rightarrow 0
$ which is strongly non-Maxwell behavior \cite{Einstein1}.

This sentence reads that a regular electrically charged
structure does not compatible with the Maxwell weak field limit
$ \mathcal{L} \rightarrow 0, \mathcal{L}_{F} \rightarrow 1, $ as $
F \rightarrow 0 $, in the center.

However, if the density does not vanish as $ r \rightarrow 0 $,
then $ \mathcal{L} $ must not vanish there, although $ F $
vanishes in all cases of the regular center. Moreover, for
solutions satisfying the Weak Energy Condition, density
takes maximum there, since the WEC requires, by
Eq.\ref{WECcondition} and Eq.\ref{sourceterm2},
$ \rho' \le 0 $. Then $ \rho $ is maximal at the center, and
one cannot expect validity of the weak field limit in the region
of maximal energy density of the field.

Let us fix the basic properties of electrical charged NED
configurations obligatory for any Lagrangian $ \mathcal{L}( F ) $:

i) First is the fundamental observation of Bronnikov's paper
\cite{Einstein1}-that $ F $ must vanish as $ r \rightarrow 0 $ to
guarantee regularity, and the electric field strength is zero in
the center of any regular electrical charged NED structure.

ii) Second is another observation from \cite{Einstein1}-since $ F
$ vanishes at both zero and infinity where it should follow the
Maxwell weak field limit, $ F $ must have at least one minimum in
between where an electrical field strength has a maximum. This
leads to branching of $ \mathcal{L}( F ) $ as a function of $ F $.
This inevitable feature of electrical charged solutions can create
problems in an effective geometry whose geodesics are world lines
of NED photons \cite{Novello}. They can be avoided for
electro vacuum solitons satisfying dominant energy condition
\cite{ChargedFP}.

iii) Third is the existence of surface of zero gravity at which
Strong Energy Condition is violated. For all electric NED
configurations this reads $ 2 p_{ \bot } = - \mathcal{L} \ge 0 $,
and SEC  is violated at the surface $ \mathcal{L} = 0 $
\cite{ChargedFP}.

\subsection {NED structures ratifying WEC }

The Weak Energy Condition requires density be non-zero and
maximal in the origin, since with $ \rho \ge 0 $ and
$ \rho' \le 0 $ , a density cannot decrease beyond zero
being obliged to be non-negative. Combined with the first
property this raises the question - whose energy density is
maximal in the center of structures where electric field
tension vanishes?

The basic feature of all solutions of class
Eq.\ref{sourceterm1} is de Sitter behavior at approaching the
regular center \cite{Irina10}. Indeed, regularity of
$ \rho ( r ) $ requires $ r \rho'/2 \rightarrow 0 $ as
$ r \rightarrow 0 $ ( which is easily to check by taking
$ r \rho' = const $ and calculating $ \rho $ ). With
$ | \rho' | < \infty $ the equation of state, by
Eq.\ref{sourceterm2} , tends to
$ p_{r} = p_{\bot} = - \rho $ as $ r \rightarrow 0 $ which
gives de Sitter asymptotic

\begin{equation}
       g(r) = 1 - \frac{\Lambda}{3} r^{2}
\label{NED1}
\end{equation}

with cosmological constant $ \Lambda = 8 \pi \rho ( 0 ) $.
For electrical NED structure Lagrangian $ \mathcal{L}(F)
\rightarrow 2 \rho ( 0 ) $ as $ r \rightarrow 0 $, by
Eq.\ref{Basic4}, so that Lagrangian is positive and takes its
maximal value at the center which testifies that the
limiting density as $ r \rightarrow 0 $ of electromagnetic origin.

Here we are able to answer the question { \it whose density
is maximal as $ r \rightarrow 0 $ where electric field
vanishes }. The $ T^{0}_{0} $ component of electromagnetic
stress-energy tensor does not vanish ( neither diverges ) as
$ r \rightarrow 0 $ and provides an effective cutoff on
self-interaction by relating it, through Einstein equation,
with cosmological constant $ \Lambda $ corresponding to
energy density of a vacuum, in this case the
electromagnetic vacuum Eq.\ref{sourceterm1}.

The WEC requirement $ \rho + p_{\bot} \ge 0 $ leads to
$ - F \mathcal{L}_{F} \ge 0 $. It gives $ \mathcal{L}_{F} \ge 0 $.
It gives also a constrain on a Lagrangian
$ \mathcal{L} \ge 2 F \mathcal{L}_{F} $ as its obligatory low
boundary.

The DEC requirement $ \rho - p_{\bot} \ge 0 $, is satisfied in the
region  where $ \mathcal{L} \ge F \mathcal{L}_{F} $. With $ F
\mathcal{L}_{F} \le 0 $ by WEC, this constraint is satisfied in
the whole region surrounding the center including a certain region
outside the surface of violation of the strong energy condition at
which $ \mathcal{L} = 0 $.

Electrically charged solutions are typically found in the
alternative form of NED obtained by the Legendre transformation:
one introduces the tensor $ P_{\mu \nu } = \mathcal{L}_{F} F_{\mu
\nu } $ with its invariant $ P = P_{\mu \nu} P^{\mu \nu} $ and
consider Hamiltonian-like function $ \mathcal{H}(P) = 2 F
\mathcal{L}_{F} - \mathcal{L} $ as a function of $ P $; the theory
is then reformulated in terms of $ P $ and specified by $
\mathcal{H}(P) $ \cite{Salzar}. $ P $ frame is related with $ F $
frame by \cite{Salzar}

\begin{equation}
       \mathcal{L} = 2 P \mathcal{H}_{P} - \mathcal{H}; ~~~
       \mathcal{L}_{F} \mathcal{H}_{P} = 1; ~~~
       F = P \mathcal{H}_{P}^{2}
\label{NED2}
\end{equation}

Here $ \mathcal{H}_{P} = d \mathcal{H} / dP $. The electric
invariant is

\begin{equation}
       P =-2 P_{01} P^{01} = - \frac{2 q^{2}}{r^{4}}
\label{NED3}
\end{equation}

The metric in P frame is calculated from Eq.\ref{Basic7}
with

\begin{equation}
       \rho (r) = - \frac{1}{2} \mathcal{H}
\label{NED4}
\end{equation}

FP duality coincides with conventional electric-magnetic
duality only in the Maxwell limit where
$ \mathcal{L} = F = P = \mathcal{H} $ \cite{Einstein1}.
Interpretation of the results obtained in P framework depends
essentially on transformation to F framework where Lagrangian
dynamics is specified. The two frames are equivalent only
when the function $ F(P) $ is monotonic \cite{Einstein1}.

The function $ F(P) $ which vanishes at both center and
infinity has at least one minimum in which

\begin{equation}
       \mathcal{L}_{FF} = \frac{1}{2} \left(
                          \frac{\mathcal{H}_{P}}{F_{P}} -
                          \mathcal{L}_{F} \right )
\label{NED5}
\end{equation}

tends to infinities of opposite sign and $ \mathcal{L}(F) $
suffers branching. Addition branching is related to extrema
of the function $  \mathcal{H}(P) $ \cite{Einstein1}.

While the first kind of branching is inevitable, the second is
avoided by WEC, since $ \mathcal{L}_{F} \ge 0 $ results in $
\mathcal{H}_{P} \ge 0 $. When $ \mathcal{H}(P) $ is monotonic
function, the function $ \mathcal{L}(F) $ has only two branches
related to one minimum of F \cite{Einstein1}. This looses problems
with restoring F-frame Lagrangian dynamics fro P-frame results.
With one cusp interpretation is transparent and inevitable cusp
become the source of information about most interesting behavior
of electrically charged NED structures which displays in
propagation of photon in an effective geometry. Typical behavior
of Lagrangian $ \mathcal{L} $ as a function of $ F $ is depicted
in Fig.{\ref{TypLagrangian}.
\begin{figure}[htbp]
\begin{center}
\rotatebox{-89}{
 \includegraphics[width=8.0cm,height=10.0cm]{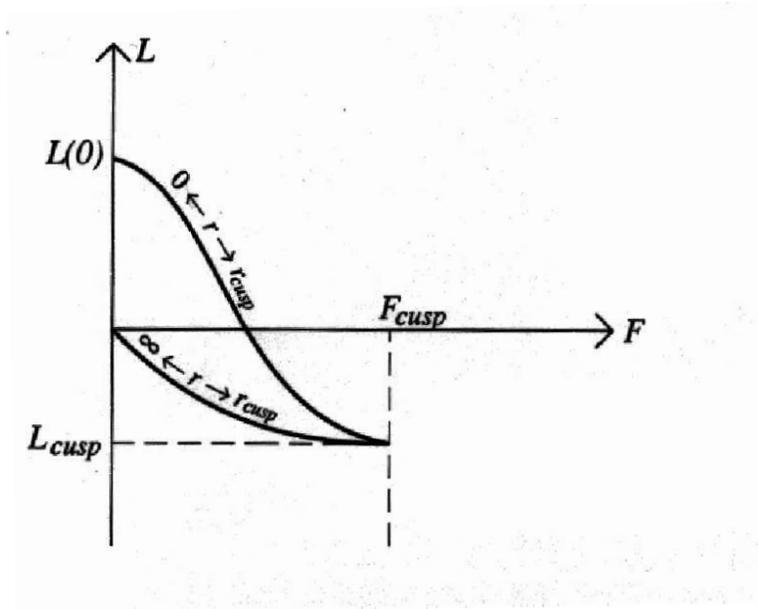}}
\end{center}
\caption{ Typical behavior of a Lagrangian $ \mathcal{L} ( F ) $. }
\label{TypLagrangian}
\end{figure}
At the cusp surface $ r = r_{cusp} $ the invariant $ F $ has
a minimum as a function of $ P $ and a function of $ r $
( since $ P(r) $ is monotonic). The Lagrangian $ r $-derivative
there $ \mathcal{L}' = \mathcal{L}_{F} F' = 0 $ and
$ \mathcal{L} $ takes its minimum value. The Lagrangian
$ \mathcal{L}(F) $ which is monotonic function of $ F $
( $ \mathcal{L}_{F} \ge 0 $ ), first decreasing smoothly along
the first branch from its maximal value $ \mathcal{L}(0) $ to
$ \mathcal{L}_{cusp} $ as $ F $ decreases from $ F = -0 $
at $ r = 0 $ to $ F_{min} = F_{cusp} $ , then the Lagrangian
increases along the second brach from its minimal value
$ \mathcal{L}_{cusp} < 0 $ to its Maxwell limit
$ \mathcal{L} \rightarrow F \rightarrow -0 $ as $ F $
increases from $ F_{cusp} $ to $ F \rightarrow -0 $ as
$ r \rightarrow \infty $.
At the cusp the electric field $ E^{2}(r) = -F(r)/2 $
achieves its maximum.

Actually there are two Lagrangians here, one, ${\cal L}_1$, for
the region from the origin to the cusp, and the other, ${\cal
L}_2$ (asymptotically Maxwellian), for the region from the cusp to
infinity. The problem can be reformulated in terms of ${\cal
L}={\cal L}_1+{\cal L}_2$, in the way correct for mathematical
physics, with the same results as outlined above \cite{ID2007}.

The tangential pressure is maximal at the cusp surface,
where $ p'_{\bot} = -\mathcal{L}'/2 = 0 $. In one-cusp configuration
tangential pressure has one extremum, this is actually
dictated by WEC which defines also the number of horizons.
The function $ g(r) $ has only one minimum and geometry
described by the metric Eq.\ref{Basic6} can have not more than
two horizons \cite{Irina10}.

With $ \mathcal{H}_{P} \ge 0 $ the electric susceptibility
$ \epsilon = 1/\mathcal{H}_{P} $ is  everywhere positive. When
$ P \rightarrow -\infty $ at the center $ \mathcal{H}_{P}
\rightarrow +0 $ ( since $ \mathcal{L}_{F} \rightarrow \infty $
there),
this leads to $ \epsilon \rightarrow + \infty $, electric
susceptibility is divergent, so that electrical charged NED
configurations demonstrate ideal conduction behavior at approaching
the regular center where the electric field tension vanishes.

Summarizing we conclude that {\it regular electrical charged NED
structures satisfying Weak Energy Condition, have de Sitter
center, not more than two horizons and precisely one cup of
$ \mathcal{L}(F) $ where the electric field strength achieves
its only maximum. }

\subsection{ New exact electric NED solution }

The just discussed NED structure leads us to choose
the function $ \mathcal{H}(P) $ in the form

\begin{equation}
       \mathcal{H}(P) = \frac{P}{(1+\alpha\sqrt{-P})^2}
\label{ExactNED1}
\end{equation}

where $ \alpha $ is the characteristic parameter of the
NED theory. Then we get

\begin{equation}
       \mathcal{H}(P) = \frac{1}{(1+\alpha\sqrt{-P})^3}
\label{ExactNED2}
\end{equation}

and

\begin{equation}
       F = \frac{P}{(1+\alpha\sqrt{-P})^6};~~~
       F_{P} = \frac{(1-2\alpha\sqrt{-P})}{(1+\alpha\sqrt{-P})^7}
\label{ExactNED3}
\end{equation}

With $ P $ defined by Eq.\ref{NED3} this gives

\begin{equation}
       \mathcal{H} = - \frac{ 2q^{2}}{(r^{2} + r_{0}^{2})^{2}};~~~
       \mathcal{H}_{P} = \frac{r^{6}}{(r^{2} + r_{0}^{2})^{3}}
\label{ExactNED4}
\end{equation}

The parameter $ r_{0}^{2} = \alpha \sqrt{2q^{2}} $ is fixed by
integrating Eq.\ref{Basic7} with the density Eq.\ref{NED4}
which connects $ r_{0} $
with the total mass $ m = \mathcal{M} ( r \rightarrow \infty ) $.
This gives

\begin{equation}
       r_{0} = \frac{\pi}{8}\frac{q^{2}}{m}
\label{ExactNED5}
\end{equation}

as classical electromagnetic radius modified by numerical
coefficient of chosen particular NED model Eq.\ref{ExactNED1}.

The only minimum of $ F (P) $ is at $ 2\alpha \sqrt{-P} = 1 $ and
the cusp surface is given by

\begin{equation}
       r_{cusp} = \sqrt{2} r_{0}
\label{ExactNED6}
\end{equation}

The density and pressure are ( up to $ 8\pi $ mentioned above )

\begin{equation}
       \rho ( r ) =  \frac{ q^{2}}{(r^{2} + r_{0}^{2})^{2}};~~~
       p_{\bot} = \frac{q^{2}(r^{2} - r_{0}^{2})}{(r^{2} + r_{0}^{2})^{3}}
\label{ExactNED7}
\end{equation}

Function $ \rho ( r ) $ is monotonically decreasing, function
$ p_{\bot}(r) $ achieves maximum at the cusp surface.

The electric field is given by

\begin{equation}
       F =  - \frac{ 2 q^{2} r^{8}}{(r^{2} + r_{0}^{2})^{6}};~~~
       E^{2} = \frac{q^{2} r ^{8}}{(r^{2} + r_{0}^{2})^{6}}
\label{ExactNED8}
\end{equation}

It achieves its maximum at the cusp surface

\begin{equation}
       E_{max} = \frac{4}{27} \frac{q}{r_{0}^{2}}
\label{ExactNED9}
\end{equation}

and Maxwell limit $ E \rightarrow 0 $ as $ r \rightarrow \infty $.

Lagrangian and its derivative are

\begin{equation}
       \mathcal{L} = \frac{ 2 q^{2}(r^{2}_{0} - r^{2})}
                      {(r^{2} + r_{0}^{2})^{3}};~~~
       \mathcal{L}_{F} = \frac{ (r^{2} + r_{0}^{2})^{3}}
                      {r^{6}}
\label{ExactNED10}
\end{equation}

The scalar curvature for this Lagrangian is given by

\begin{equation}
       R = \frac{4 q^{2} r_{0}^{2}}{(r^{2} + r_{0}^{2})^{3}}
\label{ExactNED11}
\end{equation}

It is positive everywhere, and the Dominant Energy Condition is
satisfied ( although we did not impose it ) which is a good feature
, since e.g., propagation of NED  photons in an effective
geometry resembles propagation inside a dielectric medium
\cite{Novello}, and DEC makes it free of effects produced by
speed of sound exceeding speed of light.

Integrating Eq.\ref{Basic7} with the density profile
Eq.\ref{ExactNED7} we get the metric

\begin{equation}
       g(r) = 1 - \frac{4m}{\pi r} \left( \arctan \frac{r}{r_{0}} -
              \frac{r r_{0}}{r^{2} + r_{0}^{2}}\right)
\label{ExactNED12}
\end{equation}

For $ r \gg r_{0} $ it reduces to

\begin{equation}
       g(r) = 1 - \frac{2m}{r} + \frac{q^{2}}{r^{2}}
              - \frac{2}{3} \frac{q^{2} r_{0}^{2}}{r^{4}}
\label{ExactNED13}
\end{equation}

and has Reissner-Nordstr\"om limit as $ r \rightarrow \infty $.

At small values of $ r $, $ r << r_{0} $ we get de Sitter
asymptotic Eq.\ref{NED1} with the cosmological constant

\begin{equation}
       \Lambda = \frac{q^{2}}{r_{0}^{4}}
\label{ExactNED14}
\end{equation}

which gives proper expression for a cutoff on self-energy
density by finite value of electromagnetic density
$ T_{0}^{0} ( r \rightarrow ) ) $ related to the
cosmological constant $ \Lambda = 8 \pi \rho (0) $ which
appears at the regular center.

The mass, of electromagnetic origin, is related to this
cutoff by $ m = \pi^{2} \rho(0) r_{0}^{3} $, where $ r_{0} $
is the classical electromagnetic radius.

Characteristic parameters which decides if a solution
describes a regular electrical charged black hole either
self-gravitating particle-like structure with de Sitter
vacuum inside, is given by

\begin{equation}
       \beta = \frac{8}{\pi^{2}}\left(\frac{2m}{q}\right)^{2}
             = \frac{2}{\pi}\frac{r_{g}}{r_{0}}
\label{ExactNED15}
\end{equation}

where $ r_{g} = 2 m $ is the characteristic Schwarzschild radius.

For $ \beta > \beta_{crit} = 2.816 $ solution describes a black hole.
For $ r_{g} >> r_{0} $, two horizons are

\begin{equation}
       r_{-} \simeq r_{s} \left(1 + 1.4 \frac{r_{0}}{r_{g}}\right);~~~
       r_{+} \simeq r_{g} \left(1 + 1.3 \frac{r_{0}}{r_{g}}\right)
\label{ExactNED16}
\end{equation}

Internal horizon in this limits is close to de Sitter horizon
$ r_{s} = \sqrt{3/\Lambda} $, and an event horizon to the
Schwarzschild horizon $ r_{g} $.

For $ \beta = \beta_{crit} $ there is a double horizon

\begin{equation}
       r_{\pm} = 1.825 r_{0}
\label{ExactNED17}
\end{equation}

The global structure of space-time with horizons is precisely the
same as for de Sitter-Schwarzschild geometry \cite{particle}. It
differs from Reissner-Nordstr\"om case only in that the time-like
surface $ r = 0 $ is regular.

In terms of $ q/2m $ black hole exists for $ q/2m \le 0.536 $, and
for $ q/2m > 0.536 $ we have electrically charged electro vacuum
soliton.

\subsection { Numerical comparison of the electrical charged self-
                 gravitating particle-like NED structure with the
                 ETAMFP model }

The NED structure describes a spherical spinless particle with a mass core
surrounded by a shell of an electric field. The geometrical shape
of the density distribution $ \rho $ of Eq.\ref{ExactNED7} and the field
distribution $ E $ of Eq.\ref{ExactNED8} is a function of the radius $ r $
and full symmetric in the
angle $ \theta $ and $ \phi $ ( in spherical coordinates  ).
Such a structure would be a candidate
for a charged scalar in the SUSY model e.g.\cite{ChargedHiggs}.

If we follow the ETAMFP model in the geometrical approach
of Fig.{\ref{SchematicDevFP} and the stability discussion 
in Fig.{\ref{ForcesSM-BBM}
we would expect a clustering of $ \rho $ and $ E $ in the angle
$ \theta $ and $ \phi $. The two electric charges $ 1/3 \times e $
located in point ( SP I ) at $ r = r(SP I)~,\theta = 90^{\circ}~,
\phi = 0^{\circ} $ and $ r = r(SP I)~,\theta = 90^{\circ}~,
\phi = 180^{\circ} $
would implement a cluster in the radius and angle dependence of $ \rho $ and
$ E $. 

A link between NED structure and ETAMFP model 
on a level of magnitude exist in the
radius dependence of the density and electric field
distribution of the possible particle structures. 
The most spherical FP in the ETAMFP model
is the electron. It is for this reason interesting
to study this radius dependence in more detail.

The NED structure contains no spin, we compare for this reason
the NED soliton with the non-rotating ETAMFP model.
We consider the ETAMFP model from an observer located in the center
of the FP rotating with an angular velocity which is in accordance
with the spin of the FP. In the case for the electron is $ \omega
= 5.169 \times 10^{20} $ (1/s) as displayed in Tab.
\ref{electronPARAMETERSdipol}. If we further ignore the 
cluster problem it is possible to calculate 
numerical for an electron the
geometrical distribution of the electric field and the mass
in the ETAMFP model and the NED structure.

The electron has a very small mass for this reason is the parameter
$ q/2m = 1.1 \times 10^{21} >> 0.536 $ much bigger as the limit for
a black hole. We have to consider an electrical charged self-gravitating
particle-like NED  structure.

The dynamical equation
Eq.\ref{Basic2} yields to the charge $ q $ in Eq.\ref{Basic8}. The
charge $ q $ is the constant of integration by asymptotic behavior
in the weak field limit. In this sense is the character of the charge
a constant of integration of the field $ F $ for an outside observer.
Important is that this field is not a $\delta$-function, it is
geometrical extended and the charge is a self interaction phenomenon.
In the classical experiment from Millikan 1911 \cite{Millikan} the charges
are measured fare away from the particle. The character of the charge was
a geometrical point-like aggregation of Coulomb. The particle-like NED
structure gives a more detailed picture of the character of the charge
because the character depents from the distance of the observer
measuring the charge. For a distance observer behaves the charge
like a point in the classical Millikan experiment, for an observer
close to the charge the charge behaves like an geometrical extended
object generated from an electrical charged self-gravitating
mechanism.

In the case of the electron is the charge in both models for an
observer fare outside $ q = 1 $. As discussed in 
chap.\ref{sec:The geometrical approach}
of the ETAMFP model describes the NED structure
the geometrical region outside the residual of the Planck Kernel of the FP
( See Fig.{\ref{BasicSCHEMEFP2006}}). This sheathing between the
residual radius of the Planck Kernel and the position of the charges
( $ Q $ , $ T_{3} $ and $ C $ ) is dominated from a mass density
which tends to zero at the position of the point-like charges
of the ETAMFP model. Following the position of an close observer
of the ETAMFP model it would be possible to use only the
charge $ 2/3 \times q $ for the comparison of both models for
an electron. This would lower the classical electromagnetic radius
according Eq.\ref{ExactNED5} and the mass density distribution
according Eq.\ref{ExactNED7} by a factor $ ( 2/3 )^{2} $. The
electric field distribution would be lowered according
Eq.\ref{ExactNED8} by a factor $ ( 2/3 ) $. In our study
is it only possible to study the magnitude of the
of the field and mass density distribution. For the reason
we did not take into account this refinement and
chose for the calculation of the field and mass density distribution
of the NED structure $ q = 1 $.

The comparison of both models requested to use the ETAMFP model
as a non rotating object because the NED structure does not include
a spin. For this reason we used in the calculation of the
classical electromagnetic radius of Eq.\ref{ExactNED5} only
the non rotating mass $ 0.8335 \times m_{0}(electron) $ from
Tab. \ref{electronPARAMETERSdipol} to calculate
$ r_{0}(NED) = 1.33 \times 10^{-13} $ cm.

In the Fig.{\ref{E-mass-field-NED electron}} upper part we
show the electric field of the NED structure as function of
the radius $ r $ according Eq.\ref{ExactNED8}  ( black solid
line ) and the classical electric Coulomb field for
comparison ( black broken line ). In the lower part of
Fig.{\ref{E-mass-field-NED electron}} we display the
mass density distribution of the NED structure
according Eq.\ref{ExactNED7} ( solid black line ).

\begin{figure}
\begin{center}
 \epsfig{file=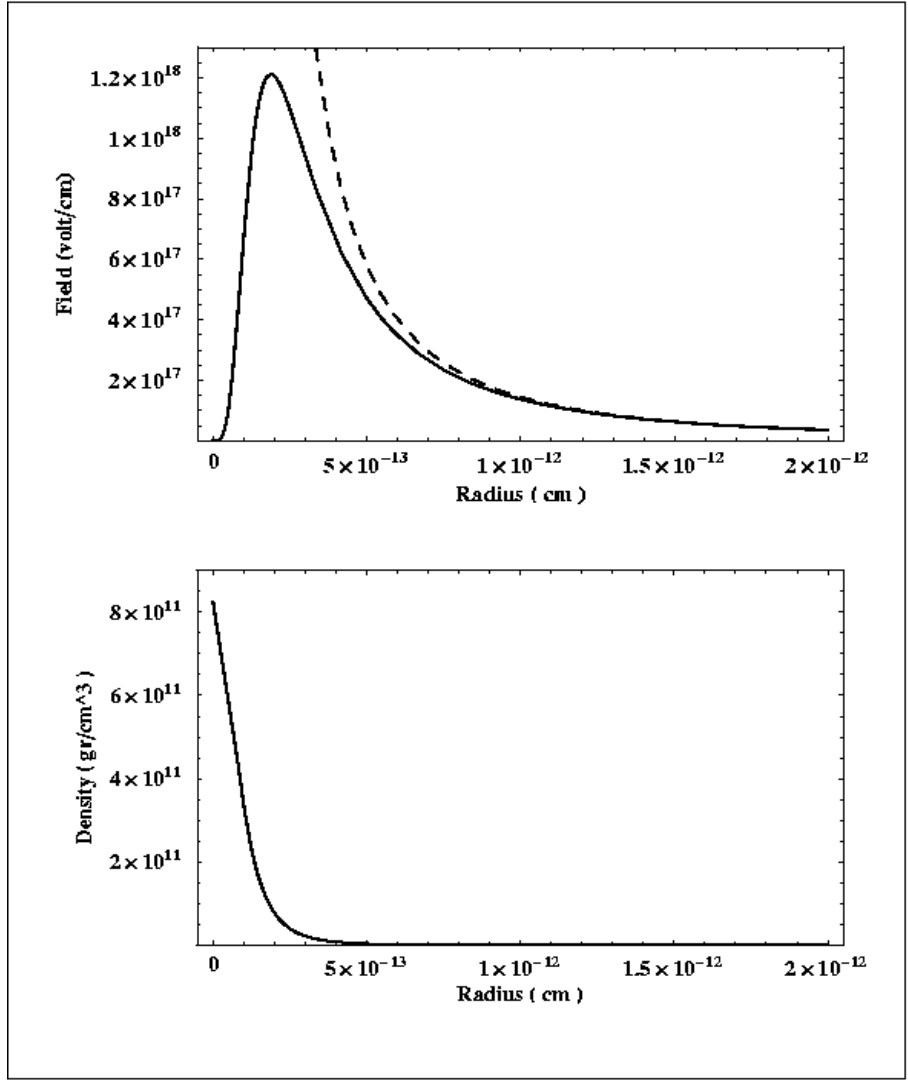,width=12.0cm,height=14.40cm}
\end{center}
\caption{ Electromagnetic field and mass density distribution
          of the NED structure for a mass of an electron. }
\label{E-mass-field-NED electron}
\end{figure}


The $ r $-dependence of the electromagnetic field of
Eq.\ref{ExactNED8} in Fig.{\ref{E-mass-field-NED electron}}
is for $ r >> r_{0} $ in perfect
agreement with the Coulomb field. It highlights that
the character of the charge depents about the distance
of the observer measuring the charge. In the Coulomb
field limit $ r >> r_{0} $ the charge behaves
like a point in the NED limit $ r \sim r_{0} $ like a self
interacting geometrical extended object.

In the ETAMFP model
we estimated in Fig.{\ref{SchematicDevFP}}
a mass shell with a density $ \rho_{ s } $ 
which is the residual of the range between the
Planck Era and GUT Era. Following this ansatz at the
GUT Era the three geometrical separated interaction
strong, electromagnetic and weak appear. The source of
this fields are the according charges color, electric-
and weak-charge. The exact electric
NED solution demonstrates the important relation
between mass and charge in the Era between Planck
and GUT, and after the GUT Era. Following this ansatz
generates mass at the GUT scale an electromagnetic
field outside the center of the FP. This field is
a self interacting phenomenon, which is characterized
be a charge according Eq.\ref{Basic8}.

The field in Fig.{\ref{E-mass-field-NED electron}} has
its maximum at about $ r \sim 2 \times 10^{-13} $ cm
close to $ r_{0}(NED) $ outside $ r = 0 $. Important is the
fact, that the mass density distribution according
Eq.\ref{ExactNED7} has his maximum at $ r = 0 $
and drops at the maximum of the field by about
a factor $ 8 $ strong decreasing with increasing
radius $ r $.

\begin{figure}[htbp]
\begin{center}
\rotatebox{0}{
 \includegraphics[width=17.0cm,height=13.0cm]{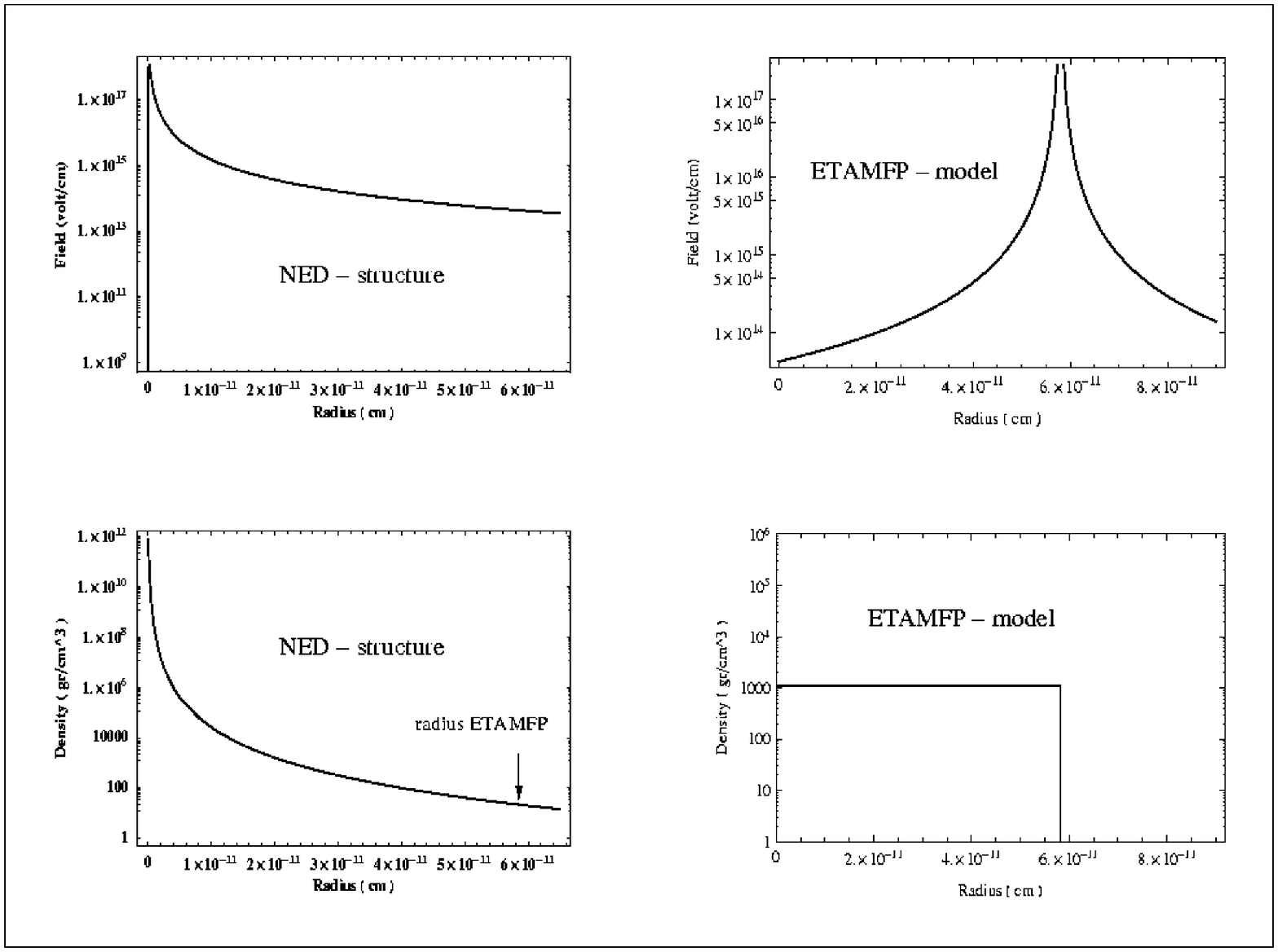}}
\end{center}
\caption{Comparison of mass density and electromagnetic field
         distribution of ETAMFP- and NED-structure
         for a paticle with the mass of an electron. }
\label{ETAMFP-NED-comparison}
\end{figure}

In the Fig.{\ref{ETAMFP-NED-comparison}} upper left part
( black solid line ) the electric field distribution of the
NED structure is displayed. The lower left part shows the
density distribution ( black solid  line ) of the NED structure.
The NED distribution of the electric field and mass
is the same as in
Fig.{\ref{E-mass-field-NED electron}} but in a log-scale.

In the Fig.{\ref{ETAMFP-NED-comparison}} upper right part
( black solid line ) the electric field distribution 
of the ETAMFP model is shown. We used the $ r $ 
dependence of a classical Coulomb field of a point-like charge
	located at $ r = r_{0} $. The lower right part
( black solid line ) the density distribution of the
ETAMFP model is displayed. We used the in
chap.\ref{sec:Mass and charge radius of electron as Lorentz
                contracted gyroscope}
introduced rigid sphere, with a
constant mass density between $ r_{kernel} $ and
$ r_{0}(ETAMFP) = 5.799 \times 10^{-11} $ cm. For simplicity
we ignored the kernel and set the mass density of the
electron according its total mass to $ \rho_{electron}
= 1.1157 \times 10^{3} $ (gr / cm$^{3}$).

Inspecting Fig.{\ref{E-mass-field-NED electron}}
and Fig.{\ref{ETAMFP-NED-comparison}} on level of global structure
shows an agreement of the $ r $ dependence
of the field and mass distribution of the NED structure and 
the ETAMFP model. In both, the NED structure
in Fig.{\ref{E-mass-field-NED electron}} and
the ETAMFP model in Fig.{\ref{ETAMFP-NED-comparison}}, is
the location of the mass distribution
in the center of the object and the charge
located at the outside, even if the NED structure
is much less extended in the radius $ r $ as
ETAMFP model.

The charge in the ETAMFP model is a point like cluster
of Coulomb at $ r = r_{0} ( ETAMFP ) $. This is replaced
by a self interacting electromagnetic field which is zero
at $ r = 0 $, maximal close to $ r = r_{0} ( NED ) $
and four magnitudes down at $ r = r_{0} ( ETAMFP ) $.
The maximum of the field is located fare inside the
radius $ r = r_{0} ( ETAMFP ) $. This is very important
for our experimental knowledge of the charge which
behaves like a point for a distance observer.
The nature of the charge $ q $ is for a close observer
according Eq.\ref{Basic8} an integral over the field.
The point like cluster of Coulomb in the ETAMFP model
has changed to more natural geometrical extended object
in the NED structure.

The density distribution in the ETAMFP model is constant
between $ r = 0 $ and $ r = r_{0} ( ETAMFP ) $ and we set this
distribution in a step function to zero at $ r = r_{0} ( ETAMFP ) $.
The NED structure replaces this simple assumption by a much more
natural strong decreasing density distribution as function of
$ r $. The density is at $ r = r_{0} ( ETAMFP ) $ about
10 magnitudes smaller at the maximum at $ r = 0 $ 
( See arrow in Fig.{\ref{ETAMFP-NED-comparison}} lower part
left side ).
This outcome supports our estimation discussed in
chap.\ref{sec:Mass and charge radius of electron as Lorentz
              contracted gyroscope} and
Fig.{\ref{Spheremassbeta} to neglect
the Lorentz correction of $ m_{0} $ for the electron.

\subsection {Conclusion to electrical charged particle-like
             structure in Nonlinear Electrodynamics
             coupled to General Relativity }

In chap.\ref{sec:Self gravitating particle-like 
structure with de Sitter vacuum core}
the Einstein equations Eq.\ref{eq.7b}, Eq.\ref{eq.7c} and
Eq.\ref{eq.7d} contain the gravitational stress-energy tensor $
T_{\mu\nu} $; in 
chap.\ref{sec:Electrical charged particle-like structure
          coupled to General Relativity}
the Einstein
equations Eq.\ref{Basic3} contain the stress-energy tensor $
T^{\mu}_{\nu} $ of a spherical symmetric electromagnetic field. As
discussed in the ETAMFP model describes the stress-energy tensor
$ T_{\mu\nu} $ a residual of the gravitational dominated regime whereas
$ T^{\mu}_{\nu} $ describes a residual of the strong,
electro magnetic and weak
interaction dominated regime. The most important outcome of the
NED structure is, that it demonstrates in an exact solution, that a
spherical mass aggregation generates a regular spherical
electromagnetic field. The mass of electromagnetic origin,
obtained by integration of electromagnetic field density over the
whole space, involves generically, de Sitter vacuum in the
origin. The integration of dynamic field equations gives the
charge of the object (as the constant of integration). In the
ETAMFP model is this transition obviously linked to the residual of the GUT Era
shown in Fig.{\ref{SchematicDevFP}}. At the GUT-scale splits the
mass in the three interactions strong, electro-magnetic and weak.
The NED structure is linked in the ETAMFP model to this scale.

Both the ETAMFP model and the NED structure agree in the global geometrical
structure of the $ r $ dependence of the mass and field 
distribution. The NED structure is more
natural in the mass and field distribution. The charge of an object,
is depending about the distance of the
observer from the center of the object, a point or an geometrical
extended self interacting phenomenon.

\section {CONCLUSION}

All experimental investigations of the
strong and electroweak interaction
searching for exited fermions or contact interaction
lead to lower and upper limits. The values of these limits
are summarized in Fig.\ref{SizeLimits}.
All the limits show that the Compton wavelength
\( \lambda\!\!\!\!-_c\) is bigger as the
characteristic size $r_s$ of the interaction area.
This highlights the experimental fact, that the size
of fundamental particles is smaller or in the
range of $ 10^{-18} $ cm.
The investigations of the pure electromagnetic interaction
$ \EEGG $ using the complete set of differential cross
sections available from VENUS, TOPAZ and LEP lead to
a $ 5 \times \sigma $ effect for the hypothesis
of an excited electron or an effective contact interaction.
This is the first experimental signal of a non point like effect
of fundamental particles. 
The increase of the significance compared to previous
analysis is simple the increase of statistic,
because we used in our standard $ \chi^{2} $ test the
most extensive data set what was
available at time of the analysis.
The statistic of data to not allows to distinguish between
both possibilities. It is interesting to notice that the
interaction size of $ 15.7 \times 10^{-18} $ cm of the $ \EEGG $
reaction is bigger as the lowest limit of the electro weak
reaction with  $ 0.9 \times 10^{-18} $ cm. This triggers the
question, why is the
signal not visible also in the electro weak case?
For example in the Bhabha scattering is the s-channel involved and
the Bhabha cross section is much bigger as the $ \EEGG $ cross section.
The annihilation reaction $ \EEGG $ which is sensitive to the signal
is suppressed by the big Bhabha channel and for this reason
hard to detect it in an experiment. 

The interpretation of the just discussed experimental result
in the Standard Theory
is in strong conflict with the point like FPs of the SM theory.
To overcome this conflict we discuss a scenario to extend the
Standard Theory and compare it with the Big Bang model.
Both theories are
extensively experimental tested and proven to describe our nature.
The quantized SM model is applicable in particular at small
distances describing the fundamental particles and the interaction
between them, whereas the Big Bang Model
describes distances from the Planck scale up to size of the universe
today. We like to stress that both theories can describe only one
nature. The logically consequence about this fact is that between
both theories similarities or links must exist.
We located in principle five links between SM- and BBM-model, both
models use the same space, energy, vacuum, charges and interactions.

The experimental paradox to measure very small distances
initiated the ETAMFP model.
To verify in an experiment a FP is a point, according
to the SM model
it would be necessary to measure distances down to zero.
This unambiguously request infinite high test energies,
the CM energy of the experiment must cross the
GUT-scale and Planck-scale. The logic consequence is that
the whole physic conditions, according to the energy scale
of the SM- or BBM-model, change to the physics laws existing
at this scales. If we assume all physic processes are time revers 
invariant up to the Planck scale, nature must be able to store
the history of this scale effect. An experiment up or beyond
the Planck scale is not feasible but nature itself performed
this experiment described in the Big Bang model.
Applied to the creation of the FPs in the BBM model in the
FP Era, request this time revers invariant,
that remains of the history of the Big Bang are
stored in the FPs.
If we use
the common ground of the time development of the SM- and BBM-model
from time $ t = 0 $ to the time of the FP Era, it is possible
to predict the overall geometrical structure of the FPs
from a radius $ r = 0 $ to a radius of the FP Era in three major
steps. At the time window $ 0 < t < t_{ Planck } $ 
the remains of a condensate of
mass and the three charges color, electromagnetic and weak
get created, between $ t_{ Planck } < t < t_{ GUT } $ the remains of the mass
is created and finally at the window $ t_{ GUT } < t < t_{ FP } $
the FPs get equipped with the three charges color,
color, electromagnetic and weak according the three
adjacent interactions strong, electro magnetic and weak.
In this sense is the
geometrical extension of the FPs a direct consequence of the
time development of the SM- or BBM-model from $ t = 0 $ to
$ t = t_{ FP-Era } $.

After establishing the ETAMFP model we tested the reliability
of the model to describe the experimental known parameters
of the FPs.

First we tested, is it possible to sort all known FPs in a
scheme of geometrical extended fundamental particles and anti-particles.
It was possible to demonstrate the ETAMFP model 
allows to sort
all FPs in a common scheme. The scheme predict the general tendency for
the different moments for strong, electromagnetic and weak interaction.
The scheme shows that the link between geometrical space and time
is of essential importance. 
The three families are in this philosophy the possibility of the
mass ( link time ) to oscillate in these three space coordinates.
The scheme also shows that the most simple charged highly symmetrical
geometrical object is the electron, what serves as ground state
for all leptons. 

Second we estimated the size of an electron
determined from its experimental measured parameters.
The classical approach to estimate the size of an electron
determined from its experimental measured parameters is an
important benchmark test of the ETAMFP model, because
both the BBM- and SM-model contribute to this estimation.
The approach follows the philosophy to use as simple as possible
solutions for the problems in question and refine them step by step.

We assume first, the electron what carries a spin,
behaves like a classical gyroscope. The charge attached to electron
at a distance $ r_{e} $ from the center will have under this
circumstances a speed what exceeded the velocity of light. 

Next we assume the velocity of light $ c $, as
absolute limit for a group velocity, must be respected
and the path of the charges circling the electron
center is free of radiation loss.
As the speet of the charge is close to the speed of light
the length of the path equivalent the size of
circumference will be contacted by the Lorentz equations.
Comparing the measurement of the magnetic moment of the
electron and the electric dipole moment with this contraction
shows, that the experiment of the magnetic moment is
sensitive to the un-contracted radius $ r_{0} $ where the
charges $ 1/3 \times e $ are located, whereas in the case of the
electric dipole moment the experiment is sensitive to the
contracted radius $ r_{e} $. This leads to the explanation
why the magnetic moment is a precise measured numerical
value and the measurement of the electric dipole moment is
so challenging. The Lorentz contraction pushes the measurable
dipole moment in the LAB-system to zero, if the charge velocity
approaches the speed of light. Assuming $ \beta = 1 $ for
the speet of the charge it is possible to estimate the important
parameters of the the electron,the angular velocity of charge, the rest mass,
the rotational energy and the un-contracted radius $ r_{0} $ where the charge
is located shown in Tab. \ref{modelbeta1}.

Including in the next step in the model an ansatz about the shape
of the mass core of the electron, as a rigid sphere and taking
into account the possibility of a non rotating inner mass kernel
implements an important limit in the model. It is only possible
to find a solution for the parameters of the electron if
only $ 2/3 \times e $ of the charge generate the
magnetic moment. The solution
for $ 1 \times e $ get excluded. It also turns out that a non
rotating inner mass kernel could exist.

In the last step of the model development so far, we include
the limit of the electric dipole moment in the ansatz. This confirms
a possible inner mass kernel and the parameters, angular velocity,
rest mass, rotational energy and the un-contracted charge radius of the
electron, already calculated for the case $ \beta = 1 $.
In particular it highlights the fact, that the numerical value
of $ \beta $ get pushed from the measurement of the electric dipole
moment of the electron extreme close to $ \beta = 1 $.

An further benchmark test about our whole model discussion is
the comparison of the size of the electron with experimental data
generated from complete different experiments. 
We insert the parameters of the electron measured
at low energies in the
ETAMFP model and calculate the parameters of an extended
electron in
in particular the radius $ r_{0} $ and $ r_{e} $.
We compare these radii with the radius
measured in the direct contact term reaction $ \EEGG $.
These measurements are complete
independent from the low energy data from the electron.
As shown in Tab. \ref{CoparisonElectronExperiment} is
the radius measured from the high energy $ \EEGG $ reaction
numerical very close the radius calculated via the ETAMFP
model. The agreement of both radius supports
the philosophy to use for a first approximation 
in the ETAMFP model a combination
of SM model, BBM model and standard classical considerations
to develop a microscopic model for a geometrical
extended fundamental particle. 

Third we compared the ETAMFP model with particle-like structure
related to gravity.
In this comparison we approximated the inner kernel
of the FP from the point of view of a self gravitating particle-like
structure with a de Sitter vacuum core and compare this ansatz
with our findings out of the experiment and ETAMFP model. 

In the framework modeling of FPs by de Sitter-Schwarzschild
geometry with vacuum interior governed by a Higgs scalar field the condition
\( \lambda \!\!\!\!-_c \approx R \) estimates a self-coupling
of corresponding potential to \( \lambda \approx \pi / 16 \).
If the scale of the generation of the FP masses
is the electroweak scale 
the size of the kernel is approximately
$ r_{c} \approx 1.7 \times 10^{-19} $ cm. This number is
in agreement with magnitude of the ETAMFP model
estimation of the kernel radius between 
$ 5.0 \times 10^{-20} < r_{K} < 1.1 \times 10^{-18} $ cm.

It is important to notice the
mechanism just under discussion  
is not only able generate the
the inner mass kernel of a FP. It would be also possible to
generate a particle like structure what carries no charge, no spin
only mass. In the SM-model is this a scalar ( Higgs particle ).
According \( \lambda \approx \pi / 16 \) would be
the mass of the corresponding
scalar $ m_{scalar} \approx 154 $ GeV.

Self-gravitating particle-like structure with de Sitter core
is generic. It is obtained from the Einstein equations with the boundary
conditions of the de Sitter vacuum at $r=0$ and Minkowski vacuum at
the infinity. The timing and the process to generate mass for the
FPs could follow the spontaneous symmetry braking ansatz of the SM
model, but would also allow a direct process to get mass to the FPs
via any process what is able to install energy in vacuum and
lifting a virtual particle on the mass shell.

Considering our discussion about the geometrical approach in
Fig.{\ref{SchematicDevFP}} is the inner mass kernel of the FPs 
a residual of the
time development of the universe in the time interval between
$ t = 0 < t < t_{ Planck Era } $.
We tested successfully to sort all known FPs
after the ETAMFP model and it was possible to estimate
the size of this inner mass kernel.
Following these experimental guided findings
contains the kernel a mixture of mass and charges. A
candidate of the content would be an energy condensate of the
four charges strong, EM, weak and pseudo charge mass described by the
de Sitter-Schwarzschild geometry. 

Fourth we compared the ETAMFP model with a
electrical charged particle-like structure in Nonlinear
Electrodynamics coupled to General Relativity.
In the section self gravitating particle-like structure with
de Sitter vacuum core the Einstein equation, contains the
gravitational stress-energy tensor $ T_{\mu\nu} $, in the
electrical charged NED structure the Einstein equation
contains the stress-energy tensor $ T^{\mu}_{\nu} $ of a
spherical symmetric electromagnetic field. As discussed in the
ETAMFP model describes the stress-energy tensor $ T_{\mu\nu} $
a gravitational dominated regime whereas $ T^{\mu}_{\nu} $
describes a from the strong, electro magnetic and weak
interaction dominated regime.
The most important outcome of the NED structure is,
it demonstrates in an exact solution, that a spherical
mass aggregation couples to a spherical
electromagnetic field. The mass is in the center of
the object and the field outside.
This radius dependence of mass and electric field
of the NED structure is on the level of magnitude
in agreement with ETAMFP model.
In the ETAMFP model is this outcome
of the NED structure
obviously linked to the residual of the GUT Era. 

The nature of the charge is in the NED structure
a constant of
the integral over the electric field. For an observer
close to the object is this a self interacting
phenomenon, which behaves like a point like charged
aggregate of Coulomb for an observer
far away from the NED structure.

The comparison of the ETAMFP model, with
the de Sitter-Schwarzschild geometry and
the NED structure opens a perspective to improve
this classical ansatz of the ETAMFP model.
The inner non rotating mass kernel of the FPs
would be linked to a de Sitter-Schwarzschild geometry.
It would be necessary to enlarge the
de Sitter-Schwarzschild geometry that the strong, electro magnetic and weak
charge could be included.
The mass torus of the FPs could be modeled with the
help of a NED structure to include the strong, weak charge
and in particular the spin.

\section {ACKNOWLEDGMENT}

We like to acknowledge the essential effort of A.~Sakharov and I.G.~Dymnikova
to perform this project.
We also like to thank Z.~Berezhiani and INFN for the hospitality and discussions
to write this article.

%
%

\end{document}